\documentclass[a4paper,twoside,12pt]{article}
\newif\ifdraft \global\drafttrue
\def\production{\global\draftfalse}

\production
\usepackage[T1]{fontenc}
\usepackage{a4wide}
\usepackage{times}
\usepackage{graphicx}
\usepackage{theorem}
\usepackage{fourier}
\usepackage[colorlinks=true,hyperindex=true]{hyperref}
\usepackage{color}
\usepackage{fancyhdr}
\usepackage{amsmath}
\usepackage{bbm}
\usepackage{dsfont}
\usepackage{amsfonts}
\usepackage{enumerate}
\usepackage{cancel}
\ifdraft
\fi
\newtheorem{Theo}{Theorem}[section]
\newtheorem{Prop}[Theo]{Proposition}
\newtheorem{Lem}[Theo]{Lemma}
\newtheorem{Def}[Theo]{Definition}
\newtheorem{Coro}[Theo]{Corollary}
\theoremstyle{plain}
\theorembodyfont{\upshape}
\newtheorem{Exa}{Example}[section]
\newtheorem{Rem}{Remark}[section]
\makeatletter
\newcounter{@hypctr}
\newenvironment{hyp}{%
\begin{enumerate}[{\bf (H1)}]
\setcounter\@enumctr{\value{@hypctr}}
}{
\setcounter{@hypctr}{\value{\@enumctr}}
\end{enumerate}
}
\makeatother
\newcommand{\Href}[1]{({\color{red}H\ref{#1}})}

\newcounter{smallarabics}
\newenvironment{arabicenumerate}
{\begin{list}{{\normalfont\textrm{(\arabic{smallarabics})}}}
{\usecounter{smallarabics}\setlength{\itemindent}{0cm}
\setlength{\leftmargin}{5ex}\setlength{\labelwidth}{4ex}
\setlength{\topsep}{0.75\parsep}\setlength{\partopsep}{0ex}
\setlength{\itemsep}{0ex}}}
{\end{list}}

\def\bel{\begin{lemma}}
\def\eel{\end{lemma}}
\def\bec{\begin{corollary}}
\def\eec{\end{corollary}}
\def\bet{\begin{theorem}}
\def\eet{\end{theorem}}
\def\bed{\begin{definition}}
\def\eed{\end{definition}}
\def\bep{\begin{proposition}}
\def\eep{\end{proposition}}
\def\ben{\begin{arabicenumerate}}  
\def\een{\end{arabicenumerate}}
\def\beq{\begin{equation}}
\def\eeq{\end{equation}}

\def\cqfd{\hfill $\rlap{$\sqcap$}\sqcup$}
\newcommand{\e}{\mathrm{e}}

\renewcommand{\i}{\mathrm{i}}

\renewcommand{\d}{\mathrm{d}}
\def\ds{\displaystyle}
\newcommand{\sign}{{\mathrm{sign\,}}}
\def\qedbox{$\rlap{$\sqcap$}\sqcup$}
\def\ad{{\rm ad}}
\def\nn{{\mathbb N}}
\def\zz{{\mathbb Z}}

\def\rr{{\mathbb R}}
\def\cc{{\mathbb C}}
\def\C{{\mathbb C}}
\def\t{{\tau}}
\def\b{{\beta}}
\def\l{{\lambda}}
\def\G{{\Gamma}}
\def\CAR{{\rm CAR}}

\def\fh{\mathfrak{h}}

\def\cL{\mathcal{L}}

\def\B{\mathcal{B}}
\def\cA{\mathcal{A}}
\def\cS{\mathcal{S}}

\def\cK{\mathcal{K}}
\def\cH{\mathcal{H}}
\newcommand{\HH}{{\widetilde{\mathcal H}}}
\def\cG{\mathcal{G}}

\def\cR{\mathcal{R}}
\def\cO{\mathcal{O}}
\def\bPt{\bar P^{\kern1pt t}}
\def\bQt{\bar Q^{\kern1pt t}}

\newcommand{\esup}{\mathop{\rm{ess-supp}}}
\newcommand{\essup}{\mathop{\rm{ess-sup}}}

\def\fh{\mathfrak{h}}
\def\Dom{{\rm Dom}}
\def\O{\Omega}
\def\bbeta{{\boldsymbol\beta}}

\def\bmu{{\boldsymbol\mu}}

\newcommand{\union}{\cup}

\def\Im{\mathrm{Im}\,}
\def\Re{\mathrm{Re}\,}
\def\Ran{\mathrm{Ran}\,}
\def\Ker{\mathrm{Ker}\,}
\def\Gr{\mathrm{Gr}\,}
\def\ac{\mathrm{ac}}
\def\pp{\mathrm{pp}}
\def\spec{\mathrm{Sp}}
\def\Res{\mathrm{Res}}

\def\supp{{\rm supp}}
\def\tr{\mathrm{tr}}

\def\Ent{{\rm Ent}}

\def\cB{{\cal B}}

\newcommand{\slim}{\mathop{\mathrm{s-lim}}\limits}
\newcommand{\wlim}{\mathop{\mathrm{w-lim}}\limits}
\newcommand{\starslim}{\mathop{\rm s^\ast-lim}}
\newcommand{\wstarlim}{\mathop{\rm w^\ast-lim}}
\newcommand{\alim}{\mathop{\rm Abel-lim}}
\def\bar{\overline}

\def\cal{\mathcal}

\renewcommand{\atop}[2]{\genfrac{}{}{0pt}{1}{#1}{#2}}
\begin{document}
\thispagestyle{empty}
\title{\bf\Huge A Geometric Approach to the Landauer-B\"uttiker Formula\\\vskip 1cm}
\author{
\sc R. Ben S\^aad$^{a}$, C.-A. Pillet$^{b}$\\ \\ \\
\small$^a$Laboratoire Math\'ematiques et Applications, INSAT Centre Urbain Nord BP 676, 1080 Tunis Cedex\\ \\
\small $^b$Aix-Marseille Universit\'e, CNRS, CPT, UMR 7332, Case 907, 13288 Marseille, France\\
\small Universit\'e de Toulon, CNRS, CPT, UMR 7332, 83957 La Garde, France\\
\small FRUMAM}
\def\today{}
\maketitle

\tableofcontents

\vskip 1cm
\begin{quote}
{\bf Abstract.} We consider an ideal Fermi gas confined to a geometric structure consisting of
a central region -- the sample -- connected to several infinitely extended ends -- the reservoirs.
Under physically reasonable assumptions on the propagation properties of the one-particle dynamics 
within these reservoirs, we show that the state of the Fermi gas relaxes to a steady
state. We compute the expected value of various current observables in this steady state and express
the result in terms of scattering data, thus obtaining a geometric version of the celebrated
Landauer-B\"uttiker formula.
\end{quote}
\vskip 1cm
\section{Introduction}

\begin{figure}
\centering
\includegraphics[scale=0.85]{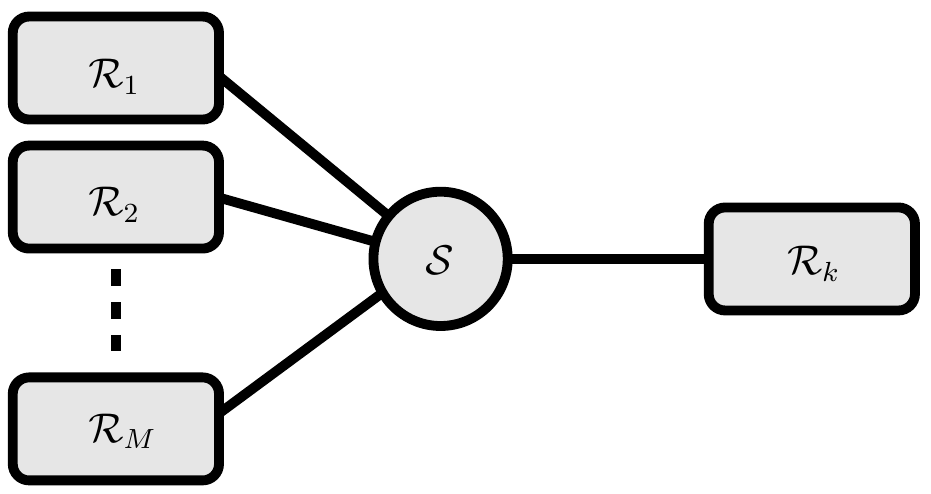}
\caption{A sample $\cS$ coupled to $M$ reservoirs $\cR_1,\ldots,\cR_M$.}
\label{OSysFig}
\end{figure}

The study of transport phenomena in the quantum regime has attracted a lot of interest over
the last decades, especially within the realm of condensed matter physics. The main efforts have 
been devoted to the development of computational tools for the calculation of steady state properties
of a confined quantum system (the sample) driven out of thermal equilibrium by mechanical or 
thermodynamical forces. This physical setup is conveniently described by an open-system model 
where the sample $\cS$ is coupled to large (eventually infinitely extended) heat and particle 
reservoirs $\cR_1,\cR_2,\ldots$ (see Figure~\ref{OSysFig}). Thermodynamical forces are
implemented by the initial state of the joint system $\cS+\cR_1+\cR_2+\cdots$. More precisely,
each reservoir $\cR_k$ is prepared in a thermal equilibrium state with its own intensive 
thermodynamic parameters: inverse temperature $\beta_k$, chemical potential $\mu_k$,\dots
In the physics literature, this is sometimes called ``the partitioned scenario'', reflecting the fact that 
each reservoir has to be prepared individually before being connected to the sample.
Mechanical forcing is obtained by imposing (possibly time dependent) potential bias in the reservoirs,
the initial state of the system being a joint thermal equilibrium state of the coupled system
$\cS+\cR_1+\cR_2+\cdots$. This is the so called ``partition free scenario'', see \cite{CCNS,C}.

Whether such an open system, prepared in a given initial state, actually relaxes to a steady state 
is a more delicate question which can not be treated by formal arguments and requires a precise 
control of quantum dynamics. To the best of our knowledge, the first rigorous results on this 
fundamental problem of nonequilibrium quantum statistical mechanics were obtained by Lebowitz 
and Spohn \cite{Sp1,Sp2,Sp3,LS} in the case of thermodynamical forcing. Besides providing simple 
and efficient criteria ensuring relaxation to a steady state in the van~Hove scaling limit (weak coupling),
they have also studied the basic thermodynamic properties of these steady states: strict positivity of 
entropy production and linear response theory. In the same limit, Davies and Spohn have studied the 
linear response of confined quantum systems to mechanical drives \cite{DSp}. These works rely on 
Davies' results on the weak coupling limit \cite{D1,D2} (see also the recent extension of Davies'
theory by Derezi\'nski and de~Roeck \cite{DdR1,DdR2}) and therefore only provide a coarse time 
resolution of transport phenomena.

In these notes we shall consider the simplest case, beyond the weak coupling limit, amenable to 
rigorous analysis: the transport properties of an ideal Fermi gas (e.g., of an electronic gas in the 
approximation of independent electrons commonly used in solid state physics). Due to the absence 
of interactions, the nonequilibrium properties of such a gas can be derived from the quantum dynamics 
of a one-particle system. We shall concentrate more specifically on the Landauer-B\"uttiker formalism 
which relates the steady currents through a sample connected to several fermionic reservoirs at 
different chemical potentials to the scattering data associated with the coupling of the sample to the 
reservoirs (we shall provide a more detailed discussion of the Landauer-B\"uttiker formalism in 
Section~\ref{LBsection1}).

Relaxation to a nonequilibrium steady state (NESS) for an ideal Fermi gas in the partitioned scenario 
was first obtained by Araki and Ho \cite{AH}. These Authors studied the large time asymptotics of the 
isotropic XY spin chain prepared in a state with different temperatures on its left and and right ends 
(the XY chain can be mapped to an ideal Fermi gas on a 1D lattice by a Jordan-Wigner transformation).
Their result has been extended to the anisotropic XY chain in \cite{AP} using a different approach, 
advocated by Ruelle \cite{R4}, and based on scattering theory. In Ruelle's approach, the NESS is
expressed in terms of the initial state of the gas and the M\o ller operator describing the
scattering of a particle from the reservoirs by the sample (see Section~\ref{Sect-NESS}). 
However, to derive the Landauer-B\"uttiker formula which 
expresses the steady state currents in terms of transmission probabilities (i.e., scattering matrix) 
requires further work. This was first achieved in \cite{AJPP2,N} within the stationary formalism of 
scattering theory and for more general classes of ideal Fermi gases driven by thermodynamical forces
(see also Section~7 in \cite{AJPP1} and \cite{CNWZ}).

In the case of mechanical forcing (in the partition-free scenario), a linearized Landauer-B\"ut\-tiker 
formula (i.e., a formula for the conductivity of the sample) was obtained by Cornean {\sl et al.} 
in \cite{CJM,CDNP}. However, relaxation to a NESS did not follow from the linear response approach 
used in these works and was first proved in \cite{CDP}\footnote{For identical intensive thermodynamic
parameters, the partitioned/partition-free scenarios lead to distinct NESS.}. 
Finally, a complete (non-linear) 
Landauer-B\"uttiker formula for the steady currents was derived in \cite{CGZ}. A unified treatment 
of the partitioned/partition-free NESS can be found in \cite{CMP2}. 

In both scenarios a necessary condition for the coupled/biased system
$\cS+\cR_1+\cdots$ to relax to a NESS is that its final, fully coupled/biased, one-particle Hamiltonian 
has empty singular spectrum. In that case, the NESS only depends on the initial states of the
reservoirs and on the final one-particle Hamiltonian. It is, in particular, independent of the initial state of the sample
and of the (possibly time-dependent) switching of the coupling/bias \cite{CNZ,CMP2}.
In fact, the presence of eigenvalues in the one-particle Hamiltonian of the fully coupled/biased system 
produces oscillations which prevent relaxation to a steady state \cite{Ste,KKSG}. These oscillations
are carried by the eigenfunctions of the Hamiltonian and hence are typically localized near the sample.
Current measurements performed deep into the reservoirs are therefore immune to this effect
\cite{CGZ}. If the  singular continuous spectrum of the final Hamiltonian is empty, then the oscillations 
induced by its eigenvalues can also be washed out by time-averaging the state of the system. 
The time-averaged state relax to a steady state which, however, depends on the initial state of the 
sample and on the history of the coupling/bias \cite{AJPP2,CGZ,CJN}.

Before turning to a detailed review of the content of these notes, let us mention some
important results in the same line of research but which will not be covered here.

Ruelle's scattering approach also works in the presence of weak local interactions
(i.e., many body interactions that are sufficiently well localized in position and momentum).
In this case, the M\o ller operator of Hilbert space scattering theory is replaced by a M\o ller morphism
acting on the $C^\ast$-algebra $\cO$ of observables of the coupled system ($\cO$ is typically
the gauge invariant part of the $C^\ast$-algebra generated by fermionic creation/annihilation 
operators satisfying the canonical anti-commutation relations, see Section~\ref{Sect-FermiGas}).
This morphism can be constructed by controlling the Dyson expansion of the interaction picture 
propagator acting on $\cO$,  using the techniques of \cite{BM1,BM2}.  
Relaxation to a NESS of a locally interacting Fermi gas in the partitioned scenario
was first proved by Fr\"ohlich {\sl et al.} \cite{DFG,FMU}. Linear response theory (including a 
central limit theorem) for such NESS was developed in \cite{JOP1,JOP2,JPP}. Using similar
techniques, a mathematical theory of basic thermodynamic processes in ideal and locally interacting 
Fermi gases has been developed in \cite{FMSU}. A unified approach to both partitioned/partition-free 
NESS of locally interacting Fermi gases was developed in \cite{CMP1,CMP2} where basic properties 
of the NESS Green-Keldysh correlation functions were also derived.

The spectral analysis of Liouvilleans provides an alternative to Ruelle's scattering approach to 
the construction of NESS. A Liouvillean for the coupled system $\cS+\cR_1+\cdots$ is an operator
$L$ acting on a Hilbert space which carries a representation of the $C^\ast$-algebra $\cO$ and such 
that the group $t\mapsto\e^{\i tL}$ implements the dynamics (see \cite{P2,DJP,JOPP}). For systems 
with finitely extended reservoirs the Liouvillean is essentially determined by the Hamiltonian. There is
however much more freedom in the choice of a Liouvillean if the reservoirs are infinitely extended.
The Liouvillean approach has been successfully used to prove return to equilibrium of a confined
system connected to a single heat bath \cite{JP1,JP2,JP3,JP4,Me,BFS3,DJ1,DJ2,FM}.
An extension of this technique to nonequilibrium situations was developed in \cite{JP6} to prove
relaxation to a steady state of a N-level system coupled to several fermionic reservoirs. 
Merkli, M\"uck and Sigal have extended this result to the technically more involved case of bosonic 
reservoirs \cite{MMS}. In these works, the steady state is characterized by a spectral resonance 
of a Liouvillean which is constructed with the help of operator algebraic techniques derived from 
the fundamental results of \cite{HHW,To,Ta}. Due to the use of spectral deformation techniques
in the resonance analysis, the method requires quite strong regularity assumptions on
the coupling of the sample to the reservoir. It does however provide a very detailed information
on the dependence of the NESS on this coupling (a convergent perturbative expansion). 
A similar approach was used by Fr\"ohlich, Merkli 
and Sigal \cite{FMS} to study the ionization process in a thermal field. We shall also mention a
series of works by Abou Salem and Fr\"ohlich \cite{AF1,AF2,AF3} who exploit the Liouvillean approach
to derive some of the basic laws of thermodynamics from microscopic quantum dynamics.
We refer the reader to the article of Schach M\o ller \cite{SM} in this volume for a detailed
exposition of the spectral theory of some important classes of Liouvilleans. 

A third approach to the relaxation problem has been developed by de Roeck and Kupiainen
in \cite{dRK1,dRK2} (see also \cite{dR1}). It uses Davies' weak coupling approximation of the dynamics as
a starting point for a systematic expansion of the true, fully coupled dynamics. The control of this 
expansion is technically more involved than the analysis required in the Liouvillean approach, but it 
is very robust and only requires minimal assumptions on the coupling to the reservoir (essentially
the existence of the Davies approximation with a spectral gap). However, the method does not
provide much information on the dependence of the NESS on the coupling.

\bigskip
 
The material presented in these notes is partly based on the PhD thesis of the first Author \cite{Sa}.
It can be read as a pedagogical introduction to some contemporary aspects of the mathematics 
of nonequilibrium quantum statistical mechanics. The main objectives are:

\begin{itemize}
\item To prove relaxation of an ideal Fermi gas under thermodynamical drive using Ruelle's
approach and geometric 
time-dependent scattering theory based on the Mourre estimate. This framework has many 
advantages over the stationary scattering theory used in the previous works on the subject.
Our main assumptions, which ensure good propagation properties at large distance from the sample,
concerns the reservoirs. They are easily checked for reservoirs with a simple geometry.  
Mourre theory gives us a 
simultaneous control over the propagation properties and the singular spectrum of the coupled system.
Finally, with the use of the two Hilbert space formalism, we avoid the decoupling of the sample
by artificial boundary conditions. The scattering matrix obtained in this way is explicitly independent 
of any decoupling scheme, which represents a serious conceptual advantage.

\item To show that the properly defined NESS expectation of a current observable can be expressed
in terms of scattering data by a geometric version of the Landauer-B\"uttiker formula.
Our approach has been deeply inspired by the works of Avron {\sl et al.} \cite{AEGS1,AEGS2,AEGS3} 
and more specifically \cite{AEGSS} who treat the similar problem of charge transport in quantum 
pumps in the adiabatic regime (and prove the B\"uttiker-Pr\^etre-Thomas formula \cite{BTP}).
\end{itemize}

Our main results are Proposition \ref{NESSexist} and Theorem \ref{NESSalt}, which guarantee
the existence and uniqueness of the NESS under physically reasonable conditions. Under
a few additional assumptions, we prove a geometric version of the Landauer-B\"uttiker formula 
in Theorem \ref{LBTheo}.

\bigskip
\noindent The organization of these notes is as follows:

\begin{itemize}
\item In Section \ref{Setup} we describe the necessary mathematical background for our work. 
The goal here is essentially to introduce the basic tools and the notation that will be used
in the following sections.

\item Section \ref{MSHEChap} is a brief introduction to nonequilibrium statistical mechanics of 
open quantum systems, and more specifically, -free fermionic systems. We introduce
the concept of NESS and describe Ruelle's scattering method for its construction. 

\item Section \ref{MourreChapter} is a thorough discussion of commutators of Hilbert space
operators and their use in spectral analysis. It introduces the elements of Mourre theory which will be 
necessary for controlling the singular spectrum and the propagation properties of quasi-free 
fermionic systems.

\item Section \ref{NESSChapter} is dedicated to the construction of NESS using the geometric theory 
of multichannel scattering and propagation estimates.

\item In Section \ref{LBForm} we discuss current observables and compute their 
expectation values in the NESS, deriving the geometric Landauer-B\"uttiker formula. 

\item The appendices \ref{HTraceProof} and \ref{LocalisationProof} contain a few technical proofs 
that we deemed appropriate to be separated from the main part of these notes.
\end{itemize}

\noindent{\bf Acknowledgments.} This work was partly supported by ANR (grant 09-BLAN-0098).
We are grateful to Y.~Barsheshat for his help in translating parts of the PhD thesis of the first
Author included in this notes. C.-A.P. is also grateful to V. Jak\v si\'c, to the Department of 
Mathematics and Statistics at McGill University and to CRM (CNRS - UMI 3457) for hospitality 
and generous support during his stay in Montreal where parts of this work were done.

\section{Mathematical background}
\label{Setup}
In this section we briefly review the necessary mathematical background. The main purpose is to
setup our notation. The covered material is very basic and the exposition is in telegraphic style, 
without formal proofs. The readers familiar with spectral analysis in Hilbert spaces and operator 
algebras can safely jump over to the next section.

In Subsection~\ref{SpectralIntro} we introduce the fundamentals of spectral analysis of operators on 
a Hilbert space, paying particular attention to self-adjoint operators and to the scattering theory of the
associated unitary groups. These are the common tools used in the mathematical study of 
quantum dynamics, i.e., solutions to the Schr\"odinger equation, either time-dependent or
time-independent. A\-mong the numerous techniques developed to study the properties of the
solutions to this equation, those based on the work of Mourre will play a central role in these notes. 
These techniques will be the object of a more detailed discussion in Section~\ref{MourreChapter}.

Subsection~\ref{CstarSection} is a brief introduction to the theory of operator algebras and more 
particularly $C^\ast$-algebras. From the perspective of the material covered in these notes, the relevance
of this subject is marginal. It does however play an important role in the more general context of the 
mathematical theory of quantum statistical mechanics. As we have already noted in the general 
introduction, the development of this theory saw a revival in the last decade, essentially revolving 
around transport problems in nonequilibrium systems. These recent developments were built upon 
the foundations of the algebraic approach to equilibrium quantum statistical mechanics developed 
in the 1960s and 1970s.

\subsection{Spectral analysis and scattering theory}
\label{SpectralIntro}
In this section we recall some fundamental results of spectral analysis of self-adjoint operators
on a Hilbert space, as well as the basics of scattering theory. The material covered in this section 
is treated in full detail in \cite{RS1}--\cite{RS4}.

\subsubsection{Closed operators and bounded operators}
\label{OperaSubSect}

If $A,B$ are non-empty sets we denote by $\langle a,b\rangle$ the elements of the Cartesian product 
$A\times B$ so as to not generate confusion with the following notation.

Let $\mathcal H$ be a Hilbert space. We denote by
$$
\begin{array}{ccc}
\mathcal H\times\mathcal H&\longrightarrow&\mathbb C\\
\langle u,v\rangle&\mapsto&(u,v),
\end{array}
$$
the inner product of $\mathcal{H}$, which is anti-linear in its first argument and linear in its second one.
Riesz' representation theorem guarantees that any continuous linear functional
$\ell:\mathcal{H}\to\cc$ can be written in the form $\ell(v)=(u,v)$ for some $u\in\mathcal{H}$.
The ortho-complement of a subset $V\subset\cH$ is defined by $V^\perp=\{u\in\cH\,|\,(v,u)=0\text{ for all }v\in V\}$.
It is a closed subspace of $\cH$ and $V^{\perp\perp}$ is the smallest closed subspace of $\cH$
containing $V$. An automorphism of $\cH$ is a linear and isometric bijection from $\cH$ onto itself.
$\cH\times\cH$ equipped with its natural vector space structure and the inner product
$(\langle u,v\rangle,\langle u',v'\rangle)=(u,u')+(v,v')$ is a Hilbert space and 
$\mathbb{K}:\langle x,y\rangle\mapsto\langle y,x\rangle$ and
$\mathbb{J}:\langle x,y\rangle\mapsto\langle -y,x\rangle$ define automorphisms
of $\mathcal{H}\times\mathcal{H}$. A net $u_\alpha$ in $\cH$ converges weakly to $u\in\cH$
if the net $(v,u_\alpha)$ converges to $(v,u)$ for all $v\in\cH$. In this case we write
$$
\wlim_\alpha u_\alpha=u.
$$

An operator on $\mathcal H$ is a linear map $A:\mathcal{D}\to\mathcal{H}$, where 
$\mathcal{D}$ is a subspace of $\mathcal{H}$. We say that $\mathcal{D}$ is the domain 
of $A$ which we denote by $\Dom(A)$. $A$ is densely defined if its domain is dense in $\cH$.
The range and the kernel of $A$ are the subspaces $\Ran(A)\equiv\{Au\,|\,u\in\Dom(A)\}$ and 
$\Ker(A)\equiv\{u\in\Dom(A)\,|\,Au=0\}$ respectively. $A$ is surjective if $\Ran(A)=\cH$ and
injective if $\Ker(A)=\{0\}$.

The graph of an operator $A$ on $\cH$ is the the subspace
$$
\Gr(A)\equiv\{\langle u,Au\rangle\,|\,u\in\Dom(A)\},
$$ 
of $\mathcal H\times\mathcal H$. The graph norm of $A$ is the norm defined by 
$\|u\|_A=\|u\|+\|Au\|$ on $\Dom(A)$.
An operator $A$ is completely characterized by its graph.
Moreover, a subspace $\cG\subset\cH\times\cH$ is the graph of an operator iff
$\langle 0,v\rangle\in\cG$ implies $v=0$. If $A$ and $B$ are two operators such that 
$\Gr(A)\subset\Gr(B)$ we say that $B$ is an extension of  $A$ and we write $A\subset B$. 
An operator $A$ is closed if its graph is closed in $\mathcal{H}\times\mathcal{H}$, and
this is the case iff $\Dom(A)$, equipped with the graph norm of $A$, is a Banach space.
If $A$ is both closed and bijective, then $\Gr(A^{-1})=\mathbb{K}\Gr(A)$ and thus $A^{-1}$ 
is also closed. If the closure $\Gr(A)^\mathrm{cl}$ of the graph of $A$ in $\mathcal H\times\mathcal H$ is a graph we 
say that $A$ is closable and we define its closure as the operator $A^\mathrm{cl}$ such that 
$\Gr(A^\mathrm{cl})=\Gr(A)^\mathrm{cl}$. It is clear that $A^\mathrm{cl}$ is the smallest closed 
extension of $A$, that is to say that if $B$ is closed and $A\subset B$, then $A^\mathrm{cl}\subset B$.
An operator $A$ is densely defined iff $\mathbb{J}(\Gr(A)^\perp)$ is a graph. In this case, 
the adjoint of $A$ is the operator $A^\ast$ defined by $\Gr(A^\ast)=\mathbb{J}(\Gr(A)^\perp)$.
$A^\ast$ is closed and its domain is given by
$$
\Dom(A^\ast)=\{u\in\cH\,|\,\sup_{v\in\Dom(A),\|v\|=1}|(u,A v)|<\infty\}.
$$ 
$(A^\ast u,v)=(u,Av)$ holds for all $\langle u,v\rangle\in\Dom(A^\ast)\times\Dom(A)$,
in particular $\Ker(A^\ast)=\Ran(A)^{\perp}$. $A$ is closable iff $A^\ast$ is densely defined. 
In this case $A^\mathrm{cl}=A^{\ast\ast}$ and $A^{\mathrm{cl}\,\ast}=A^\ast$. 

An operator $A$ is bounded if there exists a constant $C$ such that
$\Gr(A)\subset\{\langle u,v\rangle\,|\,\|v\|\le C\|u\|\}$. One easily verifies that $A$ 
is continuous as a map from $\Dom(A)$ to $\cH$ iff it is bounded. A bounded operator is obviously
closable and its closure coincide with its unique continuous extension to the closure of $\Dom(A)$.
In particular a bounded densely defined operator $A$ has a unique continuous extension $A^\mathrm{cl}$
with domain $\Dom(A^\mathrm{cl})=\mathcal H$. $A^\mathrm{cl}$ is closed and bounded. The collection of all 
bounded operators with domain $\mathcal H$ is denoted by $\B(\mathcal H)$. It is a Banach algebra 
(actually a $C^\ast$-algebra, see Section \ref{CstarSection}) with norm 
\begin{equation}
\|A\|\equiv\sup_{u\in\mathcal H, \|u\|=1}\|Au\|.
\label{OpNormDef}
\end{equation}
By the closed graph theorem, an operator $A$ with domain $\cH$ is bounded iff it is closed.
If A is bounded and densely defined, then $\Dom(A^\ast)=\mathcal{H}$ and $A^\ast$ is bounded. 
Furthermore, $\|A^\ast\|=\|A\|$ and $\|A^\ast A\|=\|A\|^2$.

A bounded net $A_\alpha$ in $\B(\mathcal H)$ is strongly (resp. weakly) convergent if 
the net $A_\alpha u$ is convergent (resp. weakly  convergent) for every $u\in\cH$. In this case, there
exists $A\in\B(\mathcal H)$ such that $\lim_\alpha A_\alpha u=Au$ for all $u\in\cH$ (resp.
$\lim_\alpha(u,A_\alpha v)=(u,Av)$ for all $u,v\in\cH$), and we write
$\mathrm{s-lim}_\alpha A_\alpha=A$ (resp. $\mathrm{w-lim}_\alpha A_\alpha=A$). 
If $A_\alpha^{(1)},\ldots,A_\alpha^{(n)}$
are bounded nets in $\B(\mathcal H)$ and $\mathrm{s-lim}_\alpha A_\alpha^{(j)}=A^{(j)}$ 
for all $j$ then $\mathrm{s-lim}_\alpha A_\alpha^{(1)}\cdots A_\alpha^{(n)}=A^{(1)}\cdots A^{(n)}$.

The resolvent set of a closed operator $A$ is defined by
$$
\Res(A)=\{z\in\mathbb C\,|\,\Ker(A-z)=\{0\}\ \text{and}\ \Ran(A-z)=\mathcal H\},
$$
thus $z\in\Res(A)$ if and only if $(A-z):\Dom(A)\to\mathcal H$ is a bijection. In this case, the operator
$R_A(z)\equiv(A-z)^{-1}:\mathcal H\to\Dom(A)$ is called the resolvent of $A$ at $z$.
It is a closed operator with domain $\mathcal H$, and is thus bounded. It satisfies the functional
equation $R_A(z)-R_A(z')=(z-z')R_A(z)R_A(z')$ (so called first resolvent equation) for all 
$z,z'\in\Res(A)$. If follows that for $z_0\in\Res(A)$
$$
R_A(z)=\sum_{n=0}^\infty R_A(z_0)^{n+1}(z-z_0)^n,
$$
this Neumann series being norm convergent for $|z-z_0|<\|R_A(z_0)\|^{-1}$. Thus, the resolvent set of $A$
is open, and the mapping $z\mapsto R_A(z)$ is an analytic function from $\Res(A)$ to 
$\B(\mathcal H)$. The closed set $\spec(A)\equiv\mathbb C\setminus\Res(A)$ is called the spectrum 
of $A$. A point $a\in\spec(A)$ is an eigenvalue of $A$ if there exists a non-zero vector $u\in\Dom(A)$ 
such that $Au=au$. We say that $u$ is an eigenvector of $A$ associated to the eigenvalue $a$. 

If $\mathcal{H}$ and $\mathcal{K}$ are two Hilbert spaces, most of the preceding facts easily 
generalize to linear maps $A$ from $\Dom(A)\subset\mathcal H$ to $\mathcal K$. We denote by 
$\B(\mathcal H,\mathcal K)$ the Banach space of continuous operators from $\mathcal{H}$ to 
$\mathcal{K}$ equipped with the norm (\ref{OpNormDef}).

\subsubsection{Self-adjoint operators}
\label{SAOpSect}

An operator $A$ is called symmetric if $A\subset A^\ast$, self-adjoint if $A=A^\ast$, and essentially
self-adjoint if $A^{\ast\ast}=A^\ast$. An essentially self-adjoint operator $A$ is closable and its closure
$A^\mathrm{cl}=A^{\ast\ast}$ is self-adjoint. In this case, we say that $\Dom(A)$ is a core of 
$A^\mathrm{cl}$.

An operator $A$ is symmetric if and only if $(u,Au)\in\mathbb R$ for all $u\in\Dom(A)$. Such an operator
is self-adjoint iff $\Ran(A\pm\i)=\mathcal H$ and it is essentially self-adjoint iff 
$\Ran(A\pm\i)$ is dense in $\mathcal H$.

If $\cK$ is a closed subspace of $\cH$ then $\cH=\cK\oplus\cK^\perp$, i.e., any $u\in\cH$ has a 
unique representation $u=x+y$ with $x\in\cK$ and $y\in\cK^\perp$. Moreover the
Pythagoras theorem $\|u\|^2=\|x\|^2+\|y\|^2$ holds. The decomposition $u=x+y$ defines a bounded
operator $P:u\mapsto x$ satisfying $P=P^2=P^\ast$. We call $P$ the orthogonal projection onto 
$\cK$. Note that $Q=I-P$ is the orthogonal projection onto $\cK^\perp$. Reciprocally, if
$P\in\cB(\cH)$ satisfies $P=P^2=P^\ast$ then it is the orthogonal projection onto the closed subspace
$\Ran(P)=\Ker(I-P)$ and $I-P$ is the orthogonal projection onto $\Ker(P)=\Ran(I-P)$.

If $\mathcal K$ is a closed subspace of $\mathcal H$ and $J$ is an operator with domain 
$\mathcal H$ such that $\|Ju\| = \|u\|$ for all $u\in\mathcal K$ and $Ju = 0$ for all 
$u\in\mathcal K^\perp$ then $\Ker(J) = \mathcal K^\perp$ and $\mathcal{R }\equiv\Ran(J)$ is a 
closed subspace of $\mathcal H$. $J$ is thus a isometric bijection from $\mathcal K$ into $\mathcal R$. 
We say that $J$ is a partial isometry with initial space $\mathcal K$ and final space $\mathcal R$.
One verifies that $JJ^\ast$ is the orthogonal projection onto $\mathcal R$ and $J^\ast J$ is the 
orthogonal projection onto $\mathcal K$. If $\mathcal K=\mathcal R=\mathcal H$ then
$JJ^\ast=J^\ast J=I$ and $J$ is unitary.

If $A$ is self-adjoint, then $\spec(A)\subset\mathbb R$. If we also have that
$\spec(A)\subset[0, \infty[$, then $A$ is called positive and we write $A\ge 0$. A self-adjoint operator 
is positive if and only if $(u,Au)\ge 0$ for all $u\in\Dom(A)$. If $C$ is a closed operator then 
$C^\ast C$ with the domain $\Dom(C^\ast C)=\{u\in\Dom(C)\,|\,Cu\in\Dom(C^\ast)\}$ is positive. 
Conversely, every positive operator is of this form.

Every closed operator $A$ has a unique polar decomposition $A=J|A|$ where $|A|\ge0$ and
$J$ is a partial isometry with initial space ${\Ran(A^\ast)}^\mathrm{cl}=\Ker(A)^\perp$ 
and final space ${\Ran(A)}^\mathrm{cl}=\Ker(A^\ast)^\perp$. Moreover, $|A|$ is the square 
root of the positive operator $A^\ast A$ constructed with the help of functional calculus which we 
shall now describe.

\bigskip
{\bf\noindent Spectral theorem 1.}
Let $B_b(\mathbb R)$ be the algebra of bounded Borel functions from $\mathbb R$ to $\mathbb C$. 
If $A$ is self-adjoint, there exists a unique morphism $\phi_A:B_b(\mathbb R)\to\B(\mathcal H)$ such 
that
\begin{enumerate}[(i)]
\item $\phi_A(\bar f)=\phi_A(f)^\ast$.
\item $\|\phi_A(f)\|\le\sup_{a\in\spec(A)}|f(a)|$.
\item If $\lim_n f_n(a)=f(a)$ for all $a\in\spec(A)$ and $\sup_{n,a\in\spec(a)}|f_n(a)|<\infty$
then $$\lim_n\phi_A(f_n)u=\phi_A(f)u,$$ for all $u\in\mathcal H$.
\item If $f\ge0$ then $\phi_A(f)\ge0$.
\item If $Au=au$ then $\phi_A(f)u=f(a)u$.
\item If $z\in\Res(A)$ and $f(a)=(a-z)^{-1}$ then $\phi_A(f)=R_A(z)$.
\end{enumerate}
We call this morphism the functional calculus associated with $A$ and we write $f(A)=\phi_A(f)$. 
We say that a bounded operator $B$ commutes with $A$ if $Bf(A) = f(A)B$ for all 
$f\in B_b(\mathbb R)$. A subspace $\mathcal K\subset\mathcal H$ is $A$-invariant 
if $f(A)\mathcal K\subset\mathcal K$ for all $f\in B_b(\mathbb R)$. It reduces $A$ 
if in addition $\mathcal K^\perp$ is also $A$-invariant. If $\mathcal K$ reduces $A$ we define
the part of $A$ in $\cK$ to be the self-adjoint operator on $\Dom(A)\cap\mathcal K$ obtained
by restricting $A$ to this subspace. We also define the part of the spectrum of $A$ in $\mathcal K$ 
as $\spec(A|\mathcal K)\equiv\spec(A|_{\mathcal K\cap\Dom(A)})$.

{\bf\noindent Spectral theorem 2.}
It follows from the functional calculus that for all $u\in\mathcal H$ the map $f\mapsto(u,f(A)u)$ is a 
continuous linear functional on $C_\infty(\mathbb R)$, the Banach space of continuous functions 
from $\mathbb R$ to $\mathbb C$ which tend to $0$ at infinity, equipped with the norm 
$\|f\|_\infty\equiv\sup_{x\in\mathbb R}|f(x)|$. The Riesz-Markov theorem implies that there
exists a finite measure $\mu_u$, with $\mu_u(\mathbb R)=\|u\|^2$, and such that
$$
(u,f(A)u)=\int f(a)\,\d\mu_u(a).
$$
$\mu_u$ is the spectral measure of $A$ associated with $u$.

Let $u\in\mathcal H$ and $\mathcal H_u={\{f(A)u\,|\,f\in C_\infty(\mathbb R)\}}^\mathrm{cl}$. 
The map  $f\mapsto f(A)u$ from $C_\infty(\mathbb R)\to\mathcal H_u$ satisfies 
$\|f(A)u\|=\|f\|_{L^2(\mathbb R,\d\mu_u)}$. It extends continuously to a unitary operator 
$U_u:L^2(\mathbb R,\d\mu_u)\to\mathcal H_u$ such that, if $M_g$ denotes the multiplication 
operator $f\mapsto gf$ on $L^2(\mathbb R,\d\mu_u)$, $U_uM_g=g(A)U_u$. If $\mathcal H$ is 
separable, one can easily show that there exists a countable family $(u_n)_{n\in N}\subset\mathcal H$ 
such that $\mathcal H=\oplus_{n\in N}\mathcal H_{u_n}$. In this way we obtain a unitary map 
$U:\oplus_{n\in N}L^2(\mathbb R,\d\mu_n)\to\mathcal H$ such that, if $g$ denotes the operator 
$\oplus_n u_n\mapsto\oplus_n gu_n$, then $Ug=g(A)U$. Alternatively stated, $A$ is unitarily 
equivalent to the operator of multiplication by the variable $a$ in the space 
$\oplus_{n\in N}L^2(\mathbb R,\d\mu_n(a))$.

One can show that $\mathcal H=\mathcal H_\pp(A)\oplus\mathcal H_\ac(A)
\oplus\mathcal H_\mathrm{sc}(A)$
where 
\begin{align*}
\mathcal H_\pp(A)&\equiv\{u\in\mathcal H\,|\,\mu_u\ \text{is purely atomic}\}\cup\{0\},\\
\mathcal H_\ac(A)&\equiv\{u\in\mathcal H\,|\,\mu_u\ \text{is Lebesgue-absolutely continuous}\}
\cup\{0\},\\
\mathcal H_\mathrm{sc}(A)&\equiv\{u\in\mathcal H\,|\,\mu_u\ 
\text{is Lebesgue-singular without atoms}\}\cup\{0\},
\end{align*}
are mutually orthogonal subspaces reducing $A$. We denote by  $P_\pp(A)$, $P_\ac(A)$ and 
$P_\mathrm{sc}(A)$ the orthogonal projections onto these subspaces and we define 
$A_\pp$, $A_\ac$, and $A_\mathrm{sc}$ to be the parts of $A$ in each of these subspaces. 
$\mathcal H_\pp(A)$ is the subspace spanned by the eigenvectors of $A$. The pure point, 
absolutely continuous, and singular spectra of $A$ are defined by
\begin{align*}
\spec_\pp(A)&\equiv\{a\in\mathbb R\,|\,a\ \text{is an eigenvalue of }\ A\},\\
\spec_\ac(A)&\equiv\spec(A_\ac),\\
\spec_\mathrm{sc}(A)&\equiv\spec(A_\mathrm{sc}),
\end{align*}
and we have that
 $\spec(A)={\spec_\pp(A)}^\mathrm{cl}\cup\spec_\ac(A)\cup\spec_\mathrm{sc}(A)$. 

The singular
spectrum of $A$ is $\spec_{\rm sing}(A)=\spec_\pp(A)^{\rm cl}\cup\spec_\mathrm{sc}(A)$. Its
discrete spectrum is the set $\spec_{\rm disc}(A)$ of all its isolated eigenvalues $a$ having finite
multiplicity, i.e., such that the corresponding eigenspace $\Ker(A-a)$ is finite dimensional. The
essential spectrum of $A$ is $\spec_{\rm ess}(A)=\spec(A)\setminus\spec_{\rm disc}(A)$.

\bigskip
{\bf\noindent Spectral theorem 3.} If $1_\Delta$ is the indicator function of a Borel set 
$\Delta\subset\mathbb R$, then  $E_A(\Delta)\equiv 1_\Delta(A)$ is an orthogonal projection. It is 
the spectral projection of $A$ associated with $\Delta$. Its image reduces $A$ and we have that 
$\spec(A|\Ran E_\Delta(A))=\spec(A)\cap\Delta\subset\Delta$ and 
$\spec(A|\Ker E_\Delta(A))\cap\Delta$ is empty. The family
$\{E_\Delta(A)\,|\,\Delta\subset\mathbb R\ \text{measurable}\}$ is called the spectral family of $A$. Stone's formula relates the spectral family to the resolvent of $A$: for all $u\in\mathcal H$ one has
\beq
\frac{1}{2}(E_A([a,b])+E_A(]a,b[))u=\lim_{\varepsilon\downarrow0}\frac{1}{2\pi\i}\int_a^b\left(
R_A(a+\i\varepsilon)u-R_A(a-\i\varepsilon)u\right)\,\d a,
\label{stoneform}
\eeq
and, in particular, if $a,b$ are not eigenvalues of $A$,
$$
E_A([a,b])u=E_A(]a,b[))u=\lim_{\varepsilon\downarrow0}\frac{1}{2\pi\i}\int_a^b\left(
R_A(a+\i\varepsilon)u-R_A(a-\i\varepsilon)u\right)\,\d a.
$$
Alternatively, the spectral family of $A$ can be interpreted as a measure with values in the orthogonal 
projections of $\mathcal H$. It is thus related to the spectral measures previously introduced by writing
$\d\mu_u(a) = (u,\d E_A(a)u)$ and we can formulate the functional calculus as
$$
(u,f(A)v)=\int f(a)(u,\d E_A(a)v).
$$
We also use the conventional notation $F(A\in\Delta)=E_\Delta(A)$ and, by extension, 
$F(A\ge a)=E_{[a,\infty[}(A)$, {\sl etc...}

The following criterion for the absence of singular spectrum is often useful.
Let $\Delta$ be a bounded open interval in $\rr$ and assume that there exists a dense set
$\mathcal D\subset\cH$ such that
$$
\sup_{\Re(z)\in\Delta,\Im(z)\not=0}|(f,R_A(z)f)|<\infty,
$$
for all $f\in\mathcal D$. It follows that $\spec_{\rm sing}(A)\cap\Delta=\emptyset$, the spectrum of $A$ in $\Delta$ is
purely absolutely continuous.

\bigskip
{\bf\noindent Spectral theorem 4.}  For $n\in\overline{\mathbb N}\equiv\mathbb N\cup\{\infty\}$ 
we write
$$
[1\!:\!n]\equiv\left\{
\begin{array}{ccc}
\emptyset&\text{if}&n=0;\\
\{1,\ldots,n\}& if& n\in\mathbb N;\\
\mathbb N^\ast& if&n=\infty.
\end{array}
\right.
$$
A function  $n:\mathbb R\to\overline{\mathbb N}$ is measurable if $n^{-1}(\{k\})$ is measurable for all 
$k\in\overline{\mathbb N}$. A family of separable Hilbert spaces $(\mathfrak h_a)_{a\in\mathbb R}$ is 
measurable if $n(a)\equiv\dim\mathfrak h_a\in\overline{\mathbb N}$ defines a measurable function. 
Let $\mu$ be a $\sigma$-finite Borel measure on $\mathbb R$. Suppose that for $\mu$-almost every 
$a\in\mathbb R$, $(e_n(a))_{n\in[1:n(a)]}$ is an orthonormal basis of $\mathfrak h_a$. 
By setting  $e_n(a)=0$ when $n>n(a)$ and when the basis $(e_n(a))_{n\in[1:n(a)]}$ 
is not defined, we can assume that $e_n(a)$ is defined for every $a\in\mathbb R$, $n\in{\mathbb N}$
(such a family is called a measurable orthonormal basis). Let $X_0$ be the set of functions 
$u:a\mapsto u(a)$ defined $\mu$-almost everywhere on $\mathbb R$, with values in 
$\cup_{a\in\mathbb R}\mathfrak h_a$, such that $u(a)\in\mathfrak h_a$ for $\mu$-almost all 
$a\in\mathbb R$ and $a\mapsto(e_n(a),u(a))_{\mathfrak h_a}$ are measurable for all 
$n\in\overline{\mathbb N}$. If $u,v\in X_0$ is it clear that $a\mapsto(u(a),v(a))_{\mathfrak h_a}$ is 
also measurable. Two functions $u,v\in X_0$ are equivalent if they agree $\mu$-almost everywhere. 
The collection of equivalence classes of elements of $X_0$ such that 
$\|u\|^2\equiv\int\|u(a)\|^2_{\mathfrak h_a}\,\d\mu(a)<\infty$ is a separable Hilbert space with the inner 
product $(u,v)=\int(u(a),v(a))_{\mathfrak h_a}\,\d\mu(a)$.  This space is independent of choice of the
family $(e_n(a))_{n\in[1:n(a)]}$, up to an isomorphism. We call it the direct integral of the family  
$(\mathfrak h_a)_{a\in\mathbb R}$ and we denote it by 
\begin{equation}
\int^\oplus\mathfrak h_a\d\mu(a).
\label{directint}
\end{equation}
The spaces $\mathfrak h_a$ are called the fibers of this space. If one assumes that
$\mathfrak h^{k}\equiv\ell^2([1:k])$, the Hilbert space of dimension $k$, and 
$\Delta_k\equiv\{a\,|\,\dim\mathfrak h_a=k\}$ for $k\in\overline{\mathbb N}$, one can show that the 
space (\ref{directint}) is isomorphic to the space 
$$
\bigoplus_{k\in\overline{\mathbb N}}L^2(\Delta_k,\d\mu)\otimes\mathfrak h^k.
$$
If $t(a)\in\B(\mathfrak h_a)$ for $\mu$-almost all  $a\in\mathbb R$ with 
$C\equiv\mu-\mathrm{esssup}_{a\in\mathbb R}\|t(a)\|<\infty$ and if $(u(a),t(a)v(a))$ is measurable 
for all measurable functions $u,v$, we say that $t(\cdot)$ is a $\mu$-measurable family of bounded 
operators. In this case, $(Tu)(a)\equiv t(a)u(a)$ defines a bounded operator $T$ on the Hilbert 
space (\ref{directint}) and $\|T\|=C$.
We refer to Chapter 7 of \cite{BS} for more details.

If $A$ is a self-adjoint operator on the separable Hilbert space $\cH$, then there exists a measure 
$\mu$, a measurable family of Hilbert spaces $(\mathfrak h_a)_{a\in\mathbb R}$ and a unitary map
$$
U:\mathcal H\to\int^\oplus\mathfrak h_a\d\mu(a),
$$
such that
$$
\Dom(A)=\{u\in\cH\,|\,\int a^2\|(Uu)(a)\|^2_{\mathfrak h_a}\,\d\mu(a)<\infty\},
$$
and, for all  $u\in\Dom(A)$, $(UAu)(a)=a(Uu)(a)$ for $\mu$-almost all $a\in\mathbb R$.

If the spectrum of $A$ is pure point, then the measure $\mu$ is purely atomic. Its atoms are the 
eigenvalues $a$ of $A$ and the fibers $\mathfrak h_a$ are the corresponding eigenspaces of $A$. 
If the spectrum of $A$ is purely absolutely continuous, one can choose $\mu$ to be the Lebesgue 
measure. In this case the set 
$$
\{u\in\cH\,|\,\essup_{a\in\rr}\|(Uu)(a)\|_{\mathfrak h_a}<\infty\},
$$
is a dense subspace of $\cH$. This applies in particular to the operators 
$A_\pp=A|_{\mathcal H_\pp(A)}$ and $A_\ac=A|_{\mathcal H_\ac(A)}$.

If $B\in\B(\mathcal H)$ commutes with $A$, there exists a $\mu$-measurable family $b(\cdot)$ of
bounded operators such that $(UBf)(a)=b(a)(Uf)(a)$ for $\mu$-almost all $a\in\mathbb R$.

\bigskip {\bf\noindent The Helffer-Sj\"ostrand Formula.}
For sufficiently smooth functions $f$, it is possible to give an explicit representation of the operator 
$f(A)$. Multiple constructions of this type exist. We will mainly use the Helffer-Sj\"ostrand formula, 
which is well adapted to the case where $f\in S(\mathbb R)$ where 
$$
S(\mathbb R)\equiv\{f\in C^\infty(\mathbb R)\,|\,\sup_{x\in\mathbb R}
\langle x\rangle^{\beta+n}|\partial_x^nf(x)|<\infty
\ \text{for some }\beta>0\text{ and all }n\ge0\},
$$
(with $\langle x\rangle\equiv(1+x^2)^{1/2}$) and in particular for $f\in C_0^\infty(\mathbb R)$, the set 
of infinitely differentiable functions which vanish outside of a compact set. We denote by $\supp\,f$ the 
support of such a function, that is to say the smallest closed set $F\subset\mathbb R$ such that 
$f=0$ on $\mathbb R\setminus F$.

For $f\in C^\infty(\mathbb R)$ and $n\in\mathbb N$, let $\tilde f:\mathbb C\to\mathbb C$ be defined 
by
\begin{equation}
\tilde f(x+\i y)\equiv\chi(y\langle x\rangle^{-1})\sum_{j=0}^{n+1}f^{(j)}(x)\frac{(\i y)^j}{j!},
\label{quasianalyticextension}
\end{equation}
where $\chi\in C_0^\infty(]-1,1[)$ is such that  $\chi(y)=1$ in a neighborhood of $y=0$. We remark 
that, apart from the factor $\chi$, (\ref{quasianalyticextension}) is a formal Taylor expansion of order 
$n$ about the point $x$ of the function $f(x+\i y)$. For functions of $z=x+\i y$ we will use the 
notation from complex analysis $\partial=(\partial_x-\i\partial_y)/2$,
$\bar\partial=(\partial_x+\i\partial_y)/2$ and $\d z=\d x+\i\d y$, $\d\bar z=\d x-\i\d y$. A simple 
calculation yields
$$
\bar\partial\partial^j\tilde f(x)=0,\qquad \partial^j\tilde f(x)=f^{(j)}(x),
$$
for all $x\in\mathbb R$ and $j\in\{0,\ldots,n\}$, and this is why $\tilde f$ is called an
almost-analytic extension of $f$ of order $n$. One easily shows that:
\begin{enumerate}[(i)]
\item There exists a constant $C$ (which depends only on $n$) such that 
\begin{equation}
\int|(\bar\partial\tilde f)(x+\i y)|\,|y|^{-1-j}\,\d y\le C\sum_{k=0}^{n+2}\langle x\rangle^{k-1-j}
|f^{(k)}(x)|,
\label{ftildeestimate}
\end{equation}
for  $j\in\{0,\ldots,n\}$.
\item If $f\in C_0^\infty(\mathbb R)$ then
$\tilde f\in C_0^\infty(\mathbb C)$ and 
$$
\supp\tilde f\subset\{z=x+\i y\,|\,x\in\supp\,f,|y|\le\langle x\rangle\}.
$$
\end{enumerate}
Moreover, the functional calculus implies that  $\|(x+\i y-A)^{-1}\|\le |y|^{-1}$. Using these properties
and starting with Stone's formula \eqref{stoneform} an integration by parts shows that
\begin{equation}
\frac{1}{j!}f^{(j)}(A)
=-\frac{1}{\pi}\int_{\mathbb C}\bar\partial\tilde f(x+\i y)(x+\i y-A)^{-1-j}\,\d x\d y
=\frac{1}{2\pi\i}\int_{\mathbb C}\bar\partial\tilde f(z)(z-A)^{-1-j}\d z\wedge\d \bar z,
\label{HSFormula}
\end{equation}
for $j\in\{0,\ldots,n\}$ and  $f\in C_0^\infty(\mathbb R)$
(see \cite{HS} and \cite{D4}, Section 2.2 for a direct approach to spectral theory from the 
Helffer-Sj\"ostrand formula). An approximation argument further shows that (\ref{HSFormula}) remains
valid if $f\in C^{n+2}(\mathbb R)$ is such that
$$
\int\langle x\rangle^{k-1}|f^{(k)}(x)|\,\d x<\infty,
$$
for  $k\in\{0,\ldots,n+2\}$ and in particular if $f\in S(\mathbb R)$.

\subsubsection{Compact operators}
An operator $C\in\B(\mathcal H)$ is compact if ${\{Cu\,|\,u\in\mathcal H,\|u\|=1\}}^\mathrm{cl}$
is a compact subset of $\mathcal H$. The set $\mathcal L^\infty(\mathcal H)$ of all compact operators 
on $\mathcal H$ is a closed two-sided $\ast$-ideal of the $C^\ast$-algebra $\cB(\mathcal H)$
(see Section \ref{CstarSection}).

Let $A$ be a self-adjoint operator on $\mathcal H$. An operator $B$ on the same
Hilbert space is called $A$-bounded (resp.\;$A$-compact) if $\Dom(A)\subset\Dom(B)$ and
there exists $z_0\in\Res(A)$ such that $B(z_0-A)^{-1}$ is bounded (resp.\;compact). In this case,
it follows from the first resolvent identity that $B(z-A)^{-1}$ is bounded (resp.\;compact) for all 
$z\in\Res(A)$. Weyl's theorem asserts that if $B$ is symmetric and $A$-compact then 
$A+B$ is self-adjoint on $\Dom(A)$ and $\spec_{\rm ess}(A+B)=\spec_{\rm ess}(A)$.

In the remaining of this subsection, we shall assume that $\mathcal H$ is separable. An operator
on $\cH$ is finite rank if $\Ran(A)$ is finite dimensional. The set $\mathcal L_\mathrm{fin}(\mathcal H)$
of all finite rank operators on $\cH$ is a $\ast$-subalgebra of $\cB(\cH)$ and is dense in 
$\mathcal L^\infty(\mathcal H)$ (in the norm topology of $\cB(\cH)$). This leads to the result that if 
$C\in\mathcal L^\infty(\mathcal H)$ and
$\wlim_\alpha u_\alpha=u$ then $\lim_\alpha Cu_\alpha=Cu$.

If $A\in\mathcal L^\infty(\mathcal H)$ is self-adjoint, then 
$\spec_\mathrm{cont}(A)=\spec_\ac(A)\cup\spec_\mathrm{sc}(A)$ is empty. Furthermore, 
$\spec_\pp(A)$ is at most countable and can only accumulate at $0$. For all 
$a\in\spec_\pp(A)\setminus\{0\}$, $\Ker(A-a)$ is finite dimensional. We can therefore deduce 
that there exists a set $N$, which is at most countable, such that  
$A=\sum_{n\in N}a_n u_n(u_n,\,\cdot\,)$ where $\{a_n\,|\,n\in N\}=\spec_\pp(A)\setminus\{0\}$ and  
$(u_n)_{n\in N}$ is an orthonormal family of eigenvectors $Au_n=a_nu_n$. More generally, 
if $A\in\mathcal L^\infty(\mathcal H)$, it follows from the polar decomposition $A=J|A|$ that
$$
A=\sum_{n\in N(A)}\kappa_n(A)\, v_n(u_n,\,\cdot\,).
$$
The numbers $\kappa_n(A)>0$ are called singular values of $A$. Their squares $\kappa_n(A)^2$ are
eigenvalues of the positive compact operator $A^\ast A$. The $u_n$ form an orthonormal family
of eigenvectors $A^\ast Au_n=\kappa_n(A)^2u_n$ while the $v_n=Ju_n$ form an orthonormal family
of eigenvectors of $AA^\ast$, $AA^\ast v_n=\kappa_n(A)^2v_n$.

A simple but very convenient compactness criterion on the Hilbert space $L^2(\rr^n)$ is due to Rellich.
Let $F$ and $G$ be two measurable functions on $\rr^n$ with the following property:
for any $K>0$ there exists $R>0$ such that $|F(x)|>K$ and $|G(x)|>K$ for almost every
$x\in\rr^n$ with $|x|>R$. Denote by $F$ and $G$ the operators of multiplication by the
corresponding functions on $L^2(\rr^n)$ and let $\mathcal{F}:L^2(\rr^n)\to L^2(\rr^n)$ denote
the Fourier transform. If $C$ is a bounded operator such that
$FC$ and $G\mathcal{F}C$ are bounded then $C$ is compact.

For $1\le p<\infty$, the von Neumann-Schatten class
$$
\mathcal L^p(\mathcal H)\equiv\left\{A\in\mathcal L^\infty(\mathcal H)\,\bigg|\,
\|A\|_p\equiv\left(\sum_{n\in N(A)}\kappa_n(A)^p\right)^{1/p}<\infty\right\},
$$
is a two-sided $\ast$-ideal of $\B(\mathcal H)$ and a Banach space equipped with the norm 
$\|\,\cdot\,\|_p$. For all $C\in\mathcal L^p(\mathcal H)$ and $B\in\B(\mathcal H)$, 
$\|BC\|_p\le\|B\|\,\|C\|_p$.  We will mainly focus on the space $\mathcal L^1(\mathcal H)$, the 
elements of which are called trace class operators on $\mathcal H$. For all 
$A\in\mathcal L^1(\mathcal H)$ and for any orthonormal basis $(u_i)_{i\in I}$ of $\mathcal H$, 
the series $\sum_{i\in I}(u_i,Au_i)$ is absolutely convergent. Furthermore, its sum is independent 
of the choice of basis, and we call this sum the trace of $A$, denoting it by $\tr(A)$. One clearly has
$$
\tr(A)=\sum_{a\in\spec(A)}a\,\dim\Ker(A-a).
$$
Moreover, the following inequality holds
\beq
|\tr(A)|\le\sum_{n\in N(a)}\kappa_n(A)
=\tr(|A|) = \|A\|_1,
\label{traceineq}
\eeq
for all $A\in\mathcal L^1(\mathcal H)$. More generally, $A\in\mathcal L^p(\mathcal H)$ if and only if 
$|A|^p\in\mathcal L^1(\mathcal H)$ and $\|A\|_p=\tr(|A|^p)^{1/p}$. If $\dim\mathcal H<\infty$ then
 $\mathcal L^p(\mathcal H)=\B(\mathcal H)$ for all $1\le p\le\infty$ and in this case it is a 
well known fact that the trace is cyclic, that is to say that $\tr(AB)=\tr(BA)$ for all 
$A,B\in\B(\mathcal H)$. In the infinite dimensional case, the cyclic property of the trace holds 
when one of the operators involved is trace class: if $A\in\mathcal L^1(\mathcal H)$ and 
$B\in\B(\mathcal H)$ then
$$
\tr(AB)=\tr(BA).
$$
If $A\in\mathcal L^1(\mathcal H)$, it follows from the estimate \eqref{traceineq} that the infinite product 
$$
\det(I+A)=\prod_{a\in\spec(A)}(1+a)^{\dim\Ker(A-a)},
$$ 
is convergent and satisfies
$$
|\det(1+A)|\le\e^{\|A\|_1}.
$$

Let $1\le p,q\le\infty$ be such that  $p^{-1}+q^{-1}=1$. If $A\in\mathcal L^p(\mathcal H)$ and 
$B\in\mathcal L^q(\mathcal H)$ then  $AB\in \mathcal L^1(\mathcal H)$ and the  H\"{o}lder inequality 
$\|AB\|_1\le\|A\|_p\,\|B\|_q$ holds. If $1<p\le\infty$, the topological dual of 
$\mathcal L^p(\mathcal H)$ is $\mathcal L^q(\mathcal H)$. The dual of  $\mathcal L^1(\mathcal H)$ 
is  $\B(\mathcal H)$. The Banach space  $\mathcal L^p(\mathcal H)$ is thus reflexive if $1<p<\infty$, 
but not if $p=1$ or $p=\infty$. In all cases the duality is given by $\langle A,B\rangle\mapsto\tr(AB)$.

Finally, we note that if $A_\alpha$ is a bounded net in  $\B(\mathcal H)$ such that 
$\slim_\alpha A_\alpha=A\in\B(\mathcal H)$ and  $B\in\mathcal L^p(\mathcal H)$ then 
$\lim_\alpha A_\alpha B=AB$ holds in $\mathcal L^p(\mathcal H)$.

\subsubsection{Unitary groups and scattering theory}
\label{firstscattering}
If $H$ is self-adjoint, the functional calculus shows that $U(t)\equiv\e^{\i tH}$, $t\in\rr$, 
defines a family of operators on $\mathcal H$ such that 
\begin{enumerate}[(i)]
\item $U(t)$ is unitary.
\item $U(0)=I$.
\item $U(t)U(s)=U(t+s)$.
\item For all $u\in\mathcal H$, $t\mapsto U(t)u$ is a continuous function from $\mathbb R$ to 
$\mathcal H$. 
\end{enumerate}
We call such a family $\{U(t)\,|\,t\in\mathbb R\}$ a strongly continuous unitary group. Stone's theorem 
states the converse; namely that if $\{U(t)\,|\,t\in\mathbb R\}$ is a strongly continuous unitary group 
on $\mathcal H$, then there exists a self-adjoint operator $H$ such that $U(t) = \e^{\i tH}$. 
Furthermore, 
$$
Hu=\lim_{t\to0}\frac{U(t)u-u}{\i t},
$$
$\Dom(H)$ being the subspace of all $u\in\cH$ such that the above limit exists.

Let $H$ be a self-adjoint operator on $\cH$. The ``core theorem'' states that if 
$\mathcal{D}\subset\Dom(H)$ is a dense subspace of $\cH$ such that 
$\e^{\i tH}\mathcal{D}\subset\mathcal{D}$ for all $t\in\rr$, then it is a core for $H$.
A special instance of such a $\cal D$ is the set $C^\omega(H)$ of vectors $u$ with
the property that the continuous function $u(t)\equiv\e^{\i tH}u$ has an entire analytic extension 
$\mathbb C\ni z\mapsto u(z)\in\cH$. The elements of the dense subspace $C^\omega(H)$
are called entire vectors of the group $\e^{\i tH}$.

If $u\in\cH_\ac(H)$, it follows from Riemann-Lebesgue's lemma that
$$
\wlim_{|t|\to\infty}\e^{\i tH}u=0.
$$
The density of $\cL_{\rm fin}(\cH)$ in $\cL^\infty(\cH)$ allows to conclude that if $C$ is a
compact operator then
\begin{equation}
\lim_{|t|\to\infty}C\e^{\i tH}P_\ac(H)u=0,
\label{RLlemma}
\end{equation}
for all $u\in\cH$.

Unitary groups play a central role in quantum dynamics. In fact, they provide the solution to the 
Cauchy problem for Schr\"{o}dinger's time dependent equation
$$
\i\partial_tu_t=Hu_t,
$$
in the form $u_t = \e^{-\i tH}u_0$. 
The dynamical properties of solutions to this equation
depend on the spectral properties of the generator $H$, the Hamiltonian of the system.
The unitary groups $U(t)= \e^{-\i tH}$ is called propagator of the system.

In this section, we review a few classical results of scattering theory in the Hilbert 
space framework (see \cite{DG,RS3,Y} for more details). We will return to the subject in more 
detail in Section \ref{ScattSect}.

Consider two strongly continuous unitary groups: $\e^{-\i tH_0}$ representing the free dynamics of the
system and $\e^{-\i tH}$ a perturbation of this free dynamics. We say that the state $u\in\mathcal H$ 
is asymptotically free as $t\to\pm\infty$ if there exists $u_\pm\in\mathcal H$ such that 
\begin{equation}
\lim_{t\to\pm\infty}\|\e^{-\i tH}u-\e^{-\i tH_0}u_\pm\|=0.
\label{Asymptote}
\end{equation}
$u_-$ ($u_+$) is the incoming (outgoing) asymptote of $u$. The condition (\ref{Asymptote}) is clearly
equivalent to any of the two following ones
\beq
\lim_{t\to\pm\infty}\|\e^{\i tH_0}\e^{-\i tH}u-u_\pm\|=0,\qquad
\lim_{t\to\pm\infty}\|\e^{\i tH}\e^{-\i tH_0}u_\pm-u\|=0.
\label{wawe-equiv}
\eeq
The fundamental problems of scattering theory are: (i) to determine the set of asymptotically free 
states, i.e., the set of $u\in\cH$ for which the limits
$$
u_\pm=\lim_{t\to\pm\infty}\e^{\i tH_0}\e^{-\i tH}u,
$$
exist; (ii) the construction of a scattering operator which maps the incoming asymptote $u_-$ 
into the corresponding outgoing one $u_+$. 

We remark that if $u$ is an eigenvector of $H$, then the above limits can only exist if $u$ is also 
an eigenvector of $H_0$ with the same eigenvalue. Since the eigenvectors of $H$ have a
particularly simple time evolution under the group $\e^{-\i tH}$ (they are stationary states), 
it is natural to restrict our attention to the subspace $\cH_\pp(H)^\perp$. This motivates the following
definition of asymptotic completeness.

\begin{Def} Let $H_0$ and $H$ be two self-adjoint operators on the Hilbert space $\mathcal H$.
\begin{enumerate}[{\rm (1)}]
\item The M\o ller operators  $\Omega^{\pm}(H,H_0)$ exist if the limits 
\beq
\O^\pm(H,H_0)u=\lim_{t\rightarrow\pm\infty}\e^{\i tH}\e^{-\i tH_0}P_{ac}(H_0)u, 
\label{Mollerdef}
\eeq
exist for all $u\in\mathcal H$. In this case, $\O^{\pm}(H,H_0)$ are clearly bounded operators on 
$\mathcal H$.
\item The M\o ller operators $\Omega^\pm(H,H_0)$  are complete if
$$
\Ran\Omega^+(H,H_0)=\Ran\Omega^-(H,H_0)=\cH_\ac(H).
$$
\item They are asymptotically complete if
$$
\Ran\Omega^+(H,H_0)=\Ran\Omega^-(H,H_0)=\mathcal{H}_\pp(H)^{\perp}.
$$
\end{enumerate}
\end{Def}

The logic behind these definitions is the following. If $\O^{\pm}(H,H_0)$ exist, then they are partial 
isometries with initial space $\cH_\ac(H_0)$ and final space $\Ran(\O^{\pm}(H,H_0))$. Since obviously 
$$
\e^{\i t H}\O^{\pm}(H,H_0)=\O^{\pm}(H,H_0)\e^{\i tH_0},
$$ 
one easily concludes that
$\Ran(\O^{\pm}(H,H_0))$ reduces $H$, that $\O^{\pm}(H,H_0)\Dom(H_0)\subset\Dom(H)$ and that 
the intertwining relation $H\O^{\pm}(H,H_0)u=\O^{\pm}(H,H_0)H_0u$ holds for all $u\in\Dom(H_0)$. 
Thus the part of $H$ in $\Ran(\O^{\pm}(H,H_0))$ is unitarily equivalent to $H_{0,\ac}$ and hence
$\Ran(\O^{\pm}(H,H_0))\subset\cH_\ac(H)$. If $\O^{\pm}(H,H_0)$ are complete, then they are unitary 
as maps from $\cH_\ac(H_0)$ to $\cH_\ac(H)$ and it follows from the equivalence of the two 
relations \eqref{wawe-equiv} that
$$
\O^\pm(H,H_0)^\ast u=\lim_{t\to\pm\infty}\e^{\i tH_0}\e^{-\i tH}P_\ac(H)u=\O^\pm(H_0,H)u,
$$
i.e., the M\o ller operators $\O^\pm(H_0,H)$ also exist and are adjoints to $\O^\pm(H,H_0)$.
Thus any $u\in\cH_\ac(H)$ has incoming/outgoing asymptotes $u_\pm=\O^\pm(H_0,H)u$.
The scattering operator $S:u_-\mapsto u_+$ is given by
$$
S=\O^+(H_0,H)\O^-(H,H_0)=\O^+(H,H_0)^\ast\O^-(H,H_0),
$$ 
and is unitary on $\cH_\ac(H_0)$. Finally, if in addition $H$ has empty singular continuous spectrum 
then asymptotic completeness holds and the set of asymptotically free states is 
$\cH_\ac(H)=\cH_\pp(H)^\perp$.

The basic method for showing the existence of the M\o ller operators $\Omega^\pm(H,H_0)$ is due to 
Cook. It is based on the fact that if a function $f$ is differentiable and if $f'\in L^1(\rr)$, then
$$
\lim_{t\to\pm\infty}f(t)=f(0)\pm\int_0^\infty f'(\pm  t)\,\d t.
$$
We thus have that
$$
\Omega^\pm(H,H_0)u=u\pm\i\int_0^\infty\e^{\pm\i tH}(H-H_0)\e^{\mp\i tH_0}u\,\d t,
$$
if $\|(H-H_0)\e^{\mp\i tH_0}u\|$ is integrable. This representation is the starting point of many
techniques used in scattering theory. In particular, if one can decompose $H-H_0=\sum_jB_j^\ast A_j$,
then the Cook representation can we rewritten as
$$
(v,\Omega^\pm(H,H_0)u)=(v,u)\pm\i\sum_j
\int_0^\infty(B_j\e^{\mp\i tH}v,A_j\e^{\mp\i tH_0}u)\,\d t,
$$
and the Cauchy-Schwarz inequality naturally leads to Kato's definition of smooth perturbation.
A closed operator $A$ is called $H$-smooth if there exists a constant $C$ such that
$$
\int_{-\infty}^\infty\|A\e^{\i tH}u\|^2\,\d t\le C\|u\|^2,
$$
for all $u\in\cH$. Smoothness is easily localized w.r.t.\;the spectrum of $H$:
$A$ is said to be $H$-smooth on the measurable subset
$\Delta\subset\rr$ if the operator $A1_\Delta(H)$ is $H$-smooth. If $\Dom(H)\subset\Dom(A)$ and
\beq
\sup_{\Re(z)\in\Delta,\Im(z)\not=0}\|A(H-z)^{-1}A^\ast\|<\infty,
\label{locsmoothcrit}
\eeq
then $A$ is $H$-smooth on $\Delta^{\rm cl}$.

\subsection{$C^\ast$-Algebras}
\label{CstarSection}

In statistical mechanics it is often useful, and sometimes necessary, to consider infinitely extended
systems with an infinite number of (classical) degrees of freedom. This is commonly referred to as the 
{\sl thermodynamic limit.} This is the case, for example, for the construction of nonequilibrium steady
states (NESS): in a confined system with a finite number of degrees of freedom there is no dissipative 
mechanism which would allow it to approach a steady state. In more technical terms, the spectrum 
of the Hamiltonian $H$ of a confined system is purely discrete and hence its propagator $\e^{\i tH}$
is an almost-periodic function of time which implies that the dynamics is recurrent.

In quantum mechanics, the structure of the algebra of observables of a system with a finite number of
degrees of freedom essentially determines the Hilbert space in which these observables are 
represented by operators (this is the content of the Stone-von Neumann theorem, see theorem 
VIII.14 in \cite{RS1}). This is the main reason why one generally ignores the algebraic structure of 
observables in such 
systems, and instead focuses attention on describing the associated Hilbert space. The situation is 
completely different when one considers systems with an infinite number of degrees of freedom. 
Such systems allow for many inequivalent representations and as such, it is necessary to precisely 
describe the 
algebra of observables. The mathematical framework necessary for implementing such an algebraic 
approach to quantum mechanics are operator algebras. Among the different operator algebras, 
$C^\ast$-algebras are particularly well suited for the fermionic systems in which we are interested. 
In this section, we introduce the basic concepts of the theory of  $C^\ast$-algebras and their 
representations. This material is treated in detail in \cite{BR1,BR2}.

\subsubsection{Definition and examples}
\label{CStarsubsect}

\begin{Def}
\begin{enumerate}[(i)]
\item A $\ast$-algebra $\mathcal A$ is a  complex algebra equipped with an involution 
$A\mapsto A^\ast$ such that 
$$ 
(A+B)^\ast=A^\ast+B^\ast,\quad (\l A)^\ast=\bar{\l}A^\ast,\quad (AB)^\ast=B^\ast A^\ast,
$$
for all $A,B\in\mathcal A$ and $\lambda\in\C$.

\item A Banach algebra  $\mathcal B$ is a complex algebra such that the underlying vector space
is a Banach space with a norm satisfying  
$$
\Vert AB\Vert\leq\Vert A\Vert\,\Vert B\Vert, 
$$
for all $A,B\in\B$. 

\item A $B^\ast$-algebra $\B$ is a Banach algebra as well as a $\ast$-algebra such that 
$\Vert A^{\ast}\Vert=\Vert A\Vert$ for all $A\in\B$.

\item A $ C^\ast$-algebra $\mathcal C$ is a $B^\ast$-algebra with a norm satisfying the 
$C^\ast$-property $$\Vert A^\ast A\Vert=\Vert A\Vert^2,$$  for all $A\in\mathcal C$.
\end{enumerate}
\end{Def}

\noindent{\bf Examples of $C^{\ast}$-algebras}

\noindent 1. $\mathcal{A}=\mathcal{B}(\mathcal{H})$, the algebra of bounded operators on a Hilbert 
space $\mathcal{H}$. In this case, the involution is the operation of adjunction, and the norm is the 
usual operator norm $\Vert A \Vert=\sup \{ \Vert A  \psi\Vert\,|\, \psi\in\mathcal{H},\Vert \psi\Vert=1 \}$. 
To verify the $C^\ast$ property of the norm, note that
$$
    \Vert A \Vert^2  =  \sup_{\|\psi\|=1}(A\psi,A\psi)=\sup_{\|\psi\|=1}(\psi,A^\ast A\psi)
     \leq  \Vert A^\ast A\Vert\leq  \Vert A^\ast\Vert \Vert A\Vert=\Vert A \Vert^2.
$$

\noindent 2. $\mathcal{A}=\mathcal{L}^\infty(\mathcal{H})$, the algebra of compact operators on a 
Hilbert space $\mathcal{H}$, is a $C^\ast$-subalgebra of $\mathcal{B}(\mathcal{H})$ (and a closed
two-sided ideal of the latter).

\noindent 3. $\mathcal{A}=C_\infty(X)$, the algebra of continuous functions on a locally compact 
space $X$ which vanish at infinity, that is to say the set of all continuous functions $f:X\to\C$
such that, for any $\epsilon>0$, there exists a compact $K\subset X$ with $|f(x)|<\epsilon$ for all 
$x\in X\setminus K$. The involution in 
this case is complex conjugation and the norm is  $\Vert f\Vert =\sup_{x\in X}|f(x)|$. Let $\mu$ be 
a regular Borel measure on $X$ such that $\mu(O)>0$ for every open $O\subset X$. By identifying 
$f\in C_\infty(X)$ with the operator of multiplication by $f$ in the Hilbert space 
$\mathcal H=L^2(X,\d\mu)$, the algebra $C_\infty(X)$ can be viewed as a commutative 
$C^\ast$-subalgebra of $\B(\mathcal H)$.

\noindent 4. A subset $\cal S$ of a $\ast$-algebra is called self-adjoint if $A\in\cal S$ implies
$A^\ast\in\cal S$. Thus, a subalgebra of a $\ast$-algebra is a $\ast$-algebra if and only if
it is self-adjoint. It follows that a subalgebra of a $B^\ast$-algebra (resp.\; $C^\ast$-algebra)
is itself a $B^\ast$-algebra (resp.\; $C^\ast$-algebra) if and only if it is closed and self-adjoint.

\bigskip
Example 1 is in some sense the most general. More precisely, any $C^\ast$-algebra is isomorphic to 
a subalgebra of $\B(\mathcal H)$ for some $\cH$. A unit in a $C^\ast$-algebra $\mathcal A$ 
is a unit for the product operation of $\mathcal A$. Such and element, $\mathds{1}$, if it exists, is 
unique and satisfies $\mathds{1}^\ast =\mathds{1}$. However, a $C^\ast$-algebra does not necessarily
contain a unit. For example, the algebra $C_\infty(X)$ has a unit if and only if $X$ is compact and the
algebra $\cL^\infty(\cH)$ of all compact operators on the Hilbert space $\cH$ has a unit if and only
if $\cH$ is finite dimensional. 
The absence of a unit can complicate the structural analysis of $\mathcal{A}$. One can avoid such
complications by embedding $\mathcal{A}$ into a larger $C^\ast$-algebra $\tilde{\mathcal{A}}$ which 
contains a unit. The following result describes the canonical construction of this extension.
\begin{Prop}
  Let $\mathcal{A}$ be a $C^\ast$-algebra without a unit and $\tilde{\mathcal{A}}=\{\langle\alpha,A\rangle\,|\,\alpha\in\C,A\in\mathcal{A} \}$ equipped with the operations
  $\langle\alpha,A\rangle+\langle\beta,B\rangle=\langle\alpha+\beta,A+B\rangle$,
  $\langle\alpha,A\rangle\langle\beta,B\rangle=\langle\alpha\beta,\alpha B+\beta A+AB\rangle$,
  $\langle\alpha,A\rangle^\ast=\langle\bar{\alpha},A^\ast\rangle$. It follows that the function
$$ 
\Vert\langle\alpha,A\rangle\Vert=\sup\{\Vert\alpha B+AB\Vert, B\in\mathcal{A},\Vert B\Vert=1\},
$$
is a $C^\ast$-algebra norm. The algebra $\mathcal{A}$ is identified with the $C^\ast$-subalgebra of 
$\tilde{\mathcal{A}}$ formed by the pairs $\langle0,A\rangle$ and the
element $\langle1,0\rangle$ is a unit of $\tilde{\mathcal{A}}$.
\end{Prop}
The majority of $C^\ast$-algebras that appear in quantum physics are naturally equipped with a
unit. In the following we will assume, without explicit mention, that all the $C^\ast$-algebras contain a unit $\mathds{1}$.

A $\ast$-morphism between two $\ast$-algebras $\mathcal{A}$ and $\mathcal{B}$ is a mapping 
$\phi:\mathcal{A}\to\mathcal{B}$ which satisfies
\begin{enumerate}[(i)]
\item $\phi(\alpha A+\beta B)=\alpha\phi(A)+\beta\phi(B)$,
\item $\phi(AB)=\phi(A)\phi(B)$,
\item $\phi(A^\ast)=\phi(A)^\ast$,
\end{enumerate} 
for all $A, B\in\mathcal{A}$, $\alpha, \beta\in\C$. A bijective $\ast$-morphism is called a 
$\ast$-isomorphism. A $\ast$-isomorphism from $\mathcal A$ onto itself is called a 
$\ast$-automorphism.

\subsubsection{Spectral theory} 
An element $A$ of a $C^\ast$-algebra $\mathcal A$ is invertible if there exists an element 
$A^{-1}\in\mathcal{A}$ such that
$$ 
A^{-1}A=AA^{-1}=\mathds{1}. 
$$
These elements form a group (w.r.t.\;the product operation of $\mathcal A$), called the group of units 
of $\mathcal A$. We call
$$
\Res(A)\equiv\{z\in\C\,|\, (z\mathds{1}-A) \ \text{is \ invertible}\},
$$
the resolvent set of $A$ and
$$ \spec (A)\equiv\C\setminus\Res(A),$$ 
the spectrum of $A$. If $\mathcal C\subset\mathcal A$ is a $C^\ast$-subalgebra and $C\in\mathcal C$, 
the spectrum of $C$, when regarded as an element of $A$, coincides with its spectrum when it is 
regarded as an element of $\mathcal C$. In particular, if $\mathcal A$ is a $C^\ast$-subalgebra of
$\B(\mathcal H)$, the notions of resolvent set and spectrum coincide with those introduced in 
Section \ref{OperaSubSect}. For all $A\in\mathcal A$ we have
\begin{enumerate}[(i)]
\item $\spec(A^\ast)=\overline{\spec(A)}$.
\item $\spec(A^{-1})=\spec(A)^{-1}$.
\item $\spec(P(A))=P(\spec(A))$ for any polynomial $P$.
\item $\spec(AB)\union\{0\}=\spec(BA)\union\{0\}$ for all $B\in\mathcal{A}$.
\end{enumerate}

If $|z|>\Vert A\Vert$ then the series
$$ 
\frac{1}{z}\sum_{n\in\mathbb N}\left(\frac{A}{z}\right)^n, 
$$
is norm convergent. Its sum is $(z\mathds{1}-A)^{-1}$, which implies that 
$\spec(A)\subset\{z\in\C\,|\,|z|\le\|A\|\}$. Also, if $A\in\mathcal A$ is invertible, and if 
$\|B-A\|\,\|A^{-1}\|<1$ then $B$ is invertible, and the series
$$
B^{-1}=\sum_{n\in\mathbb N}A^{-1}\left((B-A)A^{-1}\right)^n,
$$
converges in norm. The group of units of $\mathcal A$ is thus open in $\mathcal A$ and the mapping 
$A\mapsto A^{-1}$ is continuous. In particular, if $z_0\in\Res(A)$, then
$$
\{z\in\C\,|\,|z-z_0|<||(z_0\mathds{1}-A)^{-1}||^{-1}\}\subset\Res(A),
$$
and the series
$$
(z\mathds{1}-A)^{-1} =
\sum_{n\in\mathbb N}
(z_0-z)^n(z_0\mathds{1}-A)^{-n-1}, 
$$
converges in norm. We can deduce that:
\begin{enumerate}[(i)]
\item $\Res(A)$ is open;
\item the mapping $z\mapsto (z\mathds{1}-A)^{-1}$  is analytic on $\Res(A)$;
\item $\spec (A) $ is compact.
\end{enumerate}
We call
$$
r(A)\equiv\sup \{|\lambda|\,|\, \lambda\in\spec (A)\},
$$
the spectral radius of $A$. We have already noted that $r(A)\le\|A\|$. We also have that
$$
r(A)=\lim_n||A^n||^{1/n}=\inf_n||A^n||^{1/n}.
$$
An element $A$ of a $C^\ast$-algebra $\mathcal A$ is 
\begin{enumerate}[(i)]
\item normal if $ A^\ast A=AA^\ast$;
\item self-adjoint if $A=A^\ast$;
\item positive if $A=A^\ast$ and $\spec(A)\subset[0,\infty[$;
\item isometric if $A^\ast A=\mathds{1}$;
\item unitary if $A^\ast A=AA^\ast=\mathds{1}$.
\end{enumerate}
If $A$ is normal, then $r(A)=||A||$. If $A$ is self-adjoint, then $\spec(A)\subset [-||A||,||A||]$. 
If $A$ is isometric, then $r(A)=1$, and if it is unitary, then $\spec(A)\subset\{z\in\C\,|\,|z|=1 \}$. 
If $A$ is positive, we write $A\ge 0$.
By writing $A\geq B$ when $A-B\ge 0$ we introduce a partial order on $\mathcal A$.
For the self-adjoint elements of $\mathcal A$, the spectral theorem from Section \ref{SAOpSect} can
be formulated as follows.
\begin{Theo}
Let $A$ be a self-adjoint element of the $C^\ast$-algebra $\mathcal A$ and $C(\spec(A))$ denote the 
$C^\ast$-algebra of continuous functions on $\spec(A)$. There exists a unique $\ast$-morphism
\begin{eqnarray*}
C(\spec(A)) & \to & \mathcal{A} \\
f & \mapsto & f(A),
\end{eqnarray*}
that sends the function $1$ to $\mathds{1}$ and the function $\mathrm{Id}_{\spec(A)}$ to $A$. 
Furthermore, we have that $$ \spec(f(A))=f(\spec(A)),$$ for all $f\in C(\spec(A))$.
\end{Theo}

Applying this result to the functions $f_\pm(x)=(|x|\pm x)/2$, we obtain that any self-adjoint
$A\in\cA$ can be written as $A=A_+-A_-$ where $A_\pm=f_\pm(A)\in\cA$ are both positive. 
Since any $A\in\cA$ can be
written as $A=X+\i Y$ where both $X=(A+A^\ast)/2$ and $Y=(A-A^\ast)/2\i$ are self-adjoint elements
of $\cA$, we conclude that any $A\in\cA$ is a linear combination of $4$ positive elements of $\cA$.

\subsubsection{Representations and states}
\label{Sect-RepState}

In this section we discuss two key concepts of the theory of $C^\ast$-algebras: representations and 
states.

\paragraph{Representations.}
A $\ast$-morphism $\phi$ between two $C^\ast$-algebras preserves positivity. If $A\geq 0$, we have that 
$A=B^\ast B$ for some operator $B$ and thus
$$
\phi(A)=\phi(B^\ast B)=\phi(B)^\ast\phi(B)\geq 0.
$$
$\phi$ is also continuous and satisfies $||\phi(A)||\leq ||A||$ for all $A\in\mathcal A$. $\phi$ is injective 
if and only if one of the following conditions is satisfied.
\begin{enumerate}[(i)]
\item $\Ker\phi=\{0\}$, 
\item $||\phi(A)||=||A||$ for all $A\in\mathcal{A}$, 
\item $A>0$ implies $\phi(A)>0$ for all $A\in\mathcal{A}$.
\end{enumerate} 
In particular, every $\ast$-automorphism of a $C^\ast$-algebra is isometric.

\begin{Def}
A representation of a $C^\ast$-algebra $\mathcal A$ is a pair   $\langle\mathcal{H},\Pi\rangle$ where
$\mathcal H$ is a Hilbert space and  $\Pi:\mathcal{A}\to\mathcal{B}(\mathcal{H})$ is a
$\ast$-morphism. A representation is called faithful if $\,\Pi$ is injective.
\end{Def}

Let $\langle\mathcal{H},\Pi\rangle $  be a representation of a $C^\ast$-algebra $\mathcal A$ and let 
$\mathcal{H}_1\subset\mathcal H$ be a closed, $\Pi$-invariant subspace, that is to say that
$\Pi(A)\mathcal{H}_1\subset\mathcal{H}_1$ for all $A\in\mathcal A$. Let $P_1$ be the orthogonal
projection onto $\mathcal H_1$. For all $A\in\mathcal A$, we have that $\Pi(A)P_{1}=P_{1}\Pi(A)P_{1}$,
and by taking the adjoint, $P_1\Pi(A)=P_{1}\Pi(A)P_{1}$. We can then deduce that 
$\Pi(A)P_{1}=P_{1}\Pi(A)$, i.e. $P_1$ commutes with $\Pi(\mathcal A)$. Conversely, if an 
orthogonal projection commutes with $\Pi(\mathcal A)$, then its range is $\Pi$-invariant. This is the 
case of $P_2=I-P_1$, from which we deduce that $\mathcal H_2\equiv\mathcal H_1^\perp$ is 
$\Pi$-invariant. By writing $\Pi_i(A)=\Pi(A)|_{\mathcal H_i}$, we obtain two representations 
$\langle\mathcal H_i,\Pi_i\rangle $ of $\mathcal A$ and the decomposition 
$$ 
\langle\mathcal{H},\Pi\rangle =\langle\mathcal{H}_1,\Pi_1\rangle 
\oplus \langle\mathcal{H}_2,\Pi_2\rangle .
$$
More generally, for each orthogonal decomposition $\mathcal H=\oplus_\alpha\mathcal H_\alpha$ 
into $\Pi$-invariant subspaces, we associate the decomposition $\Pi=\oplus_\alpha\Pi_\alpha$.

A representation of a $C^\ast$-algebra is called trivial when $\Pi=0$. A representation 
$\langle\mathcal H,\Pi\rangle $ can be non-trivial but still have a trivial part $\mathcal{H}_0$ defined by
$$
\mathcal{H}_0\equiv\{ \psi\in\mathcal H\,|\,\Pi(A)\psi=0,\ \ \forall A\in\mathcal{A}\}.
$$
A representation is called non-degenerate if $\mathcal{H}_0 = \{0\}$.

Two representations $\langle\mathcal{H}_1,\Pi_1\rangle $ and $\langle\mathcal{H}_2,\Pi_2\rangle $ 
are called equivalent if there exists a unitary $U:\mathcal H_1\to\mathcal H_2$ such that 
$U\Pi_1(A)=\Pi_2(A)U$  for all $A\in\mathcal A$.

In the next subsection, we will investigate the concept of a state, which plays an important role in 
the construction of representations.

\paragraph{States.}
A linear functional $\omega$ on a $\ast$-algebra $\mathcal A$ is positive if $\omega(A^\ast A)\geq 0$ 
for all $A\in\mathcal A$. In this case, $\langle A,B\rangle\mapsto\omega(A^\ast B)$ is a positive
Hermitian form on $\mathcal A\times\mathcal A$. We deduce that 
$\omega(A^\ast B)=\overline{\omega(B^\ast A)}$ and the Cauchy-Schwarz inequality  
$|\omega(A^\ast B)|^2\leq \omega(A^\ast A)\omega(B^\ast B)$ holds for all $A,B\in\mathcal A$. In particular, 
if $\mathcal A$ has a unit $\mathds{1}$, then  $\omega(A^\ast)=\overline{\omega(A)}$ and 
$|\omega(A)|^2\leq \omega(A^\ast A)\omega(\mathds{1})$.

A linear functional $\omega$ on a $C^\ast$-algebra $\mathcal A$ is positive if and only if it is continuous 
and  $||\omega||=\omega(\mathds{1})$. If $\omega$ is a positive linear functional on the
 $C^\ast$-algebra $\mathcal A$ and $A\in\mathcal A$ then $\omega_A(B)\equiv\omega(A^\ast BA)$ 
defines a positive linear functional on $\mathcal A$ and
$|\omega(A^\ast BA)|\leq \omega(A^\ast A) ||B||$ for all $A,B\in\mathcal A$.

A state on a $C^\ast$-algebra $\mathcal A$ is a positive linear functional normalized by the condition
$||\omega|| =\omega(\mathds{1})=1$. The set $E(\mathcal A)$ of all states on $\mathcal A$ is clearly
convex. If $\mathcal{A}$ contains a unit then $E(\mathcal A)$ is a weakly-$\ast$ compact subset of 
the topological dual of $\mathcal A$.

We recall that a point $x$ of a convex set $K$ is extremal whenever $x=\l a+(1-\l)b$ with
$a,b\in K$ and $\l\in]0,1[$ implies $a=b=x$, i.e., $x$ cannot be decomposed in a convex
combination of other points of $K$. The extremal points of $E(\mathcal A)$ are called pure 
states.

\paragraph{\noindent Cyclic representations.}
Let $\langle\mathcal H,\Pi\rangle$ be a representation of the $C^\ast$-algebra $\mathcal A$. A vector 
$\Omega\in\mathcal H$ is called cyclic for $\Pi$ if the subspace $\Pi(\mathcal A)\Omega$ is dense in
$\mathcal H$. The representation $\langle\mathcal H,\Pi\rangle$ is cyclic if it admits a cyclic vector. 
A cyclic representation is non-degenerate. Conversely, every non-degenerate representation 
$\langle\mathcal H,\Pi\rangle$ can be decomposed into cyclic representations 
$\langle\mathcal H,\Pi\rangle =\oplus_\alpha\langle\mathcal H_\alpha,\Pi_\alpha\rangle$.

If $\langle\mathcal H,\Pi\rangle$ is non-degenerate and $\Omega\in\mathcal{H}$ is a unit vector, 
then the formula
$$ 
\omega_{\Omega}(A)=(\Omega,\Pi(A)\Omega),
$$
defines a state $\omega_{\Omega}$ on $\mathcal A$. The following theorem shows that all states 
on $\mathcal A$ are of this form.

\begin{Theo}[Gelfand, Naimark, Segal]
Let $\omega$ be a state on the $C^\ast$-algebra $\mathcal A$. There exists a cyclic representation 
$\langle\mathcal{H}_{\omega},\Pi_{\omega},\Omega_{\omega}\rangle$ of $\mathcal{A}$ such that 
for all $A\in\mathcal A$:
$$ 
\omega(A)=(\Omega_{\omega},\Pi_{\omega}(A)\Omega_{\omega}),
$$
with $||\Omega_{\omega}||^2=||\omega||=1$. Furthermore, this representation is unique up to a 
unitary transformation.
\end{Theo}

We call $\langle\mathcal{H}_{\omega},\Pi_{\omega},\Omega_{\omega}\rangle$ the GNS 
representation, or the canonical cyclic representation of $\mathcal A$ associated with $\omega$. 
It plays a very important role in quantum mechanics because it allows us to associate a Hilbert space 
framework to a state and to distinguish an important class of states associated 
with $\omega$. A trace class operator $\rho$ such that $\rho\ge 0$ and $\tr\rho=1$ is called a 
density matrix. To each density matrix $\rho\in\mathcal L^1(\mathcal H_\omega)$ we may associate 
a state $\omega_\rho\in E(\mathcal A)$ defined by
$$
\omega_\rho(A)=\tr_{\mathcal H_\omega}(\rho\Pi_\omega(A)).
$$
Such a state is called $\omega$-normal. We denote by $\mathcal N_\omega(\mathcal A)$ the 
subset of $E(\mathcal A$) formed by all $\omega$-normal states. We note that the set of vector
states $\omega_\Psi(A)=(\Psi,\Pi_\omega(A)\Psi)$ associated to unit vectors $\Psi\in\cH_\omega$ 
is dense in $\mathcal N_\omega(\mathcal A)$.

%
%

\subsubsection{$C^\ast$-dynamics}

A $C^\ast$-dynamics on a $C^\ast$-algebra $\mathcal A$ is a strongly continuous group 
$\mathbb R\ni t\mapsto\tau^t$ of $\ast$-auto\-mor\-phisms of $\mathcal A$, that is to say that for all 
$t\in\mathbb R$, $\tau^t$ is a $\ast$-automorphism of $\mathcal A$, such that for all $A\in\mathcal A$
 the mapping $t\mapsto\tau^t(A)$ is continuous, $\tau^0=\mathrm{Id}$, and for all $t,s\in\mathbb R$, 
$\tau^t\circ\tau^s=\tau^{t+s}$. The general theory of strongly continuous groups on Banach spaces 
shows that for all $A$ in a dense subspace $\Dom(\delta)\subset\mathcal A$, the limit
$$
\delta(A)=\lim_{t\to 0}\frac{\tau^t(A)-A}{t},
$$
exists and defines a closed operator on $\mathcal A$. The algebraic structure of $\mathcal A$ 
implies that $\Dom(\delta)$ is a $\ast$-subalgebra of $\mathcal A$ and that
$$
\delta(AB)=\delta(A)B+A\delta(B),\qquad\delta(A^\ast)=\delta(A)^\ast,\qquad\delta(\mathds{1})=0,
$$ 
for all $A,B\in\Dom(\delta)$. Such an operator $\delta$ is called a $\ast$-derivation on $\mathcal A$.

If $\tau^t$ is a $C^\ast$-dynamics on the $C^\ast$-algebra $\mathcal A$, there exists a dense 
$\ast$-subalgebra $\mathcal A_\tau\subset\mathcal A$ such that for all $A\in\mathcal A_\tau$, 
the function  $t\mapsto\tau^t(A)$  has an entire analytic extension $\mathbb C\ni z\mapsto\tau^z(A)$. 
The elements of $\mathcal A_\tau$ are called $\tau$-entire.

\bigskip {\bf\noindent Example.} 
Let $\mathcal H$ be a Hilbert space and $H$ a bounded self-adjoint operator on $\mathcal H$. Then 
$\t^t(A)=\e^{\i tH}A\e^{-\i tH}$ defines a $C^\ast$-dynamics on $\cal B(H)$. Its generator 
$\delta(A)=\i[H,A]$ is bounded. We note that the boundedness of $H$ is necessary to ensure the
strong continuity of $\t^t$. This fact is one major obstacle to the use of $C^\ast$-algebras in quantum 
mechanics. However, we shall see in Section \ref{SecondQsect} that for fermionic systems it is 
possible to define a $C^\ast$-dynamics even in cases where the Hamiltonian is unbounded.
For bosonic systems, it is generally preferable to work with von Neumann algebras which
avoid this problem.

\bigskip
{\noindent\bf $C^\ast$-dynamical systems} 

A $C^\ast$-dynamical system is a pair $\langle\cal A,\t\rangle$ where $\cal A$ is a $C^\ast$-algebra and 
$\t$ is a $C^\ast$-dynamics on $\mathcal A$. In the context of quantum mechanics, the elements of 
$\cal A$ describe physical observables and the group $\t$ specifies their time evolution in the Heisenberg picture, $A_t = \tau^t(A)$. A state $\omega\in E(\mathcal A)$ 
associates to each observable $A\in\mathcal A$ a number $\omega(A)$ which may be interpreted as 
the quantum mechanical expectation value of the observable $A$. It is thus natural to identify elements
of $E(\mathcal A)$ with the physical states of quantum mechanics. 
Since $\omega(A_t)=\omega(\tau^t(A))=\omega\circ\tau^t(A)$, 
the evolution of quantum states in the Schr\"{o}dinger picture 
is given by $\omega_t=\omega\circ\tau^t$. A 
state $\omega$ is $\t$-invariant if $\omega\circ\t^t=\omega$ for all $t\in\mathbb R$. The set 
$E(\mathcal A,\t)\subset E(\mathcal A)$ of all $\t$-invariant states is never empty. It is the set 
of all steady states of the system. A state $\omega\in E(\cal A,\t)$ is called $\t$-ergodic if 
\beq
\lim_{T\to\infty}\frac{1}{T}\int_{0}^{T}\nu\circ\t^{\pm t}(A) \,\d t=\omega(A),
\label{ergodef}
\eeq
for all $A\in\mathcal A$, $\nu\in\mathcal N_\omega(\mathcal A)$. It is called $\t$-mixing if 
\beq
\lim_{t\to\infty}\nu\circ\t^{\pm t}(A)=\omega(A),
\label{mixdef}
\eeq
for all $A\in\mathcal A$, $\nu\in\mathcal N_\omega(\mathcal A)$.

\subsubsection{KMS states}
\label{KMSSubSect}
We saw in the previous section that $C^\ast$-dynamical systems provide a mathematical framework for
quantum mechanics. In this section, we describe how to characterize thermal equilibrium states in 
the language of $C^\ast$-algebras. We shall content ourselves with a very elementary introduction
to the KMS condition. The interested reader should consult Chapter 5 of \cite{BR2} for a more detailed
discussion as well as \cite{HHW,BF} for a deeper insight into the algebraic structure induced by 
equilibrium states.

We consider a $N$-level quantum system described by a Hamiltonian $H$ on the $N$-dimensional
Hilbert space $\mathcal H$. The associated $C^\ast$-dynamical system is
$\langle\B(\mathcal H),\tau\rangle$, where $\tau^t(A)=\e^{\i tH}A\e^{-\i tH}$.
The Gibbs-Boltzmann Ansatz for the canonical ensemble a temperature $T$ is the 
density matrix
$$
\rho_{\b}=Z^{-1}_{\b}\e^{-\b H},
$$
where $\beta=1/k_BT$, $k_B$ being the Boltzmann constant. The normalization factor $Z_{\b}=\tr(\e^{-\b H})$ is the canonical partition function. 
The state $\omega\in E(\B(\mathcal H))$ corresponding to this density matrix is
$$
\omega(A)=Z^{-1}_{\b}\tr(\e^{-\b H}A)=\frac{\tr(\e^{-\b H}A)}{\tr(\e^{-\b H})}.
$$
The equilibrium correlation function
\begin{equation}
\label{KMS1}
F(A,B;t)\equiv\omega(A\t^t(B)),
\end{equation}
is an analytic function of $t\in\mathbb C$. The cyclic property of the trace yields the following 
identity
$$
\tr(\e^{-\b H}A\e^{\i tH}B\e^{-\i tH})=\tr(\e^{-\beta H}\e^{\i (t-\i\beta)H} B\e^{-\i (t-\i\b)H}A),
$$
from which we conclude that
\begin{equation}
\label{KMS2}
F(A,B;t+\i\b)=\omega(\t^t(B)A). 
\end{equation}
The relations (\ref{KMS1}) and (\ref{KMS2}) represent the values of the analytic function 
$F(A,B;z)$ along the boundary of the strip $S_{\b}\equiv\{z\in\C\,|\,0<\mathrm{Im}z<\b\}$. 
These relations are called the Kubo-Martin-Schwinger (KMS) boundary conditions.

\begin{Def}\label{KMSdef}
Let $\langle\cal A,\t\rangle$ be a $C^\ast$-dynamical system. $\omega\in E(\mathcal A)$ is a 
$\langle\t,\b\rangle$-KMS state for a $\b\ge0$ if, for all $A,B\in\cal A$, there exists a function 
$F(A,B;z)$, analytic on the strip $S_{\b}$, bounded and continuous on its closure 
and satisfying the KMS boundary conditions (\ref{KMS1}) and (\ref{KMS2}).
\end{Def}
{\bf\noindent Remarks.}
\\ 1. A $\langle\tau,0\rangle$-KMS state is also called tracial since it satisfies the cyclicity condition 
$\omega(AB)=\omega(BA)$.
\\ 2. For $\beta>0$, a $\langle\t,\b\rangle$-KMS state represents the thermodynamic state of a 
system at temperature $T=(k_B\b)^{-1}$ where $k_B$ is the Boltzmann constant.
\\ 3. $\langle\t,\b\rangle$-KMS state for $\beta<0$ can be defined by a straightforward modification
of Definition~\ref{KMSdef}. If $\omega$ is $\langle\t^t,\b\rangle$-KMS and $\gamma\not=0$ then 
$\omega$ is also $\langle\t^{\gamma t},\b/\gamma\rangle$-KMS. However, there is no simple 
relation between two KMS states at different temperatures for the same dynamics $\t^t$.
\\ 4. The abstract definition of a KMS state masks its fundamental property: Every 
$\langle\t,\b\rangle$-KMS state with $\beta>0$ is $\t$-invariant.
\\5. In practice, it suffices to check the KMS boundary condition on a large enough subalgebra of 
$\mathcal A$: let $\mathcal C$ be a dense $\ast$-subalgebra of the $\ast$-algebra
$\mathcal A_\tau$ of $\tau$-entire elements. If $\omega(A\tau^{\i\beta}(B))=\omega(BA)$
holds for all $A,B\in\mathcal{C}$ then $\omega$ is a $\langle\tau,\beta\rangle$-KMS state.
\\ 6. The set of all $\langle\t,\b\rangle$-KMS states on $\mathcal A$ is obviously a convex subset of 
$E(\mathcal{A})$. Moreover, this subset is weakly-$\ast$ closed. If $\mathcal A$ contains a unit then 
any  $\langle\t,\b\rangle$-KMS state is a convex combination of extremal
$\langle\t,\b\rangle$-KMS states. These extremal $\langle\t,\b\rangle$-KMS states are interpreted 
in statistical mechanics as pure thermodynamical phases (see Theorem 5.3.30 in \cite{BR2} for 
a more precise formulation of this property).

\bigskip 
The following example shows that KMS states of quantum systems with a finite number of 
degrees of freedom coincide with the usual notion of equilibrium states from statistical mechanics.

\paragraph{Example: Finite quantum systems.}
Let $H$ be a self-adjoint operator on the finite-dimen\-sional Hilbert space $\cal{H}$. We consider the 
$C^\ast$-dynamical system $\langle\mathcal{B}(\mathcal{H}),\t\rangle$ where
$\tau^t(A)=\e^{\i tH}A\e^{-\i tH}$. Any state on $\B(\mathcal H)$ is of the form $\omega(A)=\tr(\rho A)$ 
where $\rho$ is a density matrix on $\cH$. Suppose that the state defined by the density matrix 
$\rho$ is $\langle\t,\b\rangle$-KMS. The relations (\ref{KMS1}), (\ref{KMS2}) applied to $A\equiv\phi(\psi,\,\cdot\,)$ 
give
$$
(\psi,\t^{\i\b}(B)\rho\phi)=(\psi,\rho B\phi).
$$
Since this holds for all $\psi,\phi\in\cH$ we must have $\e^{-\b B}B\e^{\b H}\rho=\rho B$ for all 
$B\in\B(\mathcal H)$. Setting $B\equiv \phi(\psi,\,\cdot\,)$ we further obtain
$$
\rho\phi=\frac{(\e^{\b H}\psi,\rho\psi)}{(\psi,\psi)}\e^{-\b H}\phi,
$$
from which we conclude that $\rho=Z_\b^{-1}\e^{-\b H}$. Thus the $C^\ast$-dynamical system 
$\langle\mathcal{B}(\mathcal{H}),\t\rangle$ admits a unique $\langle\tau,\beta\rangle$-KMS state and this
state is given by the Gibbs-Boltzmann formula.

\subsubsection{Perturbation theory}
\label{sect-perturb}
Time-dependent perturbation theory is an essential tool in the construction of quantum dynamical 
systems. As far as $C^\ast$-dynamics are concerned, it is fairly elementary application of well
known techniques. However, the discussion of the KMS states of perturbed dynamical system
due to Araki \cite{A3} (see also \cite{BR2,DJP}) which we summarize in this section is more subtle
and has important application in equilibrium as well as non-equilibrium statistical mechanics.

Le $\langle \cA,\tau\rangle$ be a $C^\ast$-dynamical system and denote by $\delta$ the 
$\ast$-derivation generating the group $\tau$. For any self-adjoint $V\in\cA$,
$$
\delta_V=\delta+\i[V,\,\cdot\,],
$$
generates a dynamics $\tau_V^t$ on $\cA$ given by the Schwinger-Dyson expansion
$$
\tau_V^t(A)=\tau^t(A)+\sum_{n=1}^\infty\int\limits_{0\le s_1\le\cdots\le s_n\le t}
\i[\tau^{s_1}(V),\i[\tau^{s_2}(V),\cdots
\i[\tau^{s_n}(V),\tau^t(A)]\cdots]]\,\d s_1\cdots\d s_n,
$$
which converges in the norm of $\cA$ for any $t\in\rr$ and $A\in\cA$.

Let $\Gamma_V^t$ denote the solution of the initial value problem
$$
\frac{\d\ }{\d t}\Gamma_V^t=\i\Gamma_V^t\tau^t(V),\qquad \Gamma_V^0=I,
$$
that is, the time-ordered exponential
$$
\Gamma_V^t=I+\sum_{n=1}^\infty\i^n\int\limits_{0\le s_1\le\cdots\le s_n\le t}
\tau^{s_1}(V)\cdots\tau^{s_n}(V)\,\d s_1\cdots\d s_n,
$$
which, for any $t\in\rr$, converges towards a unitary element of $\cA$. One easily checks that 
\beq
\tau_V^t(A)=\Gamma_V^t\tau^t(A)(\Gamma_V^t)^{-1},
\label{tauVform}
\eeq
and the cocycle relation 
\beq
\Gamma_V^{t+s}=\Gamma_V^t\tau^t(\Gamma_V^s)
\label{Gammacocycle}
\eeq
are satisfied for all $t,s\in\rr$.

If $V\in\cA_\tau$, then the function $t\mapsto\Gamma_V^t$ has an entire analytic extension given by
the convergent expansion
$$
\Gamma_V^z=I+\sum_{n=1}^\infty(\i z)^n\int\limits_{0\le s_1\le\cdots\le s_n\le 1}
\tau^{zs_1}(V)\cdots\tau^{zs_n}(V)\,\d s_1\cdots\d s_n.
$$
It follows that $\Gamma_V^z\in\cA_\tau$ and that relations~\eqref{tauVform} (with $A\in\cA_\tau$) 
and \eqref{Gammacocycle} extend to the complex domain. Moreover, the unitarity relation
$\Gamma_V^t\Gamma_V^{t\ast}=I$ extends to $\Gamma_V^z\Gamma_V^{\bar z\ast}=I$,
i.e., $(\Gamma_V^z)^{-1}=\Gamma_V^{\bar z\ast}$.

If $\omega\in E(\cA)$ is a $\langle\tau,\beta\rangle$-KMS state, a simple calculation shows that
$$
\omega^V(A)=\frac{\omega(A\Gamma_V^{\i\beta})}{\omega(\Gamma_V^{\i\beta})}
=\frac{\omega(\Gamma_V^{\i\beta/2\ast}A\Gamma_V^{\i\beta/2})}{\omega(\Gamma_V^{\i\beta/2\ast}\Gamma_V^{\i\beta/2})},
$$
satisfies the KMS condition $\omega^V(A\tau_V^{\i\beta}(B))=\omega^V(BA)$ for any
$A,B\in\cA_\tau$, and hence $\omega_V$ is a $\langle\tau_V,\beta\rangle$-KMS state.
Let $\langle\cH_\omega,\Pi_\omega,\Omega_\omega\rangle$ denotes the GNS representation of 
$\cA$ induced by $\omega$ and set
$$
\Omega^V=\frac{\Pi_\omega(\Gamma_V^{\i\beta/2})\Omega_\omega}
{\|\Pi_\omega(\Gamma_V^{\i\beta/2})\Omega_\omega\|}.
$$
It follows that $\omega^V(A)=(\Omega^V,\Pi_\omega(A)\Omega^V)$ and in particular that
$\omega^V\in\mathcal{N}_\omega(\cA)$. 

By approximating an arbitrary self-adjoint $V\in\cA$ by a sequence $V_n\in\cA_\tau$
one can show (and this is the delicate point in the analysis) that the sequence $\Omega^{V_n}$ 
converges to a vector $\Omega^V$ such that $\omega^V(A)=(\Omega^V,\Pi_\omega(A)\Omega^V)$
is $\langle\tau_V,\beta\rangle$-KMS. The map $\omega\mapsto\omega^V$ obtained in this way
is a bijection from the set of $\langle\tau,\beta\rangle$-KMS states to the set of 
$\langle\tau_V,\beta\rangle$-KMS states. Moreover, the set $\{\omega^V\,|\,V\in\cA_\tau, V=V^\ast\}$
is dense in the set $\mathcal{N}_\omega(\cA)$ of all $\omega$-normal states.

\subsubsection{Liouvilleans and quantum Koopmanism}

Given a $C^\ast$-dynamical system $\langle\cA,\t\rangle$ and a representation $\Pi:\cA\to\B(\cH)$,
we say that the dynamics is unitarily implemented in the Hilbert space $\cH$ if there exists a 
self-adjoint operator $L$ on $\cH$ such that $\Pi(\tau^t(A))=\e^{\i tL}\Pi(A)\e^{-\i tL}$ holds for all 
$t\in\rr$ and $A\in\cA$. Such an implementation allows to reduce a number of dynamical properties 
of the system $\langle\cA,\t\rangle$ to the spectral analysis of the operator $L$. In this section,
we shall see that the unitary implementation of the dynamics always exists in the GNS representation
associated to an invariant state. Moreover, if this state is modular (see Definition~\ref{modulardef} 
below) then its ergodicity and mixing 
properties (Eq.~\eqref{ergodef} and~\eqref{mixdef}) can be deduced from the spectral properties 
of the operator $L$. This is the quantum version of the Koopman approach to the ergodic theory
of classical dynamical systems (see, e.g., Section VII.4 in~\cite{RS1}).

Let $\langle\cal A,\t\rangle$ be a $C^\ast$-dynamical systems, $\omega\in E(\cA,\t)$ and
$\langle\cH_\omega,\Pi_\omega,\Omega_\omega\rangle$ the induced GNS representation of $\cA$.
Since $\Omega_\omega$ is a cyclic vector and
$\|\Pi_\omega(\tau^t(A))\Omega_\omega\|^2=\omega(\tau^t(A^\ast A))=\omega(A^\ast A)=
\|\Pi_\omega(A)\Omega_\omega\|^2$, the map
$$
\Pi_\omega(\cA)\Omega_\omega\ni\Pi_\omega(A)\Omega_\omega
\mapsto\Pi_\omega(\tau^t(A))\Omega_\omega,
$$
is well defined on a dense subspace of $\cH_\omega$ and extends to an isometry 
$U^t:\cH_\omega\to\cH_\omega$. Observing that $U^{t}U^{s}=U^{t+s}$ and $U^0=I$, we conclude that
$U^t$ is a unitary group. Finally it follows from the estimate
$\|(U^t-I)\Pi_\omega(A)\Omega_\omega\|^2=\omega((\tau^t(A)-A)^\ast(\tau^t(A)-A))\le\|\tau^t(A)-A\|^2$
and the continuity of the map $t\mapsto\tau^t(A)$ that $U^t$ is strongly continuous. By Stone's theorem,
there exists a self-adjoint operator $L_\omega$ on $\cH_\omega$ such that
$U^t=\e^{\i tL_\omega}$. Since 
$\e^{\i tL_\omega}\Omega_\omega=\Pi_\omega(\tau^t(I))\Omega_\omega=\Omega_\omega$
one has $\Omega_\omega\in\Dom(L_\omega)$ and $L_\omega\Omega_\omega=0$.  The identity
$$
\e^{\i tL_\omega}\Pi_\omega(A)\e^{-\i tL_\omega}\Pi_\omega(B)\Omega_\omega
=\e^{\i tL\omega}\Pi_\omega(A\tau^{-t}(B))\Omega_\omega
=\Pi_\omega(\tau^t(A))\Pi_\omega(B)\Omega_\omega,
$$
holds for any $A,B\in\cA$. The cyclicity of $\Omega_\omega$ allows us to conclude that 
$\e^{\i tL_\omega}\Pi_\omega(A)\e^{-\i tL_\omega}=\Pi_\omega(\tau^t(A))$ for all $A\in\cA$.
Thus, $L_\omega$ implements the dynamics in the GNS representation. Let $L$ be a self-adjoint
operator implementing the dynamics $\tau$ on $\cH_\omega$ and such that $L\Omega_\omega=0$.
It follows that
$$
\e^{\i tL}\Pi_\omega(A)\Omega_\omega=\e^{\i tL}\Pi_\omega(A)\e^{-\i tL}\Omega_\omega
=\Pi_\omega(\tau^t(A))\Omega_\omega,
$$
from which we conclude that $L=L_\omega$. Thus, the operator $L_\omega$ is completely and
uniquely determined by the two conditions that it implements the dynamics $\tau$ on $\cH_\omega$ 
and annihilates the cyclic vector $\Omega_\omega$. We shall call $L_\omega$ the Liouvillean of the
dynamical system $\langle\cal A,\t,\omega\rangle$.

\bigskip
\begin{Def}\label{modulardef}
A state $\omega\in E(\cA)$ is called modular if there exist a dynamics $\tau$ and a $\beta\not=0$ such
that $\omega$ is $\langle\tau,\beta\rangle$-KMS.
\end{Def}
By Remark 3 of Section~\ref{KMSSubSect}, we can assume that $\beta=1$. The dynamics 
$\sigma_\omega$ such that $\omega$ is $\langle\sigma_\omega,1\rangle$-KMS is called the 
modular group of $\omega$ (the convention $\beta=-1$ has been used in the mathematical 
literature since the works of Tomita and Takesaki \cite{To,Ta}. This convention is however immaterial). 
The GNS representation induced by a modular state has a rich structure which
was unveiled by Haag, Hugenholtz and Winnink \cite{HHW} (following the pioneering works of
Araki, Woods and Wyss \cite{AW1,AW2} on the representations of canonical (anti-)commutation 
relations), and was fully developed by Tomita and Takesaki to the modular theory of von Neumann 
algebras. We shall only need the following property (see, e.g., \cite{BR1}).

\begin{Prop}
If $\omega$ is a modular state on $\cA$ and $\langle\cH_\omega,\Pi_\omega,\Omega_\omega\rangle$
the induced GNS representation then the subspace
$$
\Pi_\omega(\cA)'\Omega_\omega=\{B\Omega_\omega\,|\,B\in\B(\cH_\omega), [B,\Pi_\omega(A)]=0
\text{ for all }A\in\cA\},
$$
is dense in $\cH_\omega$.
\end{Prop}

Let $\omega\in E(\cA,\t)$ be modular and let $L_\omega$ be the Liouvillean implementing the dynamics
$\t$ in the GNS representation $\langle\cH_\omega,\Pi_\omega,\Omega_\omega\rangle$. For any
unit vector $\Psi\in\Pi_\omega(\cA)'\Omega_\omega$ one has
\begin{align*}
\omega_\Psi(\tau^t(A))&=(\Psi,\Pi_\omega(\t^t(A))\Psi)
=(B\Omega_\omega,\Pi_\omega(\t^t(A))B\Omega_\omega)\\
&=(B^\ast B\Omega_\omega,\Pi_\omega(\t^t(A))\Omega_\omega)
=(B^\ast B\Omega_\omega,\e^{\i tL_\omega}\Pi_\omega(A)\e^{-\i tL_\omega}\Omega_\omega)\\
&=(B^\ast B\Omega_\omega,\e^{\i tL_\omega}\Pi_\omega(A)\Omega_\omega),
\end{align*}
for some $B\in\Pi_\omega(\cA)'$ and it follows that
\beq
\frac1T\int_0^T\omega_\Psi(\tau^t(A))\,\d t
=\frac1T\int_0^T(B^\ast B\Omega_\omega,\e^{\i tL_\omega}\Pi_\omega(A)\Omega_\omega)\,\d t.
\label{ergobzz}
\eeq
To evaluate the limit $T\to\infty$ of the right hand side of this identity, let us define the linear
map
$$
E_T\Phi=\frac1T\int_0^T\e^{\i tL_\omega}\Phi\,\d t,
$$
where, due to the strong continuity of the group $\e^{\i tL_\omega}$, we can take the integral 
in Riemann sense. Since $\|E_T\Phi\|\le\|\Phi\|$, the maps $E_T$ form a uniformly continuous
family. For $\Phi\in\Dom(L_\omega)$ one has
$$
\e^{\i tL_\omega}L_\omega\Phi=-i\frac{\d\ }{\d t}\e^{\i tL_\omega}\Phi,
$$
so that
$$
\lim_{T\to\pm\infty}E_TL_\omega\Phi=\lim_{T\to\pm\infty}\frac\i T(I-\e^{\i TL_\omega})\Phi=0.
$$
The uniform continuity of the maps $E_T$ allows us to conclude that $\lim_{T\to\pm\infty}E_T\Phi=0$
holds for all $\Phi\in\Ran(L_\omega)^{\rm cl}=\Ker(L_\omega)^\perp$. Since $E_T\Phi=\Phi$ for
$\Phi\in\Ker(L_\omega)$, one has
$$
\slim_{T\to\pm\infty}E_T=P_0(L_\omega),
$$
where $P_0(L_\omega)$ denotes the orthogonal projection on $\Ker(L_\omega)$ (this is
von Neumann's mean ergodic theorem).

Going back to the identity~\eqref{ergobzz}, using the fact that $\Omega_\omega\in\Ker(L_\omega)$
to write $P_0(L_\omega)=\Omega_\omega(\Omega_\omega,\,\cdot\,)+Q$ where $Q$ is the 
orthogonal projection on $\Ker(L_\omega)\ominus\cc\Omega_\omega$, we get
\beq
\begin{split}
\lim_{T\to\pm\infty}\frac1T\int_0^T\omega_\Psi(\tau^t(A))\,\d t
&=(B^\ast B\Omega_\omega,\Omega_\omega)(\Omega_\omega,\Pi_\omega(A)\Omega_\omega)
+(B^\ast B\Omega_\omega,Q\Pi_\omega(A)\Omega_\omega)\\
&=(\Psi,\Psi)(\Omega_\omega,\Pi_\omega(A)\Omega_\omega)
+(B^\ast B\Omega_\omega,Q\Pi_\omega(A)\Omega_\omega)\\[4pt]
&=\omega(A)+(B^\ast B\Omega_\omega,Q\Pi_\omega(A)\Omega_\omega).
\end{split}
\label{stillwaiting}
\eeq
If $\omega$ is ergodic, then we must have 
$(B^\ast B\Omega_\omega,Q\Pi_\omega(A)\Omega_\omega)=0$ for all $B\in\Pi_\omega(\cA)'$ and
all $A\in\cA$. The cyclicity of $\Omega_\omega$ implies that $QB^\ast B\Omega_\omega=0$
for all $B\in\Pi_\omega(\cA)'$. One easily checks that $\Pi_\omega(\cA)'$ is a closed self-adjoint 
subalgebra of $\B(\cH_\omega)$, and hence a $C^\ast$-algebra (in fact a von Neumann algebra). 
Since any element of this algebra is a linear combination of positive elements, it follows that
$Q\Psi=0$ for all $\Psi\in\Pi_\omega(\cA)'\Omega_\omega$. This vectors forming a dense
subspace of $\cH_\omega$, one finally concludes that $Q=0$, i.e., 
$\Ker(L_\omega)=\cc\Omega_\omega$. Reciprocally, if $\Ker(L_\omega)=\cc\Omega_\omega$, then
$Q=0$ and we deduce from Eq.~\eqref{stillwaiting} that
$$
\lim_{T\to\pm\infty}\frac1T\int_0^T\omega_\Psi(\tau^t(A))\,\d t=\omega(A),
$$
for all $\Psi\in\Pi_\omega(\cA)'\Omega_\omega$. Since $\|\omega_\Psi-\omega_\Phi\|\le2\|\Psi-\Phi\|$,
this extends by density/continuity to all $\Psi\in\cH_\omega$.  Finally, any density matrix $\rho$ on
$\cH_\omega$ has a spectral decomposition $\rho=\sum_np_n\Phi_n(\Phi_n,\,\cdot\,)$ with
$p_n\in[0,1]$, $\sum_np_n=1$ and $\|\Phi_n\|=1$ so that, by Fubini's theorem
$$
\frac1T\int_0^T\tr_{\cH_\omega}\left(\rho\Pi_\omega(\tau^t(A))\right)\,\d t
=\sum_np_n\frac1T\int_0^T\omega_{\Phi_n}(\tau^t(A))\,\d t.
$$
The dominated convergence theorem allows us to conclude that the right hand side of this identity 
converges to $\omega(A)$ as $T\to\pm\infty$. In conclusion, we have shown that $\omega$ is ergodic
if and only if $\Ker(L_\omega)=\cc\Omega_\omega$.

Invoking similar arguments, one shows that $\omega$ is mixing if and only if
$$
\lim_{t\to\pm\infty}(\Psi,\e^{\i tL_\omega}\Phi)=(\Psi,\Omega_\omega)(\Omega_\omega,\Phi),
$$
holds for $\Psi\in\Pi_\omega(\cA)'\Omega_\omega$ and $\Phi\in\Pi_\omega(\cA)\Omega_\omega$.
The density of these two subspaces of $\cH_\omega$ implies that this is equivalent to
$$
\wlim_{t\to\pm\infty}\e^{\i tL_\omega}=\Omega_\omega(\Omega_\omega,\,\cdot\,).
$$
Moreover, the Riemann-Lebesgue lemma implies that this condition is satisfied provided the spectrum
of $L_\omega$ on the orthogonal complement of $\cc\Omega_\omega$ is purely absolutely continuous.
We have proven the following:

\begin{Prop}\label{Koopman}
Let $\langle\cA,\t\rangle$ be a $C^\ast$-dynamical system and denote by
$\langle\cH_\omega,\Pi_\omega,\Omega_\omega\rangle$ the GNS representation of $\cA$ induced
by the modular state $\omega\in E(\cA,\t)$. Denote by $L_\omega$ the corresponding Liouvillean.
\begin{enumerate}[(i)]
\item $\omega$ is ergodic if and only if $\,\Ker(L_\omega)=\cc\Omega_\omega$.
\item $\omega$ is mixing if and only if $\,\wlim_{t\to\pm\infty}\e^{\i tL_\omega}=\Omega_\omega(\Omega_\omega,\,\cdot\,)$.
\item If, apart from a simple eigenvalue $0$, the spectrum of $L_\omega$ is purely absolutely continuous,
then $\omega$ is mixing.
\end{enumerate}
\end{Prop}

\section{Elements of nonequilibrium quantum statistical mechanics}
\label{MSHEChap}


\subsection{Systems of identical particles}

\subsubsection{Bosons and fermions}

In this section we consider a system of $n$ identical particles. In quantum mechanics each one of 
these particles is described by a separable Hilbert space $\mathfrak h$. The observables associated to 
this particle are the elements of the $C^\ast$-algebra $\B(\mathfrak h)$. The Hilbert space of the entire 
system is the tensor product $\mathfrak h^{\otimes n}=\mathfrak h\otimes\cdots\otimes\mathfrak h$ 
and its observables are the elements of the $C^\ast$-algebra $\B(\mathfrak h^{\otimes n})$.

The natural action of $S_n$, the permutation group of $n$ objects, given by
$$
\pi\cdot(\phi_1\otimes\cdots\otimes\phi_n)=\phi_{\pi^{-1}(1)}\otimes\cdots
\otimes\phi_{\pi^{-1}(n)},
$$
for $\pi\in S_n$, $\phi_1,\ldots,\phi_n\in\mathfrak h$, extends uniquely to a representation 
$\psi\mapsto\pi\cdot\psi$ of $S_n$ in $\mathfrak h^{\otimes n}$. When the $n$ particles are 
indistinguishable, which is the case if they are identical, the only states that can be realized 
physically are those that are invariant under this action. In other words, a vector 
$\psi\in\mathfrak h^{\otimes n}$ represent a physical states of a system of $n$ identical particles iff
$$
\pi\cdot\psi=\l(\pi)\psi,
$$
for all $\pi\in S_n$, where $\l(\pi)$ is a phase factor ($|\lambda(\pi)|=1$). 
Thus, the subspace spanned by $\psi$ carries a one dimensional sub-representation of $S_n$.
They are only two such representations: the trivial ($\l^+(\pi)=1$) and the alternate
($\l^-(\pi)=\varepsilon(\pi)$, the sign of $\pi$). It is an experimental fact (and the so-called spin-statistic
theorem in quantum field theory, see \cite{SW,J,F}) that particles with integer spin (bosons) transform according to the trivial representation while particles with half-integer spin (fermions) transform according to the alternate representation.

To construct the Hilbert space of a system of $n$ bosons/fermions, we introduce the symmetrization
operators
$$
P_n^\pm:\psi\mapsto\frac{1}{n!}\sum_{\pi\in S_n}\lambda^\pm(\pi)\pi\cdot\psi.
$$
One can easily show that these are orthogonal projections and that 
$\pi\cdot P_n^\pm\psi=\lambda^\pm(\pi)P_n^\pm\psi$ for all $\psi\in\mathfrak h^{\otimes n}$, 
$\pi\in S_n$. The Hilbert space of the system of $n$ bosons/fermions is the subspace
$$
\Gamma_n^\pm(\mathfrak h)
\equiv P_n^\pm\mathfrak h^{\otimes n}=\Ran P_n^\pm\subset
\mathfrak h^{\otimes n}.
$$
We note that if $\dim\fh=d<\infty$ then $\dim\Gamma_n^+(\mathfrak h)=\frac{d^n}{n!}$,
$\dim\Gamma_n^-(\mathfrak h)={d\choose n}$ for $n\le d$ and 
$\Gamma_n^-(\mathfrak h)=\{0\}$ for $n>d$.

\subsubsection{Fock space}
It is often more convenient to work with an indefinite number of particles (this is in fact necessary
to describe the grand canonical ensemble). To do so, we set
$\Gamma_0^\pm(\mathfrak h)\equiv\mathbb C$ and we define
$$
\G^\pm(\mathfrak{h})=\bigoplus_{n\in\mathbb N}\Gamma_n^\pm(\mathfrak h).
$$
The vectors of this space are sequences $\Psi=\{\psi_n\}_{n\in\mathbb N}$ with
$\psi_n\in\Gamma_n^\pm(\mathfrak h)$ and such that $\|\Psi\|^2=\sum_n\|\psi_n\|^2<\infty$. 
Equipped with the scalar product
$$
(\Psi,\Phi)\equiv\sum_{n\in\mathbb N}(\psi_n,\phi_n),
$$
$\G^\pm(\mathfrak{h})$ is a Hilbert space which we call bosonic/fermionic Fock space over 
$\mathfrak h$. The state associated to the unit vector 
$\Psi=\{\psi_n\}_{n\in\mathbb N}\in\G^\pm(\mathfrak{h})$ is interpreted in the following manner.
The probability to find $n$ particles in the system is $\|\psi_n\|^2$ (note that
$1=\|\Psi\|^2=\sum_n\|\psi_n\|^2$). The vector $\Omega=\{1,0,0,\ldots\}$ thus describes the state 
in which there are no particles in the system, and is referred to as the {\sl vacuum state.} The vector 
$\Psi=\{0,\ldots,0,\psi_n,0,\ldots\}$ describes a system with $n$ particles that are in 
the state associated to $\psi_n$. In general one can write
$$
\Psi=\{\psi_n\}_{n\in\mathbb N}
=\sum_{n\in\mathbb N}\|\psi_n\|\,\{0,\ldots,0,\psi_n/\|\psi_n\|,0,\ldots\},
$$
which is a coherent superposition of states with a definite number of particles. The construction 
of the Fock space clearly shows that the subspace
$$
\Gamma^\pm_\mathrm{fin}(\mathfrak h)\equiv
\left\{\Psi=\{\psi_n\}_{n\in\mathbb N}\in\Gamma^\pm(\mathfrak h)\,\big|\,\text{the set }\{n\in\mathbb N\,|\,\psi_n\not=0\}
\ \text{is finite}\right\},
$$
is dense in $\Gamma^\pm(\mathfrak h)$.

\subsubsection{Second quantization}
\label{SecondQsect}

For all $f\in\mathfrak h$ we define the {\sl creation operator} of a boson/fermion in the state $f$ by
\begin{eqnarray}
a_\pm^\ast(f):\Gamma_n^\pm(\mathfrak h)&\rightarrow&\Gamma_{n+1}^\pm(\mathfrak h)\label{creatdef}\\
P_n^\pm \phi_1\otimes \phi_2\otimes\cdots\otimes \phi_{n}
 &\mapsto&
\sqrt{n+1}P_{n+1}^\pm \phi_1\otimes \phi_2\otimes\cdots\otimes \phi_{n}\otimes f.\nonumber
\end{eqnarray}
Apart from the factor $\sqrt{n+1}$, its interpretation is clear. This operator is extended by linearity 
to the dense subspace $\Gamma_\mathrm{fin}^\pm(\mathfrak h)$. An elementary calculation 
shows that $\Gamma_\mathrm{fin}^\pm(\mathfrak h)\subset\Dom(a_\pm^\ast(f)^\ast)$, that 
$a_\pm^\ast(f)^\ast\Omega=0$ and that
$$
a_\pm^\ast(f)^\ast P_n^\pm \phi_1\otimes \phi_2\otimes\cdots\otimes \phi_{n}
=\sum_{j=1}^n(\pm1)^j(f,\phi_j)P_{n-1}^\pm\phi_1\otimes\cdots\otimes
\cancel{\phi_j}\otimes\cdots\phi_n.
$$
From this, we deduce that $a_\pm^\ast(f)$ is closable. We shall use the same notation for its 
closure and we note that $a_\pm(f)=a_\pm^\ast(f)^\ast$ is closed as well. For
obvious reason, we call this last operator the {\sl annihilation operator} of a boson/fermion in the 
state $f$. We note that $f\mapsto a_\pm^\ast(f)$ is linear but that $f\mapsto a_\pm(f)$ is anti-linear. 
In the following, $a_\pm^{\sharp}$ will denote either $a_\pm$ or $a_\pm^\ast$.

One can easily verify that for all $f,g\in\mathfrak h$ the relations
\begin{equation}
[a_\pm(f),a_\pm(g)]_\pm=[a_\pm^\ast(f),a_\pm(g)^\ast]_\pm=0,
\quad[a_\pm(f),a_\pm^\ast(g)]_\pm=(f,g),
\stepcounter{equation}\label{CCRCAR}\tag{$\theequation_\pm$}
\end{equation}
holds on $\Gamma_\mathrm{fin}^\pm(\mathfrak h)$. There $[A,B]_\pm=AB\mp BA$
denotes the commutator/anti-commutator of $A$ and $B$. These relations play a fundamental 
role in the quantum mechanics of systems with an infinite number of degrees of freedom. 
They are known as 
the canonical commutation/anti-commutation relations (CCR/CAR). The factor of $\sqrt{n+1}$ from
definition (\ref{creatdef}) is there to ensure the validity of these relations.

In the bosonic case, the operators $a_+^\sharp(f)$ are unbounded. Indeed, assuming $a=a_+(f)$ to
be bounded, the CCR imply $aa^\ast=a^\ast a+\|f\|^2$ and hence
$\|a^\ast\Psi\|^2=\|a\Psi\|^2+\|f\|^2$ for all $\Psi\in\Gamma^+(\mathfrak h)$. From this 
we conclude that $\|a^\ast\|^2=\|a\|^2+\|f\|^2$ which implies $f=0$. In the fermionic case, if $a=a_-(f)$ 
then the CAR imply $a^2=0$ and $a^\ast a\|f\|^2=a^\ast[a,a^\ast]_-a=(a^\ast a)^2$. 
We conclude that $\|a\|^2=\|a^\ast a\|=\|f\|^2$ and hence $\|a\|=\|a^\ast\|=\|f\|$.
We note in particular that $a_-^\ast(f)^2=0$ 
shows that it is impossible to create two fermions in the same state. This fundamental property 
of fermions is known as the Pauli exclusion principle. 

Let $T\in\B(\mathfrak h)$ be a contraction, i.e., $\|T\|\le1$. We define the operator $\Gamma_n(T)$ 
acting  on $\mathfrak h^{\otimes n}$ by the formula $\Gamma_n(T)(\phi_1\otimes\cdots\otimes\phi_n)
=T\phi_1\otimes\cdots\otimes T\phi_n$. Clearly, $\Gamma_n(T)$ is a contraction which commutes with
the projections $P_n^\pm$. Its restriction to $\Gamma_n^\pm(\mathfrak h)$ is thus a contraction as 
well as the operator 
$\Gamma(T)\equiv\oplus_n\Gamma_n(T):\Gamma^\pm(\mathfrak h)
\rightarrow\Gamma^\pm(\mathfrak h)$. We say that $\Gamma(T)$ is the second quantization of $T$. 
We clearly have that $\Gamma(T)^\ast=\Gamma(T^\ast)$ and 
$\Gamma(TS)=\Gamma(T)\Gamma(S)$. It immediately follows that from definition (\ref{creatdef}) 
that $\Gamma(T)a_\pm^\ast(f)=a_\pm^\ast(Tf)\Gamma(T)$. By adjunction we obtain 
$a_\pm(f)\Gamma(T)=\Gamma(T)a_\pm(T^\ast f)$. If $T\in\mathcal L^1(\mathfrak h)$ is a 
contraction and if, in the bosonic case, $\|T\|_1<1$, then an elementary calculation shows that 
$\Gamma(T)\in\mathcal L^1(\Gamma^\pm(\mathfrak h))$ and that we have
$\|\Gamma(T)\|_{\Gamma^\pm(\mathfrak h)}\le\det(I\mp|T|)^{\mp1}$ and 
\beq
\tr_{\Gamma^\pm(\mathfrak h)}(\Gamma(T))=\det(I\mp T)^{\mp1}.
\label{FockTrace}
\eeq
The second quantization of 
unitary operators is a particularly important case. If $U$ is unitary, then so is $\Gamma(U)$ and we 
have $\Gamma(U)a_\pm^\sharp(f)\Gamma(U)^\ast=a_\pm^\sharp(Uf)$. If $U(t)=\e^{\i tH}$ is a 
strongly continuous unitary group then so is its second quantization. Stone's theorem implies the 
existence of a self-adjoint operator $\d\Gamma(H)$ on $\Gamma^\pm(\mathfrak h)$ such that 
$\Gamma(e^{\i tH})=\e^{\i t\d\Gamma(H)}$. The operator $\d\Gamma(H)$ is called the differential 
second quantization of $H$. The dense subspace $\Gamma_\mathrm{fin}^\pm(\Dom(H))$ is a core of 
$\d\Gamma(H)$ and for all $\Psi=\{\psi_n\}_{n\in\mathbb N}\in\Gamma_\mathrm{fin}^\pm(\Dom(H))$, 
we have
$$
(\d\Gamma(H)\Psi)_n=(H\otimes I\otimes\cdots\otimes I+
I\otimes H\otimes\cdots\otimes I+\cdots+I\otimes I\otimes\cdots\otimes H)\psi_n.
$$
If both $H$ and $A$ are self-adjoint then
$$
\Gamma(\e^{\i tH})\Gamma(\e^{\i sA})\Gamma(\e^{-\i tH})
=\Gamma(\e^{\i tH}\e^{\i sA}\e^{-\i tH})
=\Gamma(\e^{\i s\e^{\i tH}A\e^{-\i tH}}),
$$
which implies $\Gamma(\e^{\i tH})\d\Gamma(A)\Gamma(\e^{-\i tH})=d\Gamma(\e^{\i tH}A\e^{-\i tH})$, 
and thus $[\d\Gamma(H),\d\Gamma(A)]=\d\Gamma([H,A])$.

\subsubsection{The $C^\ast$-algebra $\CAR(\mathfrak h)$}

In what follows, we will only consider the fermionic case. To simplify the notation we set
$a^\sharp=a_-^\sharp$.

In the fermionic case, the operators $a^\sharp(f)$ being bounded, we can define 
$\CAR(\mathfrak{h})$ as the $C^\ast$-subalgebra of $\B(\Gamma^-(\mathfrak h))$ generated by 
$\{a^{\sharp}(f)\,|\,f\in\mathfrak{h}\}$. $\CAR(\mathfrak h)=\B(\Gamma^-(\fh))$ iff $\fh$ is finite
dimensional. One can show that up to a $\ast$-isomorphism it is the unique $C^\ast$-algebra 
generated by elements $a(f)$ satisfying the canonical anti-commutation relations ($\theequation_-$). 
If $\mathfrak h$ is infinite dimensional and separable, then 
$\CAR(\mathfrak h)$ is also infinite dimensional and separable. If $(e_n)_{n\in N}$ is an orthonormal 
basis of $\mathfrak h$ and $a_n^\sharp\equiv a^\sharp(e_n)$ then the $\ast$-algebra of all polynomials in 
$a_n^\sharp$ is dense in $\CAR(\mathfrak h)$. By using the anti-commutation relations, one can 
show that $\d\Gamma(f(g,\,\cdot\,))=a^\ast(f)a(g)\in\CAR(\mathfrak h)$ for all $f,g\in\mathfrak h$. 
The canonical decomposition of a compact operator allows us to deduce that
$\d\Gamma(C)\in\CAR(\mathfrak h)$ for all $C\in\mathcal L^1(\mathfrak h)$.

If $U$ (resp. $V$) is a bounded linear (resp. anti-linear) operator on $\mathfrak h$ and if 
$V^\ast U+U^\ast V=UV^\ast+VU^\ast=0$ and $U^\ast U+V^\ast V=UU^\ast+VV^\ast=I$ then there 
exists a unique $\ast$-automorphism $\tau$ of $\CAR(\mathfrak h)$ such that 
$\tau(a(f))=a(Uf)+a^\ast(Vf)$ for all $f\in\mathfrak h$. We call $\tau$ the Bogoliubov automorphism 
associated wit the pair $\langle U,V\rangle$. In particular, if $U$ is unitary and $V=0$ the above 
conditions are satisfied and $\tau(a(f))=a(Uf)$. If $U(t)=\e^{\i tH}$ is a strongly continuous unitary 
group on $\mathfrak h$, we may thus associate to it a group of Bogoliubov automorphisms 
$\tau_H^t(a^\sharp(f))=a^\sharp(\e^{\i tH}f)$. Note that
$\tau_H^t(a^\sharp(f))-a^\sharp(f)=a^\sharp(\e^{\i tH}f-f)$ implies
$\|\tau_H^t(a^\sharp(f))-a^\sharp(f)\|=\|\e^{\i tH}f-f\|$, which shows that
$t\mapsto\tau_H^t(a^\sharp(f))$ is continuous. Since $\tau_H^t$ is a $\ast$-morphism, we infer that 
$t\mapsto\tau_H^t(A)$ is continuous for all polynomials $A$ in $a^\sharp(\,\cdot\,)$. 
Finally, these polynomials being dense in $\CAR(\mathfrak h)$ and $\tau_H^t$ being isometric, 
we conclude that $t\mapsto\tau_H^t(A)$ is continuous for all $A\in\CAR(\mathfrak h)$. 
$\langle\CAR(\mathfrak h),\tau_H\rangle$ is a $C^\ast$-dynamical system.

\subsection{The ideal Fermi gas}
\label{Sect-FermiGas}

The simplest thermodynamic models describe systems of many non-interacting particles, also known as
ideal gases. In this section we discuss the ideal Fermi gas which is main object of interest in this notes.

\subsubsection{The $C^\ast$-dynamical system $\langle\CAR(\mathfrak h),\tau_H\rangle$}

As seen in the previous section, the Hilbert space of a system of $n$  indistinguishable 
fermions with one-particle Hilbert space $\fh$ is the completely anti-symmetric tensor product 
$\Gamma_n^-(\mathfrak h)$. If the fermions are non-interacting then their Hamiltonian
is given by
$$
H_n=H\otimes I\otimes\cdots\otimes I+\cdots+I\otimes I\otimes\cdots\otimes H,
$$
where $H$ denotes the one-particle Hamiltonian. The propagator of the system is thus
$$
U_n(t)=\e^{-\i tH_n}=\Gamma_n(\e^{-\i tH}).
$$
It is now easy to describe the system with an indefinite number of fermions. Its Hilbert space is the
fermionic Fock space $\Gamma^-(\mathfrak h)=\oplus_n\Gamma_n^-(\mathfrak h)$ and its 
propagator is the second quantization $$\Gamma(\e^{-\i tH})=\oplus_n\Gamma_n(\e^{-\i tH}).$$
Therefore, the time evolution of observables of the system, in the Heisenberg picture, is given by  
$A_t=\Gamma(\e^{\i tH})A\Gamma(\e^{-\i tH})$. In particular we have
$\Gamma(\e^{\i tH})a^\sharp(f)\Gamma(\e^{-\i tH})=a^\sharp(\e^{\i tH}f)=\tau_H^t(a^\sharp(f))$. 
We can thus describe an ideal Fermi gas by the $C^\ast$-dynamical system
$\langle\CAR(\mathfrak h),\tau_H^t\rangle$, where $\fh$ is the one-particle Hilbert space and
$H$ the one-particle Hamiltonian.

\subsubsection{Gauge invariance}
Gauge invariance is a fundamental symmetry of quantum mechanics. It arises from the fact that the
density matrix $\rho_\phi=\phi(\phi,\,\cdot\,)$ which describes the state associated to the vector 
$\phi\in\mathfrak h$  is invariant under the gauge transformation $\phi\mapsto\e^{\i\theta}\phi$. 
In other words, the phase of the wave function $\phi$ is not observable. The strongly continuous 
unitary group $\theta\mapsto\e^{\i\theta I}$ on $\mathfrak h$ is called the gauge group. This group is 
isomorphic to the circle $\mathbb R/2\pi\mathbb Z$. Gauge invariance is manifest in the Heisenberg 
picture since observables are invariant under transformations by elements of the gauge group, 
$\e^{\i\theta I}A\e^{-\i\theta I}=A$ for all $A\in\B(\mathfrak h)$. 

To understand the consequences of 
this invariance on the algebraic description of the Fermi gas, we note that the gauge group 
$\e^{\i\theta I}$ induces the unitary group $\Gamma(\e^{\i\theta I})$ in the Fock space 
$\Gamma^-(\mathfrak h)$. The generator of this group is the self-adjoint operator 
$N\equiv\d\Gamma(I)$ which, in the $n$-particle subspace $\Gamma_n^-(\mathfrak h)$, 
acts as multiplication by the number $n$. $N$ is aptly called the {\sl number operator} and gauge 
invariance in Fock space expresses the fact that the total number of particles is conserved. The 
observables of a Fermi gas must be invariant under the action of the gauge group. On the
$C^\ast$-algebra $\CAR(\mathfrak h)$, this action is described by the Bogoliubov group
$$
\vartheta^{\theta}(A)=\Gamma(\e^{\i\theta I})A\Gamma(\e^{-\i\theta I})
=\e^{\i\theta N}A\e^{-\i\theta N},
$$
which clearly commutes with the dynamical group $\tau_H^t$. We note that the linearity/anti-linea\-rity
properties of $f\mapsto a^\sharp(f)$ imply $\vartheta^\theta(a^\ast(f))=\e^{\i\theta}a^\ast(f)$ and 
$\vartheta^\theta(a(f))=\e^{-\i\theta}a(f)$. A monomial $A$ in $a^\sharp$ containing $n$ factors of 
$a^\ast$ and $m$ factors of $a$ transforms as $\vartheta^\theta(A)=\e^{\i\theta(n-m)}A$. Thus, $A$
is invariant under $\vartheta$ iff $n=m$, i.e. iff $A$ preserves the number of fermions. It is 
evident that a polynomial in $a^\sharp$ is invariant under $\vartheta$ if and only if all its monomials 
terms are invariant themselves. We conclude from this that the gauge invariant $C^\ast$-subalgebra
$$
\CAR_\vartheta(\mathfrak h)\equiv\{A\in\CAR(\mathfrak h)\,|\,\vartheta^\theta(A)=A\ 
\text{for all}\ \theta\in\mathbb R\},
$$
is the $C^\ast$-algebra generated by $\{a^\ast(f)a(g)\,|\,f,g\in\mathfrak h\}\cup\{I\}$.  The
$C^\ast$-dynamical system describing an ideal Fermi gas is thus 
$\langle\CAR_\vartheta(\mathfrak h),\tau_H\rangle$. It is however more convenient to work with 
the system $\langle\CAR(\mathfrak h),\tau_H\rangle$. To do this, one introduces the following notion.

A state $\omega\in E(\CAR(\mathfrak h))$ is gauge invariant if it is $\vartheta$-invariant, i.e.,
$\omega\circ\vartheta^\theta=\omega$ for all $\theta\in\rr$.

We then note that $\CAR(\mathfrak h)=\oplus_{k\in\mathbb Z}\CAR_\vartheta^k(\mathfrak h)$ where
$$
\CAR_\vartheta^k(\mathfrak h)
\equiv\{A\in\CAR(\mathfrak h)\,|\,\vartheta^\theta(A)=\e^{\i k\theta}A\ 
\text{for all}\ \theta\in\mathbb R\},
$$
are $\vartheta$-covariant subspaces. Evidently, we have that 
$\CAR_\vartheta^m(\mathfrak h)^\ast=\CAR_\vartheta^{-m}(\mathfrak h)$ and 
$$\CAR_\vartheta^n(\mathfrak h)\CAR_\vartheta^m(\mathfrak h)=\CAR_\vartheta^{n+m}(\mathfrak h).$$
If $\omega\in E(\CAR(\mathfrak h))$ is gauge invariant one can conclude that 
$\omega|_{\CAR_\vartheta^k(\mathfrak h)}=0$ for $k\not=0$ while 
$\omega|_{\CAR_\vartheta(\mathfrak h)}$ is a state on $\CAR_\vartheta(\mathfrak h)$. Conversely, a state 
$\omega\in E(\CAR_\vartheta(\mathfrak h))$ extends uniquely to a gauge invariant state on 
$\CAR(\mathfrak h)$ by setting $\omega(\oplus_{k\in\mathbb Z} A_k)=\omega(A_0)$. There thus 
exists a bijection between the states of $\CAR_\vartheta(\mathfrak h)$ and the gauge invariant 
states of $\CAR(\mathfrak h)$. The dynamical systems 
$\langle\CAR_\vartheta(\mathfrak h),\tau_H\rangle$ and $\langle\CAR(\mathfrak h),\tau_H\rangle$ 
are clearly equivalent if we restrict the latter one to its gauge invariant states.

\subsubsection{$\langle\tau_H,\beta\rangle$-KMS states on $\CAR_\vartheta(\mathfrak h)$}
\label{QFKMSsubsect}

The first issue which arises naturally after the discussion of the previous section is to characterize 
the gauge invariant states on $\CAR(\mathfrak h)$ which correspond to the $\langle\tau_H,\beta\rangle$-KMS 
states on $\CAR_\vartheta(\mathfrak h)$. This problem was solved in a very general setting by 
Araki (see Section 5.4.3 in \cite{BR2}). If $\omega$ is an extremal $\langle\tau_H,\beta\rangle$-KMS state on 
$\CAR_\vartheta(\mathfrak h)$ with $\beta>0$ there exists $\mu\in\mathbb R$ such that $\omega$ 
is a $\langle\gamma_\mu,\beta\rangle$-KMS state on $\CAR(\mathfrak h)$, where the $C^\ast$-dynamics
$\gamma_\mu$ is defined by $\gamma_{\mu}^t\equiv\tau^t\circ\vartheta^{-\mu t}$. Such a state 
represents the thermal equilibrium of an ideal Fermi gas at inverse temperature $\b$ and at 
chemical potential $\mu$ ({\sl c.f.} Section \ref{TDLimit} below). We note that
$\gamma_\mu$ is the Bogoliubov group associated with the operator $K_\mu\equiv H-\mu I$.

We now determine the gauge invariant $\langle\tau_K,\beta\rangle$-KMS states on $\CAR(\mathfrak h)$, 
for a self-adjoint operator $K$ on an arbitrary Hilbert space $\mathfrak h$. Let $\omega$ be such a
state. Then, $\langle f,g\rangle\mapsto\omega(a^\ast(g)a(f))$ is a sesquilinear form on $\mathfrak{h}$. 
Furthermore
$$
0\le\omega(a^\ast(f)a(f))\leq \|a^\ast(f)\|\,\|a(f)\|=\|f\|^2,
$$
shows that it is positive and bounded above. Thus, there exists an operator $T$ on $\mathfrak h$ 
such that $0\le T\le I$ and
\begin{equation}
\label{f2point}
\omega(a^\ast(g)a(f))=(f,Tg).
\end{equation}
For $t\in\rr$ we have $F(t)\equiv\omega(a^\ast(g)\tau_K^t(a(f)))=(\e^{\i tK}f,Tg)$ and, if 
$f\in C^\omega(K)$,
$$
F(0+\i\b)=(\e^{\b K}f,Tg).
$$
The CAR give us that $a(f)a^\ast(g)=-a^\ast(g)a(f)+(f,g)$ and thus, taking into 
account~(\ref{f2point}), 
$$
\omega(a(f)a^\ast(g))=(f,g)-\omega(a^\ast(g)a(f))=(f,(I-T)g).
$$
For $t=0$, the KMS boundary conditions $F(\i\beta)=\omega(a(f)a^\ast(g))$ imply
$$
(\e^{\b K}f,Tg)=(f,(I-T)g),
$$
that is to say $T(I+\e^{\b K})f=f$ for all $f\in C^\omega(K)$. Since $C^\omega(K)$ is dense and 
$(I+\e^{\b K})^{-1}C^\omega(K)\subset C^\omega(K)$ we may deduce that 
\beq
T=(1+\e^{\b K})^{-1}.
\label{TKlink}
\eeq
We now consider the function
\begin{equation}
W_{m,n}(g_1,\ldots,g_m;f_1,\ldots,f_n)=\omega(a^\ast(g_m)\cdots a^\ast(g_1)a(f_1)\cdots a(f_n)).
\label{WnmDef}
\end{equation}
Gauge invariance implies that $ W_{m,n}=0$ if $m\not=n$. If $m=n$, the KMS condition allows 
us to write 
$$ 
W_{n,n}(g_1,\ldots,g_n;f_1,\ldots,f_n)=\omega(a(\e^{-\b K}f_n)a^\ast(g_n)\cdots a^\ast(g_1)
a(f_1)\cdots a(f_{n-1})).
$$
By using the CAR, we can bring the first factor on the right hand side to its original 
position, thus obtaining
\begin{align*}
W_{n,n}(g_1,\ldots,g_n;&f_1,\ldots,f_n)=-W_{n,n}(g_1,\ldots,g_n;f_1,\ldots,\e^{-\beta K}f_n)\\
&+\sum_{j=1}^n(-1)^{n+j}(\e^{-\b K}f_n,g_j)
\omega(a^\ast(g_n)\cdots \cancel{a^\ast(g_j)}\cdots a^\ast(g_1)a(f_1)\cdots a(f_{n-1})).
\end{align*}
By using the multilinearity of $W_{n,n}$ we can rewrite this identity as
\begin{align*}
W_{n,n}(g_1,\ldots,g_n;&f_1,\ldots,(I+\e^{-\b K})f_n)\\
&=\sum_{j=1}^n(-1)^{n+j}(\e^{-\b K}f_n,g_j)
W_{n-1,n-1}(g_1,\ldots,\cancel{g_j},\ldots g_n;f_1,\ldots,f_{n-1}).
\end{align*}
Since $\Ran(I+\e^{-\b K})=\mathfrak h$, we may replace $f_n$ by $(I+\e^{-\b K})^{-1}f_n$ in this last 
formula to obtain
$$
W_{n,n}(g_1,\ldots,g_n;f_1,\ldots,f_n)
=\sum_{j=1}^{n+j}(-1)^{n+j}W_{1,1}(g_j,f_n)
W_{n-1,n-1}(g_1,\ldots,\cancel{g_j},\ldots g_n;f_1,\ldots,f_{n-1}).
$$
One recognizes this last expression as the Laplace expansion for the determinant of the
$n\times n$ matrix 
$[W_{1,1}(g_j,f_i)]_{i,j=1,\ldots,n}$ along the $n$th row. We are led to conclude that
$$
\omega(a^\ast(g_m)\ldots a^\ast(g_1)a(f_1)\ldots a(f_n))=\delta_{n,m}\det[(f_j,Tg_k)].
$$

\begin{Def} A gauge invariant state $\omega\in E(\CAR(\mathfrak h))$ is called quasi-free if 
there exists a self-adjoint operator $T$ on $\mathfrak h$ such that $0\leq T\leq I$ and 
\begin{equation}
\label{quasilibre}
\omega(a^\ast(g_m)...a^\ast(g_1)a(f_1)...a(f_n))=\delta_{n,m}\det[(f_j,Tg_k)]_{j,k=1,\ldots,n},
\end{equation}
for all $n,m\in\mathbb N$ and all $f_1,...,f_n,g_1,...,g_m\in\mathfrak h$. In this case we say that 
$\omega$ is the gauge invariant quasi-free state generated by $T$ and that $T$ is the generator of 
$\omega$.
\end{Def}

{\bf\noindent Remarks.} 1. Since polynomials in $a^\sharp$ are dense in $\CAR(\mathfrak h)$, it is 
clear that a state on this $C^\ast$-algebra is completely determined by its correlation 
functions~(\ref{WnmDef}).
\\2. One can show that for any self-adjoint operator $T$ on $\mathfrak h$ such that $0\leq T\leq I$, 
Formula (\ref{quasilibre}) defines a gauge invariant state on $\CAR(\mathfrak h)$. 
\\3. If $C\in\mathcal L^1(\mathfrak h)$ we have already noted that 
$\d\Gamma(C)\in\CAR(\mathfrak h)$. If $C=\sum_k\kappa_kf_k(g_k,\,\cdot\,)$ denotes the canonical 
decomposition of $C$, then
$$
\d\Gamma(C)=\sum_k\kappa_k\,a^\ast(f_k)a(g_k),
$$
and if $\omega\in E(\CAR(\mathfrak h))$ is gauge invariant quasi-free generated by $T$, then 
$$
\omega(\d\Gamma(C))=\sum_k\kappa_k\,\omega(a^\ast(f_k)a(g_k))
=\sum_k\kappa_k\,(g_k,Tf_k)=\tr(TC).
$$
\\4. If $\fh$ is finite dimensional, it follows from the final example in Section \ref{KMSSubSect} and
the calculation of this section that
$$
\frac{\tr_{\Gamma^-(\fh)}(\e^{-\d\Gamma(K)}A)}{\tr_{\Gamma^-(\fh)}(\e^{-\d\Gamma(K)})}
=\omega_T(A),
$$
where $\omega_T$ is the gauge invariant quasi-free state generated by $T=(1+\e^K)^{-1}$.
Since $\e^{-\d\Gamma(K)}=\Gamma(\e^{-K})=\Gamma(T(I-T)^{-1})$, Eq.~\eqref{FockTrace}
leads to
\beq
\omega_T(\Gamma(S))
=\frac{\tr_{\Gamma^-(\fh)}(\Gamma(T(I-T)^{-1}S)}
{\tr_{\Gamma^-(\fh)}(\Gamma(T(I-T)^{-1}))}
=\frac{\det(I+T(I-T)^{-1}S)}{\det(I+T(I-T)^{-1})}=\det(I+T(S-1)).
\label{QFGamma}
\eeq
This formula remains valid for infinite dimensional $\fh$ provided $S-I$ is trace class.
\\5. The gauge invariant quasi-free state $\omega_T$ is modular iff $\Ker(T)=\Ker(I-T)=\{0\}$. In 
this case, inverting Eq.~\eqref{TKlink} yields that the modular group of $\omega_T$ is the Bogoliubov 
group $\tau_K$ generated by the Hamiltonian $K=\log\left(T(I-T)^{-1}\right)$.

\bigskip
We thus have the following result
\begin{Theo}\label{QFKMS} For all $\beta\in\mathbb R$ there exists a unique gauge invariant 
$\langle\tau_K,\beta\rangle$-KMS state on $\CAR(\mathfrak h)$. It is the gauge invariant quasi-free 
state generated by $(1+\e^{\b K})^{-1}$.
\end{Theo}

and its corollary,

\begin{Coro}\label{QFKMScor} If $\omega$ is an extremal  $\langle\tau_H,\beta\rangle$-KMS state on 
$\CAR_\vartheta(\mathfrak h)$ for $\beta>0$ there exists $\mu\in\mathbb R$ such that $\omega$ is 
the restriction to  $\CAR_\vartheta(\mathfrak h)$ of the gauge invariant quasi-free state on 
$\CAR(\mathfrak h)$ generated by $(1+\e^{\b(H-\mu)})^{-1}$.
\end{Coro}

\subsubsection{The Araki-Wyss representation}
\label{AWsubsect}

A GNS representation $\langle\cH_{\omega_T},\Pi_{\omega_T},\Omega_{\omega_T}\rangle$
of $\CAR(\mathfrak h)$ induced by the gauge invariant quasi-free state $\omega_T$ was explicitly 
constructed by Araki and Wyss in \cite{AW2} (see also \cite{De2}). The Hilbert space is a Fermionic
Fock space
$$
\cH_{\omega_T}=\Gamma^-(\fh\oplus\fh),
$$
and the cyclic vector $\Omega_\omega$ is its the vacuum vector. The
morphism $\Pi_\omega:\CAR(\fh)\to\B(\cH_\omega)$ is completely determined by
$$
\Pi_{\omega_T}(a(f))=a((I-T)^{1/2}f\oplus 0)+a^\ast(0\oplus \bar T^{1/2}\bar f)
$$
where $\bar{\,\cdot\vphantom{f}\,}$ denotes an arbitrary complex conjugation on $\fh$
and $\bar T$ is defined  by $\bar T f=\overline{T\bar f}$. 

If $H$ is a self-adjoint operator on $\fh$ commuting with $T$ and
$$
L=\d\Gamma(H\oplus0-0\oplus \bar H),
$$
one easily check that
$$
\e^{\i tL}\Pi_{\omega_T}(a(f))\e^{-\i tL}=\Pi_{\omega_T}(a(\e^{\i tH}f)),
$$
and since $L\Omega_\omega=0$, we conclude that $L$ is the Liouvillean associated
to the Bogoliubov dynamics $\tau_H$. In particular, $\omega_T\in E(\CAR(\fh),\tau_H)$.

If $H$ has an eigenvalue $\varepsilon$ with eigenvector $f\not=0$, then 
$\Psi=a^\ast(f\oplus0)a^\ast(0\oplus\bar f)\Omega_\omega\not=0$,
$$
\e^{\i tL}\Psi=a^\ast(\e^{\i tH}f\oplus0)a^\ast(0\oplus\e^{-\i t\bar H}\bar f)\Omega_\omega
=\e^{\i t(\varepsilon-\varepsilon)}a^\ast(f\oplus0)a^\ast(0\oplus\bar f)\Omega_\omega=\Psi,
$$
and hence $L\Psi=0$ so that $0$ is not a simple eigenvalue of $L$. If on the contrary $H$
has no eigenvalues, so does $\widehat H=H\oplus0-0\oplus \bar H$.
In particular,
if $H$ has purely absolutely continuous spectrum, then the spectrum of $L$ consists
of a simple eigenvalue $0$ with eigenvector $\Omega_\omega$ and the purely absolutely
continuous spectrum
$$
\left\{\sum_{\varepsilon\in\spec(H)\cup\spec(-H)}n_\varepsilon\varepsilon\,\bigg|\,n_\varepsilon\in\{0,1\},
\sum_{\varepsilon\in\spec(H)\cup\spec(-H)}n_\varepsilon<\infty\right\}^{\rm cl}.
$$
However, if $H$ has some singular spectrum, so does $L_\omega$. Thus, the following proposition is a direct consequence of Proposition \ref{Koopman}.

\begin{Prop}\label{QFErgoProp}
Let $T$ be the generator of a gauge invariant quasi-free state on $\CAR(\fh)$
such that $\Ker(T)=\Ker(I-T)=\{0\}$ (so that $\omega_T$ is modular). Then $\omega_T$ is ergodic if and only if
$H$ has no eigenvalue. If in addition $H$ has empty singular continuous spectrum then $\omega_T$
is mixing.
\end{Prop}

\subsubsection{Gauge group and chemical potentials}

In this section, we generalize the previous result to more general Abelian gauge groups.

Suppose that there exists a family of self-adjoint operators $\{Q^{(1)},\ldots,Q^{(N)}\}$ on 
$\mathfrak h$ such that 
$$
[\e^{\i tQ^{(j)}},\e^{\i sQ^{(k)}}]=0,\qquad[\e^{\i tH},\e^{\i sQ^{(k)}}]=0,
$$ 
for all 
$s,t\in\mathbb R$ and all $j,k\in\{1,\ldots,N\}$. These operators may be interpreted as the generators 
of a symmetry group of the system. In fact
$$
\mathbb R^N\ni s=(s_1,\ldots,s_N)\mapsto U(s)
=\e^{\i\sum_{j=1}^Ns_jQ^{(j)}}
=\prod_{j=1}^N\e^{\i s_jQ^{(j)}},
$$
is a strongly continuous, faithful, unitary representation of the Abelian group $G\equiv\mathbb R^N/K$
in $\mathfrak h$, where
$$
K\equiv\{s\in\mathbb R^N\,|\,U(s)=I\},
$$
is a subgroup of $\mathbb R^N$. We will always assume that $G$ is compact. Since
$U(s)HU^\ast(s)=H$, for all $s\in\mathbb R^N$, the Bogoliubov group 
$\vartheta^s(A)=\Gamma(U(s))A\Gamma(U(s))^\ast$
commutes with the dynamics $\tau_H^t$. Thus, the $C^\ast$-subalgebra
$$
\CAR_G(\fh)\equiv\{A\in\CAR(\fh)\,|\,\vartheta^s(A)=A \ \text{for all}\ s\in\mathbb R^N\},
$$
is invariant under $\tau_H$. As a general rule, only the elements of this subalgebra are physically 
observable. In the presence of a gauge group $G$ the $C^\ast$-dynamical system corresponding 
to our system is thus $(\CAR_G(\fh),\tau_H)$. Since $\vartheta$ acts trivially on $\CAR_G(\fh)$, 
it is clear that for all $\mu=(\mu_1,\ldots,\mu_n)\in\mathbb R^N$ the unique $\beta$-KMS state 
on $\CAR(\fh)$ for the group $t\mapsto\tau_H^t\circ\vartheta^{-t\mu}$ is a $\beta$-KMS state for 
the restriction of $\tau$ to $\CAR_G(\fh)$. Conversely, every extremal $\langle\tau_H^t,\beta\rangle$-KMS state 
on $\CAR_G(\fh)$ is the restriction to this subalgebra of the
$\langle\tau_H^t\circ\vartheta^{-t\mu},\beta\rangle$-KMS state on $\CAR(\fh)$ for some
$\mu=(\mu_1,\ldots,\mu_n)\in\mathbb R^N$. This state is quasi-free, generated by
$$
T_{\beta,\mu}\equiv\left(1+\e^{\beta(H-\sum_{j=1}^N\mu_jQ^{(j)})}\right)^{-1}.
$$
We call the operators $Q^{(1)},\ldots,Q^{(N)}$ charges. The parameters $\mu_1,\ldots,\mu_N$ are 
the chemical potentials associated with these charges.

\begin{Exa} The following example is typical and provides an illustration of the previous discussion. 
Suppose that the fermions in our system are of two distinct types (and thus distinguishable), say 
red and blue. In this case, the one-particle Hilbert space can be written in the form 
$\mathfrak h=\mathfrak h_r\oplus\mathfrak h_b$ where $\mathfrak h_r$ and $\mathfrak h_b$ are 
two copies of the same space, one for the red fermions and the other for the blue ones. The wave 
function $f\oplus 0$ describe a red fermion while $0\oplus f$ describes a blue fermion in the same 
state. In this case the charges are identified with the colors
$$
Q^{(r)}\equiv\left(\begin{array}{cc}1&0\\0&0\end{array}\right),\qquad
Q^{(b)}\equiv\left(\begin{array}{cc}0&0\\0&1\end{array}\right),\qquad
$$
and the gauge group is the two-dimensional torus $G=S^1\times S^1$. Gauge invariance expresses
the fact that the phase of each component of the wave function $f\oplus g$ is not measurable, i.e., 
$\psi=f\oplus g$ and $\e^{\i(s_rQ^{(r)}+s_bQ^{(b)})}\psi=\e^{\i s_r}f\oplus\e^{\i s_b}g$ describe the 
same state. The $C^\ast$-algebra $\mathcal O=\mathrm{CAR}(\mathfrak h)$ is generated by the 
operators $r(f)=a(f\oplus 0)$ and $b(f)=a(0\oplus f)$ and their adjoints $r^\ast(f)$ (which creates a 
red fermion in the state $f$)
and $b^\ast(f)$ (which creates a blue fermion in the same state). The action of the gauge group $G$
on $\mathcal O$ is given by the $\ast$-automorphisms $\vartheta^s$, $s=(s_r,s_b)\in\mathbb R^2$ 
such that $\vartheta^s(r(f))=\e^{-\i s_r}r(f)$, $\vartheta^s(b(f))=\e^{-\i s_b}b(f)$. A monomial in the 
operators $r^\#$ and $b^\#$ is thus invariant under $\vartheta$ if it contains the same number of 
$r$ and $r^\ast$ factors, as well as the same number of $b$ and $b^\ast$ factors. 
The subalgebra $\mathcal O_G$ is generated by the operators $r^\ast(f)r(g)$, $b^\ast(f)b(g)$ and $I$.
We denote by $\mathcal O_r$ the subalgebra generated by the operators $r^\ast(f)r(g)$ and $I$, and
we denote by $\mathcal O_b$ the subalgebra generated by the operators  $b^\ast(f)b(g)$ and $I$.
It follows from the fact that $\{r(g),b^\ast(f')\}=(f\oplus0,0\oplus f')=0$ and 
$\{b(g'),r^\ast(f)\}=(0\oplus g',f\oplus 0)=0$ that we have
\begin{align*}
[r^\ast(f)r(g),b^\ast(f')b(g')]&=r^\ast(f)r(g)b^\ast(f')b(g')-b^\ast(f')b(g')r^\ast(f)r(g)\\
&=-r^\ast(f)b^\ast(f')r(g)b(g')+b^\ast(f')r^\ast(f)b(g')r(g)\\
&=-b^\ast(f')r^\ast(f)b(g')r(g)+b^\ast(f')r^\ast(f)b(g')r(g)\\
&=0.
\end{align*}
We conclude that the elements of $\mathcal O_r$ commute with the elements of $\mathcal O_b$, 
which reflects the fact that the red fermions are distinguishable from the blue fermions. Since 
$\mathcal O_G$ is clearly generated by $\mathcal O_r$ and $\mathcal O_b$ we deduce that 
$\mathcal O_G=\mathcal O_r\otimes\mathcal O_b$.

If $H=H_r\oplus H_b$ is the one-particle Hamiltonian, the dynamics on $\mathcal O$ is given by 
$\tau^t(r(f))=r(\e^{\i tH_r}f)$ and $\tau^t(b(f))=b(\e^{\i tH_b}f)$ and commutes with the gauge group 
$\vartheta$. The extremal $\langle\tau,\beta\rangle$-KMS states on $\mathcal O_G$ are quasi-free generated by
$$
T_{\beta,\mu_r,\mu_b}=\left(1+\e^{\beta(H-\mu_rQ^{(r)}-\mu_bQ^{(b)})}\right)^{-1}.
$$
\end{Exa}

\subsubsection{Thermodynamic limit}
\label{TDLimit}

In Section \ref{KMSSubSect} we identified the thermal equilibrium states of a $C^\ast$-dynamical 
system with the KMS states associated with this system. In Section \ref{QFKMSsubsect} we 
introduced the chemical potential associated with a gauge symmetry group of the dynamical system. 
In this section, we discuss the relationship between equilibrium states and the grand canonical 
ensemble often used in equilibrium statistical mechanics.

Suppose we are given a net a closed subspaces $(\mathfrak h_v)_{v\in V}$ of the one-particle
Hilbert $\mathfrak h$, as well as a corresponding family of self-adjoint operators $H_v$ on 
$\mathfrak h_v$. In the usual presentation of the thermodynamic limit in statistical mechanics,
\begin{enumerate}[(i)]
\item $\mathfrak h$ is the Hilbert space of a fermion in infinite volume, for example 
$\mathfrak h=L^2(\mathbb R^3)$.
\item $H$ is its Hamiltonian. For example $H=-\Delta$, the Laplacian on $\mathbb R^3$.
\item $V$ is a directed family of finite boxes. For example, the cubes  
  $$v=v(L)\equiv\{(x_1,x_2,x_3)\in\mathbb R^3\,|\, -L\le x_i\le
  L,\ i=1,2,3\},
  $$
  with $L>0$.
\item $\mathfrak h_v$ is the Hilbert space of a fermion confined to the box $v$. For example, 
$$
\mathfrak h_v=\{u\in L^2(\mathbb R^3)\,|\,\supp\,u\subset v\},
$$
which we will identify with $L^2(v)$. 
\item $H_v$ is its Hamiltonian. For example, $H_v=-\Delta_{v,N}$, the Laplacian on the cube $v$ 
with Neumann boundary conditions.
\end{enumerate}
Let $p_v$ be the orthogonal projection of $\mathfrak h$ onto $\mathfrak h_v$. We shall suppose that
\begin{enumerate}[(a)]
\item $\slim_v p_v=I$.
\item $\lim_v(H_v-z)^{-1}f=(H-z)^{-1}f$ for all $z\in
\mathbb C\setminus\mathbb R$ and $f\in\cup_{v\in V}\mathfrak h_v$.
\item $\e^{-\beta H_v}\in\mathcal L^1(\mathfrak h_v)$ for all $v\in V$ and $\beta>0$.
\end{enumerate}

We set $\mathcal A_v=\CAR(\mathfrak h_v)$ for all $v\in V$ and 
$$
\mathcal A_\mathrm{loc}\equiv\bigcup_{v\in V}\mathcal A_v.
$$
The injections 
$\mathfrak h_v\subset\mathfrak h_{v'}\subset\mathfrak h$ naturally induce the injections 
$\Gamma^-(\mathfrak h_v)\subset\Gamma^-(\mathfrak h_{v'})\subset\Gamma^-(\mathfrak h)$ and
$\mathcal A_v\subset\mathcal A_{v'}\subset\mathcal A$ for $v\le v'$. We thus have that 
$\mathcal A_\mathrm{loc}\subset\mathcal A$ and the condition $(a)$ implies that
$\mathcal A_\mathrm{loc}$ is dense in $\mathcal A$.

Condition (c) implies that the spectrum of $H_v$ is pure point, of finite multiplicity, bounded from
below and can only accumulate at $+\infty$. For $\Lambda>0$, we define 
$\mathfrak h_{v\Lambda}\equiv F(H_v<\Lambda)\mathfrak h_v$, 
$\mathcal A_{v\Lambda}\equiv\CAR(\mathfrak h_{v\Lambda})$ and 
$P_{v\Lambda}\equiv\Gamma(F(H_v<\Lambda))$ the orthogonal projection of  
$\Gamma^-(\mathfrak h_v)$ onto the subspace $\Gamma^-(\mathfrak h_{v\Lambda})$. 
We note that $\mathfrak h_{v\Lambda}$, $\Gamma^-(\mathfrak h_{v\Lambda})$ and 
$\mathcal A_{v\Lambda}$ are all of finite dimension. Furthermore 
$\cup_{\Lambda>0}\mathfrak h_{v\Lambda}$, being dense in 
$\mathfrak h_{v}$, $\cup_{\Lambda>0}\mathcal A_{v\Lambda}$ is also dense in 
$\mathcal A_{v}$ and $\slim_{\Lambda\to\infty}P_{v\Lambda}=P_v=\Gamma(p_v)$.

If we denote $\mathcal K_{\mu,v}\equiv\d\Gamma(H_v-\mu I)$ then $\e^{-\beta\mathcal K_{\mu,v}}$
is trace class,
$$
\e^{-\beta\mathcal K_{\mu,v}}-
\e^{-\beta\mathcal K_{\mu,v}}P_{v\Lambda}
=\e^{-\beta\mathcal K_{\mu,v}}(I-P_{v\Lambda}),
$$
and we deduce that 
$$
\lim_{\Lambda\to\infty}\e^{-\beta\mathcal K_{\mu,v}}P_{v\Lambda}
=\e^{-\beta \mathcal K_{\mu,v}},
$$
in the norm of $\mathcal L^1(\Gamma^-(\mathfrak h_v))$.

The grand canonical ensemble in the box $v$, at inverse temperature $\beta$ with chemical 
potential $\mu$, given by $\omega_{\beta,\mu,v}(A)=\tr(\rho_{\beta,\mu,v}A)$ where
$$
\rho_{\beta,\mu,v}=
\frac{\e^{-\beta\mathcal K_{\mu,v}}}
{\tr_{\Gamma_-(\mathfrak h_v)}(\e^{-\beta\mathcal K_{\mu,v}})},
$$
defines a state on $\mathcal A_v$. For all $A\in\cup_{\Lambda>0}\mathcal A_{v,\Lambda}$ we have
$$
\omega_{\beta,\mu,v}(A)=\lim_{\Lambda\to\infty}\omega_{\beta,\mu,v,\Lambda}(A),
$$
where
$$
\omega_{\beta,\mu,v,\Lambda}(A)=
\frac{\tr_{\Gamma^-(\mathfrak h_{v\Lambda})}(\e^{-\b\mathcal K_{\mu,v}}A)}
{\tr_{\Gamma^-(\mathfrak h_{v\Lambda})}(\e^{-\b\mathcal K_{\mu,v}})}.
$$

The group of $\ast$-automorphisms 
$\gamma^t_{\mu,v}(A)=\e^{\i t\mathcal K_{\mu,v}}A\e^{-\i t\mathcal K_{\mu,v}}$ on 
$\B(\Gamma^-(\mathfrak h_v))$ leaves the $C^\ast$-algebra $\mathcal A_{v\Lambda}$ invariant. 
Example 1 from Section \ref{KMSSubSect} shows that $\omega_{\beta,\mu,v,\Lambda}$ is the 
unique $\langle\gamma^t_{\mu,v},\beta\rangle$-KMS state on $\mathcal A_{v\Lambda}$. Theorem \ref{QFKMS} 
implies that $\omega_{\beta,\mu,v,\Lambda}$ is gauge invariant quasi-free generated by
$$
T_{\beta,\mu,v,\Lambda}=(I+\e^{\beta(H_v-\mu I)})^{-1}|_{\mathfrak h_{v\Lambda}}.
$$
For all $f_1,\ldots,g_1,\ldots\in\mathfrak h_{v\Lambda}$, we thus have
\begin{align*}
\omega_{\beta,\mu,v}(a^\ast(g_m)\cdots a^\ast(g_1)a(f_1)\cdots a(f_n))
&=\lim_{\Lambda'\to\infty}
\omega_{\beta,\mu,v,\Lambda'}(a^\ast(g_m)\cdots a^\ast(g_1)a(f_1)\cdots a(f_n))\\
&=\lim_{\Lambda'\to\infty}
\delta_{nm}\det[(f_i,T_{\beta,\mu,v,\Lambda'}g_j)]_{i,j=1,\ldots n}.
\end{align*}
We deduce that 
$$
\omega_{\beta,\mu,v}(a^\ast(g_m)\cdots a^\ast(g_1)a(f_1)\cdots a(f_n))
=\delta_{nm}\det[(f_i,T_{\beta,\mu,v}g_j)]_{i,j=1,\ldots n}.
$$
where
$$
T_{\beta,\mu,v}=(I+\e^{\beta(H_v-\mu)})^{-1}.
$$
This last formula extends continuously to all $f_1,\ldots,g_1,\ldots\in\mathfrak h_{v}$. 
$\omega_{\beta,\mu,v}$  is thus the gauge invariant quasi-free state generated by $T_{\beta,\mu,v}$. 
Condition (b) implies that
$$
\lim_{v'}T_{\beta,\mu,v'}f=T_{\beta,\mu}f\equiv(I+\e^{\beta(H-\mu)})^{-1}f,
$$
for all $f\in\mathfrak h_v$. We deduce that
\begin{align*}
\omega_{\beta,\mu}(a^\ast(g_m)\cdots a^\ast(g_1)a(f_1)\cdots a(f_n))
&\equiv\lim_{v'}
\omega_{\beta,\mu,v'}(a^\ast(g_m)\cdots a^\ast(g_1)a(f_1)\cdots a(f_n))\\
&=\lim_{v'}
\delta_{nm}\det\{(f_i,T_{\beta,\mu,v'}g_j)\}_{i,j=1,\ldots n}\\
&=\delta_{nm}\det\{(f_i,T_{\beta,\mu}g_j)\}_{i,j=1,\ldots n},
\end{align*}
for all $f_1,\ldots,g_1,\ldots\in\mathfrak h_{v}$. This shows that $\omega_{\beta,\mu}$, the gauge 
invariant quasi-free state on $\mathcal A$ generated by $T_{\beta,\mu}$, is the thermodynamic 
limit of the grand canonical ensemble
$$
\omega_{\beta,\mu}(A)=\lim_{v'}\omega_{\beta,\mu,v'}(A),
$$
for all $A\in\mathcal A_\mathrm{loc}$.

\subsection{Open quantum systems}
\label{OQS-sect}
This section is a brief introduction to the $C^\ast$-algebraic description of open quantum systems 
and to the nonequilibrium statistical mechanics of these systems. For a more detailed discussion, 
we refer the reader to \cite{JP7,AJPP1}. 

\subsubsection{Algebraic description}
\label{OSYSDef}
A system is called open if it interacts with some environment. A typical example of an open quantum 
system is an atom (or a molecule) whose charged constituents interact with the electromagnetic field. If 
we neglect these interactions, an atom generally allows for a series of steady states, corresponding to
the eigenvalues of its Hamiltonian. Taking into account the interaction of the electrons
with the electromagnetic field, only the ground state remain stable. The excited steady states
turn into metastable states with a finite lifetime (see \cite{BFS1,BFS2} for a rigorous treatment of 
this fundamental effect).

To model an open system, one generally considers that the system is made up of a small, confined 
system $\mathcal S$, with a finite number of classical degrees of freedom, as well as one or several
reservoirs $\mathcal R_1,\ldots,\mathcal R_M$, each of those being an extended system with a 
large number of degrees of freedom. While the small system $\mathcal S$  may have a complex 
internal structure, the reservoirs are generally simple systems, for example ideal gases. In terms of 
Hamiltonians, the dynamics of such a system is determined, at least formally, by the sum 
$$
H^\mathrm{tot}=H_{\mathcal S}+\sum_{j=1}^MH_{\mathcal R_j}
+\sum_{j=1}^MH^\mathrm{int}_{\mathcal S,\mathcal R_j},
$$
where $H_{\mathcal S}$ is the Hamiltonian of the small system, $H_{\mathcal R_j}$ is that of 
reservoir $\mathcal R_j$, and $H^\mathrm{int}_{\mathcal S,\mathcal R_j}$ is the Hamiltonian 
representing the interaction between the small system and reservoir $\mathcal R_j$.

In a mathematically rigorous approach to open systems it is often convenient to idealize the 
reservoirs and consider them to be infinitely extended (this is in particular the case for the
construction of non\-equilibrium steady states). In such situations the algebraic formulation of 
quantum dynamics provides a more appropriate framework than the familiar Hilbert space/Hamiltonian
approach. The coupled system $\mathcal S+\mathcal R_1+\cdots\mathcal R_M$ is described by a 
$C^\ast$-dynamical system $\langle \mathcal O,\tau\rangle$ which has the following structure
\begin{enumerate}[(i)]
\item There exist $C^\ast$-subalgebras
$\mathcal{O_S,O_{R_{\mathrm 1}},\ldots O_{R_{\mathrm M}}\subset O}$, such that
$\cO_\cS\cap\cO_{\cR_j}=\cO_{\cR_k}\cap\cO_{\cR_j}=\cc I$ for $j\not=k$ and
$$
\mathcal{O=O_S\vee O_{R_{\mathrm 1}}\vee\cdots\vee O_{R_{\mathrm M}}},
$$
that is to say that $\mathcal O$ is generated by these subalgebras.
$\cO_\cS$ is the algebra of observables of the small system, and 
${\mathcal O_{\mathcal R_j}}$ is that of the $j$-th reservoir.

\item For each $\alpha\in\{\cS,\cR_1,\ldots,\cR_M\}$ there exists a $C^\ast$-dynamical system
$\langle \mathcal{O},\tau_\alpha\rangle$ such that $\tau_\alpha^t(\cO_\alpha)$ $\subset\cO_\alpha$
and $\tau_\alpha^t(A)=A$ for $A\in\cO_\beta$ and $\beta\not=\alpha$. The $C^\ast$-dynamical 
systems $\langle\cO_\alpha,\tau_\alpha|_{\cO_\alpha}\rangle$ describe the com\-po\-nents of the 
system without interactions between them.

\item  $\tau_\cS^t=\e^{t\delta_{\mathcal S}}$ where the $\ast$-derivation $\delta_\cS$ is inner, i.e.,
there exists a self-adjoint element $H_\mathcal{S}\in\mathcal O_{\mathcal S}$ such that 
$\delta_{\mathcal S}=\i[H_\mathcal{S},\,\cdot\,]$.

\item Let $\delta,\delta_{\mathcal R_j}$ be the $\ast$-derivations generating the $C^\ast$-dynamics 
$\tau,\tau_{\mathcal R_j}$. There exist self-adjoint elements 
$V_j\in\mathcal O_{\mathcal S}\vee\mathcal O_{\mathcal R_j}$ such that
$$
\delta(A)=\delta_{\mathcal S}(A)+\sum_{j=1}^M\delta_{\mathcal R_j}(A)
+\sum_{j=1}^M\i[V_j,A].
$$
$V_j$ thus describes the interaction between the small system and the $j$-th reservoir.
\end{enumerate}

The dynamics of the system admits a perturbative expansion, the Schwinger-Dyson 
series 
\begin{align*}
\tau^t(A)=\tau_0^t(A)\!+\!\sum_{n=1}^\infty\,
\int\limits_{0\le s_1\le\cdots\le s_n\le t}\!\!\!\!\!\!
\i[\tau_0^{s_1}(V),\i[\cdots,\i[\tau_0^{s_n}(V),\tau_0^t(A)]\cdots]]
\,\d s_1\cdots\d s_n,
\end{align*}
where  $\tau_0^t=\e^{t\delta_0}$ with 
$\delta_0=\delta_{\mathcal S}+\sum_{j=1}^M\delta_{\mathcal R_j}$ and $V=\sum_{j=1}^MV_j$. 
This series converges in the norm of $\mathcal O$ for all $t\in\mathbb R$ and $A\in\cO$.

\subsubsection{Non-equilibrium steady states (NESS)}
\label{NESSSubSect}
Let $\omega\in E(\mathcal O)$ be the initial state of the system. If this state is close enough to a 
thermodynamic equilibrium $\omega_{\mathrm eq}$ it is expected that the system will relax towards 
equilibrium,
$$
\lim_{t\to\infty}\omega\circ\tau^t(A)=\omega_\mathrm{eq}(A),
$$
for all $A\in\mathcal O$. On the other hand, if $\omega$ is sufficiently far from a thermodynamic 
equilibrium state, the system may evolve to a nonequilibrium steady state.

Following Ruelle (\cite{R2,R3}) we define a nonequilibrium steady state (NESS) associated with 
the initial state $\omega$ as a limit point, in the weak-$\ast$ topology of $E(\mathcal O)\subset\cO'$, 
of the net
$$
\langle\omega\rangle_t\equiv \frac{1}{t}\int_0^t\omega\circ\t^s ds,
$$
with $t>0$. We denote by $\Sigma^+(\omega)$ the set of all NESS associated to $\omega$. 
We thus have $\omega^+\in\Sigma^+(\omega)$ if and only if there exists a net $t_\alpha\to+\infty$ 
such that 
\begin{equation}
\lim_\alpha\,\langle\omega\rangle_{t_\alpha}(A)=\omega^+(A),
\label{NESSConverg}
\end{equation}
for all $A\in\mathcal O$. One can easily show that all elements of $\Sigma^+(\omega)$ are 
$\tau$-invariant states. Furthermore, since $E(\mathcal O)$ is weak-$\ast$ compact, 
$\Sigma^+(\omega)$ is never empty. The fundamental problem of nonequilibrium statistical 
mechanics of the system $\langle\mathcal O,\tau\rangle$ is the study of the properties of these NESS. 
We are particularly interested in showing that $\Sigma^+(\omega)=\{\omega^+\}$ (there is only one 
NESS), and that 
\begin{equation}
\lim_{t\to\infty}\omega\circ\tau^t(A)=\omega^+(A),
\label{NESSUniq}
\end{equation}
the unique NESS, is an attractor.

\subsubsection{Scattering theory of $C^\ast$-dynamical systems}
\label{CstarScattSubSect}

As already remarked in the introduction, three methods have been implemented to study the limit in 
equation~(\ref{NESSUniq}):
\begin{enumerate}[-]
\item Ruelle's scattering approach \cite{R4} uses scattering theory of $C^\ast$-dynamical systems to
construct a unique NESS. This method operates directly on the algebra $\mathcal O$ without passing 
to a representation.
\item The spectral method of Jak\v si\'c-Pillet \cite{JP6} (see also \cite{MMS}). It reduces the problem
to the analysis of the complex resonances of a Liouvillean, a non self-adjoint generator of the dynamics
in the canonical cyclic representation of the $C^\ast$-algebra $\mathcal O$ associated with the initial 
state $\omega$.
\item The de Roeck-Kupiainen cluster expansion technique \cite{dRK1,dRK2} which operates directly on the 
sample subalgebra $\cO_\cS$.
\end{enumerate}
We shall apply the first approach in these notes. In the following, we provide a general 
description of this method. We refer the reader to \cite{R4,R5,JP7,AJPP1} for further discussion and to
\cite{DFG,FMU,FMSU,JOP1,JPP2,CMP1,CMP2} for examples of application of the method.

\bigskip
The $C^\ast$-algebraic scattering theory is inspired by the Hilbert space scattering theory introduced
in Section~\ref{firstscattering} and relies on the existence of the strong limits
\begin{equation}
\label{moller}
\gamma^\pm=\slim_{t\to\pm\infty}\t_0^{-t}\circ\t^t,
\end{equation}
(compare with Eq.~\eqref{Mollerdef}, and observe that the order of the free and perturbed groups 
is reversed since we are working here in the Heisenberg picture whereas the Hilbert space scattering 
is formulated in the Schr\"odinger picture). 
The groups $\tau_0$ and $\tau$ being isometric, these limits, 
if they exist, define injective $\ast$-endomorphisms of $\mathcal O$ such that
$$
\gamma^\pm\circ\t^t=\t_0^t\circ\gamma^\pm,
$$
for all $t\in\mathbb R$. We call them M\o ller morphisms. As in the Hilbert spaces theory, the 
$\tau_0$-invariant $C^\ast$-subalgebras $\mathcal O^\pm\equiv\gamma^\pm(\mathcal O)$ play a 
central role. For all $A\in\mathcal O$, we have
$$
0=\lim_{t\to\pm\infty}\|\gamma^\pm(A)-\t_0^{-t}\circ\t^t(A)\|=
\lim_{t\to\pm\infty}\|\t_0^t(\gamma^\pm(A))-\t^t(A)\|,
$$
that is to say that the evolution of $A$ under $\tau$ is asymptotically that of $\gamma^\pm(A)$ under
$\t_0$ when $t\to\pm\infty$. We may thus define a $\ast$-isomorphism
$$
\sigma\equiv\gamma^+\circ(\gamma^-)^{-1}:\mathcal O^-\to\mathcal O^+,
$$
which transforms the incoming asymptote $\gamma^-(A)$ into the outgoing asymptote $\gamma^+(A)$.
It is evidently the equivalent to the scattering operator in the Hilbert space approach.

For all $A\in\mathcal O$ one has
$$
0=\lim_{t\to\pm\infty}\|\t_0^t(\gamma^\pm(A))-\t^t(A)\|
=\lim_{t\to\pm\infty}\|\t^{-t}\circ\t_0^t(\gamma^\pm(A))-A\|,
$$
which shows that the strong limits
\begin{equation}
\alpha^\pm=\slim_{t\to\pm\infty}\t^{-t}\circ\t_0^t|_{\mathcal O^\pm},
\label{AlphaPM}
\end{equation}
exists and that $\alpha^\pm=(\gamma^\pm)^{-1}$.

In the context of open systems described in the preceding section we expect
\begin{equation}
\mathcal O^-=\mathcal O^+=\mathcal O_{\mathcal R}\equiv\bigvee_{j=1}^M\mathcal O_{\mathcal R_j},
\label{OPMOSYS}
\end{equation}
where $\mathcal O_{\mathcal R}$ is the $C^\ast$-subalgebra of the reservoirs. In fact, the small 
system $\mathcal S$ being confined, the spectrum of its Hamiltonian $H_\mathcal{S}$ will be pure 
point. In this case $\delta_{\mathcal S}=\i[H_{\mathcal S},\,\cdot\,]$ will also have a pure point 
spectrum and we thus do not expect that the limit in Eq.~(\ref{AlphaPM}) exists 
on $\mathcal O_\mathcal{S}$.

We note in particular that if (\ref{OPMOSYS}) is verified then $\gamma^\pm$ provide 
$\ast$-isomorphisms between the $C^\ast$-dynamical systems $\langle\mathcal O,\tau\rangle$ and 
$\langle\mathcal O_{\mathcal R},\tau_{\mathcal R}\rangle$ where 
$\tau_{\mathcal R}=\tau_0|_{\mathcal O_R}$ denotes the free
dynamics of the reservoirs. If the initial state 
$\omega$ is $\t_0$-invariant, then
$$
\omega\circ\tau^t=\omega\circ\tau_0^{-t}\circ\tau^t,
$$
and
$$
\lim_{t\to\pm\infty}\omega\circ\tau^t(A)=\lim_{t\to\pm\infty}
\omega\circ\tau_0^{-t}\circ\tau^t(A)=\omega|_{\mathcal O_{\mathcal R}}\circ\gamma^\pm(A),
$$
for all $A\in\mathcal O$. From this we get that $\Sigma^+(\omega)=\{\omega^+\}$ and the unique
NESS associated with $\omega$ is independent of the initial state of the small system 
$\omega|_{\mathcal O_{\mathcal S}}$. If $\omega|_{\mathcal O_{\mathcal R}}$ has ergodic 
properties (which is typically the case for ideal reservoirs) we can say even more. 
For all $\nu\in E(\mathcal O)$, we have
$$
|\nu\circ\tau^t(A)-\nu\circ\tau_0^t(\gamma^+(A))|\le\|\tau^t(A)-\tau_0^t(\gamma^+(A))\|
=\|\tau_0^{-t}\circ\tau^t(A)-\gamma^+(A)\|,
$$
and thus
$$
\lim_{t\to\infty}|\nu\circ\tau^t(A)-\nu\circ\tau_0^t(\gamma^+(A))|=0.
$$
If $\omega|_{\mathcal O_{\mathcal R}}$ is $\tau_{\mathcal R}$-ergodic and
if $\nu|_{\mathcal O_{\mathcal R}}$ is $\omega|_{\mathcal O_{\mathcal R}}$-normal, we deduce that
$$
\lim_{T\to\infty}\frac{1}{T}\int_0^T\nu\circ\tau^t(A)\,\d t
=\lim_{T\to\infty}\frac{1}{T}\int_0^T\nu\circ\tau_0^t(\gamma^+(A))\,\d t=
\omega(\gamma^+(A)),
$$
that is to say $\Sigma^+(\nu)=\{\omega^+\}$. Similarly, if $\omega|_{\mathcal O_{\mathcal R}}$ is 
$\tau_{\mathcal R}$-mixing and if 
$\nu|_{\mathcal O_{\mathcal R}}$ is $\omega|_{\mathcal O_{\mathcal R}}$-normal,
$$
\lim_{t\to\infty}\nu\circ\tau^t(A)
=\lim_{t\to\infty}\nu\circ\tau_0^t(\gamma^+(A))=
\omega(\gamma^+(A)).
$$
Ruelle's approach may thus be summarized by the following proposition
\begin{Prop}\label{AlgNESSProp}
Suppose that the M\o ller morphism $\gamma^+$ defined in Eq.~(\ref{moller}) exists and is such that
$\gamma^+(\mathcal O)=\mathcal O_{\mathcal R}$. If $\omega$ is $\t_0$-invariant, then, for all 
$A\in\cal O$,
$$
\lim_{t\to\infty}\omega\circ\t^t(A)=\omega^+(A),
$$
where $\omega^+=\omega|_{\mathcal O_{\mathcal R}}\circ\gamma^+$. In particular, we have 
$\Sigma^+(\omega)=\{\omega^+\}$ and the unique NESS $\omega^+$ associated with $\omega$ is 
independent of the initial state of the small system $\mathcal S$. If 
$\omega|_{\mathcal O_{\mathcal R}}$ is $\tau_{\mathcal R}$-ergodic, we have 
$\Sigma^+(\nu)=\{\omega^+\}$ for all $\nu\in E(\mathcal O)$ such that 
$\nu|_{\mathcal O_{\mathcal R}}$ is $\omega|_{\mathcal O_{\mathcal R}}$-normal. If, furthermore, 
$\omega|_{\mathcal O_{\mathcal R}}$ is $\tau_0|_{\mathcal O_{\mathcal R}}$-mixing, then
$$
\lim_{t\to\infty}\nu\circ\tau^t(A)=\omega^+(A),
$$
holds for all $A\in\cal O$ and all $\nu\in E(\mathcal O)$ such that 
$\nu|_{\mathcal O_{\mathcal R}}$ is $\omega|_{\mathcal O_{\mathcal R}}$-normal.
\end{Prop} 

\subsubsection{Entropy production}

Another central notion in nonequilibrium statistical mechanics is entropy production. 
The general definition of entropy production is problematic since the concept of 
entropy itself -- fundamental for equilibrium thermodynamics -- does not have a satisfying 
generalization outside of equilibrium (the reader interested in this problematic should read 
the following enlightening discussions by Gallavotti and Ruelle~\cite{G1,G2,R5,R6,R7,R8}).

Following Ruelle \cite{R5} and Jak\v si\'c-Pillet \cite{P1,JP5,JP4} we can however give a satisfactory 
definition of entropy production for a large class of NESS. This definition is based on the concept of 
relative entropy (we refer to \cite{O,OHI,LS,Sp3} for similar considerations and to \cite{AF2} for a 
careful analysis of entropy production in the framework of cyclic processes). 

The relative entropy of two density matrices $\rho$ and $\omega$ on a Hilbert space $\mathcal H$ is 
defined, analogously to the relative entropy of two measures, by the formula
$$
\Ent(\rho|\omega)\equiv \tr(\rho(\log\omega-\log\rho)). 
$$ 
Let $(\varphi_i)_i$ be an orthonormal basis of 
eigenvectors of $\rho$ and let $p_i$ be the associated eigenvalues. Then $p_i\in [0,1]$ and 
$\sum_ip_i=1$. Let $q_i\equiv(\varphi_i,\omega\varphi_i)$. We thus have that $q_i\in [0,1]$ and 
$\sum_iq_i=\tr(\omega)=1$. By applying Jensen's inequality twice we get (with the convention
$0\log 0=0$)
\begin{align*}
\Ent(\rho|\omega) &= \sum_ip_i((\varphi_i,\log\omega\varphi_i)-\log p_i) \\
& \leq \sum_ip_i(\log q_i-\log p_i) \leq \log\sum_iq_i=0.
\end{align*}
We thus have that $\Ent(\rho|\omega)\leq 0$. We can also show that $\Ent(\rho|\omega)=0$ if and 
only if $\rho=\omega$. Araki extended this definition to states of a $C^\ast$-algebra \cite{A1,A2} 
(see also \cite{OP,BR2}). We will not go into 
the details of this extension, which is based on the modular theory of Tomita and Takesaki. The only 
property of this extension of interest to us is precisely the one which we describe in the 
following result (\cite{JP5}).

\begin{Theo} Let $\omega\in E(\mathcal O)$ be a $\tau_0$-invariant state. Suppose that $\omega$ 
is $\langle\sigma_\omega,1\rangle$-KMS for a group $t\mapsto\sigma_\omega^t$ of $\ast$-automorphisms of 
$\mathcal O$. We denote by $\delta_\omega$ the $\ast$-derivation generating the group 
$\sigma_\omega$. If $V\in\Dom(\delta_\omega)$ then
$$
\Ent(\nu\circ\tau^t|\omega)=\Ent(\nu|\omega)+\int_0^t\nu\circ\tau^s(\delta_\omega(V))
\,\d s,
$$
for all $\nu\in E(\mathcal O)$.
\label{EPRODTheo}
\end{Theo}

We shall use this result to define the entropy production rate of a NESS. In order 
to get a convincing physical interpretation of this definition we will restrict ourselves to initial 
states $\omega$ which are close enough to ``product states'' in which each reservoir is in thermal 
equilibrium.

Let $\bbeta=(\beta_1,\ldots,\beta_M)\in\rr_+^M$. We say that $\omega_\bbeta$ is a
$\bbeta$-KMS state if it is a $\langle\sigma_\bbeta,1\rangle$-KMS state, where $\sigma_\bbeta$ denotes 
the group of $\ast$-automorphisms of $\cO$ generated by
$$
\delta_{\bbeta}=\sum_{j=1}^M\beta_j\delta_{\cR_j}.
$$
To simplify our exposition we shall always assume here that such a state exists and is unique
(this is the case if the reservoirs are ideal Fermi gases, which is the situation that will prevail in
the remaining parts of these notes).
Remark 3 of Section \ref{KMSSubSect} shows that the restricted state 
$\omega_\bbeta|_{\cO_{\cR_j}}$ is a $\beta_j$-KMS state for $\tau_{\cR_j}$, i.e., in the state 
$\omega_\bbeta$ each reservoir is in thermal equilibrium. However, if the $\beta_j$ are not all equal 
then the joint reservoir system $\cR$ is not in a global equilibrium state. 

Applying the results of Section~\ref{sect-perturb}, for any self-adjoint $K\in\cO$
$$
\delta_\bbeta^{(K)}=\delta_{\bbeta}+\i[K,\,\cdot\,],
$$
generates a dynamics $\sigma_\bbeta^{(K)}$ with a unique $\langle\sigma_\bbeta^{(K)},1\rangle$-KMS state 
$\omega_\bbeta^{(K)}\in E(\cO)$. The set $E_\bbeta(\cO)$ of all states obtained in this way is 
dense in the set $\mathcal{N}_{\omega_\bbeta}$ of all $\omega_\bbeta$-normal states.
Moreover, one has the estimates \cite{A3,A4}
\beq
\omega_\bbeta^{(K)}(K)-\omega_\bbeta(K)\le\Ent(\omega_\bbeta^{(K)}|\omega_\bbeta)
\le\omega_\bbeta^{(K)}(K)+\log\omega_\bbeta(\e^{-K}).
\label{PBGT}
\eeq

Let $\omega\in E_\bbeta(\cO)$ and $\omega^+\in\Sigma^+(\omega)$ so that Eq.~\eqref{NESSConverg}
holds for a net $t_\alpha$. We define the entropy production rate of $\omega^+$ by
\beq
\mathrm{Ep}(\omega^+)=-\lim_\alpha \frac1{t_\alpha}\Ent(\omega\circ\tau^{t_\alpha}|\omega_\bbeta).
\label{EpDef}
\eeq

Assuming that $V_j\in\Dom(\delta_{\mathcal R_j})$ for all $j$, Theorem \ref{EPRODTheo} allows 
us to write
$$
\frac1{t_\alpha}\Ent(\omega\circ\tau^{t_\alpha}|\omega_\bbeta)
=\frac1{t_\alpha}\Ent(\omega|\omega_\bbeta)
+\frac{1}{t_\alpha}\int_0^{t_\alpha}\omega\circ\tau^s(\delta_{\bbeta}(V))\,\d s,
$$
and  we deduce from Eq.~\eqref{NESSConverg} and \eqref{PBGT} that
\begin{equation}
\mathrm{Ep}(\omega^+)\equiv\omega^+(-\delta_{\bbeta}(V)).
\label{EBal}
\end{equation}
We will come back to the physical interpretation of this relation in the following section. Meanwhile, 
we note that the inequality
$$
\mathrm{Ep}(\omega^+)\ge0,
$$
is a consequence of the fact that the relative entropy of two states is never positive.

\subsubsection{First and second laws of thermodynamics}
\label{ThermoSect}

To legitimate Definition (\ref{EpDef}) and interpret Relation \eqref{EBal} we discuss in this section 
the first two laws of thermodynamics 
in the framework of open quantum systems. To do this, we must identify the observables $\Phi_j$ 
which describe the energy flux leaving the reservoirs $\mathcal R_j$ and entering the small system 
$\mathcal S$.

As in the preceding section, we shall assume that $V_j\in\Dom(\delta_{\mathcal R_j})$ for
$j=1,\ldots, M$. The total energy flux leaving the reservoirs is given by
$$
\frac{\d\ }{\d t}\t^t(H_{\mathcal{S}}+V)=\t^t(\delta(H_{\mathcal{S}}+V)).
$$
Since
$$
\delta(H_{\mathcal{S}}+V)=\i[H_\mathcal{S}+V,H_\mathcal{S}+V]
+\sum_{j=1}^M\delta_{\mathcal R_j}(H_\mathcal{S}+V),
$$
and $\delta_{\mathcal R_j}(H_\mathcal{S})=0$, we have
$$
\frac{\d\ }{\d t}\t^t(H_{\mathcal{S}}+V)=\sum_{j=1}^M\t^t(\delta_{\mathcal{R}_j}(V)).
$$
We may thus identify
$$
\Phi_j=\delta_{\mathcal{R}_j}(V)=\delta_{\mathcal{R}_j}(V_j),
$$
as the observable describing the energy flux leaving the $j$-th reservoir. The identity
$$
\sum_{j=1}^M\Phi_j=\delta(H_{\mathcal{S}}+V),
$$
expresses the conservation of energy -- the first law of thermodynamics: for any $\t$-invariant state 
$\nu$,
$$
\sum_{j=1}^M\nu(\Phi_j)=\left.\frac{\d\ }{\d t}\nu\circ\tau^t(H_\mathcal{S}+V)\right|_{t=0}=0. 
$$
Relation (\ref{EBal}) can now be written as
$$
\mathrm{Ep}(\omega^+)=-\sum_{j=1}^M\beta_j\omega^+(\Phi_j),
$$
which can be interpreted as an entropy balance equation. Its right hand side is the phenomenological 
expression of the entropy flux leaving the system $\cS$ and must coincide with the entropy
produced within this system (see for example \cite{DGM}).
In particular, for all NESS $\omega^+\in\Sigma^+(\omega)$, we obtain
$$
\sum_{j=1}^M\beta_{j}\omega^+(\Phi_j)=-{\rm Ep}(\omega_+)\leq 0,
$$
which is an expression of the second law of thermodynamics.

\subsection{Open fermionic systems}
\label{FQS-sect}

In this section, we show how to adapt the description of open systems developed in
Section~\ref{OQS-sect} to the special case of quasi-free fermionic systems. We shall see that 
scattering theory takes a particularly simple form in this case.

\subsubsection{The one-particle setup}
\begin{figure}
\centering
\includegraphics[scale=0.35]{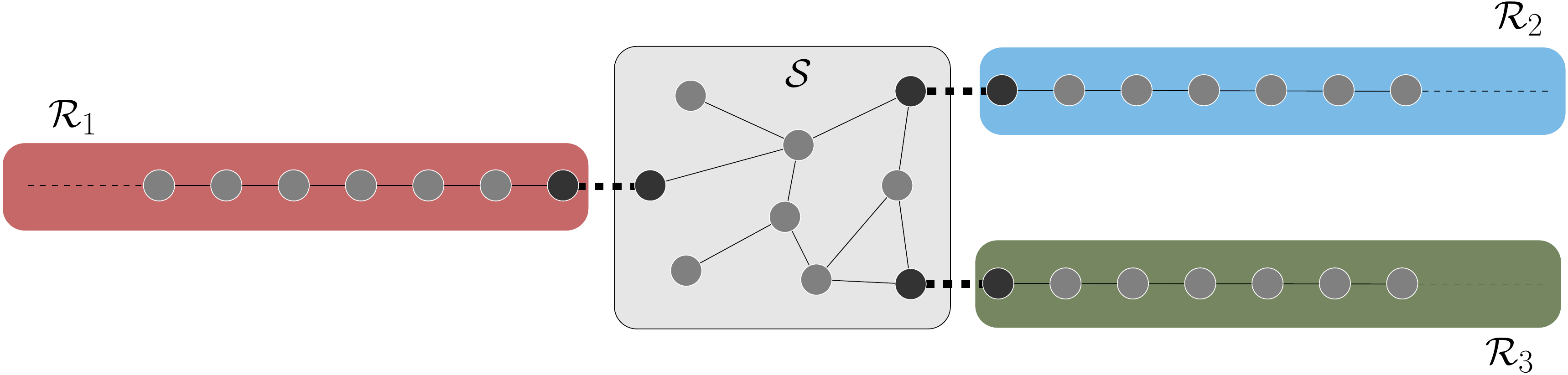}
\caption{A discrete structure $\mathfrak{M}=\mathcal{S}\cup\mathcal{R}_1\cup\mathcal{R}_2\cup\mathcal{R}_3$.}
\label{DiscreteFig}
\end{figure}

In order to simplify the presentation and avoid unnecessary technical difficulties we consider an 
ideal Fermi gas on a connected discrete structure $\mathfrak M$ which is the disjoint union of a finite
set $\cS$ and of $M$ semi-infinite one-dimensional lattices $ \cR_1,\ldots,\cR_M$ (see 
Figure~\ref{DiscreteFig}). This situation is typical of the tight binding approximation widely used
in solid state physics.
The one-particle Hilbert space admits the following decomposition
$$
\fh=\fh_\cS\oplus\fh_\cR,\qquad\fh_\cR=\bigoplus_{k=1}^M\fh_k,
$$
where $\fh_\cS=\ell^2(\cS)$ and $\fh_k=\ell^2(\cR_k)$ with $\cR_k=\nn$. Let $H_\cS$ be a 
self-adjoint operator on $\fh_\cS$ which describes the internal structure of the sample $\cS$.
For each $k$ denote by $H_k$ a copy of the discrete Laplacian on $\nn$ with Dirichlet boundary 
condition, i.e., the operator on $\ell^2(\nn)$ defined by
$$
(Lu)(x)=\frac12\sum_{|x-y|=1}u(y).
$$
It is the standard tight binding Hamiltonian for an electron in a single band of a one-dimen\-sional lead. 
One easily checks that
\beq
(Uu)(\varepsilon)=\sqrt{\frac2{\pi\sqrt{1-\varepsilon^2}}}
\sum_{x\in\nn}u(x)\sin(\arccos(\varepsilon)(x+1)),
\label{Udef}
\eeq
defines 	a unitary operator from $\ell^2(\nn)$ to $L^2([-1,1],\d\varepsilon)$ such that
$(ULu)(\varepsilon)=\varepsilon(Uu)(\varepsilon)$. Thus, $H_k$ has purely absolutely 
continuous spectrum $\spec(H_k)=\spec_\ac(H_k)=[-1,1]$, $\spec_{\rm sing}(H_k)=\emptyset$.

The one-particle Hamiltonian is given by
$$
H=(H_\cS\oplus H_\cR)+V,\qquad
H_\cR=\bigoplus_{k=1}^MH_k,
$$
with a coupling term
$$
V=\sum_{k=1}^M\left(\chi_k(\delta_{0_k},\,\cdot\,)+\delta_{0_k}(\chi_k,\,\cdot\,)\right),
$$
where $\chi_k\in\fh_\cS$ and $\delta_{0_k}\in\fh_{\cR_k}$ denotes the Kronecker delta function at site 
$0$. Since $H_\cS+V$ is compact (in fact finite rank), it follows from Weyl's theorem that
$\spec_{\rm ess}(H)=\spec_{\rm ess}(H_\cR)=[-1,1]$.
To simplify our discussion, we shall assume that $H$ has purely absolutely continuous spectrum.
In the so called fully resonant case, i.e., when $\spec(H_\cS)\subset]-1,1[$, this condition is verified 
provided the coupling strength $\max_k\|\chi_k\|$ is small enough. We will discuss the effect of 
singular spectra in Sections \ref{NESSChapter} and \ref{LBForm}.

\subsubsection{Quasi-free NESS}

We define the $C^\ast$-algebras
$$
\mathcal O\equiv\CAR(\mathfrak h),\quad
\mathcal O_\mathcal{S}\equiv\CAR(\mathfrak h_\mathcal{S}),\quad
\mathcal O_\mathcal{R}\equiv\CAR(\mathfrak h_\mathcal{R}),\quad
\mathcal O_{\mathcal R_k}\equiv\CAR(\mathfrak h_k),
$$
and denote by $\tau$ the $C^\ast$-dynamics on $\mathcal O$ associated to $H$, i.e., 
$\tau^t(a^\#(f))=a^\#(\e^{\i tH}f)$. Let $T$ be the generator of a gauge invariant quasi-free state 
$\omega_T\in E(\mathcal O)$. For all $f_1,\ldots,g_1,\ldots\in\mathfrak h$ we have
\begin{align*}
\omega_T\circ\tau^t(a^\ast(g_m)\cdots a^\ast(g_1)a(f_1)\cdots a(f_n))
&=\omega_T(a^\ast(\e^{\i tH}g_m)\cdots a^\ast(\e^{\i tH}g_1)
a(\e^{\i tH}f_1)\cdots a(\e^{\i tH}f_n))\\
&=\delta_{nm}\det\{(\e^{\i tH}f_i,T\e^{\i tH}g_j)\}_{i,j=1,\ldots,n}\\
&=\delta_{nm}\det\{(f_i,T_tg_j)\}_{i,j=1,\ldots,n}\\
&=\omega_{T_t}(a^\ast(g_m)\cdots a^\ast(g_1)a(f_1)\cdots a(f_n)),
\end{align*}
where $T_t\equiv\e^{-\i tH}T\e^{\i tH}$. We conclude that
\beq
\omega_T\circ\tau^t=\omega_{T_t}.
\label{TtStuf}
\eeq
Furthermore if 
\begin{equation}
T^+\equiv\wlim_{t\to+\infty}T_t,
\label{wlimT}
\end{equation}
exists, then
\begin{align*}
\lim_{t\to+\infty}\omega_T\circ\tau^t(a^\ast(g_m)&\cdots a^\ast(g_1)a(f_1)\cdots a(f_n))
=\omega_{T^+}(a^\ast(g_m)\cdots a^\ast(g_1)a(f_1)\cdots a(f_n)).
\end{align*}
The mapping $A\mapsto\omega_{T_t}(A)$, being uniformly continuous in $t\in\mathbb R$ and
the monomials $a^\ast(g_m)\cdots$ $a(f_n)$ forming a total subset of $\mathcal O$, 
we can conclude that
\begin{equation}
\Sigma^+(\omega_T)=\{\omega_{T^+}\}.
\label{SigmaPlusForm}
\end{equation}

\subsubsection{Multi-channel scattering}

In the preceding subsection, we reduced the problem of the existence and uniqueness of the 
NESS associated with a gauge invariant quasi-free state on $\mathcal O$ to the existence of the 
weak limit (\ref{wlimT}). To control this limit we will, in this subsection, implement the theory of 
multi-channel scattering. We present here a simplified version. A more detailed 
discussion will be made in Section \ref{ScattSect}.

Let $\tau_{\mathcal R_k}$ denotes the $C^\ast$-dynamics generated by $H_k$ on 
$\mathcal O_{\mathcal R_k}$. Since the canonical injections 
$J_k:\mathfrak h_k\to\mathfrak h$ are partial isometries, we have 
$J_j^\ast J_k=\delta_{jk}I_{\mathfrak h_k}$ and 
$J_kJ_k^\ast=1_k$ is the orthogonal 
projection of $\mathfrak h$ onto the subspace $\mathfrak h_k$.

Since $HJ_k-J_kH_k=(H_\cS+V)J_k$ is trace class, it follows from
Pearson's theorem (Theorem XI.7 in \cite{RS3}) that the partial M\o ller operators
\beq
\Omega_k^\pm\equiv\slim_{t\to\pm\infty}\e^{\i tH}J_k\e^{-\i tH_k}P_\ac(H_k),
\label{FridayMorning}
\eeq
exist. Like the ordinary M\o ller operators, they satisfy the intertwining relations
$f(H)\Omega_k^\pm=\Omega_k^\pm f(H_k)$ which imply in particular that 
$\Ran(\O_k^\pm)\subset\fh_\ac(H)$. For any $u,v\in\fh$ one has
\begin{equation}
\begin{split}
(\Omega_j^\pm u,\Omega_k^\pm v)
&=\lim_{t\to\pm\infty}(\e^{\i tH}J_j\e^{-\i tH_j}P_\ac(H_j)u,\e^{\i tH}J_k\e^{-\i tH_k}P_\ac(H_k)v)\\
&=\lim_{t\to\pm\infty}(J_j\e^{-\i tH_j}P_\ac(H_j)u,J_k\e^{-\i tH_k}P_\ac(H_k)v)\\
&=\lim_{t\to\pm\infty}\delta_{jk}(\e^{-\i tH_k}P_\ac(H_k)u,\e^{-\i tH_k}P_\ac(H_k)v)\\
&=\delta_{jk}(u,P_\ac(H_k)v),
\end{split}
\label{OstarO}
\end{equation}
from which we conclude that $\Omega_k^\pm$ is a partial isometry with initial space 
$\mathfrak h_{k,\ac}(H_k)$ and final space $\Ran(\Omega_k^\pm)$. Moreover,
the subspaces $\Ran(\Omega_k^\pm)$ are orthogonal to each other.

Pearson's theorem also implies the existence of the strong limits
\beq
W_k^\pm\equiv\slim_{t\to\pm\infty}\e^{\i tH_k}J_k^\ast\e^{-\i tH}P_\ac(H).
\label{Ostarslim}
\eeq
Repeating the arguments of the previous paragraph, we obtain that $W_k^\pm$ is a partial
isometry with initial space $\fh_\ac(H)$ and final space $\Ran(W_k^\pm)\subset\fh_\ac(H_k)$.
Thus, one has
\begin{align*}
(W_k^\pm u,v)&=\lim_{t\to\pm\infty}(\e^{\i tH_k}J_k^\ast\e^{-\i tH}P_\ac(H)u,P_\ac(H_k)v)\\
&=\lim_{t\to\pm\infty}(P_\ac(H)u, \e^{\i tH}J_k\e^{-\i tH_k}P_\ac(H_k)v)=(u,\Omega_k^\pm v),
\end{align*}
which shows that $W_k^\pm=\Omega_k^{\pm\ast}$.

We note that $\sum_kJ_kJ_k^\ast=\sum_k1_k=I-1_\cS$, where $1_\cS$ denotes the orthogonal 
projection of $\fh$ onto $\fh_\cS$. Since $\fh_\cS$ is finite dimensional,
$1_\cS$ is compact and it follows from the Riemann-Lebesgue
lemma, Eq.~\eqref{RLlemma}, that
\begin{align*}
\sum_k(u,\Omega_k^\pm\Omega_k^{\pm\ast}v)=\sum_k(W_k^\pm u,W_k^\pm v)
&=\sum_k\lim_{t\to\pm\infty}(\e^{\i tH_k}J_k^\ast\e^{-\i tH}P_\ac(H)u,
\e^{\i tH_k}J_k^\ast\e^{-\i tH}P_\ac(H)v)\\
&=\sum_k\lim_{t\to\pm\infty}(J_k^\ast\e^{-\i tH}P_\ac(H)u,J_k^\ast\e^{-\i tH}P_\ac(H)v)\\
&=\sum_k\lim_{t\to\pm\infty}(\e^{-\i tH}P_\ac(H)u,J_kJ_k^\ast\e^{-\i tH}P_\ac(H)v)\\
&=(u,P_\ac(H)v)-\lim_{t\to\pm\infty}(\e^{-\i tH}P_\ac(H)u,1_\cS\e^{-\i tH}P_\ac(H)v)\\
&=(u,P_\ac(H)v).
\end{align*}
Thus, one has
\beq
\sum_k\Omega^{\pm}_k\Omega^{\pm\ast}_k=P_\ac(H),
\label{OOstar}
\eeq
which implies that the full M\o ller operators
$$
\Omega^\pm:\fh\ni u\mapsto\sum_k\Omega_k^\pm 1_k u,
$$
is a partial isometry with initial space $\fh_\ac(H_\cS\oplus H_\cR)=\oplus_k\fh_\ac(H_k)$ and
final space $\fh_\ac(H)=\fh$. The scattering operator
$S=\Omega^{+\ast}\Omega^-:\oplus_k\mathfrak h_{\ac}(H_k)\to\oplus_k\mathfrak h_{\ac}(H_k)$
has a block matrix structure $S=[S_{jk}]$ where
$$
S_{jk}=\Omega^{+\ast}_j\Omega_k^{-}:\mathfrak h_{\ac}(H_k)\to\mathfrak h_{\ac}(H_j).
$$
It follows from Eq.~\eqref{OstarO} and \eqref{OOstar} that
$$
(S^\ast S)_{jk}
=\sum_l(\Omega^{+\ast}_l\Omega_j^{-})^\ast \Omega^{+\ast}_l\Omega_k^{-}
=\Omega_j^{-^\ast}\left(\sum_l\Omega^{+}_l\Omega^{+\ast}_l\right)\Omega_k^{-}
=\Omega_j^{-^\ast}\Omega_k^{-}=\delta_{jk}P_\ac(H_k),
$$
which shows that $S$ is unitary. 

\subsubsection{The NESS}

Fix $\bbeta=(\beta_1,\ldots,\beta_M)\in\rr_+^M$, $\bmu=(\mu_1,\ldots,\mu_M)\in\rr^M$ and
set $T_k=(I+\e^{\beta_k(H_k-\mu_k)})^{-1}$. Let $T_\cS\in\cB(\fh_\cS)$ be such that $0<T_\cS<1$. Then
\beq
T=T_\cS\oplus T_\cR=T_\cS\oplus\left(\bigoplus_{j=1}^MT_k\right),
\label{Tinit}
\eeq
generates a modular, gauge invariant quasi-free state on $\mathcal O$ such that
$\omega_T(A)=\omega_{T_k}(A)$ for all $A\in\mathcal O_{\mathcal R_k}$.
It follows from Theorem~\ref{QFKMS} that $\omega_T$ describes a physical state
of the joint system $\cS+\cR$ in which each reservoir $\cR_k$ is in thermal equilibrium at 
inverse temperature $\beta_k$ and chemical potential $\mu_k$.

Since $T_k$ commutes with $\e^{\i tH_k}$ we may write
\begin{align*}
T_t=\e^{-\i tH}T\e^{\i tH}&=\e^{-\i tH}T_\cS\e^{\i tH}
+\sum_k\e^{-\i tH}J_k T_k J_k^\ast\e^{\i tH}\\
&=\e^{-\i tH}T_\cS\e^{\i tH}
+\sum_k\e^{-\i tH}J_k\e^{\i tH_k}T_k\e^{-\i tH_k} J_k^\ast\e^{\i tH},
\end{align*}
Since $T_\cS$ is compact, it follows from the Riemann-Lebesgue lemma,
Eq.~\eqref{FridayMorning} and~\eqref{Ostarslim} (recall that we assumed $\fh_\ac(H)=\fh$) that
\beq
\slim_{t\to+\infty}T_t=\sum_k\Omega_k^-T_k\Omega_k^{-\ast}=\Omega^-T_\cR\Omega^{-\ast}.
\label{slimform}
\eeq
so that Relation \eqref{SigmaPlusForm} holds with $T^+=\Omega^-T_\cR\Omega^{-\ast}$.

Let us now make connection with the $C^\ast$-scattering approach of Section \ref{CstarScattSubSect}.
For $f\in\fh$, one has
$$
\tau_0^{-t}\circ\tau^t(a^\#(f))=a^\#(J_\cS\e^{-\i tH_\cS}J_\cS^\ast\e^{\i tH}f)
+\sum_k a^\#(J_k\e^{-\i tH_k}J_k^\ast\e^{\i tH}f),
$$
where $J_\cS$ denotes the canonical injection $\fh_\cS\to\fh$. The same argument as before
and the continuity of the map $f\mapsto a^\#(f)$ yield
$$
\lim_{t\to\infty}\tau_0^{-t}\circ\tau^t(a^\#(f))=
\sum_k a^\#(J_kW_k^-f)=a^\#(\Omega^{-\ast}f)
=\Gamma(\Omega^-)^\ast a^\#(f)\Gamma(\Omega^-).
$$
The uniform continuity of the $\ast$-automorphisms $\tau_0^{-t}\circ\tau^t$ and the density of
polynomials in $a^\#$ in $\CAR(\fh)$ imply that
$$
\lim_{t\to\infty}\tau_0^{-t}\circ\tau^t(A)=\Gamma(\Omega^-)^\ast A\Gamma(\Omega^-),
$$
holds for any $A\in\CAR(\fh)$. Thus, the M\o ller morphism $\gamma^+$ exists and is given by
the Bogoliubov morphism
$$
\gamma^+(A)=\Gamma(\Omega^-)^\ast A\Gamma(\Omega^-).
$$
Its range is $\CAR(\Ran\Omega^{-\ast})=\CAR(\fh_\cR)=\cO_\cR$. The NESS can be written as
$$
\omega_{T^+}(A)=\omega_{T_\cR}(\gamma^+(A)),
$$
and it follows from Propositions \ref{QFErgoProp}, \ref{AlgNESSProp} that
$$
\lim_{t\to\infty}\nu(\tau^t(A))=\omega_{T^+}(A),
$$
holds for all $A\in\cO$ and all $\nu\in\mathcal{N}_{\omega_T}$. Note in particular that the NESS is 
independent of the initial state $T_\cS$ of the sample. Note also that the
above arguments extend without modification to the more general class of initial states $\omega_T$ such that $0<T<I$ and $J_k^\ast TJ_k=T_k$.

\subsubsection{Flux observables}
\label{Sect-cc}

In the framework of quasi-free fermionic systems, the total energy of reservoir $\mathcal R_k$ 
can be identified with the operator $\d\Gamma(H_k)$ (here and in the following, an operator $A_k$ 
acting on $\fh_k$ is identified with the operator $J_kA_kJ_k^\ast$ which acts on $\fh$). 
The energy flux leaving reservoir $\mathcal R_k$ is thus given by 
$$
\Phi_k^{\rm e}=
-\left.\frac{\d\ }{\d t}\e^{\i t\d\Gamma(H)}\d\Gamma(H_k)\e^{-\i t\d\Gamma(H)}\right|_{t=0}
=-\d\Gamma(\i[H,H_k]).
$$
Besides energy fluxes, we can also introduce particle fluxes. The number of particles in 
reservoir $\mathcal R_k$ being given by $\d\Gamma(1_k)$, the particle flux leaving this reservoir is
$$
\Phi_k^{\rm p}=
-\left.\frac{\d\ }{\d t}\,
\e^{\i t\d\Gamma(H)}\d\Gamma(1_k)\e^{-\i t\d\Gamma(H)}\right|_{t=0}
=-\d\Gamma(\i[H,1_k]).
$$
We note that all these flux observables have the same structure: each of them has the form
\beq
\Phi_k=\d\Gamma(\phi_k),
\label{fluxobs}
\eeq
where $\phi_k=-\i[H,Q_k]$ for a self-adjoint operator $Q_k$ commuting with $H_\cR$ and $H_\cS$.
Even though the second quantized charge $\mathcal{Q}_k=\d\Gamma(Q_k)$ does 
not belong to the algebra $\cO$, one has
$$
\phi_k=-\i[H_\cS+H_\cR+V,Q_k]=-\i[V,Q_k]
=\left((\chi_k,\,\cdot\,)\varphi_k+(\varphi_k,\,\cdot\,)\chi_k\right),
$$
where $\varphi_k=\i Q_k\delta_{0_k}$. Thus 
$$
\Phi_k=a^\ast(\varphi_k)a(\chi_k)+a^\ast(\chi_k)a(\varphi_k),
$$
is a self-adjoint element of $\cO$. 

Conservation of energy and particle number are expressed by the identities 
$$
\sum_k\Phi_k^{\rm e}=\d\Gamma(\i[H,H_\cS+v]),\qquad
\sum_k\Phi_k^{\rm p}=\d\Gamma(\i[H,1_\cS]).
$$
Indeed, since $\d\Gamma(H_\cS+V)$ and $\d\Gamma(1_\cS)$ belong to $\cO$, one has
\beq
\sum_k\nu(\Phi_k^{\rm e})=\left.\frac{\d\ }{\d t}\nu(\tau^t(\d\Gamma(H_\cS+V)))\right|_{t=0}=0,\qquad
\sum_k\nu(\Phi_k^{\rm p})=\left.\frac{\d\ }{\d t}\nu(\tau^t(\d\Gamma(1_\cS)))\right|_{t=0}=0,
\label{ECONS}
\eeq
for any $\tau$-invariant state $\nu$.

\subsubsection{Entropy production}

Since the NESS $\omega_{T^+}$ is independent of the initial state of the sample $\cS$,
let us assume that 
$$
T_\cS=(1+\e^{-K_\cS})^{-1},
$$
for some self-adjoint operator $K_\cS$ commuting with $H_\cS$. Then we can write 
$T=(1+\e^{-K})^{-1}$ with $K=K_\cS\oplus K_\cR$ and 
$$
K_\cR=-\bigoplus_j \beta_j(H_j-\mu_j1_j).
$$
It follows that $\omega_T$ is a KMS state at inverse temperature $\beta=-1$ for the group
of Bogoliubov automorphisms $\sigma^t(A)=\e^{\i t\d\Gamma(K)}A\e^{-\i t\d\Gamma(K)}$.
By Theorem~\ref{EPRODTheo}, we have
$$
\Ent(\omega_{T_t}|\omega_T)=\int_0^t\omega_{T_s}(\d\Gamma(\i[K,V]))\,\d s.
$$
In the limit $t\to+\infty$, we obtain
\begin{align*}
0\le-\lim_{t\to+\infty}\frac1t\Ent(\omega_{T_t}|\omega_T)
&=-\omega_{T^+}(\d\Gamma(\i[K,V]))\\
&=-\omega_{T^+}(\d\Gamma(\i[K_\cS,V]))
+\sum_{j=1}^M\omega_{T^+}(\d\Gamma(\i[\beta_j(H_j+\mu_j1_j),V]))\\
&=-\omega_{T^+}(\d\Gamma(\i[K_\cS,V]))
-\sum_{j=1}^M\omega_{T^+}(\d\Gamma(\i[H,\beta_j(H_j+\mu_j1_j)]))\\
&=-\omega_{T^+}(\d\Gamma(\i[K_\cS,V]))
-\sum_{j=1}^M\beta_j\omega_{T^+}(\Phi_j^{\rm h}),
\end{align*}
where $\Phi_j^{\rm h}=\Phi_j^{\rm e}-\mu_j\Phi_j^{\rm p}$ denotes the heat flux leaving
the $j$-th reservoir. Since the resulting inequality is valid for any $K_\cS$, we can conclude that
$\omega_{T^+}(\d\Gamma(\i[K_\cS,V]))$ vanishes for any choice of $K_\cS$. Thus, the total entropy flux $\sum_j\beta_j\Phi_j^{\rm h}$ entering 
the sample satisfies
$$
\sum_{j=1}^M\beta_j\omega_{T^+}(\Phi_j^{\rm h})\le0.
$$
It follows that the entropy production in the steady state is also independent 
of the choice of the initial state of the sample and satisfies the entropy balance equation
$$
\mathrm{Ep}(\omega_{T^+})=-\sum_{j=1}^M\beta_j\omega_{T^+}(\Phi_j^{\rm h}).
$$

\subsubsection{The Landauer-B\"uttiker formula}
\label{LBsection1}

Until the end of the next section $T$ is given by Eq. \eqref{Tinit}, $Q_k$ stands for
either $H_k$ or $1_k$ and $\phi_k=-\i[H,Q_k]$.

Formula (\ref{slimform}) allows us to compute the expectation of the current 
observable~\eqref{fluxobs}, associated to the charge $Q_k$, in the NESS $\omega_{T^+}$,
\begin{equation}
\omega_{T^+}(\Phi_k)=\tr_{\mathfrak h}(T^+\phi_k)
=\sum_{j=1}^M\tr_{\mathfrak h_j}(T_j\Omega_j^{-\ast}\phi_k\Omega_j^{-}).
\label{ScattNESS}
\end{equation}

The celebrated Landauer-B\"uttiker formula expresses the right hand side of this identity in terms of 
the scattering matrix $S=[S_{jk}]$. To write down this formula, we shall now describe the spectral 
representations of the reference Hamiltonians $H_k$.

Eq.~\eqref{Udef} defines unitary operators
\beq
U_k:\fh_k\to L^2([-1,1],\d\varepsilon),
\label{Ukdef}
\eeq
such that $(U_kH_ku)(\varepsilon)=\varepsilon(U_ku)(\varepsilon)$. It follows that
$(U_kT_ku)(\varepsilon)=(1+\e^{\beta_k(\varepsilon-\mu_k)})^{-1}(U_ku)(\varepsilon)$. Since the 
scattering matrix $S=[S_{jk}]$ satisfies $H_jS_{jk}=S_{jk}H_k$, there exists measurable functions 
$s_{jk}(\varepsilon)$ such that  $(U_jS_{jk}u)(\varepsilon)=s_{jk}(\varepsilon)(U_ku)(\varepsilon)$ 
for almost all $\varepsilon\in[-1,1]$. The matrix $S(\varepsilon)=[s_{jk}(\varepsilon)]$ is
the so called {\sl on-shell scattering matrix} at energy $\varepsilon$.

The Landauer-B\"uttiker formula for the energy currents reads
\beq
\omega_{T^+}(\Phi_{k}^{\rm e})=\sum_{j=1}^M\int_{-1}^1\mathcal{T}_{kj}(\varepsilon)\varepsilon
\left(\frac1{1+\e^{\beta_k(\varepsilon-\mu_k)}}-\frac1{1+\e^{\beta_j(\varepsilon-\mu_j)}}\right)
\frac{\d\varepsilon}{2\pi},
\label{firstLBe}
\eeq
where $\mathcal{T}_{kj}(\varepsilon)=|\delta_{kj}-s_{kj}(\varepsilon)|^2$ is the so called
{\sl transmittance matrix.} A similar formula holds for the particle current
\beq
\omega_{T^+}(\Phi_{k}^{\rm p})=\sum_{j=1}^M\int_{-1}^1\mathcal{T}_{kj}(\varepsilon)
\left(\frac1{1+\e^{\beta_k(\varepsilon-\mu_k)}}-\frac1{1+\e^{\beta_j(\varepsilon-\mu_j)}}\right)
\frac{\d\varepsilon}{2\pi}.
\label{firstLBc}
\eeq

It is instructive to recover the energy conservation identity~\eqref{ECONS} from Eq.~\eqref{firstLBe}.
To this end, we remark that the unitarity of the $S$-matrix, $\sum_m\bar{s}_{mj}(\varepsilon)s_{mk}(\varepsilon)=\delta_{jk}$, implies the sum rule
$$
\sum_j\left(\mathcal{T}_{kj}(\varepsilon)-\mathcal{T}_{jk}(\varepsilon)\right)=0,
$$
for almost every $\varepsilon\in[0,1]$ and every $k\in\{1,\ldots,M\}$.
The Landauer-B\"uttiker formula~\eqref{firstLBe} thus yields
\beq
\sum_k\omega_{T^+}(\Phi_{k}^{\rm e})=\sum_{j,k=1}^M\int_{-1}^1
\left(\mathcal{T}_{kj}(\varepsilon)-\mathcal{T}_{jk}(\varepsilon)\right)
\varepsilon
\frac1{1+\e^{\beta_k(\varepsilon-\mu_k)}}
\frac{\d\varepsilon}{2\pi}=0.
\label{eCons}
\eeq
The particle number conservation identity
\beq
\sum_k\omega_{T^+}(\Phi_{k}^{\rm p})=0,
\label{cCons}
\eeq
follows similarly from Eq.~\eqref{firstLBc}.

A formula expressing the electric current through a sample connected to two electronic reservoirs  in 
terms of scattering data was first proposed by Landauer \cite{L1,L2}. Similar formulas for more than 
two reservoirs were obtained later by Fischer and Lee \cite{FL}, Langreth and Abrahams \cite{LA} 
and B\"uttiker and his coworkers \cite{BILP,B1,B2}.
Anderson and Engquist \cite{AE} and Sivan and Imry \cite{SI} have also considered the case of
energy transport. We refer to \cite{Da,I,IL} for more exhaustive references to the enormous
physical literature on the subject. 

We note however that physicists usually assume relaxation to a unique NESS and derive their
formula from  this assumption. Mathematical proofs of relaxation to a unique NESS and of the
related Landauer-B\"uttiker formula first appeared in \cite{AJPP1} for a simple special case and in
\cite{AJPP2,N} in more general settings.

The Landauer-B\"uttiker formula can be used to compute the conductance matrix, and more generally 
the Onsager matrix which expresses the steady state energy/particle currents in terms of temperature 
and chemical potential differentials to first order in these differentials (linear response theory). 
Let $\bar\beta$ and $\bar\mu$ denote equilibrium values of the inverse temperature and
chemical potential and denote by
\beq
X^{\rm e}_j=\bar\beta-\beta_j,\qquad X^{\rm p}_j=\beta_j\mu_j-\bar\beta\bar\mu,
\label{XYdef}
\eeq
the thermodynamic forces which describe departures from the equilibrium situation.
The Onsager matrix
$L=[L_{kj}^{{\rm a}\,{\rm b}}]_{{\rm a},{\rm b}\in\{{\rm e},{\rm p}\}; j,k\in\{1,\ldots,M\}}$ is defined by
$$
L_{kj}^{{\rm a}\,{\rm b}}=\left.\partial_{X^{\rm b}_j}\omega_{T^+}(\Phi_k^{\rm a})\right|_{X=0},
$$
where we have set $X=(X^{\rm e}_1,\ldots,X^{\rm e}_M,X^{\rm p}_1,\ldots,X^{\rm p}_M)\in\rr^{2M}$.
Thus, linear response to the thermodynamic forces $X$ is given by
$$
\omega_{T^+}(\Phi_k^{\rm a})=\sum_{{\rm b},j}L_{kj}^{{\rm a}\,{\rm b}}X^{\rm b}_j
+\mathcal{O}(|X|^2).
$$
Since the energy/particle number conservation identities~\eqref{eCons}/\eqref{cCons} imply
$\sum_k L_{kj}^{{\rm a}\,{\rm b}}=0$, one has 
$$
L_{jj}^{{\rm a}\,{\rm b}}=-\sum_{k\not=j}L_{kj}^{{\rm a}\,{\rm b}},
$$
and it is a simple exercise to differentiate Eq.~\eqref{firstLBe}, \eqref{firstLBc} to obtain, for $j\not=k$,
\beq
\begin{array}{rclcrcl}
\ds L_{kj}^{{\rm e}\,{\rm e}}&=&\ds\int_{-1}^1\mathcal{T}_{kj}(\varepsilon)
\varepsilon^2
f_{\rm eq}(1-f_{\rm eq})
\frac{\d\varepsilon}{2\pi},&\qquad&
\ds L_{kj}^{{\rm e}\,{\rm p}}&=&\ds\int_{-1}^1\mathcal{T}_{kj}(\varepsilon)
\varepsilon
f_{\rm eq}(1-f_{\rm eq})
\frac{\d\varepsilon}{2\pi},\\[12pt]
\ds L_{kj}^{{\rm p}\,{\rm e}}&=&\ds\int_{-1}^1\mathcal{T}_{kj}(\varepsilon)
\varepsilon
f_{\rm eq}(1-f_{\rm eq})
\frac{\d\varepsilon}{2\pi},&\qquad&
\ds L_{kj}^{{\rm p}\,{\rm p}}&=&\ds\int_{-1}^1\mathcal{T}_{kj}(\varepsilon)
f_{\rm eq}(1-f_{\rm eq})
\frac{\d\varepsilon}{2\pi},
\end{array}
\label{LLB}
\eeq
where
$$
f_{\rm eq}(\varepsilon)=\frac1{1+\e^{\bar\beta(\varepsilon-\bar\mu)}},
$$
is the equilibrium Fermi-Dirac distribution. A first rigorous proof of these linearized Lan\-dauer-B\"uttiker 
formulas was obtained in \cite{CJM}. 

The open system $\cS+\cR$ is time reversal invariant (TRI) if there exists an anti-unitary involution 
$\theta_\cS$ on $\fh_\cS$ such that $\theta_\cS H_\cS\theta_\cS^\ast=H_\cS$ and 
$\theta_\cS\chi_k=\chi_k$. 
Let $\theta_\cR=\oplus_k\theta_k$ where $\theta_k$ is the complex conjugation on $\fh_k=\ell^2(\nn)$.
Then $\theta_kH_k\theta_k^\ast=H_k$ and $\theta=\theta_\cS\oplus\theta_\cR$ satisfies 
$\theta H\theta^\ast=H$.  Thus  $\theta_k\e^{\i tH_k}\theta_k^\ast=\e^{-\i tH_k}$ and
 $\theta\e^{\i tH}\theta^\ast=\e^{-\i tH}$ from which we conclude that
$$
\theta\Omega_k^\pm\theta_k^\ast
=\slim_{t\to\pm\infty}\theta\e^{\i tH}J_k\e^{-\i tH_k}\theta_k^\ast
=\slim_{t\to\pm\infty}\e^{-\i tH}J_k\e^{\i tH_k}=\Omega_k^\mp,
$$
and hence
$$
\theta_jS_{jk}\theta_k^\ast=\theta_j\Omega_j^{+\ast}\theta^\ast\theta\Omega_k^-\theta_k^\ast
=(\theta\Omega_j^+\theta_j^\ast)^\ast\theta\Omega_k^-\theta_k^\ast
=\Omega_j^{-\ast}\Omega_k^+=S_{jk}^\ast.
$$
Since the unitary map $U_k$ defined in Eq. \eqref{Udef}, \eqref{Ukdef} satisfies 
$(U_k\theta_ku)(\varepsilon)=\overline{(U_ku)(\varepsilon)}$, we conclude that the on-shell 
$S$-matrix $s(\varepsilon)=[s_{jk}(\varepsilon)]$ is symmetric, i.e., 
$s^\ast(\varepsilon)=\overline{s(\varepsilon)}$ for almost every $\varepsilon\in[-1,1]$.

If the system $\cS+\cR$ is TRI, then the transmittance matrix 
$\mathcal{T}(\varepsilon)=[\mathcal{T}_{jk}(\varepsilon)]$
is also symmetric, and the
linearized Landauer-B\"uttiker formulas \eqref{LLB} imply the Onsager reciprocity relations
\beq
L_{kj}^{{\rm a}\,{\rm b}}=L_{jk}^{{\rm b}\,{\rm a}}.
\label{OnsagRel}
\eeq

We shall give a proof of the Landauer-B\"uttiker formulas~\eqref{firstLBe}, \eqref{firstLBc} based
on the Levitov formula in the next section.
In Section \ref{LBForm} we shall prove a more general form of these formulas under appropriate
but physically reasonable hypotheses.
As opposed to the proofs in \cite{AJPP2,N} which use the abstract stationary approach
to scattering theory, we shall work within the framework of geometric, time-dependent scattering 
theory.

\subsubsection{Full counting statistics}
\label{Sect-FCS}

The total charge transferred from reservoir $\cR_k$ to the sample $\cS$ during the time interval $[0,t]$
can be expressed as an integral of the corresponding current
$$
\delta\mathcal{Q}_k(t)=-\left(\tau^t(\mathcal{Q}_k)-\mathcal{Q}_k\right)
=\int_0^t \tau^s(\Phi_k)\,\d s.
$$
Note that even though the second quantized charge $\mathcal{Q}_k$ does not belong to the 
algebra $\cO$ (and $\omega_T(\tau^t(\mathcal{Q}_k))$ is generally infinite for all $t$),
the charge transfer $\delta\mathcal{Q}_k(t)$ is a self-adjoint element of $\cO$ and
\beq
\omega_T(\delta\mathcal{Q}_k(t))=\int_0^t \omega_T(\tau^s(\Phi_k))\,\d s,
\label{ctrans}
\eeq
is finite for all $t$ and satisfies 
\begin{align*}
\lim_{t\to\infty}\frac1t\omega_T(\delta\mathcal{Q}_k(t))
=\lim_{t\to\infty}\frac1t\int_0^t\omega_T(\tau^s(\Phi_k))\,\d s
&=\lim_{t\to\infty}\omega_T(\tau^t(\Phi_k))\\
&=\omega_{T^+}(\Phi_k)
=\lim_{t\to\infty}\frac{\d\  }{\d t}\omega_T(\delta\mathcal{Q}_k(t)).
\end{align*}
We note however that since the observable $\tau^s(\Phi_k)$ and $\tau^{s'}(\Phi_k)$ do not commute
for $s\not=s'$, there is no obvious measurement process for the observable $\delta\mathcal{Q}_k(t)$.
Moreover, $\delta\mathcal{Q}_k(t)$ does not commute with $\delta\mathcal{Q}_j(t)$ for $k\not=j$,
making an exact joint measurement of the components of
$\delta\mathcal{Q}(t)=(\delta\mathcal{Q}_1(t),\ldots,\delta\mathcal{Q}_M(t))$ impossible.

In this section we discuss a more satisfactory approach to the charge transfer problem and derive
the Levitov formula which provides a complete description of the transport statistics
in the long time limit. The reader should consult \cite{BN,EHM} for a pedagogical introduction
and references to the physics literature. More mathematically oriented discussions and applications
to other models can be found in \cite{dR1,dR2,DdRM,JOPP,JPP2}.

\paragraph{Finite size approximation.}
Let us assume for a while that the sets $\cR_k$ are finite lattices $\{0,1,\ldots,R\}$
so that the one-particle Hilbert spaces $\fh_k=\ell^2(\cR_k)$ and hence $\fh$ are finite dimensional.
We also replace the one-particle Hamiltonians $H_k$ with the discrete Dirichlet Laplacians on 
$\ell^2(\cR_k)$. In this setup the second quantized charges
form a commuting family $\mathcal{Q}=(\mathcal{Q}_1,\ldots,\mathcal{Q}_M)$
of self-adjoint elements of $\cO=\CAR(\fh)=\cB(\Gamma^-(\fh))$. Thus, one can analyze charge transport by measuring the 
vector observable $\mathcal{Q}$ at time $0$ and at the later time $t$. The results of these two 
measurements are elements $q$ and $q'$ of the joint spectrum
$\spec(\mathcal{Q})=\spec(\mathcal{Q}_1)\times\cdots\times\spec(\mathcal{Q}_M)\subset\rr^M$.
The probability distribution of the charge differences $\delta q=q-q'$ obtained with this protocol is called full 
counting statistics (FCS) of the 
charge transport. To compute this distribution, we note that since the components of $\mathcal{Q}$ 
commute, there exists a spectral family $\{P_q\}_{q\in\spec(\mathcal{Q})}$ of orthogonal projections
such that
$$
f(\mathcal{Q})=\sum_{q\in\spec(\mathcal{Q})}f(q)P_q,
$$
for all functions $f:\spec(\mathcal{Q})\to\cc$. The probability for the measurement of $\mathcal{Q}$
at time $0$ to yield the result $q$ is given by $\omega_T(P_q)$. After the measurement, the state
of the system is given by 
$$
\omega(\,\cdot\,)=\frac{\omega_T(P_q\,\cdot\,P_q)}{\omega_T(P_q)},
$$
so that the probability for a subsequent measurement of $\mathcal{Q}$ at the later time $t$ to yield $q'$ is
$$
\omega(\tau^t(P_{q'}))=\frac{\omega_T(P_q\tau^t(P_{q'})P_q)}{\omega_T(P_q)}.
$$
Hence, the joint probability distribution of the pair $(q,q')$ is given by the Bayes formula
$$
\mathbb{P}_t(q,q')=\omega_T(P_q)\frac{\omega_T(P_q\tau^t(P_{q'})P_q)}{\omega_T(P_q)}
=\omega_T\left(P_q\tau^t(P_{q'})P_q\right).
$$
Following the argument leading to Eq.~\eqref{TtStuf}, one shows that for any $\alpha\in\rr^M$,
$$
\omega_T(\e^{\i\alpha\cdot\mathcal{Q}}A\e^{-\i\alpha\cdot\mathcal{Q}})=\omega_{T_\alpha}(A),
$$
where $T_\alpha=\e^{-\i\alpha\cdot Q}T\e^{\i\alpha\cdot Q}$, $Q=(Q_1,\ldots,Q_M)$,
$\alpha\cdot Q=\alpha_1Q_1+\cdots+\alpha_MQ_M$, and 
$\alpha\cdot\mathcal{Q}=\d\Gamma(\alpha\cdot Q)$.
Since $Q_k$ commutes with $T_k$  for all $k$ one has $T_\alpha=T$ and it follows that 
$\omega_T(\e^{\i\alpha\cdot\mathcal{Q}}A\e^{-\i\alpha\cdot\mathcal{Q}})=\omega_T(A)$.
Thus,
$$
\e^{\i\alpha\cdot(q-q')}\omega_T(P_qAP_{q'})
=\omega_T(\e^{\i\alpha\cdot\mathcal{Q}}P_qAP_{q'}\e^{-\i\alpha\cdot\mathcal{Q}})
=\omega_T(P_qAP_{q'}),
$$
and hence $\omega_T(P_qAP_{q'})=0$ for $q\not=q'$. Since $\sum_qP_q=I$, we get
\beq
\omega_T(P_qAP_{q})=\omega_T(P_qA)=\omega_T(AP_q),
\label{PAP}
\eeq
which allows us to write
$$
\mathbb{P}_t(q,q')=\omega_T\left(P_q\tau^t(P_{q'})\right).
$$
If $f$ is a polynomial in $M$ variables, then
$$
\mathbb{E}_t(f(\delta q))=\sum_{q,q'\in\spec(\mathcal{Q})}\mathbb{P}_t(q,q') f\left(q-q'\right)
=\omega_T\left(\mathbf{T}f\left(\mathcal{Q}-\tau^t(\mathcal{Q})\right)\right),
$$
where the time ordered observable $\mathbf{T}f\left(\mathcal{Q}-\tau^t(\mathcal{Q})\right)$ 
is obtained by substituting $X_k$ by $\mathcal{Q}_k$ and $Y_k$ by $\tau^t(\mathcal{Q}_k)$ in the 
expansion
$$
f\left(X-Y\right)=\sum_{\alpha,\beta}f_{\alpha,\beta}X_1^{\alpha_1}\cdots X_M^{\alpha_M}
Y_1^{\beta_1}\cdots Y_M^{\beta_M}.
$$
It follows in particular that
\beq
\mathbb{E}_t(\delta q_k)=\omega_T\left(\delta\mathcal{Q}_k(t)\right)
=\int_0^t\omega_T(\tau^s(\Phi_k))\,\d s,
\label{ctrans2}
\eeq
and, taking Eq.~\eqref{PAP} into account,
\beq
\mathbb{E}_t(\delta q_k \delta q_j)=
\omega_T\left(\delta\mathcal{Q}_k(t)\delta\mathcal{Q}_j(t)\right).
\label{ctrans3}
\eeq
Thus the moments of order one and two of the family of random variables $\delta q_k$ coincide
with the corresponding moments of the observables $\delta\mathcal{Q}_k(t)$ in the state $\omega_T$.
We stress however that this is no more the case for higher moments,
as shown by a simple calculation.

The full counting statistics is the distribution of $\delta q=q-q'$, that is
$$
\mathbb{P}_t(\delta q)=\sum_{\atop{q,q'\in\spec(\mathcal{Q})}{q-q'=\delta q}}\mathbb{P}_t(q,q').
$$
Its Laplace transform is given by
\beq
\begin{split}
\rr^M\ni\alpha\mapsto
\chi_t(\alpha)
&=\sum_{\delta q\in\spec(\mathcal{Q})-\spec(\mathcal{Q})}\mathbb{P}_t(\delta q)
\e^{\alpha\cdot\delta q}\\
&=\sum_{q,q'\in\spec(\mathcal{Q})}\omega_T\left(P_q\tau^t(P_{q'})\right)\e^{-\alpha\cdot(q'-q)}
=\omega_T(\e^{\alpha\cdot\mathcal{Q}}\tau^t(\e^{-\alpha\cdot\mathcal{Q}})).
\end{split}
\label{lastsunday}
\eeq
The function $\chi_t(\alpha)$ is the moment generating function of the random variable $\delta q$,
i.e.,
$$
\mathbb{E}_t(\delta q_k)
=\left.\frac{\partial\chi_t}{\partial\alpha_k}\right|_{\alpha=0},\qquad
\mathbb{E}_t(\delta q_k\delta q_j)=
\left.\frac{\partial^2\chi_t}{\partial\alpha_k\partial\alpha_j}\right|_{\alpha=0},
$$
etc.

Writing $\e^{\alpha\cdot\mathcal{Q}}\tau^t(\e^{-\alpha\cdot\mathcal{Q}})
=\Gamma(\e^{\alpha\cdot Q}\e^{\i t H}\e^{-\alpha\cdot Q}\e^{-\i t H})$, it follows from
Eq.~\eqref{QFGamma} that
$$
\chi_t(\alpha)=\det(I+T(\e^{\alpha\cdot Q}\e^{-\alpha\cdot Q_t}-I)),
$$
where we have set $Q_t=\e^{\i tH}Q\e^{-\i tH}$ (with an obvious abuse of notation).
Using the fact that $Q$ commutes with $T$, some elementary algebraic manipulations lead to
$\chi_t(\alpha)=\det(I+X_t(\alpha))$ where
$$
X_t(\alpha)=T^{1/2}\e^{\alpha\cdot Q/2}\left(
\e^{-\alpha\cdot Q_t}-\e^{-\alpha\cdot Q}
\right)\e^{\alpha\cdot Q/2}T^{1/2}.
$$
Note that $X_t(\alpha)$ is self-adjoint. Moreover, integrating its derivative w.r.t.\;$t$ yields the integral representation
\beq
X_t(\alpha)=\int_0^t
T^{1/2}\e^{\alpha\cdot Q/2}
\e^{\i sH}\i[V,\e^{-\alpha\cdot Q}]\e^{-\i sH}
\e^{\alpha\cdot Q/2}T^{1/2}\,\d s.
\label{finiteLev}
\eeq

\paragraph{Thermodynamic limit.}
At this point, we can investigate the thermodynamic limit $R\to\infty$ of our model. We use a
superscript ${\,}^{(R)}$ to denote the objects pertaining to the system with finite reservoirs of size $R$,
e.g., $\fh^{(R)}$ is the one-particle Hilbert space of the system with finite reservoirs. Let $J_R$ be the 
canonical injection of $\fh^{(R)}$ into the one-particle Hilbert space $\fh$ of the system with infinite 
reservoirs. We first note that the coupling $V^{(R)}$ is such that $J_RV^{(R)}J_R^\ast=V$.
We also observe that 
$$
\slim_{R\to\infty}J_RQ_k^{(R)}J_R^\ast=Q_k,\qquad(Q_k=H_k\text{ or }Q_k=1_k),
$$
from which we easily conclude that
$$
\slim_{R\to\infty}J_R\e^{\alpha\cdot Q^{(R)}}J_R^\ast=\e^{\alpha\cdot Q},\qquad
\slim_{R\to\infty}J_R\e^{\i tH^{(R)}}J_R^\ast=\e^{\i tH},\qquad
\slim_{R\to\infty}J_RT^{(R)}J_R^\ast=T,
$$
for any $\alpha\in\cc^M$ and $t\in\rr$. Rewriting Eq.~\eqref{finiteLev} as
$$
J_RX_t^{(R)}(\alpha)J_R^\ast=\int_0^t
J_RT^{(R)1/2}\e^{\alpha\cdot Q^{(R)}/2}
\e^{\i sH^{(R)}}\i[V,\e^{-\alpha\cdot Q^{(R)}}]\e^{-\i sH^{(R)}}
\e^{\alpha\cdot Q^{(R)}/2}T^{(R)1/2}
J_R^\ast\,\d s,
$$
inserting the identity $I^{(R)}=J_R^\ast J_R$ between each factors of the right hand side of this
formula and using the fact that $V$ is finite rank we obtain that
\beq
\begin{split}
\lim_{R\to\infty}J_R X_t^{(R)}(\alpha)J_R^\ast
&=\int_0^t
T^{1/2}\e^{\alpha\cdot Q/2}
\e^{\i sH}\i[V,\e^{-\alpha\cdot Q}]\e^{-\i sH}
\e^{\alpha\cdot Q/2}T^{1/2}\,\d s\\
&=T^{1/2}\e^{\alpha\cdot Q/2}\left(
\e^{-\alpha\cdot Q_t}-\e^{-\alpha\cdot Q}
\right)\e^{\alpha\cdot Q/2}T^{1/2}=X_t(\alpha),
\end{split}
\label{latethursday}
\eeq
holds in trace norm. This implies in particular that the right hand side of this identity is trace class, and it 
follows from the continuity property of the determinant (see e.g., Theorem 3.4 in \cite{S}) that
$$
\chi_t(\alpha)=
\lim_{R\to\infty}\chi_t^{(R)}(\alpha)
=\det(I+X_t(\alpha)).
$$
Since the function $\alpha\mapsto\chi_t(\alpha)$ is continuous, this pointwise convergence implies that the FCS $\mathbb{P}_t^{(R)}$ converges weakly to a probability measure $\mathbb{P}_t$ on $\rr^M$ such that
\beq
\int\e^{\alpha\cdot \delta q}\d\mathbb{P}_t(\delta q)=\chi_t(\alpha),
\label{chiform}
\eeq
for all $\alpha\in\cc^M$ (see, e.g., Theorem 26.3 and its Corollary in \cite{Bi}). We call $\mathbb{P}_t$ the FCS of the model with infinite reservoirs.
Note that Eq.~\eqref{ctrans2} and \eqref{ctrans3} also survive the thermodynamic limit, i.e.,
\begin{eqnarray}
\ds\int \delta q_k\,\d\mathbb{P}_t(\delta q)=&\ds\left.\frac{\partial\chi_t}{\partial\alpha_k}\right|_{\alpha=0}
&\ds=\omega_T(\delta\mathcal{Q}_k(t)),\label{ctrans4}\\[12pt]
\ds\int \delta q_k\delta q_j\,\d\mathbb{P}_t(\delta q)=&\ds\left.
\frac{\partial^2\chi_t}{\partial\alpha_k\partial\alpha_j}\right|_{\alpha=0}
&\ds=\omega_T(\delta\mathcal{Q}_k(t)\delta\mathcal{Q}_j(t)).
\label{ctrans5}
\end{eqnarray}

\paragraph{Large time limit.}
We shall now consider the large time asymptotics of the FCS. For $\alpha\in\rr^M$, $t\in\rr$ and
$u\in\fh$ one has
\begin{align*}
(u,(I+X_t(\alpha))u)&=(u,(I-T)u)
+(u,T^{1/2}\e^{\alpha\cdot Q/2}\e^{\i tH}\e^{-\alpha\cdot Q}\e^{-\i tH}\e^{\alpha\cdot Q/2}T^{1/2}u)\\
&\ge(u,(I-T)u)+\e^{-\sup\,\spec(\alpha\cdot Q)}(u,T^{1/2}\e^{\alpha\cdot Q}T^{1/2}u)\\
&\ge(u,(I-T)u)+\e^{-\sup\,\spec(\alpha\cdot Q)+\inf\,\spec(\alpha\cdot Q)}(u,Tu)\\
&\ge(u,(I-(1-\kappa(\alpha))T)u)\\
&\ge\|u\|^2-(1-\kappa(\alpha))(u,Tu)\ge\|u\|^2-(1-\kappa(\alpha))\|u\|^2= \kappa(\alpha)\|u\|^2,
\end{align*}
where $ \kappa(\alpha)=\e^{\inf\,\spec(\alpha\cdot Q)-\sup\,\spec(\alpha\cdot Q)}\in]0,1]$.
It follows that $I+\gamma X_t(\alpha)\ge\kappa(\alpha)>0$ for $\gamma\in[0,1]$ and hence 
that
$$
\frac{\d\ }{\d\gamma}\log\,\det(I+\gamma X_t(\alpha))
=\frac{\d\ }{\d\gamma}\tr\,\log(I+\gamma X_t(\alpha))
=\tr\left((I+\gamma X_t(\alpha))^{-1}X_t(\alpha)\right).
$$
Using Eq.~\eqref{latethursday}, the cyclicity of the trace and a change of integration variable allow 
us to write
\begin{align*}
\frac1t\log\chi_t(\alpha)&=\frac1t\int_0^1\d\gamma\int_0^t\d s\,\tr\left(
(I+\gamma X_t(\alpha))^{-1}T^{1/2}\e^{\alpha\cdot Q/2}
\e^{\i sH}\i[V,\e^{-\alpha\cdot Q}]\e^{-\i sH}
\e^{\alpha\cdot Q/2}T^{1/2}\right)\\
&=\int_0^1\d\gamma\int_0^1\d s\,\tr\left(Y_t(s,\alpha,\gamma)\i[V,\e^{-\alpha\cdot Q}]\right),
\end{align*}
where $Y_t(s,\alpha,\gamma)=\e^{-\i tsH}\e^{\alpha\cdot Q/2}T^{1/2}
(I+\gamma X_t(\alpha))^{-1}T^{1/2}\e^{\alpha\cdot Q/2}\e^{\i tsH}$ is easily seen to satisfy the
estimate
\beq
\|Y_t(s,\alpha,\gamma)\|\le\e^{2\sup\,\spec(\alpha\cdot Q)-\inf\,\spec(\alpha\cdot Q)}.
\label{Ybound}
\eeq
Elementary manipulations further yield
\begin{align*}
Y_t(s,\alpha,\gamma)&=\left(\e^{-\i tsH}\e^{-\alpha\cdot Q/2}T^{-1/2}
(I+\gamma X_t(\alpha))T^{-1/2}\e^{-\alpha\cdot Q/2}\e^{\i tsH}\right)^{-1}\\
&=\left(\e^{-\i tsH}\e^{-\alpha\cdot Q}(T^{-1}-\gamma)\e^{\i tsH}
+\gamma\e^{\i t(1-s)H}\e^{-\alpha\cdot Q}\e^{-\i t(1-s)H}\right)^{-1},
\end{align*}
and repeating the argument leading to Eq.~\eqref{slimform}, we can write
\begin{align*}
\slim_{t\to+\infty}\left(\e^{-\i tsH}\e^{-\alpha\cdot Q}(T^{-1}-\gamma)\e^{\i tsH}\right.
&\left.+\gamma\e^{\i t(1-s)H}\e^{-\alpha\cdot Q}\e^{-\i t(1-s)H}\right)\\
&=\Omega^-\e^{-\alpha\cdot Q}(T_\cR^{-1}-\gamma)\Omega^{-\ast}
+\gamma\Omega^+\e^{-\alpha\cdot Q}\Omega^{+\ast}\\
&=\Omega^-\left(\e^{-\alpha\cdot Q}(T_\cR^{-1}-\gamma)
+\gamma\Omega^{-\ast}\Omega^+\e^{-\alpha\cdot Q}\Omega^{+\ast}\Omega^-\right)\Omega^{-\ast}\\
&=\Omega^-\left(\e^{-\alpha\cdot Q}(T_\cR^{-1}-\gamma)
+\gamma S^\ast\e^{-\alpha\cdot Q}S\right)\Omega^{-\ast},
\end{align*}
for $s\in]0,1[$. The estimate \eqref{Ybound} allows us to conclude that
$$
\slim_{t\to+\infty}Y_t(s,\alpha,\gamma)
=\Omega^-\left(\e^{-\alpha\cdot Q}(T_\cR^{-1}-\gamma)
+\gamma S^\ast\e^{-\alpha\cdot Q}S\right)^{-1}\Omega^{-\ast}
=\Omega^-Y(\alpha,\gamma)\Omega^{-\ast},
$$
and an elementary calculation shows that
$$
Y(\alpha,\gamma)=\left(\e^{-\alpha\cdot Q}(T_\cR^{-1}-\gamma)
+\gamma S^\ast\e^{-\alpha\cdot Q}S\right)^{-1}
=\left(I+\gamma T_\cR(\e^{\alpha\cdot Q}S^\ast\e^{-\alpha\cdot Q}S-I)\right)^{-1}T_\cR\e^{\alpha\cdot Q}.
$$
It follows that
\begin{align*}
\lim_{t\to+\infty}\frac1t\log\chi_t(\alpha)
&=\int_0^1\int_0^1
\tr\left(\Omega^-Y(\alpha,\gamma)\Omega^{-\ast}\i[V,\e^{-\alpha\cdot Q}]\right)\,\d s\d \gamma\\
&=\int_0^1\tr_{\fh_\cR}\left(Y(\alpha,\gamma)\Omega^{-\ast}\i[V,\e^{-\alpha\cdot Q}]\Omega^{-}\right)
\,\d\gamma.
\end{align*}
To evaluate the right hand side of this identity we need some technical results which we shall prove
in Section~\ref{CurrentSectII}. By Theorem~\ref{TraceKernel}, the finite rank operator $C$ inside 
the trace has a integral representation
$$
(U_jJ^\ast_jCu)(\varepsilon)
=\sum_k\int_{-1}^1 c_{jk}(\varepsilon,\varepsilon')(U_kJ^\ast_ku)(\varepsilon')\,d\varepsilon',
$$
and its trace is given by
\beq
\tr_{\fh_\cR}(C)=\sum_k\int_{-1}^1 c_{kk}(\varepsilon,\varepsilon)\,\d\varepsilon.
\label{treece}
\eeq
Moreover, by Lemma~\ref{DiagKernel},
\beq
\sum_{jk}\int_{-1}^1
\overline{(U_jJ_j^\ast u)(\varepsilon)}c_{jk}(\varepsilon,\varepsilon)
(U_kJ_k^\ast w)(\varepsilon)\,\d\varepsilon=
\lim_{\eta\downarrow0}\frac{1}{2\pi}\int_{-\infty}^\infty\e^{-\eta|t|}(u,\e^{-\i tH_\cR}C\e^{\i tH_\cR}w)\,\d t,
\label{cmatrix}
\eeq
holds for $u,w$ in a dense subspace of $\fh_\cR$. The intertwining property of the M\o ller operator
and the fact that $Q$ commutes with $H_\cS\oplus H_\cR$ yield
$$
\e^{-\i tH_\cR}C\e^{\i tH_\cR}
=Y(\alpha,\gamma)\Omega^{-\ast}\e^{-\i tH}\i[V,\e^{-\alpha\cdot Q}]\e^{\i tH}\Omega^{-}
=-\frac{\d\ }{\d t}Y(\alpha,\gamma)\Omega^{-\ast}\e^{-\i tH}\e^{-\alpha\cdot Q}\e^{\i tH}\Omega^{-},
$$
so that an integration by parts leads to the Abelian mean
$$
\int_{-\infty}^\infty\e^{-\eta|t|}(u,\e^{-\i tH_\cR}C\e^{\i tH_\cR}w)\,\d t
=\eta\int_0^\infty\e^{-\eta t}\left(u,Y(\alpha,\gamma)
\Omega^{-\ast}(\e^{-\alpha\cdot Q_t}-\e^{-\alpha\cdot Q_{-t}})\Omega^{-}w\right)\,\d t.
$$
Proceeding as above, we get
$$
\slim_{t\to\pm\infty}\e^{-\alpha\cdot Q_{t}}=\Omega^\pm\e^{-\alpha\cdot Q}\Omega^{\pm\ast},
$$
and hence
\begin{align*}
\lim_{\eta\downarrow0}\int_{-\infty}^\infty\e^{-\eta|t|}(u,&\e^{-\i tH_\cR}C\e^{\i tH_\cR}w)\,\d t
=\left(u,Y(\alpha,\gamma)\Omega^{-\ast}(\Omega^+\e^{-\alpha\cdot Q}\Omega^{+\ast}
-\Omega^-\e^{-\alpha\cdot Q}\Omega^{-\ast})\Omega^{-}w\right)\\
&=\left(u,Y(\alpha,\gamma)(S^\ast\e^{-\alpha\cdot Q}S-\e^{-\alpha\cdot Q})w\right)\\
&=\left(u,\left(I+\gamma T_\cR(\e^{\alpha\cdot Q}S^\ast\e^{-\alpha\cdot Q}S-I)\right)^{-1}T_\cR
(\e^{\alpha\cdot Q}S^\ast\e^{-\alpha\cdot Q}S-I)w\right).
\end{align*}
Applying Lemma~\ref{CompareKernel}, Eq.~\eqref{cmatrix} allows us to conclude that the 
$M\times M$ matrix $c(\varepsilon)=[c_{jk}(\varepsilon,\varepsilon)]$ is given by
$$
c(\varepsilon)=\frac1{2\pi}
\left(I+\gamma t(\varepsilon)(\e^{q(\alpha;\varepsilon)}s(\varepsilon)^\ast
\e^{-q(\alpha;\varepsilon)}s(\varepsilon)-I)\right)^{-1} t(\varepsilon)
(\e^{q(\alpha;\varepsilon)}s(\varepsilon)^\ast
\e^{-q(\alpha;\varepsilon)}s(\varepsilon)-I),
$$
where $t(\varepsilon)$ is the diagonal matrix with entries 
$t_{kk}(\varepsilon)=(1+\e^{\beta_k(\varepsilon-\mu_k)})^{-1}$, $s(\varepsilon)=[s_{jk}(\varepsilon)]$ is
the on-shell scattering matrix and $q(\alpha;\varepsilon)=\sum_k\alpha_kq_k(\varepsilon)$ is the 
diagonal matrix with entries
$$
q_{kk}(\alpha;\varepsilon)=\left\{\begin{array}{ll}
\alpha_k&\text{for particle transport,}\\[4pt]
\varepsilon\alpha_k&\text{for energy transport.}
\end{array}\right.
$$
Eq.~\eqref{treece} becomes
$$
\tr_{\fh_\cR}(C)=\int_{-1}^1\tr_{\cc^M}(c(\varepsilon))\d\varepsilon
=\int_{-1}^1\frac{\d\ }{\d\gamma}\tr_{\cc^M}\log\left(
I+\gamma t(\varepsilon)(\e^{q(\alpha;\varepsilon)}s(\varepsilon)^\ast
\e^{-q(\alpha;\varepsilon)}s(\varepsilon)-I)\right)
\frac{\d\varepsilon}{2\pi},
$$
and integration over $\gamma$ yields the Levitov formula \cite{LL,LLL,Kl,ABGK,BN} 
\beq
e_+(\alpha)=
\lim_{t\to+\infty}\frac1t\log\chi_t(\alpha)=\int_{-1}^1\log\,\det{}_{\cc^M}\left(
I+ t(\varepsilon)(\e^{q(\alpha;\varepsilon)}s(\varepsilon)^\ast
\e^{-q(\alpha;\varepsilon)}s(\varepsilon)-I)\right)
\frac{\d\varepsilon}{2\pi}.
\label{LevitovForm}
\eeq
Observing that $X_t(\alpha)$ is an entire analytic function of $\alpha\in\cc^M$, it is not hard to show 
that the above limit holds for $\alpha$ in an open neighborhood of $\rr^M$ in $\cc^M$. 
Moreover, Eq.~\eqref{chiform} and H\"older's inequality imply that the function 
$\rr^M\ni\alpha\mapsto\log\chi_t(\alpha)$ is convex. Thus, the function
$\rr^M\ni\alpha\mapsto e_+(\alpha)$ is real analytic and convex. With  
$\mathbf{1}=(1,\ldots,1)\in\rr^M$ the matrix $q(\lambda\mathbf{1},\varepsilon)$ is a multiple of the 
identity for any $\lambda\in\rr$ so that the function $e_+(\alpha)$ satisfies
\beq
e_+(\alpha+\lambda \mathbf{1})=e_+(\alpha).
\label{tsym}
\eeq
This property is clearly related to the conservation of energy/particle number and is an instance of 
the translation symmetry discussed in \cite{AGMT} (see also \cite{JPW,JPP2}).

\paragraph{Applications.}
1. Our first application of Levitov formula is a derivation of the Landauer-B\"uttiker formulas.
Eq.~\eqref{ctrans4} and the convexity of the functions $t^{-1}\log\chi_t(\alpha)$ 
and $e_+(\alpha)$ imply that
$$
\lim_{t\to+\infty}\mathbb{E}_t\left(\frac{\delta q_k}t\right)=
\lim_{t\to+\infty}\frac1t\int \delta q_k\,\d\mathbb{P}_t(\delta q)
=\lim_{t\to+\infty}\left.\frac{\partial\ \ }{\partial\alpha_k}\frac1t\log\chi_t\right|_{\alpha=0}
=\left.\frac{\partial e_+}{\partial\alpha_k}\right|_{\alpha=0},
$$
(see, e.g., Theorem 25.7 in \cite{Ro}). Thus, it follows from Eq.~\eqref{ctrans2} that
$$
\omega_{T^+}(\Phi_k)=\lim_{t\to+\infty}\frac1t\int_0^t\omega_T(\tau^s(\Phi_k))
\,\d s
=\left.\frac{\partial e_+}{\partial\alpha_k}\right|_{\alpha=0}.
$$
It is a simple exercise to compute the derivative on the right hand side of this identity starting  from
the Levitov formula~\eqref{LevitovForm}. The result of this calculation
$$
\omega_{T^+}(\Phi_k)
=\int_{-1}^1\tr_{\cc^M}\left(t(\varepsilon)
(s(\varepsilon)^\ast q_k(\varepsilon) s(\varepsilon)-q_k(\varepsilon))
\right)\frac{\d\varepsilon}{2\pi},
$$
is easily recognized to be the Landauer-B\"uttiker formulas~\eqref{firstLBe}, \eqref{firstLBc}.

2. As a second application of Levitov formula, we show that it implies a large deviation principle
which gives quantitative estimates for fluctuations of order $1$ of the charge transfer rates
$\delta q_k/t$ around their mean values $\omega_{T^+}(\Phi_k)$ for large time.
Applying the G\"artner-Ellis theorem (see, e.g., Theorem 2.3.6 in \cite{DZ}), we conclude that the 
family of FCS $\{\mathbb{P}_t\}_{t\ge0}$ satisfies a large deviation principle, i.e., for any Borel set 
$A\subset\rr^M$, one has
$$
-\inf_{q\in A^{\rm int}}I(q)\le\liminf_{t\to+\infty}\frac1t\log\,\mathbb{P}_t\left(\frac{\delta q}t\in  A\right)\le
\limsup_{t\to+\infty}\frac1t\log\,\mathbb{P}_t\left(\frac{\delta q}t\in  A\right)\le-\inf_{q\in A^{\rm cl}}I(q),
$$
where $A^{\rm int}$/$A^{\rm cl}$ denotes the interior/closure of the set $A$ and
the rate function $I:\rr^M\to[0,\infty[$ is the Legendre transform of $e_+$,
$$
I(q)=\sup_{\alpha\in\rr^M}\left(\alpha\cdot q-e_+(\alpha)\right).
$$
It follows from Eq.~\eqref{tsym} that $I(q)=+\infty$ unless $\mathbf{1}\cdot q=0$, i.e., for any $a>0$
the probability $\mathbb{P}_t\left(\frac{\delta q}t\in  A\right)$ decays more rapidly than $\e^{-a t}$ as $t\to\infty$
unless the closure of $A$ intersects the hyperplane $\mathcal{X}=\{q\in\rr^M\,|\,\mathbf{1}\cdot q=0\}$
where the energy/particle number conservation is satisfied. It is therefore natural to decompose
$\rr^M=\mathcal{X}\oplus\rr\mathbf{1}$ and rewrite the large deviation principle as
$$
-\inf_{\hat q\in A^{\rm int}}I(\hat q)\le\liminf_{t\to+\infty}\frac1t\log\,\mathbb{P}_t\left(\frac{\delta q}t\in A\oplus\rr\mathbf{1}\right)\le
\limsup_{t\to+\infty}\frac1t\log\,\mathbb{P}_t\left(\frac{\delta q}t\in A\oplus\rr\mathbf{1}\right)\le-\inf_{\hat q\in A^{\rm cl}}I(\hat q),
$$
for any Borel set $A\subset\mathcal{X}$. Roughly speaking, this means that
$$
\mathbb{P}_t\left(\left\{\delta q\in\rr^M\,\bigg|\,\delta q-\left(\frac1M\sum_k\delta q_k\right)\mathbf{1}\simeq t\hat q\right\}\right)
\simeq\e^{-tI(\hat q)},
$$
for $\hat q\in\mathcal{X}$ and $t\to\infty$. One easily shows that $I(\hat q)\ge0$ with equality if and
and only if $\hat q=\omega_{T^+}(\Phi_k)$ (see, e.g. Lemma 2.3.9 in \cite{DZ}).

3. Our third application of Levitov formula links FCS to linear response
theory and more precisely to the Onsager matrix. We shall assume here that the system $\cS+\cR$ is TRI.

Applying the Levitov formula to the joint FCS $\mathbb{P}_t(q^{\rm e},q^{\rm p})$ of energy and
particle transport which corresponds to the choice of commuting family
$\mathcal{Q}=(\d\Gamma(h_1),\ldots,\d\Gamma(h_M),\d\Gamma(1_1),\ldots,\d\Gamma(1_M))$,
we obtain a generating function
$$
e_+(\alpha,\nu)=\lim_{t\to\infty}\frac1t\log\int\e^{\alpha\cdot \delta q^{\rm e}+\nu\cdot \delta q^{\rm p}}
\d\mathbb{P}_t(\delta q^{\rm e},\delta q^{\rm p}),
$$
given by
$$
e_+(\alpha,\nu)=\int_{-1}^1\log\,\det\left(
I+ t(\varepsilon)(\e^{q(\alpha,\nu;\varepsilon)}s(\varepsilon)^\ast
\e^{-q(\alpha,\nu\varepsilon)}s(\varepsilon)-I)\right)
\frac{\d\varepsilon}{2\pi},
$$
where $q(\alpha,\nu;\varepsilon)$ is the diagonal $M\times M$-matrix with entries
$q_{kk}(\alpha,\nu;\varepsilon)=\alpha_k\varepsilon+\nu_k$. It follows that the translation symmetry
$$
e_+(\alpha+\lambda\mathbf{1},\nu+\kappa\mathbf{1})=e_+(\alpha,\nu),
$$
holds for all $\alpha,\nu\in\rr^M$ and $\lambda,\kappa\in\rr$.

We note that
$$
t(\varepsilon)(I- t(\varepsilon))^{-1}=\e^{-\bar\beta(\varepsilon-\bar\mu)}
\e^{-q(X^{\rm e},X^{\rm p};\varepsilon)},
$$
where $X^{\rm e}$ and $X^{\rm p}$ are the thermodynamic forces defined in Eq.~\eqref{XYdef}.
Writing the Levitov formula as
$$
e_+(\alpha,\nu)
=\int_{-1}^1\left[\log\,\det\left(I- t(\varepsilon)\right)+
\log\,\det\left(I+t(\varepsilon)(I- t(\varepsilon))^{-1}
\e^{q(\alpha,\nu;\varepsilon)}s(\varepsilon)^\ast
\e^{-q(\alpha,\nu;\varepsilon)}s(\varepsilon)\right)
\right]\frac{\d\varepsilon}{2\pi},
$$
and using the fact that TRI implies the symmetry of the scattering matrix,
we observe that the second determinant on the right hand side is
\begin{align*}
\det\left(I+\e^{-\bar\beta(\varepsilon-\bar\mu)}\right.&\left.\e^{-q(X^{\rm e}-\alpha,X^{\rm p}-\nu;\varepsilon)}
s^\ast(\varepsilon)
\e^{-q(\alpha,\nu;\varepsilon)}s(\varepsilon)\right)\\
&=\det\left(I+\e^{-\bar\beta(\varepsilon-\bar\mu)}
\e^{-q(\alpha,\nu;\varepsilon)}s(\varepsilon)
\e^{-q(X^{\rm e}-\alpha,X^{\rm p}-\nu;\varepsilon)}s(\varepsilon)^\ast\right)\\
&=\det\left(I+
\e^{-\bar\beta(\varepsilon-\bar\mu)}
\e^{-q(\alpha,\nu;\varepsilon)}s^\ast(\varepsilon)
\e^{-q(X^{\rm e}-\alpha,X^{\rm p}-\nu;\varepsilon)}s(\varepsilon)\right)\\
&=\det\left(I+
\e^{-\bar\beta(\varepsilon-\bar\mu)}\e^{-q(X^{\rm e},X^{\rm p};\varepsilon)}
\e^{q(X^{\rm e}-\alpha,X^{\rm p}-\nu;\varepsilon)}s^\ast(\varepsilon)
\e^{-q(X^{\rm e}-\alpha,X^{\rm p}-\nu;\varepsilon)}s(\varepsilon)\right),
\end{align*}
which yields a quantum version of the generalized Evans-Searles symmetry
$$
e_+(\alpha,\nu)=e_+(X^{\rm e}-\alpha,X^{\rm p}-\nu).
$$
These symmetries play a central role in nonequilibrium statistical mechanics. The interested reader should consult \cite{MR,JPR} for reviews of the classical theory and \cite{K,dR1,DdRM,AGMT,JPW,JOPP,JPP2} for
its adaptation to the quantum world. We also refer to \cite{JOPS} for the link between the FCS of entropy
production and the hypothesis testing of the arrow of time.

Since
$$
\omega_{T^+}(\Phi_k^{\rm e})=\left.\frac{\partial e_+}{\partial\alpha_k}\right|_{\alpha=\nu=0},
\qquad
\omega_{T^+}(\Phi_k^{\rm p})=\left.\frac{\partial e_+}{\partial\nu_k}\right|_{\alpha=\nu=0},
$$
one can write the Onsager matrix as
\beq
\begin{array}{rclcrcl}
\ds L_{kj}^{{\rm e}\,{\rm e}}&=&\ds
\left.\frac{\partial^2 e_+}{\partial X^{\rm e}_j\partial\alpha_k}
\right|_{\alpha=\nu=0,X=0},&\qquad&
\ds L_{kj}^{{\rm e}\,{\rm p}}&=&\ds
\left.\frac{\partial^2 e_+}{\partial X_j^{\rm p}\partial\alpha_k}
\right|_{\alpha=\nu=0,X=0},\\[12pt]
\ds L_{kj}^{{\rm p}\,{\rm e}}&=&\ds
\left.\frac{\partial^2 e_+}{\partial X_j^{\rm e}\partial\nu_k}
\right|_{\alpha=\nu=0,X=0},&\qquad&
\ds L_{kj}^{{\rm p}\,{\rm p}}&=&\ds
\left.\frac{\partial^2 e_+}{\partial X_j^{\rm p}\partial\nu_k}
\right|_{\alpha=\nu=0,X=0}.
\end{array}
\label{Lform}
\eeq
If a function $f(x,y)$ is $C^2$ near $(x,y)=(0,0)\in\rr^M\times\rr^M$ and satisfies the symmetry
$f(x,y)=f(x,x-y)$, then  $(\partial_{y_k}f)(x,y)=-(\partial_{y_k}f)(x,x-y)$ and
$(\partial_{x_j}\partial_{y_k}f)(x,y)=-(\partial_{x_j}\partial_{y_k}f)(x,x-y)-(\partial_{y_j}\partial_{y_k}f)(x,x-y)$
so that
$$
(\partial_{x_j}\partial_{y_k}f)(0,0)=-\frac12(\partial_{y_j}\partial_{y_k}f)(0,0).
$$
Applying this result to the derivatives in Eq.~\eqref{Lform}, the
translation symmetry and the
generalized Evans-Searles symmetry yield the following linear response
formulas
$$
\begin{array}{rclcrcl}
\ds L_{kj}^{{\rm e}\,{\rm e}}&=&\ds
-\left.\frac12\frac{\partial^2 e_+}{\partial\alpha_j\partial\alpha_k}
\right|_{\alpha=\nu=0,X=0},&\qquad&
\ds L_{kj}^{{\rm e}\,{\rm p}}&=&\ds
-\left.\frac12\frac{\partial^2 e_+}{\partial\nu_j\partial\alpha_k}
\right|_{\alpha=\nu=0,X=0},\\[12pt]
\ds L_{kj}^{{\rm p}\,{\rm e}}&=&\ds
-\left.\frac12\frac{\partial^2 e_+}{\partial\alpha_j\partial\nu_k}
\right|_{\alpha=\nu=0,X=0},&\qquad&
\ds L_{kj}^{{\rm p}\,{\rm p}}&=&\ds
-\left.\frac12\frac{\partial^2 e_+}{\partial\nu_j\partial\nu_k}
\right|_{\alpha=\nu=0,X=0},
\end{array}
$$
where the right hand sides depend on the equilibrium FCS, i.e., on the function $e_+(\alpha,\nu)|_{X=0}$.

4. It is instructive to evaluate Levitov's formula for charge transport 
between 2 reservoirs $\cR_L$, $\cR_R$. In this situation the general form of the scattering matrix (i.e., of a unitary $2\times2$ matrix) is 
$$
s(\varepsilon)=\left[\begin{array}{cc}
(1-\mathcal{T})^{1/2}\e^{\i\theta}&\mathcal{T}^{1/2}\e^{\i(\kappa-\eta)}\\
\mathcal{T}^{1/2}\e^{\i(\theta+\eta)}&-(1-\mathcal{T})^{1/2}\e^{\i\kappa}
\end{array}\right]
$$
where $\mathcal{T}\in[0,1]$ is the transmittance from one reservoir to the other and
$\theta$, $\kappa$, $\eta$ are real phases, all depending on the energy $\varepsilon$. An elementary calculation leads to the following
expression of Levitov's formula
\begin{align*}
e_+(\nu_L,\nu_R)
&=\lim_{t\to\infty}\frac1t\log\int\e^{\nu_L \delta q_L^{\rm p}+\nu_R \delta q_R^{\rm p}}\,\d\mathbb{P}_t(\delta q_L^{\rm p},\delta q_R^{\rm p})\\
&=\int\log\left(p_0+p_+\e^{(\nu_L-\nu_R)}
+p_-\e^{-(\nu_L-\nu_R)}\right)\frac{\d\varepsilon}{2\pi},
\end{align*}
where
$$
p_+=t_L(1-t_R)\mathcal{T},\qquad p_-=t_R(1-t_L)\mathcal{T},
\qquad p_0=1-p_--p_+.
$$
We note that, as a consequence of the translation symmetry (i.e., charge conservation), one has
 $e_+(\nu_L,\nu_R)=\hat e_+(\nu_L-\nu_R)$ where
$$
\hat e_+(\nu)=
e_+(\nu,0)=\lim_{t\to\infty}\frac1t\log\int\e^{\nu\delta q_L^{\rm p}}\,\d\mathbb{P}_t(\delta q_L^{\rm p},\delta q_R^{\rm p}).
$$
If we further
specialize to the zero temperature case ($\beta_L=\beta_R=+\infty$)
then $t_L=1_{]-\infty,\mu_L]}$, $t_R=1_{]-\infty,\mu_R]}$ and
assuming $\mu_R<\mu_L$, we get $p_-=0$ and
$p_+=1_{]\mu_R,\mu_L]}\mathcal{T}$. It follows that
$$
\hat e_+(\nu)=\int_{\mu_R}^{\mu_L}\log\left(1-\mathcal{T}+\mathcal{T}\e^{\nu}\right)\frac{\d\varepsilon}{2\pi}.
$$
Neglecting the variation of the transmittance over the energy interval
$[\mu_R,\mu_L]$ we finally obtain
$$
\hat
e_+(\nu)=\frac{\Delta\mu}{2\pi}\log\left(1-\mathcal{T}+\mathcal{T}\e^{\nu}\right),
$$
with $\Delta\mu=\mu_L-\mu_R$.
This can be interpreted in the following way. Let $(\xi_j)_{j\in\nn^\ast}$ be a sequence of independent
identically distributed random variables with values in $\{0,1\}$ and
the law $\mathbb{P}(\xi_j=1)=\mathcal{T}$. Set
$$
\Xi_t=\sum_{j=1}^{[t/\tau]}\xi_j,
$$
with $\tau=2\pi/\Delta\mu$ (here $[\,\cdot\,]$ denotes the integer part). One easily computes
$$
\lim_{t\to\infty}\frac1t\log\mathbb{E}\left(\e^{\nu\Xi_t}\right)=\hat e_+(\nu).
$$
Thus, the Bernoulli process $\Xi_t$ and the FCS of the charge transfer $\delta q_L^{\rm p}$ shares the same large deviations. Loosely speaking,
$\xi_j$ is the total charge transferred from the left reservoir
to the right one in the time interval $[(j-1)\tau,j\tau]$ and the
binomial law
$$
\mathbb{P}_t(\delta q_L^{\rm p}=q)={t/\tau\choose q}
\mathcal{T}^q(1-\mathcal{T})^{t/\tau-q},
$$
holds for large $t$ (see, e.g., \cite{BN}).

\section{Commutators and Mourre Estimates}
\label{MourreChapter}

The commutator $[A,B]=AB-BA$ of two operators appears naturally in many problems of 
spectral theory and the use of commutators has a long history. Putnam's monograph \cite{Pu} 
is a good introduction to the first results in this domain. The works of Mourre \cite{M1,M2,M3} 
had a profound influence on the development of spectral analysis and scattering 
theory. They brought technical tools allowing for the proof of asymptotic
completeness of the $N$-body problem which had been a struggle for decades \cite{SS1,Gr,De1,SS2}. 
This section is a brief introduction to the elements of Mourre theory that we shall need in these notes.
The monograph \cite{ABG} provides a more detailed exposition (see also \cite{DG} and \cite {CFKS}).

\subsection{Commutators}
\label{CommutSect}

\subsubsection{The commutator $[\,\cdot\,,\,\cdot\,]$ on $\B(\mathcal H)\times\B(\mathcal H)$}

If $A$ and $B$ are bounded operators on the Hilbert space $\mathcal H$ then their commutator 
$[A,B]=AB-BA$ is also a bounded operator. We may thus define the operator
$$
\ad_A:B\mapsto\i[A,B],
$$
on $\B(\cH)$. We note that $\ad_A$ is bounded with $\|\ad_A\|\le2\,\|A\|$ so that
$\theta\mapsto\tau_A^\theta=\e^{\theta\ad_A}$ is an entire function and
$$
\frac{\d^k\ }{\d\theta^k}\tau_A^\theta(B)=\tau_A^\theta(\ad_A^k(B)),
$$
and
$$
\tau_A^\theta(B)=\sum_{k=0}^\infty\frac{\theta^k}{k!}\ad_A^k(B).
$$
Note that $\tau_A^\theta(B)=\e^{\i\theta A}B\e^{-\i\theta A}$. If $A\in\B(\mathcal H)$ is self-adjoint
then $\ad_A$ is a $\ast$-derivation of the $C^\ast$-algebra $\mathcal B(\mathcal H)$. In this case 
$\mathbb R\ni\theta\mapsto\e^{\i\theta\ad_A}$ is a real analytic (and thus strongly continuous) group 
of $\ast$-auto\-morphisms of $\B(\mathcal H)$.

\subsubsection{The commutator $[A,\,\cdot\,]$ on $\B(\mathcal H)$}

Domain related problems make the definition of the commutator of two unbounded operators
more delicate. However, Mourre theory which we shall need to develop the geometric scattering 
theory of quasi-free fermionic systems, is based on such commutators. In this section we discuss 
the definition of the commutator $[A,\,\cdot\,]$ with a self-adjoint operator $A$ and its relation to 
the regularity of the group of $\ast$-automorphisms of $\B(\mathcal H)$ generated by $A$.

Let $A$ be a self-adjoint operator on the Hilbert space $\mathcal H$ with domain $\Dom(A)$.
If $B\in\B(\mathcal H)$ and if there exists a constant $c$ such that
$$
|(Au,Bu)-(u,BAu)|\le c\, \|u\|^2,
$$
for all $u\in\Dom(A)$ then the sesquilinear form
$$
\Dom(A)\times\Dom(A)\ni\left<u,v\right>\mapsto
(Au,Bv)-(u,BAv),
$$
is continuous on a dense subspace of $\mathcal H\times\mathcal H$. Thus, it has a continuous
extension to $\mathcal H\times\mathcal H$ and there exists an operator $C\in\B(\mathcal H)$ such 
that
$$
(Au,Bv)-(u,BAv)=(u,Cv),
$$
for all $u,v\in\Dom(A)$. In this case we say that the commutator of $A$ and $B$ is bounded and we 
denote the operator $C$ by the symbol $[A,B]$. If it is possible to iterate this construction we write
$$
\mathrm{ad}_A^0B\equiv B,\qquad
\mathrm{ad}_A^kB\equiv\i[A,\mathrm{ad}_A^{k-1}B], (k=1,2,\ldots).
$$
\begin{Def} Let $A$ be a self-adjoint operator on the Hilbert space $\mathcal H$. For each integer 
$n\ge 0$ we define
$$
\mathcal B_A^n(\mathcal H)\equiv\{B\in\mathcal B(\mathcal H)\,|\,
\mathrm{ad}_A^k(B)\in\mathcal B(\mathcal H), k=0,1,\ldots,n\}.
$$
We remark that $\mathcal B_A^0(\mathcal H)=\mathcal B(\mathcal H)$. We also write 
$\mathcal B_A(\mathcal H)\equiv\mathcal B_A^1(\mathcal H)$.
\end{Def}

The following characterization of $\B_A(\mathcal H)$ will play an essential role.
\begin{Lem}\label{BUcomm} Let $A$ be a self-adjoint operator on the Hilbert space $\mathcal H$. 
For all $B\in\mathcal B(\mathcal H)$ the following conditions are equivalent.
\begin{enumerate}[(i)]
\item $B\in\mathcal B_A(\mathcal H)$.
\item $B\Dom(A)\subset\Dom(A)$ and there exists a constant $c$ such that
\begin{equation}\label{Comutbound}
  \|ABv-BAv\|\le c\|v\|,
\end{equation}
for all $v\in\Dom(A)$.
\item $B^\ast\Dom(A)\subset\Dom(A)$ and there exists a constant $c$ such that
\begin{equation}\label{Comutbound2}
\|B^\ast Au-AB^\ast u\|\le c\|u\|,
\end{equation}
for all $u\in\Dom(A)$.
\end{enumerate}
When one of these conditions is satisfied, we have
$$
[A,B]u=ABu-BAu,\qquad [A,B]^\ast u=B^\ast Au-AB^\ast u,
$$ 
for all $u\in\Dom(A)$.
\end{Lem}

\begin{demo} $(i)\Rightarrow(ii)$. Suppose $B\in\mathcal B_A(\mathcal H)$ and set $C=[A,B]$. 
We have that $(Au,Bv)=(u,(BA+C)v)$ for all $u,v\in\Dom(A)$. Since $A$ is self-adjoint, we deduce 
that $Bv\in\Dom(A)$ and that $(u,ABv)=(u,(BA+C)v)$. We thus have $B\Dom(A)\subset\Dom(A)$ 
and since $\Dom(A)$ is dense $Cv=ABv-BAv$ for all $v\in\Dom(A)$ and we can choose 
$c=\|C\|$ in (\ref{Comutbound}).

\noindent
$(ii)\Rightarrow(iii)$. If $B\Dom(A)\subset\Dom(A)$ and if (\ref{Comutbound}) holds, then for all 
$u,v\in\Dom(A)$ we have that
$$
(B^\ast u,Av)=(u,BAv)=(u,ABv)-(u,ABv-BAv)=(B^\ast Au,v)-(u,ABv-BAv).
$$
We deduce that $|(B^\ast u,Av)|\le \left(\|B\|\,\|Au\|+c\|u\|\right)\|v\|$, which allows us to conclude that 
$B^\ast u\in\Dom(A)$ and that $(AB^\ast u,v)-(B^\ast Au,v)=(u,ABv-BAv)$. In particular
$$
|(AB^\ast u-B^\ast Au,v)|\le c\|u\|\,\|v\|,
$$
shows that (\ref{Comutbound2}) holds.

\noindent
$(iii)\Rightarrow(i)$. If $B^\ast\Dom(A)\subset\Dom(A)$ and if (\ref{Comutbound2}) holds, then for all 
$u,v\in\Dom(A)$ we have
$$
|(Au,Bv)-(u,BAv)|=|(B^\ast Au-AB^\ast u,v)|\le c\,\|u\|\,\|v\|,
$$
and thus $B\in\mathcal B_A(\mathcal H)$. 

Furthermore,  since $\Dom(A)$ is dense in $\mathcal H$ we deduce from the fact that
$$
(u,Cv)=(u,ABv-BAv)=(B^\ast Au-AB^\ast u,v)=(C^\ast u,v),
$$
for all $u,v\in\Dom(A)$, that $Cv=ABv-BAv$ for all $v\in\Dom(A)$, and that 
$C^\ast u=B^\ast Au-AB^\ast u$ for all $u\in\Dom(A)$.\cqfd
\end{demo}

\begin{Lem}\label{commutderiv} Let $A$ be a self-adjoint operator on the Hilbert space $\mathcal H$
and let $n\ge 1$.
\begin{enumerate}[(i)]
\item $\mathcal B_A^n(\mathcal H)$ is a $\ast$-subalgebra of $\mathcal B(\mathcal H)$.
\item $\mathrm{ad}_A\colon\mathcal B_A^n(\mathcal H)\to\mathcal B_A^{n-1}(\mathcal H)$ is a 
$\ast$-derivation.
\end{enumerate} 
\end{Lem}

\begin{demo} It is obvious that $\mathcal B_A^n(\mathcal H)$ is a subspace of 
$\mathcal B(\mathcal H)$ and that $\mathrm{ad}_A$ is linear. Let $\mathcal C$ be a 
$\ast$-subalgebra of $\mathcal B(\mathcal H)$ and 
$$
\mathcal C_A\equiv\{B\in\mathcal C\,|\,\mathrm{ad}_A(B)\in\mathcal B(\mathcal H)\}.
$$
We show that $\mathcal C_A$ is a $\ast$-subalgebra of $\mathcal B(\mathcal H)$ and that 
$\mathrm{ad}_A$ is a $\ast$-derivation on $\mathcal C_A$. The facts that $B\in\mathcal C_A$ 
implies $B^\ast\in\mathcal C_A$ and that 
$$
\mathrm{ad}_A(B^\ast)=(\mathrm{ad}_A(B))^\ast,
$$
are immediate consequences of Lemma \ref{BUcomm}.

Let $B,C\in\mathcal C_A$. By Lemma \ref{BUcomm}, $B$ and $C$ preserve $\Dom(A)$. 
Thus so does $BC$. For all $u\in\Dom(A)$ we also have $ABu-BAu=[A,B]u$ and $ACu-CAu=[A,C]u$.
Thus,
$$
ABCu-BCAu=[A,B]Cu+A[B,C]u,
$$
and therefore
$$
\|ABCu-BCAu\|\le\left(\|[A,B]\|\,\|C\|+\|A\|\,\|[B,C]\|\right)\|u\|.
$$
Lemma \ref{BUcomm} allows us to conclude that $BC\in\mathcal C_A$ and that 
$\mathrm{ad}_A(BC)=\mathrm{ad}_A(B)C+B\,\mathrm{ad}_A(C)$.

The proof of the lemma now follows by induction on $n$, noting that
$$
\mathcal B_A^n(\mathcal H)
=\{B\in\mathcal B_A^{n-1}(\mathcal H)\,|\,\mathrm{ad}_A(B)\in\mathcal B(\mathcal H)\}.
$$
\cqfd
\end{demo}

The following result relates the iterated commutators $\ad_A^k$ to the regularity of the group of 
$\ast$-auto\-morphisms
$$
\tau_A^\theta(B)\equiv\e^{\i\theta A}B\e^{-\i\theta A},
$$
generated by $A$. To formulate it, we make the following definition.

\begin{Def} Let $A$ be a self-adjoint operator on a Hilbert space $\mathcal H$. An operator 
$B\in\mathcal B(\mathcal H)$ is of class $C^n(A)$ if, for all $u\in\mathcal H$, the function
$$
\begin{array}{rrcl}
\mathbb R&\to&\mathcal H\\
\theta&\mapsto&\tau_A^\theta(B)u,
\end{array}
$$
is of class $C^n(\mathbb R)$ in the norm topology of $\mathcal H$.
\end{Def}

\begin{Lem}\label{fcommut} Let $A$ be a self-adjoint operator and $B$ a bounded operator on the 
Hilbert space $\mathcal H$. The following conditions are equivalent.
\begin{enumerate}[(i)]
\item $B\in\mathcal B_A^n(\mathcal H)$.
\item $B$ and $B^\ast$ are of class $C^n(A)$.
\end{enumerate}
If one of these conditions is satisfied, then
$$
\frac{\d^k}{\d\theta^k}\tau_A^\theta(B)u=\tau_A^\theta(\ad_A^k(B))u,
$$
for all $u\in\mathcal H$ and $k\in\{0,1,\ldots,n\}$.
\end{Lem}

\begin{demo} $(ii)\Rightarrow(i)$ If $\theta\mapsto \tau_A^\theta(B)u$  and 
$\theta\mapsto \tau_A^\theta(B^\ast)u$ are of class $C^n$ for all $u\in\mathcal H$ then
$$
\lim_{h\to0}\frac{\tau_A^h(B)u-Bu}{h}=Du,\qquad
\lim_{h\to0}\frac{\tau_A^h(B^\ast)u-B^\ast u}{h}=\tilde Du,
$$
define two linear operators $D,\tilde D:\mathcal H\to\mathcal H$. Furthermore, for all 
$u,v\in\mathcal H$, we have
$$
(v,Du)=\lim_{h\to0}\frac{(v,\tau_A^h(B)u-Bu)}{h}=
\lim_{h\to0}\frac{(\tau_A^h(B^\ast)v-B^\ast v,u)}{h}=(\tilde Dv,u),
$$
that is to say that $D^\ast=\tilde D$. The Hellinger-Toeplitz theorem allows us to conclude that $D$ 
is bounded. For $u\in\Dom(A)$
$$
Du=\lim_{h\to0}\e^{\i h A}B\frac{\e^{-\i hA}u-u}{h}+\frac{\e^{\i hA}Bu-Bu}{h}
=-\i BAu+\lim_{h\to0}\frac{\e^{\i hA}Bu-Bu}{h},
$$
shows that $Bu\in\Dom(A)$ and $Du=\i ABu-\i BAu$. Lemma \ref{BUcomm} allows us to conclude 
that $B\in\mathcal B_A(\mathcal H)$ and $D=\mathrm{ad}_AB$.

If $u\in\mathcal H$ and $v=\e^{-\i\theta A}u$ then
$$
\lim_{h\to0}\frac{\tau_A^{\theta+h}(B)u-\tau_A^\theta(B)u}{h}=
\lim_{h\to0}\e^{\i\theta A}\frac{\tau_A^h(B)v-Bv}{h}=
\e^{\i\theta A}Dv=\tau_A^\theta(D)u,
$$
and it follows from our hypothesis that $\theta\mapsto\tau_A^\theta(D)u$ is of class $C^{n-1}$. 
A similar argument shows that the same is true for $\theta\mapsto\tau_A^\theta(D^\ast)u$. 
By iteration it is thus easy to conclude that
$$
\frac{\d^k\ }{\d\theta^k}\tau_A^\theta(B)u=
\tau_A^\theta\left(\mathrm{ad}_A^kB\right)u,
$$
and that $\mathrm{ad}_A^kB\in\mathcal B(\mathcal H)$ for $k=0,\ldots,n$ and in particular that 
$B\in\mathcal B_A^n(\mathcal H)$.

$(i)\Rightarrow(ii)$ Let $B\in\mathcal B_A^n(\mathcal H)$. By Lemma \ref{BUcomm}, we have 
$B\Dom(A)\subset\Dom(A)$ and thus for all $u\in\Dom(A)$ we have
$$
\tau_A^{\theta+h}(B)u-\tau_A^\theta(B)u
=\int_0^1\frac{\d\ }{\d s}\,\e^{\i(\theta+sh)A}B\e^{-\i(\theta+sh)A}u\,\d s
=h\int_0^1\e^{\i(\theta+sh)A}\ad_A(B)\e^{-\i(\theta+sh)A}u\,\d s.
$$
This relation extends by continuity to all $u\in\mathcal H$. Furthermore, the strong continuity of the 
unitary group generated by $A$ gives
\begin{align*}
\e^{\i(\theta+t)A}\ad_A(B)\e^{-\i(\theta+t)A}u
=\tau_A^\theta\left(\ad_A(B)\right)u
&+(\e^{\i(\theta+t)A}-\e^{\i\theta A})\ad_A(B)\e^{-\i\theta A}u\\
&+\e^{\i(\theta+t)A}\ad_A(B)(\e^{-\i(\theta+t) A}-\e^{-\i\theta A})u\\
&=\tau_A^\theta\left(\ad_A(B)\right)u+\mathrm{o}(1),
\end{align*}
as $t\to0$. This allows us to write, for all $u\in\mathcal H$,
$$
\frac{\tau_A^{\theta+h}(B)u-\tau_A^\theta(B)u}{h}
=\int_0^1\tau_A^{\theta+sh}\left(\mathrm{ad}_A(B)\right)u\,\d s
=\tau_A^\theta\left(\mathrm{ad}_A(B)\right)u+\mathrm{o}(1).
$$
We conclude that $\tau_A^\theta(B)u$ is of class $C^1$ and that 
$$
\frac{\d\ }{\d\theta}\tau_A^\theta(B)u=\tau_A^\theta\left(\mathrm{ad}_A(B)\right)u,
$$
with $\mathrm{ad}_A(B)\in\mathcal B_A^{n-1}(\mathcal H)$. By iterating this argument we conclude 
that $\tau_A^\theta(B)u$ is of class $C^n$. Lemma \ref{commutderiv} implies that 
$B^\ast\in\mathcal B_A^{n}(\mathcal H)$ and we conclude that $\tau_A^\theta(B^\ast)u$ is also of 
class $C^n$.
\cqfd
\end{demo}

\bigskip
The next lemma provides a strong approximation result for iterated commutators in 
$\B_A^n(\mathcal H)$.

\begin{Lem}\label{CommutApprox} Let $A$ be a self-adjoint operator on the Hilbert space 
$\mathcal H$. For $\varepsilon\in\mathbb R^\ast$, we set
$$
A_\varepsilon\equiv\frac{\e^{\i\varepsilon A}-I}{\i\varepsilon} \in\B(\mathcal H).
$$
For all $B\in\B(\mathcal H)$, the two following conditions are equivalent.
\begin{enumerate}[(i)]
\item $B\in\mathcal B_A^n(\mathcal H)$.
\item $\sup_{\varepsilon\in\mathbb{R}^\ast}\|\ad_{A_\varepsilon}^n(B)\|<\infty$.
\end{enumerate}
If one of these conditions is satisfied then
$$
\slim_{\varepsilon\to0}\,\ad_{A_\varepsilon}^n(B)=\ad_A^n(B).
$$
\end{Lem}

\begin{demo} One easily shows that
$$
\varepsilon^{-1}(\tau_A^\varepsilon-\mathrm{Id})(B)=
\ad_{A_\varepsilon}(B)\e^{-\i\varepsilon A},
$$
from which one deduces
\begin{equation}
\varepsilon^{-n}(\tau_A^\varepsilon-\mathrm{Id})^n(B)=
\ad_{A_\varepsilon}^n(B)\e^{-\i n\varepsilon A}.
\label{difftruc}
\end{equation}
Newton's formula
$$
(\tau_A^\varepsilon-\mathrm{Id})^n=\sum_{k=0}^n{n \choose k}(-1)^{n-k}
\tau_A^{k\varepsilon},
$$
thus allows us to write, for all $u,v\in\mathcal H$,
\begin{equation}
(u,\ad_{A_\varepsilon}^n(B)v)=
\sum_{k=0}^n{n \choose k}(-1)^{n-k}
\varepsilon^{-n}(u,\tau_A^{k\varepsilon}(B)
\e^{\i n\varepsilon A}v).
\label{adnform1}
\end{equation}
For all $u,v\in\Dom(A^\infty)\equiv\cap_{k>0}\Dom(A^k)$ the function
$$
\mathbb R\ni\theta\mapsto (u,\tau^{\theta}_A(B)v)=(\e^{-\i\theta A}u,B\e^{-\i\theta A}v),
$$
is of class $C^\infty$ and thus admits a Taylor expansion around $\theta=0$
\begin{equation}
(u,\tau^{\theta}_A(B)v)
=\sum_{j=1}^{n-1}\frac{\theta^j}{j!}\,a_j(u,v)
+\frac{\theta^n}{(n-1)!}\int_0^1
(1-s)^{n-1}a_n(\e^{-\i s\theta A}u,\e^{-\i s\theta A}v)\,\d s,
\label{tautaylor}
\end{equation}
where $a_j$ denotes the quadratic form defined on $\Dom(A^\infty)\times\Dom(A^\infty)$ by 
\begin{equation}
a_j(u,v)\equiv\left.\frac{\d^j}{\d\theta^j}(u,\tau^{\theta}_A(B)v)\right|_{\theta=0}
=\left.\frac{\d^j}{\d\theta^j}(\e^{-\i\theta A}u,B\e^{-\i\theta A}v)\right|_{\theta=0}
=\i^j\sum_{l=0}^j{j\choose l}(-1)^l(A^{j-l}u,BA^lv).
\label{adABform}
\end{equation}
Taking into account the fact that, for $j=0,1,\ldots,n-1$, we have
$$
\sum_{k=0}^n{n \choose k}(-1)^{n-k}k^j=\left.(x\partial_x)^j(x-1)^n\right|_{x=1}=0,
$$
we obtain, after inserting the series (\ref{tautaylor}) into (\ref{adnform1}), the formula
\begin{equation}
(u,\ad_{A_\varepsilon}^n(B)v)=
\sum_{k=1}^n{n \choose k}(-1)^{n-k}
\frac{nk^n}{n!}\!\!\int_0^1\!(1-s)^{n-1}
a_n(\e^{-\i ks\varepsilon A}u,\e^{\i(n-ks)\varepsilon A}v)\,\d s.
\label{adnform2}
\end{equation}

$(i)\Rightarrow(ii)$ If $B\in\B_A^n(\mathcal H)$ then $\ad_A^n(B)\in\B(\mathcal H)$.
Lemma \ref{fcommut} implies that $\theta\mapsto\tau_A^\theta(B)u$ is of class $C^n$ and that
$$
a_n(u,v)=(u,\ad_A^n(B)v).
$$
Formula (\ref{adnform2}) implies the bound
$$
|(u,\ad_{A_\varepsilon}^n(B)v)|\le c\, \|u\|\,\|v\|\,\|\ad_A^n(B)\|,
$$
for all  $u,v\in\Dom(A^\infty)$ and a constant $c$. We deduce that
$$
\sup_{\varepsilon\in\mathbb{R}^\ast}\|\ad_{A_\varepsilon}^n(B)\|<\infty,
$$
and that, for all $v\in\mathcal H$,
$$
\ad_{A_\varepsilon}^n(B)v=
\sum_{k=1}^n{n \choose k}(-1)^{n-k}
\frac{nk^n}{n!}\int_0^1(1-s)^{n-1}
\e^{\i ks\varepsilon A}\ad_A^n(B)\e^{\i(n-ks)\varepsilon A}v\,\d s.
$$
It is then easy to conclude that
$$
\lim_{\varepsilon\to0}\ad_{A_\varepsilon}^n(B)v=
\sum_{k=1}^n{n \choose k}(-1)^{n-k}
\frac{k^n}{n!}
\ad_A^n(B)v=\ad_A^n(B)v.
$$

$(ii)\Rightarrow(i)$  Suppose that $\sup_{\varepsilon\in\mathbb{R}^\ast}\|\ad_{A_\varepsilon}^n(B)\|=c<\infty$. 
For all $u,v\in\Dom(A^\infty)$ we have
$$
a_n(\e^{-\i ks\varepsilon A}u,\e^{\i(n-ks)\varepsilon A}v)
=\i^n\sum_{j=0}^n{n\choose j}(-1)^j
(\e^{-\i ks\varepsilon A}A^{n-j}u,B\e^{\i(n-ks)\varepsilon A}A^jv),
$$
from which we deduce
$$
\lim_{\varepsilon\to0}a_n(\e^{-\i ks\varepsilon A}u,\e^{\i(n-ks)\varepsilon A}v)=a_n(u,v).
$$
Formula (\ref{adnform2}) thus implies
$$
a_n(u,v)=\lim_{\varepsilon\to0}(u,\ad_{A_\varepsilon}^n(B)v),
$$
and therefore
$$
|a_n(u,v)|\le\,c \|u\|\,\|v\|.
$$
Also, by writing the Taylor expansion (\ref{tautaylor}) as 
$$
\sum_{j=0}^{n-1}\frac{\theta^j}{j!}a_j(u,v)=
(u,\tau_A^\theta(B)v)-
\frac{\theta^n}{(n-1)!}\int_0^1(s-1)^{n-1}a_n(\e^{-\i s\theta A}u,\e^{-\i s\theta A}v)\,\d s,
$$
we deduce that there exists constants $c_j$ such that
$$
|a_j(u,v)|\le c_j\,\|u\|\,\|v\|.
$$
In particular
$$
|a_1(u,v)|=|(Au,Bv)-(u,BAv)|\le c_1\,\|u\|\,\|v\|,
$$
implies  $B\in\B_A(\mathcal H)$ and $a_1(u,v)=(u,\ad_A(B)v)$ by Lemma \ref{BUcomm}. 
We can finish the proof by induction. If $B\in\B_A^j(\mathcal H)$ for $1\le j<n$ then
$$
|a_{j+1}(u,v)|=|(Au,\ad_A^j(B)v)-(u,\ad_A^j(B)Av)|\le c_{j+1}\|u\|\,\|v\|,
$$
shows that $B\in\B_A^{j+1}(\mathcal H)$.
\cqfd\end{demo}

\bigskip
We finish this subsection with two results concerning the expansion of commutators which will be 
very useful to us later on. The first is purely algebraic whereas the second, a simple consequence 
of the Helffer-Sj\"ostrand formula, is due to \cite{SiSo} (see also \cite{HS}).

\begin{Lem}\label{AfAinv}\label{ExactCummutExpansion} Let $A$ be a self-adjoint operator on a 
Hilbert space $\mathcal H$ and $B\in\B_A^n(\mathcal H)$. For all $z\in\Res(A)$ we have
\begin{equation}
[(A-z)^{-n},B]=\sum_{j=1}^n{n\choose j}
\i^j(A-z)^{-j}\ad_{A}^j(B)(A-z)^{-n}.
\label{ResolvCommutId}
\end{equation}
Furthermore, $B\Dom(A^n)\subset\Dom(A^n)$ and in particular, $A^nB(A-z)^{-n}\in\B(\mathcal H)$
for all $z\in\Res(A)$.
\end{Lem}
\begin{demo} We prove (\ref{ResolvCommutId}) by induction on $n$. For $n=1$ we have
$$
[(A-z)^{-1},B]=-(A-z)^{-1}[A,B](A-z)^{-1}={1\choose 1}\i(A-z)^{-1}\ad_A(B)(A-z)^{-1}.
$$
For $1\le k\le n$ we set
$$
C_k\equiv\sum_{j=1}^k{k\choose j}
\i^j(A-z)^{-j}\ad_{A}^j(B),
$$
and we assume that $[(A-z)^{-k},B]=C_k(A-z)^{-k}$ for all integers $k$ such that $1\le k<n$.

We may thus write
\begin{align*}
[(A-z)^{-k-1},B]&=[(A-z)^{-1}(A-z)^{-k},B]\\
&=(A-z)^{-1}[(A-z)^{-k},B]+[(A-z)^{-1},B](A-z)^{-k}\\
&=(A-z)^{-1}C_k(A-z)^{-k}+[(A-z)^{-1},B](A-z)^{-k}\\
&=C_k(A-z)^{-k-1}+[(A-z)^{-1},B+C_k](A-z)^{-k}\\
&=C_k(A-z)^{-k-1}-(A-z)^{-1}[A,B+C_k](A-z)^{-k-1}\\
&=\left(C_k-(A-z)^{-1}[A,B+C_k]\right)(A-z)^{-k-1},
\end{align*}
and we easily verify that
$$
C_k-(A-z)^{-1}[A,B+C_k]=C_{k+1},
$$
which finishes the induction step.

By taking the adjoint, the identity (\ref{ResolvCommutId}) becomes
$$
[B,(A-z)^{-n}]=\sum_{j=1}^n{n\choose j}
(-\i)^j(A-z)^{-n}\ad_{A}^j(B)(A-z)^{-j},
$$
from which we deduce
$$
(A-z)^nB(A-z)^{-n}=\sum_{j=0}^n{n\choose j}
(-\i)^j\ad_{A}^j(B)(A-z)^{-j}.
$$
\cqfd
\end{demo}

\begin{Theo}\label{CommutExpand} Let $A$ be a self-adjoint operator on the Hilbert space 
$\mathcal H$, $B\in\B_A^n(\mathcal H)$ and $f\in C^\infty(\mathbb R)$ such that 
\begin{equation}
\VERT f\VERT_{n,k}\equiv
\sum_{j=0}^n\sup_{x\in\mathbb R}|f^{(j)}(x)|+
\sum_{j=n}^{k+2}\int\langle x\rangle^{j-n-1} |f^{(j)}(x)|\,\d x<\infty,
\label{CommutExpandCond}
\end{equation}
for some $k\ge n$. Then
\begin{equation}
[f(A),B]=\sum_{j=1}^{n}\frac{\i^{j}}{j!}f^{(j)}(A)
\ad_A^j(B)+\mathcal R_n(f,A,B),
\label{CommutExpandForm}
\end{equation}
where the remainder is given by
\begin{equation}
\mathcal R_n(f,A,B)=
\frac{1}{2\pi\i^{n+1}}\int\bar\partial\tilde f(z)(A-z)^{-n}[\ad_A^n(B),(A-z)^{-1}]\,
\d\bar z\wedge\d z,
\label{RestnOne}
\end{equation}
where $\tilde f$ is the almost-analytic extension of $f$ of order $k$ defined by 
(\ref{quasianalyticextension}). Furthermore, if $B\in\B_A^{n+1}(\mathcal H)$, the 
remainder can also be written as
$$
\mathcal R_n(f,A,B)=
\frac{1}{2\pi\i^{n+2}}\int\bar\partial\tilde f(z)(A-z)^{-n-1}\ad_A^{n+1}(B)(A-z)^{-1}\,
\d\bar z\wedge\d z.
$$
\end{Theo}

\noindent{\bf Remark.} By taking the adjoint we obtain a similar formula
$$
[f(A),B]=-\sum_{j=1}^{n}\frac{(-\i)^{j}}{j!}
\ad_A^j(B)f^{(j)}(A)-\mathcal R_n(\bar f,A,B^\ast)^\ast.
$$

\begin{demo} We first consider $f\in C_0^\infty(\mathbb R)$. Let $\tilde f$ be the almost-analytic 
extension of order $n$ given by (\ref{quasianalyticextension}). The Helffer-Sj\"os\-trand formula
(\ref{HSFormula}) allows us to write
$$
[f(A),B]=\frac{1}{2\pi\i}\int\bar\partial\tilde f(z)[B,(A-z)^{-1}]\,\d\bar z\wedge\d z.
$$
A repeated use of the identity
\begin{align*}
[\ad_A^j(B),(A-z)^{-1}]&=(A-z)^{-1}[A,\ad_A^j(B)](A-z)^{-1}\\
&=-\i(A-z)^{-1}\ad_A^{j+1}(B)(A-z)^{-1}\\
&=-\i(A-z)^{-2}\ad_A^{j+1}(B)-\i(A-z)^{-1}[\ad_A^{j+1}(B),(A-z)^{-1}],
\end{align*}
and Formula (\ref{HSFormula}) lead to
\begin{align*}
[f(A),B]&=\sum_{j=1}^{n}\left(\frac{1}{2\pi\i}\int\bar\partial\tilde f(z)
(A-z)^{-1-j}\,\d\bar z\wedge\d z\right)(-\i)^{j}\ad_A^j(B)\nonumber\\
&+\frac{1}{2\pi\i^{n+1}}\int\bar\partial\tilde f(z)(A-z)^{-n}[\ad_A^{n}(B),(A-z)^{-1}]
\,\d\bar z\wedge\d z\\
&=\sum_{j=1}^{n}\frac{\i^{j}}{j!}f^{(j)}(A)
\ad_A^j(B)+\mathcal R_n(f,A,B),
\end{align*}
where $\mathcal R_n$ is given by (\ref{RestnOne}).

We now consider $f\in C^\infty(\mathbb R)$ satisfying (\ref{CommutExpandCond}). Let 
$\varphi_m(x)\equiv\varphi(x/m)$ with $\varphi\in C_0^\infty(]-1,1[)$ being such that $0\le\varphi\le 1$
and $\varphi(x)=1$ for all $x\in[-1/2,1/2]$. Then $f_m\equiv f\varphi_m\in C_0^\infty(\mathbb R)$, 
$\lim_m f_m^{(j)}(x)=f^{(j)}(x)$ for all $x\in\mathbb R$ and (\ref{CommutExpandCond}) imply 
that $\sup_{m,x\in\mathbb R}|f_m^{(j)}(x)|<\infty$ for all $j\in\{0,\ldots,n\}$. The functional calculus 
allows us to conclude that $\slim_mf_m^{(j)}(A)=f^{(j)}(A)$ and in particular that 
$$\slim_m[f_m(A),B]=[f(A),B].$$

The estimate (\ref{ftildeestimate}) allows us to obtain, starting from the representation of 
the remainder (\ref{RestnOne}),
\begin{align*}
\|\mathcal R_n(f,A,B)\|&\le
\frac{\|\ad_A^{n}(B)\|}{\pi}\int|\bar\partial\tilde f(x+\i y)|\,|y|^{-n-1}\d\bar z\wedge\d z\\
&\le C\sum_{j=0}^{k+2}\int\langle x\rangle^{j-n-1}|f^{(j)}(x)|\,\d x.
\end{align*}
We deduce that
$$
\|\mathcal R_n(f,A,B)-\mathcal R_n(f_m,A,B)\|
\le C\int g_m(x)\,\d x,
$$
where we have set
$$
g_m(x)=\sum_{j=0}^{k+2}\langle x\rangle^{j-n-1}
|f^{(j)}(x)-f_m^{(j)}(x)|.
$$
Starting from the expansion
$$
f^{(j)}(x)-f_m^{(j)}(x)=\sum_{l=0}^j{j\choose l}f^{(l)}(x)
(\delta_{jl}-m^{-(j-l)}\varphi^{(j-l)}(x/m)),
$$
we obtain the estimate
\begin{align*}
g_m(x)&\le
\sum_{j=0}^{k+2}\langle x\rangle^{j-n-1}
\sum_{l=0}^j{j\choose l}
|f^{(l)}(x)||\delta_{jl}-m^{-(j-l)}\varphi^{(j-l)}(x/m)|\\
&\le\sum_{l=0}^{k+2}\langle x\rangle^{l-n-1} |f^{(l)}(x)|
\sum_{j=l}^{k+2}{j\choose l}\langle x\rangle^{j-l}
|\delta_{jl}-m^{-(j-l)}\varphi^{(j-l)}(x/m)|\\
&\le\sum_{l=0}^{k+2}\langle x\rangle^{l-n-1} |f^{(l)}(x)|
\sum_{j=l}^{k+2}{j\choose l}
(\delta_{jl}+\langle m\rangle^{j-l}m^{-(j-l)}\varphi^{(j-l)}(x/m))\\
&\le C'\sum_{l=0}^{k+2}\langle x\rangle^{l-n-1} |f^{(l)}(x)|\equiv g(x),
\end{align*}
and as (\ref{CommutExpandCond}) implies that $g\in L^1(\mathbb R)$ we can apply the dominated 
convergence theorem to conclude that
$$
\lim_m \mathcal R_n(f_m,A,B)=\mathcal R_n(f,A,B).
$$
The identity (\ref{CommutExpandForm}), with $f=f_m$ is thus preserved in the limit $m\to\infty$, 
which proves (\ref{CommutExpandForm}) in the general case.

Finally, we note that if  $B\in\B_A^{n+1}(\mathcal H)$ the remainder $\mathcal R_n$ can also be 
written as
\begin{align*}
\mathcal R_n(f,A,B)&
=\frac{1}{2\pi\i^{n+1}}\int\bar\partial\tilde f(z)(A-z)^{-n}[\ad_A^{n}(B),(A-z)^{-1}]
\,\d\bar z\wedge\d z\\
&=\frac{1}{2\pi\i^{n+2}}\int\bar\partial\tilde f(z)(A-z)^{-n-1}\ad_A^{n+1}(B)(A-z)^{-1}
\,\d\bar z\wedge\d z.
\end{align*}
\cqfd
\end{demo}

\subsubsection{The commutator of two self-adjoint operators}
\label{CommutSASect}
In this subsection we shall extend the discussion to commutators of type $[A,H]$ where $A$ and 
$H$ are self-adjoint operators on $\mathcal H$. We shall start by studying commutators of the form 
$[A,(H-z)^{-1}]$.

\begin{Lem}\label{ResCommut} Let $A$ be a self-adjoint operator and $B$ a closed operator on the 
Hilbert space $\mathcal H$. Let $\Res(B)$ be the resolvent set of $B$ and $R(z)\equiv(B-z)^{-1}$
its resolvent.
\begin{enumerate}[(i)]
\item If $R(z_0)\in\mathcal B_A^n(\mathcal H)$ for some $z_0\in\Res(H)$ then 
$R(z)\in\mathcal B_A^n(\mathcal H)$ for all $z\in\Res(B)$.
\item For all $z,z_0\in\Res(B)$ we have
$$
\mathrm{ad}_A(R(z))=(I+(z-z_0)R(z))\mathrm{ad}_A(R(z_0))(I+(z-z_0)R(z)).
$$
This relation allows for the inductive calculation of $\mathrm{ad}_A^k(R(z))$ for $k=2,\ldots,n$. 
\end{enumerate}
\end{Lem}

\begin{demo} We set $R\equiv R(z)$, $R_0\equiv R(z_0)$, and $w\equiv z-z_0$. The first resolvent 
equation $R-R_0=wRR_0$ gives $(I-wR_0)^{-1}=I+wR$. With the notation from the proof of 
Lemma \ref{CommutApprox} we have, for $u\in\Dom(A)$, 
\begin{align*}
(I+wR)Au&=\lim_{\varepsilon\to0}(I+wR)A_\varepsilon u
=\lim_{\varepsilon\to0}(I+wR)A_\varepsilon (I-wR_0)(I+wR)u\\
&=\lim_{\varepsilon\to0}\left(A_\varepsilon(I+wR)u-w
(I+wR)[A_\varepsilon,R_0](I+wR)u\right).
\end{align*}
By applying Lemma \ref{CommutApprox} we obtain
$$
\lim_{\varepsilon\to0}A_\varepsilon(I+wR)u=
w(I+wR)[A,R_0](I+wR)u+(I+wR)Au,
$$
which shows that $(I+wR)u\in\Dom(A)$ and allows us to write
$$
A(I+wR)u=w(I+wR)[A,R_0](I+wR)u+(I+wR)Au,
$$
or alternatively
$$
ARu-RAu=(I+wR)[A,R_0](I+wR)u.
$$
Using Lemma \ref{BUcomm} gives
$$
[A,R]=(I+wR)[A,R_0](I+wR),
$$
which shows (i) in the particular case where $n=1$ as well as (ii). To show (i) in the general case 
we proceed by induction on $n$. Suppose assertion (i) holds for $n\le m$ and that 
$R_0\in\mathcal B_A^{m+1}(\mathcal H)$. We then have that $R_0\in\mathcal B_A^{m}(\mathcal H)$
and the induction hypothesis allows us to state that $R\in\mathcal B_A^{m}(\mathcal H)$. 
$\mathrm{ad}_A$, being a derivation, satisfies the Leibniz formula
$$
\mathrm{ad}_A^m(BC)
=\sum_{k=0}^m{m\choose k}\mathrm{ad}_A^k(B) \,\mathrm{ad}_A^{(m-k)}(C).
$$
Since $R=(I+wR)R_0$ we have
$$
\mathrm{ad}_A^m(R)
=\sum_{k=0}^m{m\choose k}\mathrm{ad}_A^k(I+wR) \,\mathrm{ad}_A^{(m-k)}(R_0),
$$
which can also be written as
$$
\mathrm{ad}_A^m(R)(I-wR_0)
=\sum_{k=0}^{m-1}{m\choose k}\mathrm{ad}_A^k(I+wR) \,\mathrm{ad}_A^{(m-k)}(R_0),
$$
or as
\begin{equation}
\mathrm{ad}_A^m(R)
=\sum_{k=0}^{m-1}{m\choose k}\mathrm{ad}_A^k(I+wR) \,\mathrm{ad}_A^{(m-k)}(R_0)
(I+wR).
\label{Resolviter}
\end{equation}
We deduce that
\begin{align*}
\mathrm{ad}_A^{m+1}(R)
=\sum_{k=0}^{m-1}{m\choose k}&\left[w\,\mathrm{ad}_A^{k+1}(R) \,\mathrm{ad}_A^{(m-k)}(R_0)
(I+wR)\right.\\
&+\mathrm{ad}_A^{k}(I+wR) \,\mathrm{ad}_A^{(m+1-k)}(R_0)(I+wR)\\[5pt]
&\left.+w\,\mathrm{ad}_A^{k}(I+wR) \,\mathrm{ad}_A^{(m-k)}(R_0)\mathrm{ad}_A(R)\right],
\end{align*}
and thus $R\in\mathcal B_A^{m+1}(\mathcal H)$, which validates the induction step.
\cqfd
\end{demo}

\begin{Lem}\label{fcommutencore}Let $A$ and $H$ be self-adjoint operators on the Hilbert space 
$\mathcal H$. If there exists $z_0\in\Res(H)$ such that $(H-z_0)^{-1}\in\mathcal B_A^n(\mathcal H)$ 
then $f(H)\in\mathcal B_A^n(\mathcal H)$ for all $f\in C_0^\infty(\mathbb R)$.
\end{Lem}

\begin{demo} We set $R(z)\equiv(H-z)^{-1}$ and $\VERT B\VERT_{A,n}\equiv\max_{k\le n}
\|\mathrm{ad}_A^kB\|$ for $B\in B_A^n(\mathcal H)$. Using Formula~(\ref{Resolviter}) 
we easily show that for $n\ge1$
$$
\left\|\mathrm{ad}_A^nR(z)\right\|\le
(1+|z-z_0|)^{n-1}\left(1+\frac{|z-z_0|}{|\mathrm{Im}z|}\right)^{n+1}
\left(n-1+\VERT R(z_0)\VERT_{A,n}\right)^n.
$$
We may thus use the Helffer-Sj\"ostrand formula to obtain
$$
\mathrm{ad}_A^n(f(H))=\int_{\mathbb C}\bar\partial\tilde f(z) 
\,\mathrm{ad}_A^nR(z)\,\frac{\d z\wedge\d\bar z}{\pi}.
$$
\cqfd
\end{demo}

\begin{Def}\label{Cnlocdef} Let $A$ be a self-adjoint operator on the Hilbert space $\mathcal H$. 
A self-adjoint operator $H$ on $\mathcal H$ is locally of class $C^n(A)$ (or of class 
$C^n_\mathrm{loc}(A)$) if $f(H)$ is of class $C^n(A)$ for all $f\in C_0^\infty(\mathbb R)$.
\end{Def}

A self-adjoint operator $H$ is thus of class $C^n_\mathrm{loc}(A)$ if, for all $u\in\mathcal H$ and 
for all $f\in C_0^\infty(\mathbb R)$ the function 
$\mathbb R\ni\theta\to\tau_A^\theta(f(H))u=f(\tau_A^\theta(H))u$ is of class $C^n$. The following 
lemma is an immediate corollary of Lemmas \ref{fcommut} and \ref{fcommutencore}. 

\begin{Lem}\label{localCn} Let $A$ and $H$ be self-adjoint operators on the Hilbert space
$\mathcal H$. The two following conditions are equivalent.
\begin{enumerate}[(i)]
\item $H$ is of class $C^n_\mathrm{loc}(A)$.
\item There exists $z_0\in\Res(H)$ such that $(H-z_0)^{-1}\in\mathcal B_A^n(\mathcal H)$.
\end{enumerate}
\end{Lem}

\bigskip 
In practice it is useful to have a criteria characterizing  $C^n_\mathrm{loc}(A)$-operators
without reference to their resolvents. Moreover, Mourre theory requires the commutator 
$[A,H]$ to be defined as an operator, at least locally (in the sense of the spectrum of $H$).
To this end we introduce a scale of Banach spaces.

\begin{Def}\label{Sobolevdef} Let $H$ be a self-adjoint operator on the Hilbert space $\mathcal H$, 
$\Lambda=(I+|H|)$, and $s\in\mathbb R$. We denote by $\mathcal H_H^s$ the Banach 
space obtained by completing $\Dom(\Lambda^s)$ equipped with the norm
$$
\|u\|_{H,s}=\|\Lambda^su\|.
$$
\end{Def}

For $s\ge0$ we have $\|u\|_{H,s}\ge\|u\|$ and the norm $\|\,\cdot\,\|_{H,s}$ is equivalent to the graph 
norm of the closed operator $\Lambda^s$. This implies that $\mathcal H_H^s=\Dom(\Lambda^s)$ and 
that $\Lambda^s$ is an isometry from $\mathcal H_H^s$ onto $\mathcal H$. For $s\le0$ the map
$u\mapsto\Lambda^{-s}u$ extends continuously to an isometry from $\mathcal H_H^{-s}$ onto 
$\mathcal H$ and 
\beq
\mathcal H_H^{s}\times\mathcal H_H^{-s} \ni\langle u,v\rangle
\mapsto(\Lambda^{s}u,\Lambda^{-s}v)\equiv (u,v),
\label{dual}
\eeq
describes the duality between $\mathcal H_H^{-s}$ and $\mathcal H_H^{s}$. We therefore obtain a 
scale of spaces
$$
\mathcal H_H^{s'}\subset\mathcal H_H^s\subset\mathcal H=\mathcal H_H^0\subset\mathcal H_H^{-s}
\subset\mathcal H_H^{-s'},
$$
for $0\le s\le s'$, all embeddings being dense and continuous. We note in particular 
that for all $s\in\mathbb R$ one has $H\in\B(\mathcal H_H^{s},\mathcal H_H^{s-1})$ and 
$(H-z)^{-1}\in\B(\mathcal H_H^{s},\mathcal H_H^{s+1})$ for $z\in\Res(H)$.

Using the duality \eqref{dual}, we can associate to each continuous sesquilinear form 
$q\colon\mathcal H_H^{s}\times\mathcal H_H^{s}\to\mathbb C$ a unique operator 
$Q\in\B(\mathcal H_H^{s},\mathcal H_H^{-s})$ such that $q(u,v)=(u,Qv)$ for all 
$u,v\in\mathcal H_H^{s}$.

\begin{Lem}\label{locform} Let $A$ and $H$ be self-adjoint operators on the Hilbert space 
$\mathcal H$, and set $R(z)\equiv(H-z)^{-1}$. If $H\in C^n_\mathrm{loc}(A)$ then the following
statements hold.
\begin{enumerate}[(i)]
\item $\Dom(A)\cap\Dom(H)$ is dense in $\mathcal H$.
\item For all $z\in\Res(H)$ and $u,v\in\Dom(A)\cap\Dom(H)$
\begin{equation}
(Au,Hv)-(Hu,Av)=-((H-z)^\ast u,[A,R(z)](H-z)v),
\label{locformeq}
\end{equation}
\item There exists a constant $c$ such that
$$
|(Au,Hu)-(Hu,Au)|\le c\,\|u\|_{H,1}^2,
$$
for all $u\in\Dom(A)\cap\Dom(H)$. 
\item The quadratic form defined on $\Dom(A)\cap\Dom(H)$  by the left hands side 
of Eq.~(\ref{locformeq}) extends continuously to a bounded quadratic form on $\mathcal H_H^1$. 
\end{enumerate}
\end{Lem}

\begin{demo}(i) Lemma \ref{localCn} shows that $H\in C^n_\mathrm{loc}(A)$ implies that
$R(z_0)\in\B_A^n(\mathcal H)$ for some $z_0\in\Res(H)$. Since $\Dom(A)$ is dense in $\cH$ and 
$R(z_0)^\ast=R(\bar z_0)$ is injective we may conclude that $R(z_0)\Dom(A)$ is dense. 
We finish the proof of assertion (i) by remarking that Lemma \ref{BUcomm} implies 
that $R(z_0)\Dom(A)\subset\Dom(A)$ while the inclusion $R(z_0)\Dom(A)\subset\Dom(H)$ is evident.

(ii) For all  $u,v\in\Dom(A)\cap\Dom(H)$ and $z\in\Res(H)$ we may write, with $A_\varepsilon$ defined 
as in Lemma \ref{CommutApprox}, 
$$
((H-z)^\ast u,[A_\varepsilon,R(z)](H-z)v)
=((H-z)^\ast u,A_\varepsilon v)-(A_\varepsilon^\ast u,(H-z)v)
=(H u,A_\varepsilon v)-(A_{-\varepsilon} u,Hv).
$$
The proof of (ii) is obtained by taking the limit $\varepsilon\to0$ and invoking 
Lemma~\ref{CommutApprox}.

(iii) is an immediate consequence of (ii) and the fact that $[A,R(z)]$ is bounded.

(iv) follows directly from (iii).
\cqfd
\end{demo}

\begin{Def} Let $A$ and $H$ be self-adjoint operators on the Hilbert space $\mathcal H$ such that 
$\mathcal D=\Dom(A)\cap\Dom(H)$ is dense in $\mathcal H$. If the sesquilinear form
$$
\mathcal D\times\mathcal D\ni\langle u,v\rangle\mapsto (Au,Hv)-(Hu,Av),
$$
extends continuously to a bounded form on $\mathcal H_H^s$, we denote by $[A,H]$ the operator 
associated with this extension, and we write $[A,H]\in\B(\mathcal H_H^s,\mathcal H_H^{-s})$.
\end{Def}

\begin{Rem} If $[A,H]\in\B(\mathcal H_H^s,\mathcal H_H^{-s})$ for some $s\ge 0$ then
the operator $f(H)\i[A,H]f(H)$ is bounded on $\mathcal H$ for all $f\in C_0^\infty(\mathbb R)$. 
In general however, we can not claim that it is self-adjoint.

\bigskip
Lemma \ref{locform} stipulates that if $H\in C^1_\mathrm{loc}(A)$ then  
$[A,H]\in\B(\mathcal H_H^1,\mathcal H_H^{-1})$. The converse is not true without an additional 
assumption. One possibility is given by the following result.
\end{Rem}

\begin{Lem}\label{loccom} Let $A$ and $H$ be self-adjoint operators on the Hilbert space 
$\mathcal H$ and set $R(z)=(H-z)^{-1}$. The two following conditions are equivalent.
\begin{enumerate}[(i)]
\item $H\in C_\mathrm{loc}^1(A)$.
\item $[A,H]\in\B(\mathcal H_H^1,\mathcal H_H^{-1})$ and there exists $z_0\in\Res(H)$ such that
\[
R(z_0)\Dom(A)\subset\Dom(A),\quad\text{and}\quad R(\bar z_0)\Dom(A)\subset\Dom(A).
\]
\end{enumerate}
If one of these conditions is satisfied, then for each real measurable function $f$ such that
$$
\sup_{E\in\spec(H)}(1+|E|)|f(E)|<\infty,
$$
the operator $f(H)\i[A,H]f(H)$ is bounded and self-adjoint on $\mathcal H$. Furthermore, if 
$g\in C_0^\infty(\mathbb R)$ is such that $g(E)=E$ for all $E\in\mathrm{supp}(f)$, then
\begin{equation}
\label{loccommutid}
f(H)\i[A,H]f(H)=f(H)\ad_A(g(H))f(H).
\end{equation}
\end{Lem}

\begin{demo}$(i)\Rightarrow(ii)$ is a direct consequence of the preceding remark and 
Lemmas~\ref{locform}, \ref{localCn}, and \ref{BUcomm}.

$(ii)\Rightarrow(i)$ By setting $C=[A,H]$ we may write
$$
(Au,(H-z_0)v)-((H-\bar z_0)u,Av)=(Au,Hv)-(Hu,Av)=(u,Cv),
$$
for all $u,v\in\Dom(A)\cap\Dom(H)$. Since $R(z_0)\Dom(A)\subset\Dom(A)\cap\Dom(H)$ and 
$R(\bar z_0)\Dom(A)\subset\Dom(A)\cap\Dom(H)$ by hypothesis, we also have
$$
(AR(\bar z_0)u,(H-z_0)R(z_0)v)-((H-\bar z_0)R(\bar z_0)u,AR(z_0)v)=(R(\bar z_0)u,CR(z_0)v),
$$
that is,
$$
(AR(\bar z_0)u,v)-(u,AR(z_0)v)=(R(\bar z_0)u,CR(z_0)v),
$$
or also
$$
(u,R(z_0)Av)-(Au,R(z_0)v)=(u,R(z_0)CR(z_0)v).
$$
It follows from $C\in\B(\mathcal H_H^1,\mathcal H_H^{-1})$ and 
$R(z_0)\in\B(\mathcal H,\mathcal H_H^{1})\cap\B(\mathcal H_H^{-1},\mathcal H)$ that 
$R(z_0)CR(z_0)\in\B(\mathcal H)$, and the last identity shows that $R(z_0)\in\B_A^1(\mathcal H)$. 
The proof is finished by invoking Lemma \ref{localCn}.

To prove the final assertions of the lemma we note that 
$f(H)\in\B(\mathcal H,\mathcal H_H^1)\cap\B(\mathcal H_H^{-1},\mathcal H)$
since $\Lambda f(H)$ is bounded. This shows that 
$f(H)[A,H]f(H)$ is bounded. By Lemma \ref{locform} we have
$$
(u,[A,H]v)=-((H-z)^\ast u,[A,R(z)](H-z)v),
$$
for all $u,v\in\mathcal H_H^1$ and $z\in\Res(H)$. Thus
$$
(f(H)u,[A,H]f(H)v)=-((H-z)^\ast f(H) u,[A,R(z)](H-z)f(H)v),
$$
for all $u,v\in\mathcal H$. Since $Ef(E)=g(E)f(E)$ for all $E\in\mathbb R$ we may, without loss of 
generality, suppose that $g$ is real and write the preceding relation as
$$
(f(H)u,[A,H]f(H)v)=-(u,f(H)(g(H)-z)[A,R(z)](g(H)-z)f(H)v).
$$
Lemma \ref{CommutApprox} allows us to write
$$
f(H)(g(H)-z)[A,R(z)](g(H)-z)f(H)v
=\lim_{\varepsilon\to0}f(H)(g(H)-z)[A_\varepsilon,R(z)](g(H)-z)f(H)v,
$$
and a simple calculation shows that 
$f(H)(g(H)-z)[A_\varepsilon,R(z)](g(H)-z)f(H)v=-[g(H),A_\varepsilon]v$. 
Since $g(H)\in\B_A(\mathcal H)$ we may once again invoke Lemma \ref{CommutApprox} to obtain
$$
\lim_{\varepsilon\to0}[g(H),A_\varepsilon]v=[g(H),A]v,
$$
and conclude that
$$
(f(H)u,[A,H]f(H)v)=(f(H)u,[A,g(H)]f(H)v),
$$
which proves (\ref{loccommutid}). Finally, since $\ad_A$ is a $\ast$-derivation 
(Lemma \ref{commutderiv}), $\ad_A(g(H))$ is self-adjoint and so is $f(H)\ad_A(g(H))f(H)$.
\cqfd
\end{demo}

\bigskip In practice it is often much easier to compute iterated commutators of $A$ with $H$ than 
$\ad_A^k((H-z)^{-1})$ and to verify the invariance of $\Dom(H)$ by the group $\e^{\i\theta A}$ than 
that of $\Dom(A)$ by the resolvent $(H-z)^{-1}$. The following results are therefore important in this 
cases.

\begin{Lem}\label{Ainvar} Let $A$ and $H$ be self-adjoint operators on the Hilbert space $\mathcal H$.
The following statements hold if $\e^{\i\theta A}\Dom(H)\subset\Dom(H)$ for all $\theta\in\mathbb R$.
\begin{enumerate}[(i)]
\item $\Dom(A^\infty)\cap\mathcal  H_H^s$ is dense in $\mathcal H_H^s$ for all $s\in[-1,1]$.
\item For all $s\in[0,1]$ and $\theta\in\mathbb R$ we have 
$\e^{\i\theta A}\mathcal H_H^s\subset\mathcal H_H^s$. The restriction of  
$\e^{\i\theta A}$ to $\mathcal H_H^s$ defines a strongly continuous, quasi-bounded group on 
$\mathcal H_H^s$. Its generator $A_s$ is 
given by $\Dom(A_s)=\{u\in\mathcal H_H^s\,|\,u\in\Dom(A), Au\in\mathcal H_H^s\}$ and
$A_su=Au$ for all $u\in\Dom(A_s)$.
\item For all $s\in[0,1]$ and $\theta\in\mathbb R$ the operator $\e^{\i\theta A}$ extends continuously 
to $\mathcal H_H^{-s}$. This extension defines a strongly continuous, quasi-bounded group on 
$\mathcal H_H^{-s}$ which we denote again by $\e^{\i\theta A}$. Its generator $A_{-s}$ is the closure 
of $A$ in $\mathcal H_H^{-s}$.
\item For all $s\in[0,1]$, $u\in\mathcal H_H^{-s}$, $v\in\mathcal H_H^{s}$
and $\theta\in\mathbb R$, $(\e^{\i\theta A}u,\e^{\i\theta A}v)=(u,v)$. In particular
$A_s^\ast=A_{-s}$.
\end{enumerate}
\end{Lem}

\begin{demo} To simplify our proof, we shall assume that $\mathcal H$ is separable and we 
refer the reader to proposition 3.2.5 of \cite{ABG} for the general case. By hypothesis, the operator 
$\Lambda\e^{\i\theta A}\Lambda^{-1}$ is defined everywhere on $\mathcal H$. We easily verify that 
its graph is closed. This operator is thus bounded, which is equivalent to saying that $\e^{\i\theta A}$ 
is bounded on $\mathcal H_H^1$. We thus have 
$\e^{\i\theta A}\in\B(\mathcal H_H^1)\cap\B(\mathcal H_H^0)$ from which we deduce 
$\e^{\i\theta A}\in\B(\mathcal H_H^s)$ for all $s\in[0,1]$ by interpolation.

Since $\mathcal H$ is separable it has a dense countable subset $\mathcal D_0$. 
The sets $\mathcal D_+\equiv\{\Lambda^{-s}v/\|v\|\,|\,v\in\mathcal
D_0,v\not=0\}\subset\mathcal H$ and $\mathcal
D_-\equiv\{v/\|\Lambda^{-s}v\|\,|\,v\in\mathcal
D_0,v\not=0\}\subset\mathcal H$ are countable and dense in the unit spheres of 
$\mathcal H_H^{\pm s}$. For all $\theta\in\mathbb R$ we thus have
$$
\|\e^{\i\theta A}\|_{\B(\mathcal H_H^s)}=\sup_{\langle u,v\rangle\in\mathcal D_-\times\mathcal D_+}
|(u,\e^{\i\theta A}v)|.
$$
Since $\mathcal D_-\times\mathcal D_+$ is countable and $\theta\mapsto|(u,\e^{\i\theta A}v)|$ is 
continuous and thus measurable for all 
$\langle u,v\rangle\in\mathcal D_-\times D_+\subset\mathcal H\times\mathcal H$, the functions 
$f_\pm(\theta)=\log\|\e^{\pm\i\theta A}\|_{\B(\mathcal H_H^s)}$ are measurable. They are 
sub-additive ($f_\pm(\theta+\theta')\le f_\pm(\theta)+f_\pm(\theta')$) and thus bounded on all 
compact intervals (see, for example, Theorem 7.4.1 in \cite{HP}). We easily conclude that for 
all $\delta>0$, $\theta\ge0$,
$$
f_\pm(\theta)\le \theta\frac{|f_\pm(\delta)|}{\delta}+\sup_{\vartheta\in[0,\delta]}f_\pm(\vartheta),
$$
which shows that the group $\e^{\i\theta A}$ is quasi-bounded on $\mathcal H_H^1$,
\begin{equation}
\|\Lambda^s\e^{\i\theta A}\Lambda^{-s}\|
=\|\e^{\i\theta A}\|_{\B(\mathcal H^s)}\le M\e^{\omega|\theta|},
\label{qbd}
\end{equation}
for some constants $M$ and $\omega$ and all $\theta\in\rr$. For $u\in\mathcal H$ and 
$\varepsilon>0$ we set
$$
u_\varepsilon\equiv\e^{-\varepsilon^2A^2/4}u=\pi^{-1/2}
\int_{-\infty}^\infty\e^{-\vartheta^2}
\,\e^{\i\varepsilon\vartheta A}u\,\d\vartheta.
$$
We easily show that, for all $u\in\mathcal H$, $\lim_{\varepsilon\to0}\|u_\varepsilon-u\|=0$. 
Furthermore, the bound (\ref{qbd}) implies $\|u_\varepsilon\|_{H,s}\le c\,\|u\|_{H,s}$ for a constant 
$c$ and $\varepsilon\in]0,1]$. Since $\mathcal H$ is dense in $\mathcal H_H^{-s}$ we may 
conclude that $u_\varepsilon$ converges to $u$ in the weak topology of $\mathcal H_H^s$.
It follows that $\mathcal D\equiv\{u_\varepsilon\,|\,u\in\mathcal H_H^s,\varepsilon>0\}$  
is dense in $\mathcal H$ and thus in $\mathcal H_H^s$. The identity
$$
(\e^{\i\theta A}-I)u_\varepsilon=\pi^{-1/2}
\int_{-\infty}^\infty\left(\e^{-(\vartheta-\theta/\varepsilon)^2}-\e^{-\vartheta^2}\right)
\,\e^{\i\varepsilon\vartheta A}u\,\d\vartheta,
$$
and the bound (\ref{qbd}) lead to
$$
\|(\e^{\i\theta A}-I)u_\varepsilon\|_{H,s}\le\pi^{-1/2}
\int_{-\infty}^\infty\left|\e^{-(\vartheta-\theta/\varepsilon)^2}-\e^{-\vartheta^2}\right|
\,M\e^{\varepsilon|\vartheta|\omega}\|u\|_{H,s}\,\d\vartheta.
$$
We deduce that $\lim_{\theta\to0}\|(\e^{\i\theta A}-I)u\|_{H,s}=0$ for all $u\in\mathcal D$. 
The bound (\ref{qbd}) and the density of $\mathcal D$ allow us to conclude that 
$\theta\mapsto\e^{\i\theta A}$ is strongly continuous on $\mathcal H_H^s$. Let $A_s$ be 
the generator of this group. If $u\in\Dom(A_s)$ then
$$
\lim_{\theta\to0}\left\|\frac{\e^{\i\theta A}u-u}{\i\theta}-A_su\right\|_{H,s}=0.
$$
We conclude that $u\in\Dom(A)$ and that $Au=A_su\in\mathcal H_H^s$. Conversely, if 
$u\in\Dom(A)\cap\mathcal H_H^s$ and $v=Au\in\mathcal H_H^s$ then the identity
$$
\frac{\e^{\i\theta A}u-u}{\i\theta}-v=\int_0^1\left(\e^{\i t\theta A}-I\right)v\,\d t,
$$
the bound (\ref{qbd}), and the continuity of $t\mapsto\e^{\i t\theta A}v$ in $\mathcal H_H^s$ allow
us to conclude that $u\in\Dom(A_s)$. Finally we note that 
$\mathcal D\subset\Dom(A^\infty)\cap\mathcal H_H^s$, which finishes the proof for assertions 
(i) and (ii) for $s\in[0,1]$.

By duality $(T^\theta u,v)\equiv(u,\e^{\i\theta A}v)$ defines a strongly continuous, quasi-bounded 
group on $\mathcal H_H^{-s}$. Furthermore, $T^\theta=\e^{-\i\theta A}$  on the dense subspace 
$\mathcal H\subset\mathcal H_H^{-s}$. $T^\theta$ is thus the unique continuous extension of
$\e^{\i\theta A}$ to $\mathcal H_H^{-s}$. Since $\Dom(A)$ is dense in $\mathcal H$ it is also 
dense in $\mathcal H_H^{-s}$. Furthermore it is invariant by $T^\theta$. It is thus a core for 
the generator $A_s$ of $T^\theta$ (see Theorem X.49 in \cite{RS2}). The same argument
applies to $\Dom(A^\infty)$. The assertions (ii) for $s\in[-1,0]$ and $(iii)$ are proven. 
Assertion (iv) is an immediate consequence of the duality.
\cqfd
\end{demo}

\bigskip With a small abuse of notation, we shall denote by $A$ the generator $A_s$ when the 
space on which it acts is clearly determined by the context. Similarly, $A_\varepsilon$ will denote 
the element of $\B(\mathcal H_H^s)$ corresponding to the restriction (for $s\ge0$) or the 
continuous extension (for $s\le0$) of the operator $(\i\varepsilon)^{-1}(\e^{\i\varepsilon A}-I)$ 
on $\mathcal H_H^s$.

\bigskip
Under the hypotheses of Lemma \ref{Ainvar}, if $B\in\B(\mathcal H_H^s,\mathcal H_H^{s'})$ with 
$s,s'\in[-1,1]$, then one has $B^\ast\in\B(\mathcal H_H^{-s'},\mathcal H_H^{-s})$ and 
$$
(Au,Bv)-(B^\ast u,Av),
$$
defines a quadratic form on $\Dom(A_{-s'})\times\Dom(A_s)$. If there exists a constant $c$ such 
that, for all $\langle u,v\rangle\in\Dom(A_{-s'})\times\Dom(A_s)$,
$$
|(Au,Bv)-(B^\ast u,Av)|\le c\,\|u\|_{H,-s'}\,\|v\|_{H,s},
$$
then there exists an operator $C\in\B(\mathcal H_H^s,\mathcal H_H^{s'})$ such that
$$
(Au,Bv)-(B^\ast u,Av)=(u,Cv).
$$
In this case we shall write $C=[A,B]$ and, when this construction can be iterated, we define 
$\ad_A^k(B)\in\B(\mathcal H_H^s,\mathcal H_H^{s'})$ as before.

\begin{Def} Let $A$ and $H$ be self-adjoint operators on the Hilbert space $\mathcal H$ 
such that, for all $\theta\in\rr$, $\e^{\i\theta A}\Dom(H)\subset\Dom(H)$. 
For all $n\in\mathbb N$, we denote
$$
\B_A^n(\mathcal H_H^s,\mathcal H_H^{s'})\equiv\{B\in\B(\mathcal H_H^s,\mathcal H_H^{s'})\,|\,
\ad_A^k(B)\in\B(\mathcal H_H^s,\mathcal H_H^{s'}), k=0,\ldots, n\}.
$$
In particular $\B(\mathcal H_H^s,\mathcal H_H^{s'})=\B_A^0(\mathcal H_H^s,\mathcal H_H^{s'})$
and $\B_A(\mathcal H_H^s,\mathcal H_H^{s'})=\B_A^1(\mathcal H_H^s,\mathcal H_H^{s'})$.
\end{Def}
\bigskip

\begin{Lem}\label{AHequiv} Let $A$ and $H$ be self-adjoint operators on $\mathcal H$ such that 
$\e^{\i\theta A}\Dom(H)\subset\Dom(H)$ for all $\theta\in\mathbb R$.
\begin{enumerate}[(i)]
\item $B\in\B_A^n(\mathcal H_H^s,\mathcal H_H^{s'})$ if and only if
$\sup_{\varepsilon\in\mathbb{R}^\ast}\|\ad_{A_\varepsilon}^n(B)\|_{\B(\mathcal H_H^s,\mathcal H_H^{s'})}
<\infty$ and in this case
$$
\ad_A^n(B)v=\lim_{\varepsilon\to0}\ad_{A_\varepsilon}^n(B)v,
$$
in $\mathcal H_H^{s'}$ for all $v\in\mathcal H_H^s$.
\item $B\in\B_A(\mathcal H_H^s,\mathcal H_H^{s'})$
if and only if $B^\ast\in\B_A(\mathcal H_H^{-s'},\mathcal H_H^{-s})$.
Furthermore
$$
\ad_A(B^\ast)=\ad_A(B)^\ast.
$$
\item If $B\in\B_A(\mathcal H_H^{s'},\mathcal H_H^{s''})$ and
$C\in\B_A(\mathcal H_H^{s},\mathcal H_H^{s'})$ then
$BC\in\B_A(\mathcal H_H^{s},\mathcal H_H^{s''})$ and
$$
\ad_A(BC)=\ad_A(B)C+B\ad_A(C).
$$
\end{enumerate}
\end{Lem}

\begin{demo} The proof of Lemma \ref{CommutApprox} is easily adapted to prove assertion $(i)$.

$(ii)$ is a direct consequence of the identity
$$
(Au,Bv)-(B^\ast u,Av)=-\overline{(Av,B^\ast u)-(Bv,Au)}.
$$

$(iii)$ For $\langle u,v\rangle\in\Dom(A_{-s''})\times\Dom(A_s)$ we have
\begin{align*}
(Au,BCv)-(C^\ast B^\ast u,Av)
&=\lim_{\varepsilon\to0}(A_{-\varepsilon}u,BCv)-(C^\ast B^\ast u,A_\varepsilon v)\\
&=\lim_{\varepsilon\to0}(u,A_{\varepsilon}BCv -BCA_\varepsilon v)\\
&=\lim_{\varepsilon\to0}(u,[A_{\varepsilon},B]Cv+B[A_\varepsilon,C]v)\\
&=-\lim_{\varepsilon\to0}([A_{-\varepsilon},B^\ast]u,Cv)
+\lim_{\varepsilon\to0}(B^\ast u,[A_\varepsilon,C]v).
\end{align*}
The assertions $(i)$ and $(ii)$ allows us to conclude that
$$
\lim_{\varepsilon\to0}[A_{-\varepsilon},B^\ast]u=[A,B^\ast]u=-[A,B]^\ast u,
$$
in $\mathcal H_H^{-s'}$ and that 
$$
\lim_{\varepsilon\to0}[A_{\varepsilon},C]v=[A,C]v,
$$
in $\mathcal H_H^{s'}$. It is then easy to finish the proof.
\cqfd
\end{demo}

\begin{Def} Let $A$ and $H$ be self-adjoint operators on the Hilbert space $\mathcal H$ such that,
for all $\theta\in\rr$,  $\e^{\i\theta A}\Dom(H)\subset\Dom(H)$. 
An operator $B\in\B(\mathcal H_H^s,\mathcal H_H^{s'})$, with $s,s'\in[-1,1]$, is of class 
$C^n(A;H;s,s')$ if, for all $v\in\mathcal H_H^s$ the function
$$
\begin{array}{rcc}
\mathbb R&\to&\mathcal H_H^{s'}\\
\theta&\mapsto&\e^{\i\theta A}B\e^{-\i\theta A}v,
\end{array}
$$
is of class $C^n$.
\end{Def}

\begin{Lem}\label{AHequivencore} Let $A$ and $H$ be self-adjoint operators on the Hilbert space 
$\mathcal H$ such that, for all $\theta\in\rr$, $\e^{\i\theta A}\Dom(H)\subset\Dom(H)$.
For an operator $B\in\B(\mathcal H_H^s,\mathcal H_H^{s'})$, the following are equivalent
\begin{enumerate}[(i)]
\item $B\in C^n(A;H;s,s')$ and $B^\ast\in C^n(A;H;-s',-s)$.
\item $B\in\B_A^n(\mathcal H_H^s,\mathcal H_H^{s'})$.
\end{enumerate}
If one of these statements holds, then
$$
\frac{\d^k}{\d\theta^k}\e^{\i\theta A}B\e^{-\i\theta A}v=\e^{\i\theta A}\ad_A^k(B)\e^{-\i\theta A}v,
$$
for $k=1,\ldots,n$ and $v\in\mathcal H_H^s$.
\end{Lem}

\begin{demo} The strategy is identical to that of the proof of Lemma \ref{fcommut}
\cqfd
\end{demo}

\begin{Theo}\label{localchar} Let $A$ and $H$ be self-adjoint operators on the Hilbert space 
$\mathcal H$ such that, for all $\theta\in\rr$, $\e^{\i\theta A}\Dom(H)\subset\Dom(H)$.
\begin{enumerate}[(i)]
\item $H\in C^1_\mathrm{loc}(A)$ if and only if 
$H\in\B_A(\mathcal H_H^1,\mathcal H_H^{-1})$.
\item If $H\in\B_A^n(\mathcal H_H^1,\mathcal H)$ then $H\in C^n_\mathrm{loc}(A)$. 
\end{enumerate}
\end{Theo}

\begin{demo}
$(i)$ By combining Lemmas \ref{CommutApprox} and \ref{AHequiv}, it suffices to show that for 
$z\in\Res(H)$,
$$
\sup_{\varepsilon\in\mathbb{R}^\ast}\|[A_\varepsilon,R(z)]\|<\infty \Longleftrightarrow
\sup_{\varepsilon\in\mathbb{R}^\ast}\|[A_\varepsilon,H]\|_{\B(\mathcal H_H^1,\mathcal H_H^{-1})}<\infty.
$$
For all $s\in\mathbb R$, $R(z)$ is an isomorphism of $\mathcal H_H^s$ into $\mathcal H_H^{s+1}$, 
with inverse $H-z$. For $u\in\mathcal H_H^1$ the identity
$$
[A_\varepsilon,H]u
=[A_\varepsilon,H-z]u
=A_\varepsilon(H-z)u-(H-z)A_\varepsilon u
=(H-z)\left[R(z)A_\varepsilon-A_\varepsilon R(z)\right](H-z)u,
$$
implies
$$
\|[A_\varepsilon,H]\|_{\B(\mathcal H_H^1,\mathcal H_H^{-1})}\le
\|H-z\|_{\B(\mathcal H_H^1,\mathcal H)}
\|[A_\varepsilon, R(z)]\|\,
\|H-z\|_{\B(\mathcal H,\mathcal H_H^{-1})},
$$
and thus $\sup_{\varepsilon\in\mathbb{R}^\ast}\|[A_\varepsilon,R(z)]\|<\infty$ $\Rightarrow$
$\sup_{\varepsilon\in\mathbb{R}^\ast}\|[A_\varepsilon,H]\|_{\B(\mathcal H_H^1,\mathcal H_H^{-1})}<\infty$. 
Conversely, we have, for $u\in\mathcal H$,
$$
[A_\varepsilon,R(z)]u=R(z)\left[HA_\varepsilon-A_\varepsilon H\right]R(z)u,
$$
which implies
$$
\|[A_\varepsilon,R(z)]\|\le
\|R(z)\|_{\B(\mathcal H_H^{-1},\mathcal H)}
\|[A_\varepsilon,H]\|_{\B(\mathcal H_H^1,\mathcal H_H^{-1})}
\|R(z)\|_{\B(\mathcal H,\mathcal H_H^1)},
$$
and thus
$\sup_{\varepsilon\in\mathbb{R}^\ast}\|[A_\varepsilon,H]\|_{\B(\mathcal H_H^1,\mathcal H_H^{-1})}<\infty$
$\Rightarrow$ $\sup_{\varepsilon\in\mathbb{R}^\ast}\|[A_\varepsilon,R(z)]\|<\infty$.

$(ii)$ Let $z\in\Res(H)$ and $R=(H-z)^{-1}$. We easily show, by induction on $k$, that
$$
\ad_{A_\varepsilon}^k(R)=\sum_{l=1}^k\sum_{k_1+\cdots+k_l=k}C^{(l)}_{k_1\ldots k_l}
R\,\ad_{A_\varepsilon}^{k_1}(H)R\,\ad_{A_\varepsilon}^{k_2}(H)R\cdots\ad_{A_\varepsilon}^{k_l}(H)R,
$$
where the $C^{(l)}_{k_1\ldots k_l}$ are numerical coefficients. Since
$$
\|\ad_{A_\varepsilon}^{k}(H)R\|\le
\|\ad_{A_\varepsilon}^{k}(H)\|_{\B(\mathcal H_H^1,\mathcal H)}\,
\|R\|_{\B(\mathcal H,\mathcal H_H^1)},
$$
Lemma \ref{AHequiv} implies that if $H\in \B_A^n(\mathcal H_H^1,\mathcal H)$ we have
$$
\sup_{\varepsilon\in\mathbb{R}^\ast}\|\ad_{A_\varepsilon}^n(R)\|<\infty,
$$
and Lemma \ref{CommutApprox} allows us to conclude that $R\in C^n(A)$. 
Finally, Lemma \ref{localCn} shows that $H\in C^n_\mathrm{loc}(A)$.
\cqfd
\end{demo}

\subsection{The Mourre estimate}
\label{MourreSect}

The following definition is due to Mourre \cite{M1}.

 \begin{Def}\label{MourreEstDef} Let $H$ and $A$ be self-adjoint operators on the Hilbert space 
$\mathcal H$. 
\begin{enumerate}[(i)]
\item $H$ satisfies a Mourre estimate at $E\in\mathbb R$ with the conjugate operator $A$ if there 
exists $\theta>0$, a function $g\in C_0^\infty(\mathbb R)$ and a compact operator $K$ such that
$0\le g\le 1$, $g(E)=1$ and 
\begin{equation}
\label{MourreIneq}
g(H)\i[H,A]g(H)\ge\theta g(H)^2 +K.
\end{equation}
\item Let $O\subset\mathbb R$ be open. $H$ satisfies a Mourre estimate on $O$ with the conjugate 
operator $A$ if, for all $E\in O$, $H$ satisfies a Mourre estimate with the conjugate operator $A$ at 
$E$.
\item If it is possible to take $K=0$ in (\ref{MourreIneq}), we say that $H$ satisfies a strict Mourre 
estimate at $E$ (respectively on $O$) with the conjugate operator $A$. 
\end{enumerate}
\end{Def}

The following lemma shows that the set of $E$ at which $H$ satisfies a (strict) Mourre estimate
with the conjugate operator $A$ is open. 

\begin{Lem}\label{localME} If $H$ satisfies a (strict) Mourre estimate at $E\in\mathbb R$, there 
exists an open interval $\Delta\ni E$ such that (\ref{MourreIneq}) is satisfied with $g=1_\Delta$, 
the indicator function of $\Delta$ (and $K=0$).
\end{Lem}

\begin{demo} By hypothesis, there exists $\theta>0$, $g\in C_0^\infty(\mathbb R)$ such that 
$0\le g\le 1$, $g(E)=1$, as well as  a compact operator $K$ satisfying (\ref{MourreIneq}). 
There thus exists $\delta>0$ such that $\Delta\equiv]E-\delta,E+\delta[\subset g^{-1}(]1/2,1])$. 
It follows that $0\le h\equiv 1_\Delta/g \le2$ and in particular that $h(H)$ is bounded. 
By multiplying (\ref{MourreIneq}) on both sides by $h(H)$ we obtain
$$
1_\Delta(H)\i[H,A]1_\Delta(H)\ge\theta 1_\Delta(H) +K',
$$
where $K'=h(H)Kh(H)$ is compact (and vanishes if $K=0$).\cqfd
\end{demo}

\bigskip
The first consequences of the Mourre estimate concern the singular spectrum of $H$ (see \cite{M1}).

\begin{Theo}\label{MourreSingSpec} We suppose that a self-adjoint operator $H$ on the Hilbert space $\mathcal H$ satisfies a Mourre estimate on the open set $O\subset\mathbb R$ with the conjugate operator $A$.
\begin{enumerate}[(i)]
\item If $H\in C^1_\mathrm{loc}(A)$ and $I\subset O$ is compact then $\spec_\pp(H)\cap I$ is finite. 
This set is empty if the Mourre estimate is strict on $I$. 
\item If $H\in C^2_\mathrm{loc}(A)$ then $\spec_\mathrm{sc}(H)\cap O$ is empty.
\end{enumerate}
\end{Theo}

\begin{demo} We essentially follow the proof in \cite{M1}.

(i) For all $E\in O$ we denote by $\Delta_E\ni E$ the interval described in Lemma \ref{localME}.
We thus have the Mourre estimate
\begin{equation}
1_{\Delta_E}(H)\i[H,A]1_{\Delta_E}(H)\ge\theta_E1_{\Delta_E}(H)+K_E,
\label{ThisMourreEstimate}
\end{equation}
for a constant $\theta_E>0$ and a compact operator $K_E$. We fix $f_E\in C_0^\infty(\mathbb R)$ 
such that $f_E(x)1_{\Delta_E}(x)=x$ for all $x\in\Delta_E$.

If $E_0\in\spec_\pp(H)\cap\Delta_E$, $Hu=E_0u$ and $\|u\|=1$ we have $1_{\Delta_E}(H)u=u$ 
and $f_E(H)u=E_0u$. Since $H\in C^1_\mathrm{loc}(A)$, $[A,f_E(H)]$ is bounded and 
Lemmas~\ref{CommutApprox} and~\ref{fcommut} imply
$$
(u,\i[f_E(H),A]u)=\lim_{\varepsilon\to0}(u,\i[f_E(H),A_\varepsilon]u)
=\i E_0\lim_{\varepsilon\to0}(u,A_\varepsilon u)-(A_\varepsilon^\ast u,u)=0.
$$
Identity (\ref{loccommutid}) and the Mourre estimate (\ref{ThisMourreEstimate}) allow us to write
$$
0=(u,\i[f_E(H),A]u)
=(u,1_{\Delta_E}(H)\i[H,A]1_{\Delta_E}(H)u)\ge\theta_E+(u,K_E \,u),
$$
and thus
\begin{equation}
0<\theta_E\le|(u,K_E\,u)|.
\label{TheContradic}
\end{equation}

Suppose now that $\spec_\pp(H)\cap\Delta_E$ is infinite. There thus exists a sequence 
$E_n\in\spec_\pp(H)\cap\Delta_E$ and a corresponding orthonormal sequence $u_n$
of eigenvectors. It follows that $\wlim_nu_n=0$ and, since $K_E$ is compact, 
$\lim_n\|K_E\,u_n\|=0$. We deduce that
$$
0<\theta_E\le|(u_n,K_E\,u_n)|\le\|K_E\,u_n\|\to0\quad(n\to\infty),
$$
a contradiction which shows that $\spec_\pp(H)\cap\Delta_E$ is finite.

We have shown that every $E\in O$ has an open neighborhood $\Delta_E$ such that 
$\Delta_E\cap\spec_\pp(H)$ is finite. If $I\subset O$ is compact there exists a finite set 
$\mathcal E\subset O$ such that
$$
I\subset\bigcup_{E\in\mathcal E} \Delta_E,
$$
and we conclude that $\spec_\pp(H)\cap I$ is finite.

If the Mourre estimate is strict on $I$ then $K_E=0$ for all $E\in I$ and (\ref{TheContradic}) leads 
to a contradiction which forces us to conclude that $\Delta_E\cap\spec_\pp(H)$ is empty and in 
particular that $E\not\in\spec_\pp(H)$.

(ii) Let $E\in O\setminus\spec_\pp(H)$. Assertion (i) implies that there exists $\delta>0$ such that 
$]E-\delta,E+\delta[\subset O\setminus\spec_\pp(H)$. By denoting 
$P_\varepsilon\equiv 1_{]E-\varepsilon,E+\varepsilon[}(H)$ we thus have 
$\wlim_{\varepsilon\to0}P_\varepsilon=0$ and consequently 
$\lim_{\varepsilon\to0}\|P_\varepsilon KP_\varepsilon\|=0$. We deduce that if $\varepsilon$ is small 
enough then $\|P_\varepsilon KP_\varepsilon\|\le\theta/2$ and 
$$
P_\varepsilon\i[H,A]P_\varepsilon\ge\theta P_\varepsilon+
P_\varepsilon KP_\varepsilon\ge\frac{\theta}{2}P_\varepsilon\ge0.
$$
In the following, we fix such an $\varepsilon$, we set $I\equiv ]E-\varepsilon/2,E+\varepsilon/2[$ and 
we denote
$$
\Omega_\gamma\equiv\{\langle z,\mu\rangle\in\mathbb C\setminus\mathbb R\times\mathbb R
\,|\,\mathrm{Re}z\in I, \sign(\mu)=\sign(\mathrm{Im}z), 0<|\mu|<\gamma\},
$$
where the constant $\gamma>0$ will be fixed later. We also choose a function 
$g\in C_0^\infty(]E-\varepsilon,E+\varepsilon[)$ such that $0\le g\le 1$ and $g=1$ on $[E-3\epsilon/4,E+3\epsilon/4]$. We thus have
\begin{equation}
Q\equiv g(H)\i[H,A]g(H)\ge\frac{\theta}{2}g(H)^2\ge0.
\label{MourreTrickery}
\end{equation}

Since $H\in C^1_\mathrm{loc}(A)$, Lemma \ref{loccom} implies that $Q$ is a bounded self-adjoint
operator. The operator $K_\mu \equiv H-\i\mu Q$ is thus closed on $\Dom(H)$. 
For $\langle z,\mu\rangle\in\Omega_\gamma$ we have
$$
\|u\|\,\|(K_\mu-z)u\|\ge|(u,(K_\mu-z)u)|
\ge|\mathrm{Im}(u,(H-\i\mu Q-z)u)|
\ge|\mathrm{Im}\,z|\,\|u\|^2,
$$
for all $u\in\Dom(H)$. We deduce
\begin{equation}
\|u\|\le|\mathrm{Im}\,z|^{-1} \|(K_\mu-z)u\|,\label{resestone}
\end{equation}
so that $\Ker(K_\mu-z)=\{0\}$. We show that this inequality also implies that 
$\Ran(K_\mu-z)$ is closed. If $v_n=(K_\mu-z)u_n$ for a sequence $u_n\in\Dom(H)$ and if 
$\lim_n v_n=v$ then (\ref{resestone}) implies that $\|u_n-u_m\|\le\|v_n-v_m\|/|\mathrm{Im(z)}|$ 
which shows that $u_n$ is Cauchy. Let $u=\lim_n u_n$. Since $K_\mu-z$ is closed, it follows that
$u\in\Dom(K_\mu)=\Dom(H)$, $v=(K_\mu-z)u$ and thus that $v\in\Ran(K_\mu-z)$.

If $u\in\Ran (K_\mu-z)^\perp$ then, for all  $v\in\Dom(H)$, we have
$$
(u,(K_\mu-z)v)=
(u,Hv)-(u,(z+\i\mu Q)v)=0,
$$
and thus
$$
|(u,Hv)|\le(|z|+|\mu|\,\|Q\|)\|u\|\,\|v\|.
$$
We deduce that $u\in\Dom(H)$ and that
$$
((H-\bar z+\i\mu Q)u,v)=0,
$$
for all $v\in\Dom(H)$. It follows that $u\in\Ker(K_{-\mu}-\bar z)$ and since 
$\langle \bar z,-\mu\rangle\in\Omega_\gamma$ it follows that $u=0$. 
We have thus shown that $\Ran(K_\mu-z)=\mathcal H$. The inverse operator
$$
G_\mu(z)\equiv(K_\mu-z)^{-1},
$$
being closed  with domain $\mathcal H$ is bounded. 
Furthermore, we clearly have $G_\mu(z)^\ast=G_{-\mu}(\bar z)$.

The Mourre estimate (\ref{MourreTrickery}) further gives
\begin{align}
G_\mu(z)^\ast g(H)^2 G_\mu(z)&\le\frac{2}{\theta}G_{\mu}(z)^\ast Q G_\mu(z)\nonumber\\
&\le\frac{2}{\theta\mu}G_{\mu}(z)^\ast(\mathrm{Im} z+\mu Q )G_\mu(z)\label{MourreDeriv}\\
&=\frac{\i}{\theta\mu}(G_{\mu}(z)^\ast -G_\mu(z)),\nonumber
\end{align}
from which we conclude that $\|g(H)G_\mu(z)\|^2\le2\|G_\mu(z)\|/\theta|\mu|$. Using
the second resolvent identity $G_\mu(z)=G_0(z)(I+\i\mu QG_\mu(z))$ we can write
$$
G_\mu(z)^\ast G_\mu(z)=G_\mu(z)^\ast g(H)^2G_\mu(z)+G_\mu(z)^\ast(1-g(H)^2)G_0(z)(I+\i\mu QG_\mu(z)).
$$
Since $G_0(z)=(H-z)^{-1}$ and $\mathrm{dist}(I,\supp(1-g^2))\ge\epsilon/4$, the functional calculus yields the
estimate
$\|(1-g(H)^2)G_0(z)\|\le4/\epsilon$ for $\Re z\in I$ and we obtain
$$
\|G_\mu(z)\|^2\le\frac2{\theta|\mu|}\|G_\mu(z)\|+\frac4\epsilon(1+|\mu|\|Q\|\|G_\mu(z)\|)\|G_\mu(z)\|.
$$
Rewriting the last inequality as
$$
\left(1-\frac{4\|Q\||\mu|}{\epsilon}\right)\|\mu G_\mu(z)\|\le\frac2{\theta}+\frac{4|\mu|}\epsilon,
$$
it is easy to conclude that
\begin{equation}
\sup_{\langle z,\mu\rangle\in\Omega_\gamma}\|\mu G_\mu(z)\|
\le\frac2\theta\frac{\epsilon+2\theta\gamma}{\epsilon-4\|Q\|\gamma}
<\infty,
\label{muGest}
\end{equation}
provided we impose $\gamma<\epsilon/4\|Q\|$.

We shall denote $T\equiv|A+\i|^{-1}$, $\Lambda\equiv(1+|H|)$, and we shall use the notation 
introduced in Definition \ref{Sobolevdef} in what follows.

Multiplying the inequality (\ref{MourreDeriv}) on both sides with $T$ gives
$$
\|g(H) G_\mu(z)T\|^2\le\frac{2}{\theta|\mu|}\|TG_\mu(z)T\|.
$$
Since $g(H)\Lambda^2g(H)\le Cg(H)^2$, it follows that there exists a constant $C_1$ such that
\begin{equation}
\|g(H) G_\mu(z)T\|_{\B(\mathcal H,\mathcal H_H^1)}
\le C_1 |\mu|^{-1/2}\|TG_\mu(z)T\|^{1/2},
\label{C1def}
\end{equation}
for all $\langle z,\mu\rangle\in\Omega_\gamma$.

Using again the second resolvent equation, we can write
$$
\|\Lambda(1-g(H))G_\mu(z)\|
\le\|\Lambda(1-g(H))G_0(z)\|\,(I+\|Q\| \|\mu G_\mu(z)\|),
$$
and the functional calculus and inequality~(\ref{muGest}) yield
\begin{equation}
C_2\equiv\sup_{\langle z,\mu\rangle\in\Omega_\gamma}
\|(I-g(H))G_\mu(z)\|_{\B(\mathcal H,\mathcal H_H^1)}<\infty.
\label{C2def}
\end{equation}

By combining this last estimate and (\ref{C1def}) we obtain
\begin{align}
\|G_\mu(z)T\|_{\B(\mathcal H,\mathcal H_H^1)}
&\le\,\|g(H) G_\mu(z)T\|_{\B(\mathcal H,\mathcal H_H^1)}
+\|(1-g(H))G_\mu(z)\|_{\B(\mathcal H,\mathcal H_H^1)}\,\|T\|\nonumber\\
&\le C_1|\mu|^{-1/2}\|TG_\mu(z)T\|^{1/2}+C_2.\label{GmuT}
\end{align}

Since $\Ran G_\mu(z)=\Dom(H)=\mathcal H_H^1$, and 
$[H,A]\in\B(\mathcal H_H^1,\mathcal H_H^{-1})$ by Lemma \ref{loccom}, the identity
$$
Q=\i[H,A]-g(H)\i[H,A](1-g(H))-(1-g(H))\i[H,A]g(H)-(1-g(H))\i[H,A](I-g(H)),
$$
may be substituted into
$$
\frac{\d\ }{\d\mu}TG_\mu(z)T=\i TG_\mu(z)QG_\mu(z)T,
$$
to obtain, using (\ref{C1def}) and (\ref{C2def}),
\begin{align}
\left\|\frac{\d\ }{\d\mu}TG_\mu(z)T\right\|&\le
\|TG_\mu(z)[H,A]G_\mu(z)T\|\label{PreDiffIneq}\\
&+C_2\,\|[H,A]\|_{\B(\mathcal H_H^{1},\mathcal H_H^{-1})}\left(C_2+2C_1\,|\mu|^{-1/2}
\|TG_\mu(z)T\|^{1/2}\right).\nonumber
\end{align}
To estimate the first term on the right hand side of this inequality, we decompose
$$
[H,A]=[K_\mu-z,A]+\i\mu[Q,A].
$$
Let $f\in C_0^\infty(\mathbb R)$ be such that $f(x)g(x)=xg(x)$. By invoking Lemma \ref{loccom} we 
remark that 
\begin{align*}
[Q,A]=&\i[A,g(H)[A,f(H)]g(H)]\\
=&\i g(H)[A,f(H)][A,g(H)]+\i[A,g(H)][A,f(H)]g(H)
+\i g(H)[A,[A,f(H)]]g(H),
\end{align*}
and since $H\in C^2_\mathrm{loc}(A)$, we can conclude that $[Q,A]$ is bounded.
The estimate (\ref{GmuT}) gives us 
\begin{equation}
\|TG_\mu(z)\i\mu[Q,A]G_\mu(z)T\|\le\|[Q,A]\|\,
(C_1\|TG_\mu(z)T\|^{1/2}+C_2|\mu|^{1/2})^2.
\label{TGQGT}
\end{equation}
Also, the identity 
$$
G_\mu(z)[K_\mu-z,A]G_\mu(z)=[A,G_\mu(z)],
$$
and the estimate (\ref{GmuT}) allows us to write
\begin{align}
\|TG_\mu(z)[K_\mu-z,A]G_\mu(z)T\|&=\|T[A,G_\mu(z)]T\|\nonumber\\
&\le\|G_\mu(z)T\|+\|TG_\mu(z)\|\label{TGKGT}\\
&\le2(C_1|\mu|^{-1/2}\|TG_\mu(z)T\|^{1/2}+C_2).\nonumber
\end{align}
By combining (\ref{TGQGT}) and (\ref{TGKGT}) in (\ref{PreDiffIneq}) we obtain, after taking into 
account (\ref{muGest}), the following differential inequality.
$$
\left\|\frac{\d\ }{\d\mu}TG_\mu(z)T\right\|\le
a+b\,\|TG_\mu(z)T\|+c\frac{\|TG_\mu(z)T\|^{1/2}}{|\mu|^{1/2}},
$$
valid for all $\langle z,\mu\rangle\in\Omega_\gamma$ and where $a,b,c$ are positive constants. 
Setting $\phi(\mu)=\|TG_\mu(z)T\|$ we can write, for $0<\mu<\mu_0<\gamma$,
\begin{align*}
\phi(\mu)&\le\phi(\mu_0)+\int_{\mu}^{\mu_0}(a+b\phi(\nu)+c\nu^{-1/2}\phi(\nu)^{1/2})\,\d\nu\\
&\le\Phi(\mu)\equiv(\phi(\mu_0)+a\mu_0)+\int_{\mu}^{\mu_0}(b\phi(\nu)
+c\nu^{-1/2}\phi(\nu)^{1/2})\,\d\nu.
\end{align*}
With $\Psi(\mu)\equiv\Phi(\mu)\e^{-b(\mu_0-\mu)}$, we easily compute
\begin{align*}
\frac{\d\ }{\d\mu}\Psi(\mu)
&=\left(b(\Phi(\mu)-\phi(\mu))-c\mu^{-1/2}\phi(\mu)^{1/2})\right)\e^{-b(\mu_0-\mu)}\\
&\ge-c\mu^{-1/2}\phi(\mu)^{1/2}\e^{-b(\mu_0-\mu)}\\
&\ge-c\mu^{-1/2}\Phi(\mu)^{1/2}\e^{-b(\mu_0-\mu)}.
\end{align*}
Introducing $\psi(\mu)\equiv c\mu^{-1/2}\e^{-b(\mu_0-\mu)/2}$ we can rewrite this last 
inequality as 
$$
\frac{\d\ }{\d\mu}\Psi(\mu)\ge-\psi(\mu)\Psi(\mu)^{1/2},
$$
and from this we get that
$$
\frac{\d\ }{\d\mu}\Psi(\mu)^{1/2}\ge-\frac{1}{2}\psi(\mu).
$$
After integration, we obtain
$$
\Psi(\mu)^{1/2}\le\Psi(\mu_0)^{1/2}+\frac{1}{2}\int_\mu^{\mu_0}\psi(\nu)\,\d\nu,
$$
and since $\Psi(\mu_0)=\Phi(\mu_0)=\phi(\mu_0)+a\mu_0$,
$$
\Psi(\mu)^{1/2}\le(\phi(\mu_0)+a\mu_0)^{1/2}+\frac{c}{2}\int_\mu^{\mu_0}\nu^{-1/2}\e^{-b(\mu_0-\nu)/2}
\,\d\nu.
$$
Finally, we get
\begin{align*}
\phi(\mu)^{1/2}&\le\Phi(\mu)^{1/2}=\e^{b(\mu_0-\mu)/2}\Psi(\mu)^{1/2}\\
&\le\e^{b\mu_0/2}(\phi(\mu_0)+a\mu_0)^{1/2}+\frac{c}{2}\int_0^{\mu_0}\nu^{-1/2}
\e^{b(\nu-\mu)/2}\,\d\nu\\
&\le\e^{b\mu_0/2}\left((\phi(\mu_0)+a\mu_0)^{1/2}+c\mu_0^{1/2}\right),
\end{align*}
and it is straightforward to check that the same estimate is valid for $-\gamma<-\mu_0<\mu<0$. 
We have thus shown that $\sup_{\langle z,\mu\rangle\in\Omega_\gamma}\|TG_\mu(z)T\|<\infty$, 
and in particular that
$$
\sup_{\Re(z)\in I,\Im(z)\not=0}\|T(H-z)^{-1}T\|<\infty.
$$
Invoking a covering argument, we easily show that, for any compact set 
$I\subset O\setminus\spec_\pp(H)$, 
\begin{equation}
C(I)\equiv\sup_{\Re(z)\in I,\Im(z)\not=0}\|T(H-z)^{-1}T\|<\infty.
\label{LAP}
\end{equation}
Since  $\Ran(T)=\Dom(A)$ is dense in $\mathcal H$, we can conclude that 
$I\cap\spec_\mathrm{sc}(H)$ is empty.
\cqfd
\end{demo}

\subsection{Propagation estimates}
\label{PropagSect}
The Mourre estimate provides a very efficient method to derive propagation estimates, which 
are essential ingredients of the time-dependent approach to scattering theory. We shall also use
these estimates in our derivation of the Landauer-B\"uttiker formula.

The first result is a simple corollary of the proof of Theorem \ref{MourreSingSpec} (see \cite{M2,M3}).

\begin{Coro}\label{Katosmooth} Let $H$ be a self-adjoint operator on the Hilbert space $\mathcal H$.
If $H$ satisfies a Mourre estimate on the open set $O\subset\mathbb R$ with the conjugate
operator $A$ and if $H\in C^2_\mathrm{loc}(A)$ then, for any function
$g\in C_0^\infty(O\setminus\spec_\pp(H))$ there exists a constant $C$ such that
\begin{equation}
\int_{-\infty}^\infty\|(1+A^2)^{-1/2}\e^{-\i tH}g(H)u\|^2\,\d t\le
C\,\|u\|^2,
\label{SmoothIntegral}
\end{equation}
for all $u\in\mathcal H$.
\end{Coro}

\begin{demo} Let $g\in C_0^\infty(O\setminus\spec_\pp(H))$ and set $I=\supp\,g$. 
The hypotheses of Theorem \ref{MourreSingSpec} being satisfied and 
$(1+|A|)^{-1}(1+A^2)^{1/2}$ being bounded, the inequality (\ref{LAP}) implies
$$
\sup_{\Re(z)\in I,\Im(z)\not=0}\|(1+A^2)^{-1/2}(H-z)^{-1}(1+A^2)^{-1/2}\|<\infty.
$$
It follows that $(1+A^2)^{-1/2}$ is $H$-smooth on $I$ (compare with Eq.~\eqref{locsmoothcrit}). 
The operator $(A^2+1)^{-1/2}g(H)$ is thus $H$-smooth, and (\ref{SmoothIntegral}) follows. \cqfd
\end{demo}

\bigskip
Our main tool in controlling the dynamics is the propagation estimate of Sigal-Soffer \cite{SiSo} 
(see also \cite{HSS} and Section 4.12 of \cite{DG}). Recall that if $A$ is a self-adjoint operator
then $F(A\le a)$ denotes the spectral projection $E_{]-\infty,a]}(A)$, etc.

\begin{Prop}\label{MainPropEst} Let $A$ and $H$ be self-adjoint operators on the Hilbert space 
$\mathcal H$ such that:
\begin{enumerate}[(i)]
\item $H\in C^n_\mathrm{loc}(A)$ for an integer $n\ge 2$. 
\item $H$ satisfies a strict Mourre estimate with the conjugate operator $A$ on the open set 
$O\subset\mathbb R$.
\end{enumerate}
Then, for all $s<n-1$ and $g\in C_0^{\infty}(O)$, there exist constants $\vartheta>0$ and $c$ 
such that
\begin{equation}
\|F(\pm A\le a-b+\vartheta t)\e^{\mp\i tH}g(H)F(\pm A\ge a)\|\le c\,\langle b+\vartheta t\rangle^{-s},
\label{TheMainPropEstim}
\end{equation}
for all $a\in\mathbb R$, $b\ge0$ and $t\ge0$.
\end{Prop}


\section{Non-equilibrium steady states}
\label{NESSChapter}

In this section, we reconsider the problem of constructing nonequilibrium steady states for an 
open system of quasi-free fermions driven by extended reservoirs. Like the Authors of \cite{AJPP2,N}, 
we follow Ruelle's scattering approach \cite{R2,R3}. The originality of the present work is
in the use of time dependent scattering theory, inspired by the approach of Avron {\sl et al.\;}to
the related problem of adiabatic charge pumping \cite{AEGSS}.

The stationary approach to scattering used in \cite{AJPP2,N} has the advantage of providing explicit 
representations of certain objects (M\o ller operators, scattering matrix). It thus allows for fairly 
simple and direct calculations. On the other hand, it requires quite strong assumptions, 
specifically some trace conditions on the coupling between the system and the 
reservoir. Furthermore, it does not provide any control on the singular continuous spectrum. 
The absence of this spectral component is part of the assumptions of \cite{AJPP2}. 
Our time-dependent approach is base on Mourre theory. It simultaneously gives us control over 
the singular spectrum, propagation estimates, and the property of local {\it Kato-Smoothness} which
provide the construction of complete M\o ller operators and unitary scattering matrix.

\subsection{Model and hypotheses}

\begin{figure}
\centering
\includegraphics[scale=0.50]{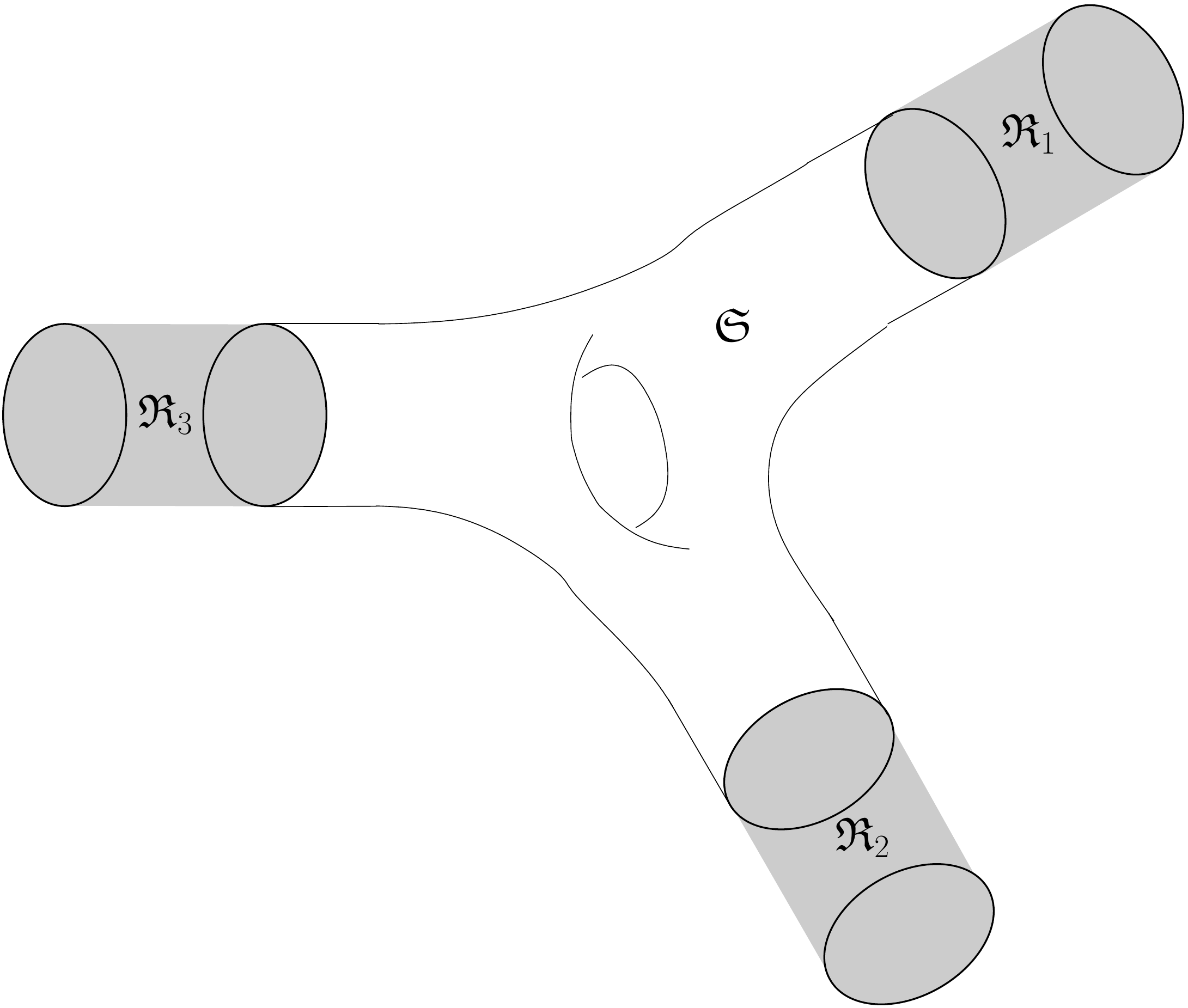}
\caption{An ideal Fermi gas on a geometric structure $\mathfrak M$ structure.}
\label{TubularFig}
\end{figure}

We consider an ideal Fermi gas confined to a connected geometric structure $\mathfrak M$. This 
structure may be a domain $\mathfrak M\subset\mathbb R^d$ or a finite dimensional Riemannian 
manifold which we shall assume is itself embedded in a Euclidean space $\mathbb R^d$. We shall 
suppose that $\mathfrak M$ is the disjoint union of a a compact subset $\mathfrak S$ and of $M$ 
infinite tubular or cylindrical branches $\mathfrak R_1,\ldots\mathfrak R_M$ 
(see figure \ref{TubularFig}).

The fermions visiting the compact part $\mathfrak S\subset\mathfrak M$ 
form the small system $\mathcal S$. The reservoirs $\mathcal R_1,\ldots,\mathcal R_M$ consist
in the fermions contained in the infinitely extended branches $\mathfrak R_1,\ldots,\mathfrak R_M$. 
In the context of mesoscopic physics, $\mathcal S$ is a sample connected to the electronic 
reservoirs $\mathcal R_k$. We denote by $\mathcal H$ the one-particle Hilbert space 
of the system, and by $H$ its one-particle Hamiltonian, a self-adjoint operator on $\mathcal H$.

The system is described by the $C^\ast$-algebra $\mathcal O\equiv\mathrm{CAR}(\mathcal H)$ 
equipped with the group of Bogoliubov automorphisms 
$$
\tau^t(a(f))=a(\e^{\i tH}f).
$$
\begin{figure}
\centering
\includegraphics[scale=0.5]{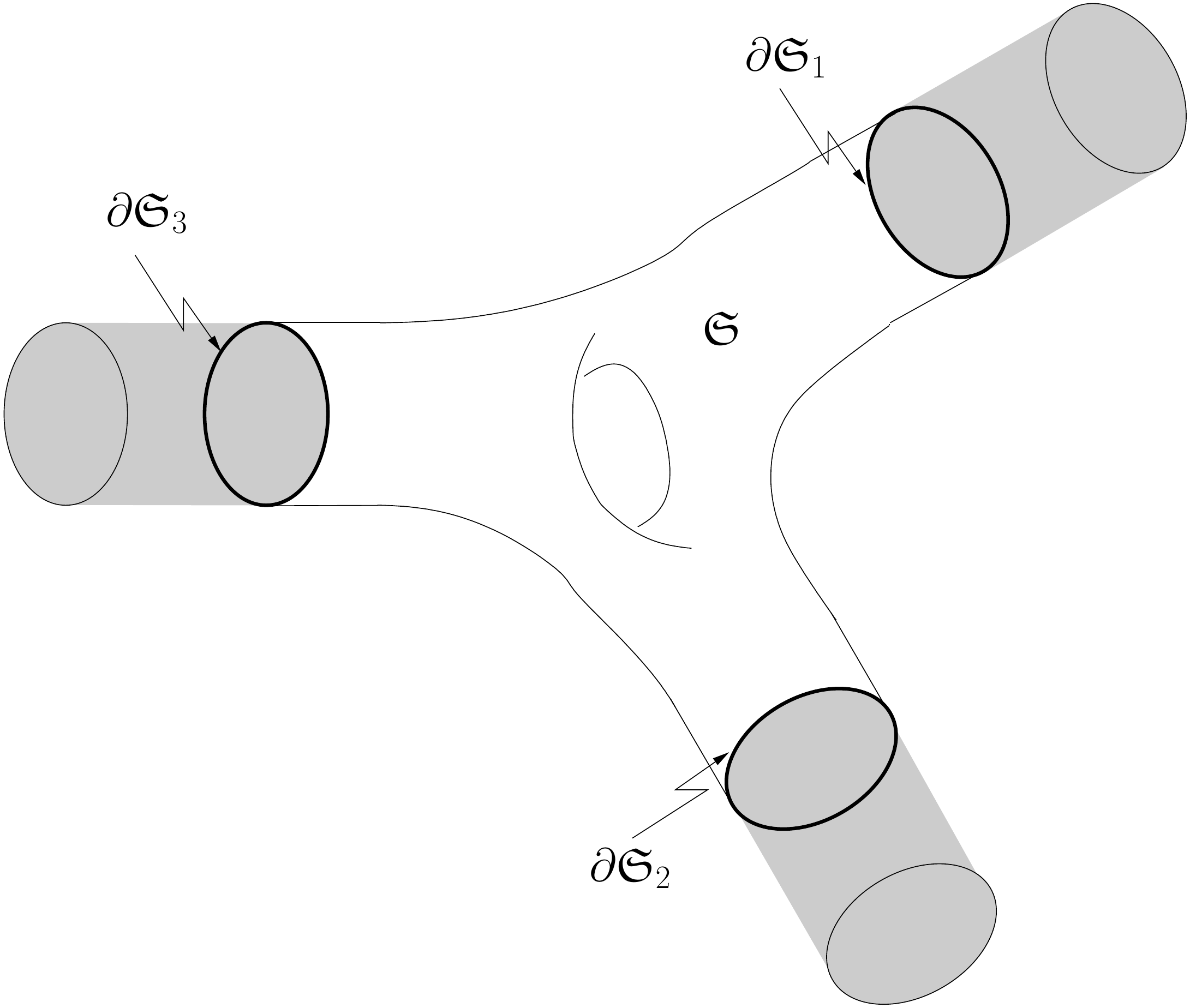}
\caption{Decoupling of the sample $\mathfrak S$. Appropriate boundary conditions are imposed 
on the fictitious walls $\partial\mathfrak S_k$.}
\label{Lambdafig}
\end{figure}

\begin{figure}
\centering
\includegraphics[scale=0.5]{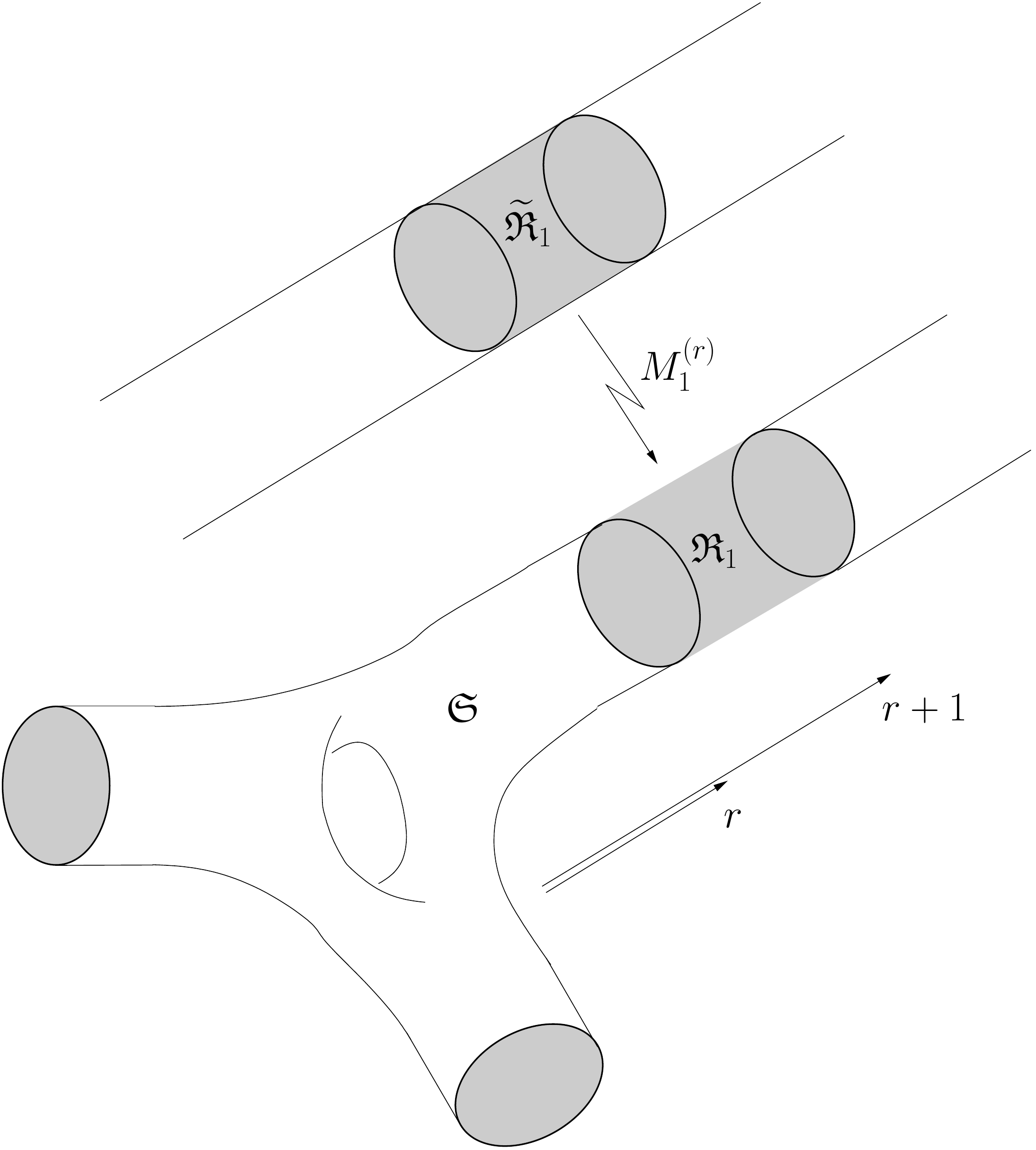}
\caption{Submersion of the super-reservoir $\widetilde{\mathfrak R}_1$ into the extended branch 
$\mathfrak R_1$ of $\mathfrak{M}$ . The parameter $r$ describes the depth of this submersion.}
\label{ImmersionFig}
\end{figure}

To apply scattering theory to this system requires the definition of a reference dynamics. A possible 
approach consists in decoupling the compact part $\mathfrak S$ from the extended branches 
$\mathfrak R_k$ by imposing boundary conditions on fictitious walls $\partial\mathfrak{S}_k$ 
surrounding $\mathfrak S$ (see Figure \ref{Lambdafig}). 
We then obtain a new Hamiltonian $H_\mathrm{ref}$ which will 
serve as reference. This is essentially the approach followed in \cite{AJPP2} for example. 
However, this method has a serious disadvantage. In fact, the scattering matrix obtained in this
way depends {\em a priori} on the largely arbitrary method used to perform the decoupling, i.e. 
the position of the decoupling walls, as well as the boundary conditions imposed at these walls. 
We shall avoid this difficulty by adopting a more geometric approach. We shall consider each reservoir 
as part of a larger system, a kind of super-reservoir, by immersing each branch $\mathfrak R_k$ 
in a reference structure $\widetilde{\mathfrak R}_k$ (see Figure \ref{ImmersionFig}). 
In this context, the two Hilbert spaces formalism of scattering theory applies. The advantage
of this method is in the fact that the scattering matrix only depends on the geometry of 
the reservoirs and not on artificial decoupling techniques. 

Let us now formulate our main hypotheses.

\begin{hyp}
\item {\bf Submersion.} For $k\in\{1,\ldots,M\}$ there exists a Hilbert space $\HH_k$ as well as a 
family of identification operators 
$\{M_k^{(r)}\,|\,r\ge0\}\subset\mathcal B(\widetilde{\mathcal H}_k,\mathcal H)$. We denote by 
$M_k^{(r)}=J_k^{(r)}\widetilde\chi_k^{(r)}$ their polar decompositions,  with
$\widetilde\chi_k^{(r)}\equiv|M_k^{(r)}|=(M_k^{(r)\ast}M_k^{(r)})^{1/2}$. We also set
$\chi_k^{(r)}\equiv(M_k^{(r)}M_k^{(r)\ast})^{1/2}$, $\tilde 1_k^{(r)}\equiv J_k^{(r)\ast}J_k^{(r)}$ and 
$1_k^{(r)}\equiv J_k^{(r)}J_k^{(r)\ast}$.
We assume that:
  \begin{enumerate}[(i)]
  \item $\|M_k^{(r)}\|=1$.
  \item $\tilde 1_k^{(r+1)}\le\widetilde\chi_k^{(r)2}$ and $1_k^{(r+1)}\le\chi_k^{(r)2}$.  
  \item $J_k^{(r)\ast}J_l^{(r)}=0$ for $k\not=l$.
  \item $\tilde 1_k^{(s)}\le\tilde 1_k^{(r)}$ if $0\le r\le s$ and 
   $\slim_{r\to\infty}\tilde 1_k^{(r)}=0$.
  \item $1_k^{(s)}\le1_k^{(r)}$ if $0\le r\le s$ and $\slim_{r\to\infty}1_k^{(r)}=0$.
  \item $J_k^{(r)}\tilde1_k^{(s)}=1_k^{(s)}J_k^{(r)}=J_k^{(s)}$ for $0\le r\le s$.
  \end{enumerate}
\label{HypReservoir}
\end{hyp}

$\HH_k$ is the Hilbert space of the super-reservoir and $M_k^{(r)}$ is the operator which 
maps one half of this super-reservoir $\widetilde{\mathcal R}_k$ into $\mathfrak M$. 
The parameter $r$ describes the ``depth'' of this submersion. One can also think of $r$ as
describing the ``position'' of a fictitious interface between the system $\mathfrak S$ and the 
reservoir $\mathcal R_k$ (see Figure \ref{ImmersionFig}).

The properties of the polar decomposition imply that the operator $J_k^{(r)}$ is a partial 
isometry with initial space
$$
\widetilde{\mathcal H}_k^{(r)}\equiv
\Ran(M_k^{(r)\ast})^\mathrm{cl}=\Ker(M_k^{(r)})^\perp
=\Ran(\widetilde\chi_k^{(r)})^\mathrm{cl}=\Ker(\widetilde\chi_k^{(r)})^\perp
=\Ran(J_k^{(r)\ast})=\Ker(J_k^{(r)})^\perp,
$$
and final space
$$
\mathcal H_k^{(r)}\equiv\Ran(M_k^{(r)})^\mathrm{cl}=\Ker(M_k^{(r)\ast})^\perp
=\Ran(\chi_k^{(r)})^\mathrm{cl}=\Ker(\chi_k^{(r)})^\perp
=\Ran(J_k^{(r)})=\Ker(J_k^{(r)\ast})^\perp.
$$
$J_k^{(r)\ast}$ is the inverse isometry. $1_k^{(r)}$ is the orthogonal projection 
of $\mathcal H$ onto $\mathcal H_k^{(r)}$ while $\widetilde{1}_k^{(r)}$ is the orthogonal projection 
of $\HH_k$ onto $\HH_k^{(r)}$. In particular, since $\widetilde\chi_k^{(r)}\ge0$ and 
$\|\widetilde\chi_k^{(r)2}\|=\|M_k^{(r)\ast}M_k^{(r)}\|=\|M_k^{(r)}\|^2=1$ one has
$0\le\widetilde\chi_k^{(r)}\le I$ and since $\Ran(\widetilde\chi_k^{(r)})^{\rm cl}=\Ran(\tilde1_k^{(r)})$
we easily verify the inequalities
\begin{equation}
0\le\widetilde\chi_k^{(r)2}\le
\widetilde\chi_k^{(r)}\le\tilde1_k^{(r)}\le I,\qquad
0\le\chi_k^{(r)2}\le\chi_k^{(r)}\le1_k^{(r)}\le I,
\label{chidomin}
\end{equation}
for $r\ge0$. If $u\in\Ran\tilde1_k^{(r+1)}$ we deduce from~(\ref{chidomin}) and Property~(ii) that
$$
0\le(u,(I-\widetilde\chi_k^{(r)})u)\le(u,(I-\widetilde\chi_k^{(r)2})u)
=(u,(\tilde1_k^{(r+1)}-\widetilde\chi_k^{(r)2})u)\le0,
$$
so that $(I-\widetilde\chi_k^{(r)2})u=0$. Since 
$(I-\widetilde\chi_k^{(r)})u=(I+\widetilde\chi_k^{(r)})^{-1}(I-\widetilde\chi_k^{(r)2})u=0$
we conclude that $\widetilde\chi_k^{(r)}\tilde 1_k^{(r+1)}=\tilde1_k^{(r+1)}
=\tilde1_k^{(r+1)}\widetilde\chi_k^{(r)}$ (the same identity is verified without the tildes). 
For $s\ge r+1\ge 1$ Properties (iv)-(v) yield $\tilde 1_k^{(r+1)}\tilde 1_k^{(s)}=\tilde 1_k^{(s)}$
(and the same relation without the tildes), hence one has
\begin{equation}
\widetilde\chi_k^{(r)}\tilde 1_k^{(s)}=\tilde 1_k^{(s)}\widetilde\chi_k^{(r)}=\tilde 1_k^{(s)},
\qquad
\chi_k^{(r)}1_k^{(s)}=1_k^{(s)}\chi_k^{(r)}=1_k^{(s)},
\label{chidomin2}
\end{equation}
Property (iii) implies that $1_k^{(r)}1_l^{(r)}=\delta_{kl}1_k^{(r)}$. 
The operator $1_0^{(r)}=I-\sum_{j=1}^M1_k^{(r)}$ is the orthogonal projection onto a neighborhood 
of ``radius'' $r$ containing the system $\mathfrak S$.

We thus obtain a partition of unity on $\mathcal H$. We shall write
$$
\mathcal H=\bigoplus_{k=0}^M\mathcal H_k^{(r)},
$$
for the corresponding decomposition. By setting $\HH_0^{(r)}\equiv\mathcal H_0^{(r)}$ and 
$J_0^{(r)}=I$ we obtain a unitary
$$
\begin{array}{rlccl}
U^{(r)}:&\HH\equiv\oplus_{k=0}^M\HH_k^{(r)}&\to&\mathcal H\\
&(u_0,\ldots,u_M)&\mapsto&\sum_{j=0}^MJ_k^{(r)}u_k,
\end{array}
$$
with inverse
$$
\begin{array}{rlccr}
U^{(r)\ast}:&\mathcal H&\to&\HH\\
&u&\mapsto&(J_0^{(r)\ast}u,\ldots,J_M^{(r)\ast}u).
\end{array}
$$

\begin{hyp}
\item {\bf Coupling.} For $k\in\{1,\ldots,M\}$ there exists a self-adjoint Hamiltonian  $\widetilde H_k$ on $\HH_k$ such that
\begin{enumerate}[(i)]
  \item $M_k^{(r)}\Dom(\widetilde H_k)\subset\Dom(H)$ and 
  $M_k^{(r)\ast}\Dom(H)\subset\Dom(\widetilde H_k)$.
  \item $H$ ``coincides'' with $\widetilde H_k$ on 
  $M_k^{(r)}\Dom(\widetilde H_k)$:
    $$
    HM_k^{(r)}u=
    J_k^{(r)}\widetilde H_k\widetilde\chi_k^{(r)}u,
    $$
  for all $u\in\Dom(\widetilde H_k)$.
  \item $(I-\tilde1_k^{(r)})\widetilde H_k\widetilde\chi_k^{(r)}u=0$ for all
   $u\in\Dom(\widetilde H_k)$ and $r\ge0$.
  \item The operator $B_k^{(r)}\equiv[\widetilde H_k,\widetilde\chi_k^{(r)}]$ is
  $\widetilde H_k$-compact.
  \item The operator $B_k^{(r)\ast}J_k^{(r)\ast}$ is $H$-compact.
  \item For all $r\ge0$, $1_0^{(r)}$ is $H$-compact.
  \item For all $r,s\ge0$, $\tilde1_k^{(r)}-\tilde1_k^{(s)}$ 
        is $\widetilde H_k$-compact.
\end{enumerate}
\label{HypResDyn}
\end{hyp}

We note that condition (iii) and Hypothesis \Href{HypReservoir} (ii) imply that if $s\ge r+1\ge 1$ 
then
\begin{equation}
\widetilde\chi_k^{(s)}\widetilde H_k(I-\widetilde\chi_k^{(r)})u
=(I-\widetilde\chi_k^{(r)})\widetilde H_k\widetilde\chi_k^{(s)}u=0,
\label{OldHyp3}
\end{equation}
for all $u\in\Dom(\widetilde H_k)$.

\bigskip
Finally our main hypothesis ensures good propagation properties in the reservoirs. We shall denote by 
$\HH_k^1$ the space $\Dom(\widetilde H_k)$ equipped with the graph norm and $\HH_k^{-1}$ its dual 
(see Definition \ref{Sobolevdef}).

\begin{hyp}
\item {\bf Mourre estimate.} 
For $k\in\{1,...,M\}$ there exists a self-adjoint operator $\widetilde A_k$ on $\HH_k$
and a closed, countable subset $\Sigma_k\subset\mathbb R$ such that for any
$E\in\mathbb R\setminus\Sigma_k$, $\widetilde{H}_k$ satisfies a Mourre estimate at $E$
with the conjugate operator $\widetilde A_k$. Furthermore:
\begin{enumerate}[(i)]
  \item $\widetilde\chi_k^{(r)}\Dom(\widetilde A_k)\subset\Dom(\widetilde A_k)$ 
        for all $r\ge0$.
  \item $[\widetilde A_k,\widetilde\chi_k^{(r)}]=0$ for $r\in[0,2]$.
  \item $(\tilde 1_k^{(0)}-\widetilde\chi_k^{(r)})\widetilde A_k=0$ for $r\in[0,2]$.
  \item $\e^{\i\theta \widetilde A_k}\Dom(\widetilde{H}_k)\subset\Dom(\widetilde{H}_k)$ for $\theta\in\mathbb R$.
  \item $\widetilde H_k\in\B_{\widetilde A_k}^n(\HH_k^1,\HH_k)$ for some integer $n\ge 2$.
\end{enumerate}\label{HypConjugate}
\end{hyp}

\subsection{A simple model}
\label{cylindricends}

In this section we illustrate our Hypotheses  \Href{HypReservoir}--\Href{HypConjugate} with
a simple non-trivial example. We also discuss various possible extensions and modifications 
of this example.

Let $\mathfrak M$ be a smooth 2-dimensional connected sub-manifold of $\rr^3$ such that
$\mathfrak{M}=\mathfrak{S}\cup\mathfrak{R}_-\cup\mathfrak{R}_+$ where
$\mathfrak{S}$ is compact with boundary $\partial\mathfrak{S}=\gamma_-\cup\gamma_+$
where
$$
\gamma_\mp=\left\{x=(x_1,x_2,\mp1)\in\rr^3\,\big|\,x_1^2+x_2^2=R^2\right\},
$$
and $\mathfrak{R}_\mp$ are semi-infinite cylinders
$$
\mathfrak{R}_\mp=\left\{x=(x_1,x_2,x_3)\in\rr^3\,\big|\,x_1^2+x_2^2=R^2,\mp x_3>1\right\},
$$
The super-reservoirs are infinite cylinders
$$
\widetilde{\mathfrak{R}}_\mp=\{x=(x_1,x_2,x_3)\in\rr^3\,|\,x_1^2+x_2^2=R^2\},
$$
and for $r\ge0$ we set 
$\mathfrak{R}_\mp^{(r)}=\{x\in\mathfrak{R}_\mp\,|\,\mp x_3\ge 1+r\}
\subset\mathfrak{M}\cap\widetilde{\mathfrak{R}}_\mp$.

The various Hilbert spaces are just the corresponding $L^2$-spaces with the induced surface 
measures, e.g., $\cH=L^2(\mathfrak{M})$. The operator $M_+^{(r)}$ is defined by
$$
(M_+^{(r)}f)(x_1,x_2,x_3)=\left\{
\begin{array}{ll}
g(x_3-1-r)f(x_1,x_2,x_3)& \text{if }(x_1,x_2,x_3)\in\mathfrak{R}_+,\\[4pt]
0&\text{otherwise},
\end{array}
\right.
$$
where $g\in C^\infty(\rr)$ is such that $0\le g(x)\le 1$, $g(x)=0$ for $x\le0$ and $g(x)=1$ for $x\ge1$.
A similar definition holds for $M_-^{(r)}$. One easily checks that $\widetilde{\chi}_+^{(r)}$ is the 
operator of multiplication by the function $g(x_3-1-r)$ on the Hilbert space 
$L^2(\widetilde{\mathfrak{R}}_+)$ while $\chi_1^{(r)}$ is the operator of multiplication 
by the function $1_{\mathfrak{R}_+}(x)g(x_3-1-r)$ on the Hilbert spaces $L^2(\mathfrak{M})$
where $1_{\mathfrak{R}_+}$ denotes the indicator function of $\mathfrak{R}_+\subset\mathfrak{M}$.
The partial isometry $J_\mp^{(r)}:L^2(\widetilde{\mathfrak{R}}_\mp)\to L^2(\mathfrak{M})$ is given by
$$
(J_\mp^{(r)}u)(x)=\left\{\begin{array}{ll}
u(x)&\text{ if }x\in\mathfrak{R}_\mp^{(r)},\\[4pt]
0&\text{ if }x\in\mathfrak{M}\setminus\mathfrak{R}_\mp^{(r)},
\end{array}
\right.
$$
and its adjoint $J_\mp^{(r)\ast}:L^2(\mathfrak{M})\to L^2(\widetilde{\mathfrak{R}}_\mp)$ acts as
$$
(J_\mp^{(r)\ast}u)(x)=\left\{\begin{array}{ll}
u(x)&\text{ if }x\in\mathfrak{R}_\mp^{(r)},\\[4pt]
0&\text{ if }x\in\widetilde{\mathfrak{R}}_1\setminus\mathfrak{R}_\mp^{(r)}.
\end{array}
\right.
$$
The orthogonal projection $1_\mp^{(r)}$ (resp. $\widetilde{1}_\mp^{(r)}$) acts on $L^2(\mathfrak{M})$
(resp. on $L^2(\widetilde{\mathfrak{R}}_\mp)$) as multiplication with the indicator function of the 
subset $\mathfrak{R}_\mp^{(r)}$. It is a simple exercise to verify Properties (i)--(vi) of 
Hypothesis~\Href{HypReservoir}.

Denote by  $\Delta_{\mathfrak{M}}$ the Laplace-Beltrami operator acting on 
$\mathcal{D}_{\mathfrak M}=C_0^\infty(\mathfrak M)$.
If $q=(q_1,q_2)$ are local coordinates on some open subset
$\mathfrak{Q}\subset\mathfrak M$ then $\Delta_{\mathfrak{M}}$ acts on $C_0^\infty(\mathfrak{Q})$
as the second order, elliptic differential operator
$$
\sum_{i,j=1}^2 g(q)^{-1/2}\partial_{q_i} g(q)^{1/2}g^{ij}(q)\partial_{q_j},
$$
where $g$ denotes the determinant of the metric tensor $[g_{ij}]$ and $[g^{ij}]$ its inverse. 
In particular $\Delta_{\mathfrak{M}}$ maps $\mathcal{D}_{\mathfrak M}$ into itself. Since the surface 
measure is given locally on $\mathfrak{Q}$ by $\d\sigma=g(q)^{1/2}\d q_1\d q_2$, one easily checks that
$\Delta_{\mathfrak{M}}$ is symmetric as an operator on $L^2(\mathfrak{M})$ with domain 
$\mathcal{D}_{\mathfrak M}$. We denote by the same symbol the dual action of 
$\Delta_{\mathfrak{M}}$ on the space $\mathcal{D}'(\mathfrak{M})$ of distributions on 
$\mathfrak{M}$ (the duality being induced by the inner product of $L^2(\mathfrak M)$).

In fact, the operator $-\Delta_{\mathfrak{M}}$ is essentially self-adjoint on 
$\mathcal{D}_{\mathfrak{M}}$ (see, e.g., \cite{Ch,Co,D5,Str}) and we denote by $H$ its self-adjoint 
extension. Explicitly, the domain of $H$ is given by
$\Dom(H)=\{u\in\cH\,|\,\Delta_{\mathfrak{M}}u\in\cH\}$ and for $u\in\Dom(H)$ one has
$Hu=-\Delta_{\mathfrak{M}}u$ in distributional sense. Moreover, $H$ is positive and its quadratic form
is the Dirichlet form
$$
(u,Hu)=\int_{\mathfrak{M}}|\nabla u|^2\d\sigma,
$$
where $\nabla$ denotes the gradient operator (in local coordinates $\nabla^i=\sum_jg^{ij}(q)\partial_{q_j}$).
The same conclusions hold for the Laplace-Beltrami operator $\Delta_{\widetilde{\mathfrak R}_\mp}$
acting on $C_0^\infty(\widetilde{\mathfrak R}_\mp)$ and we denote by $\widetilde{H}_\mp$ the self-adjoint 
extension of $-\Delta_{\widetilde{\mathfrak R}_\mp}$. It is now straightforward to verify Properties
(i)--(iii) of Hypothesis~\Href{HypResDyn}. Properties (iv)--(vii) easily follow from the fact that
the weighted Sobolev space
$$
H_w^s(\mathfrak{M})=\{u\in\Dom(H^{s/2})\,|\,wH^{s/2} u\in L^2(\mathfrak{M})\},
$$
with $s>0$ is compactly embedded in $L^2(\mathfrak{M})$ if the weight $w\in C^\infty(\mathfrak{M})$
is such that 
$$
\lim_{|x_3|\to\infty}w(x)=+\infty,
$$
(and a similar statement for 
$H_w^s(\widetilde{\mathfrak{R}}_\mp)$, see, e.g. \cite{Lo}).

We note that $\HH_\mp=L^2(\rr\times\gamma_\mp,\d x\,R\d\varphi)$, where 
$R\d\varphi$ is the arc-length measure on the circle $\gamma_\mp$, and that
$$
\widetilde{H}_\mp=-\partial_x^2+\Lambda^2,
\qquad \Lambda^2=-R^{-2}\partial_\varphi^2.
$$
Set $\Sigma_\mp=\spec(\Lambda^2)=\{\lambda_n^2=n^2/R^2\,|\,n\in\nn\}$ and
let $v\in C^\infty(\rr)$ be 
such that $v(x)=0$ for $|x|\le 5$, $v'(x)\ge0$ and $v(x)=x$ for $|x|\ge15$. Denote by $\Phi^t$ the global 
flow defined by the ODE $\dot x=v(x)$ and set $j^t=(\partial_x\Phi^t)^{1/2}$. The operators defined by
$(U^tu)(x,\varphi)=j^t(x)u(\Phi^t(x),\varphi)$ form a strongly continuous unitary group on  
$\HH_\mp$ leaving the subspaces $C_0^\infty(\widetilde{\mathfrak{R}}_\mp)$ as well as 
$\Dom(\widetilde{H}_\mp)$ invariants.
Define $\widetilde{A}_\mp$ to be its self-adjoint generator. One easily checks that
$C_0^\infty(\widetilde{\mathfrak{R}}_\mp)$ is in the domain of $\widetilde{A}_\mp$ and that
$$
\widetilde{A}_\mp=\frac1{2\i}\left(v(x)\partial_x+\partial_xv(x)\right),
$$
on this subspace. By the core theorem,  $\widetilde{A}_\mp$  is essentially self-adjoint on 
$C_0^\infty(\widetilde{\mathfrak{R}}_\mp)$ and hence acts in the same way, in the sense of
distributions, on its domain 
$\Dom(\widetilde{A}_\mp)=\{u\in\HH_\mp\,|\,\widetilde{A}_\mp u\in\widetilde\cH_\mp\}$.

\begin{figure}
\centering
\includegraphics[scale=0.5]{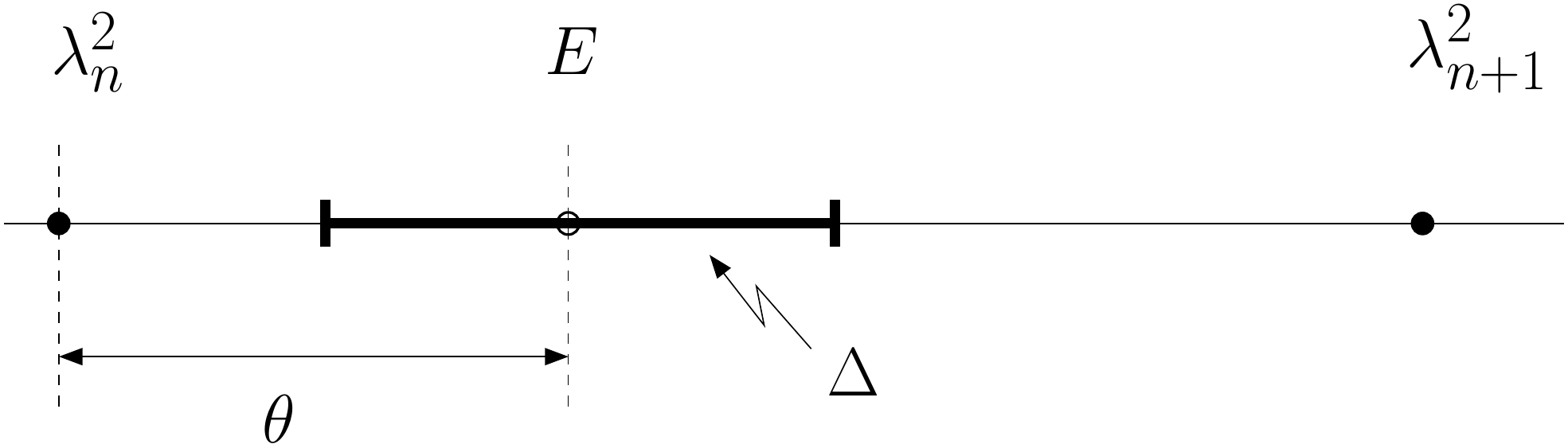}
\caption{If $E\not\in\Sigma_\mp=\{\lambda_k^2\}_{k\in\nn}$, one can chose $\Delta$ such that
${\rm dist}(\Sigma_\mp,\Delta)=\theta/2>0$.}
\label{SpedFig}
\end{figure}

A formal calculation shows that
$$
\i[\widetilde{H}_\mp,\widetilde{A}_\mp]=2(\widetilde{H}_\mp-\Lambda^2)
+2\partial_x(1-v'(x))\partial_x-\frac12 v'''(x),
$$
which defines a bounded quadratic form on $\Dom(\widetilde{H}_\mp)$. Applying 
Theorem \ref{localchar}
we conclude that $\widetilde{H}_\mp$ is of class $C^1_{\rm loc}(\widetilde{A}_\mp)$ and
Lemma \ref{loccom} yields that 
$$
1_\Delta(\widetilde{H}_\mp)\i[\widetilde{H}_\mp,\widetilde{A}_\mp]1_\Delta(\widetilde{H}_\mp)
=1_\Delta(\widetilde{H}_\mp)\left(2(\widetilde{H}_\mp-\Lambda^2)
+2\partial_x(1-v'(x))\partial_x-\frac12 v'''(x)\right)1_\Delta(\widetilde{H}_\mp),
$$
is self-adjoint for any bounded interval $\Delta\subset\rr$. Let $E\in\rr\setminus\Sigma_\mp$ so that
$\theta={\rm dist}(E,\Sigma_\mp)>0$ (see Figure~\ref{SpedFig}). With $\Delta=[E-\theta/2,E+\theta/2]$,
it follows from the functional calculus that
$$
1_\Delta(\widetilde{H}_\mp)2(\widetilde{H}_\mp-\Lambda^2)1_\Delta(\widetilde{H}_\mp)
\ge\theta 1_\Delta(\widetilde{H}_\mp).
$$
Moreover, since $1-v'$ and  $v'''$ belong to $C_0^\infty(\rr)$, Rellich's criterion yields that
$$
K_\mp=1_\Delta(\widetilde{H}_\mp)\left(
2\partial_x(1-v'(x))\partial_x-\frac12 v'''(x)\right)1_\Delta(\widetilde{H}_\mp),
$$
is compact. Thus, one has
$$
1_\Delta(\widetilde{H}_\mp)\i[\widetilde{H}_\mp,\widetilde{A}_\mp]1_\Delta(\widetilde{H}_\mp)
\ge\theta 1_\Delta(\widetilde{H}_\mp)+K_\mp,
$$
which shows that $\widetilde{H}_\mp$ satisfies a Mourre estimate with conjugate operator 
$\widetilde{A}_\mp$ at every 
$E\in\rr\setminus\Sigma_\mp$. Properties (i)--(iv) of Hypothesis \Href{HypConjugate} are now easily
verified. A simple induction argument shows that for any integer $n\ge1$ one has
$$
\rm{ad}_{\widetilde{A}_j}^n(\widetilde{H}_j)=-\partial_x(2^n+a_n(x))\partial_x+b_n(x),
$$
with $a_n,b_n\in C_0^\infty(\widetilde{\mathfrak{R}}_\mp)$. Lemma \ref{AHequivencore} allows us
to conclude that Hypothesis \Href{HypConjugate} (v) holds for any integer $n\ge0$.

\bigskip
This simple example can be modified in a number of ways by straightforward adaptations of the above
discussion:
\begin{itemize}
\item The manifold $\mathfrak{M}$ can have more that two cylindrical ends $\mathfrak{R}_j$.
\item The cylindrical ends $\mathfrak{R}_j$ can have arbitrary, smooth, compact bases $\gamma_j$
with metric $g_{\gamma_j}$.
\item The dimension of $\mathfrak{M}$ can be arbitrary, as long as this smooth manifold is the union 
of a compact piece $\mathfrak S$ and a finite number of cylindrical ends of the type 
$\mathfrak{R}_j=]1,\infty[\times\gamma_j$ with the metric
$g_{\widetilde{\mathfrak{R}}_j}=\d x^2+g_{\gamma_j}$.
\item A potential $V:\mathfrak{S}\to\rr$ can be added to the Hamiltonian $H$.
\item The metric on the cylindrical ends of the manifold $\mathfrak{M}$ can be slightly perturbed.
\end{itemize}

Scattering theory on non-compact complete Riemannian manifolds with Euclidean or hyperbolic metric 
near infinity has been intensively studied. Most of the cases covered by these studies can be
casted within our framework. We refer the reader to \cite{Hi,IKL,RT} for the development of scattering 
theory on manifolds with cylindrical ends.

A more radical change of the metric of the reservoirs leads to the concept of scattering manifold with
super-reservoirs of the type $\widetilde{\mathfrak{R}}_j=]1,\infty[\times\gamma_j$ equipped with a 
metric $g_{\widetilde{\mathfrak{R}}_j}=\d x^2+r(x)^2g_{\gamma_j}$, such that $r(x)\to\infty$ as
$x\to\infty$. The method used in these notes to derive the Lan\-dauer-B\"uttiker formula does not
apply to this class of models, for this reason we will not consider them here and refer the 
interested reader to \cite{Hi,IN,IS,Ku,MS} for discussions of the scattering 
theory. We note however that extending our results to this context is an interesting
open problem.

\subsection{The Mourre estimate}
In this section we construct a conjugate operator for $H$, in the sense of Mourre estimate.
To simplify our notation, we shall set $M_k\equiv M_k^{(1)}$, 
$J_k\equiv J_k^{(1)}$, 
$\widetilde\chi_k\equiv\widetilde\chi_k^{(1)}$,
$\chi_k\equiv\chi_k^{(1)}$,
$1_k\equiv 1_k^{(1)}$, $\tilde 1_k\equiv \tilde 1_k^{(1)}$,
$\mathcal H_k\equiv\mathcal H_k^{(1)}$, and $U\equiv U^{(1)}$ for all
$k\in\{0,\ldots,M\}$.

\begin{Lem}\label{ADefOK} If Hypotheses \Href{HypReservoir} and \Href{HypConjugate} are satisfied, the operator $A$ defined by
$$
A\equiv\sum_{k=1}^MJ_k\widetilde A_kJ_k^{\ast},
$$
on $\Dom(A)\equiv\{u\in\mathcal H\,|\,J_k^\ast u\in\Dom(\widetilde A_k),k=1,\ldots,M\}$ is self-adjoint. Furthermore, $A$ is reduced by the orthogonal projection $1_k$,
$$
[\e^{\i\theta A},1_k]=0,\quad k\in\{0,\ldots,M\}, \theta\in\mathbb R.
$$
For $k\in\{0,\ldots,M\}$ we set $A_k\equiv A1_k$. These self-adjoint operators satisfy $A_k=1_kA1_k=J_k\widetilde A_kJ_k^\ast=M_k\widetilde A_kM_k^\ast$ and $A_0=0$.
\end{Lem}

\begin{Rem}\label{AlocalRemark} The sample $\mathfrak S$, being localized in $\Ran 1_0$, 
it is also localized in $\Ran F(A=0)$. For $a>0$, $\Ran F(\pm A_k>a)\subset\Ran 1_k$, which leads to
$F(\pm A_k>a)$ being localized in the interior of the reservoir $\mathcal R_k$. This fact will be 
very useful for the calculations of currents in the reservoirs.
\end{Rem}

\begin{Theo}\label{HMourre}
Under Hypotheses \Href{HypReservoir}, \Href{HypResDyn} and \Href{HypConjugate}, we have
\begin{enumerate}[(i)]
\item $\e^{\i\theta A_k}\Dom(H)\subset\Dom(H)$ and $\e^{\i\theta A}\Dom(H)\subset\Dom(H)$
 for all $\theta\in\mathbb R$ and $k\in\{1,\ldots,M\}$.
\item $H\in \B^n_{A_k}(\mathcal H_H^1,\mathcal H)$ for $k\in\{1,\ldots,M\}$.
\item $H\in C^n_\mathrm{loc}(A)$.
\item There exists a closed and countable set $\Sigma_H\subset\mathbb R$ such that
$H$ satisfies a strict Mourre estimate on $\mathbb R\setminus\Sigma_H$
with the conjugate operator $A$.
\end{enumerate}
\end{Theo}

\begin{Coro}\label{NoSingSpec}
Under the hypotheses of Theorem \ref{HMourre} we have
\begin{enumerate}[(i)]
\item $\spec_\mathrm{sc}(\widetilde H_k)$ is empty for $k\in\{1,\ldots,M\}$.
\item $\spec_\mathrm{pp}(\widetilde H_k)\subset\Sigma_k$.
\item $\spec_\mathrm{sc}(H)$ is empty.
\item $\spec_\mathrm{pp}(H)\subset\Sigma_H$
\end{enumerate}
\end{Coro}

The remaining of this section is devoted to the proofs of these important results.

{\noindent\bf Proof of Lemma \ref{ADefOK}.} We begin by showing that for $k\in\{1,\ldots,M\}$ the 
orthogonal projection $\tilde 1_k$ reduces $\widetilde A_k$. In fact
Hypotheses \Href{HypConjugate} (i) and (ii) imply that for all $u\in\Dom(\widetilde A_k)$ and $r\in[0,2]$,
$$
\frac{\d\ }{\d\theta}\e^{\i\theta \widetilde A_k}
\widetilde\chi_k^{(r)}\e^{-\i\theta \widetilde A_k}u=0,
$$
and consequently $[\widetilde\chi_k^{(r)},\e^{\i\theta \widetilde A_k}]=0$. Hypotheses 
\Href{HypReservoir} (iv) and \Href{HypConjugate} (iii) allow us to write 
$$
\frac{\d\ }{\d\theta}(\tilde 1_k^{(r)}-\widetilde\chi_k^{(s)})\e^{\i\theta\widetilde A_k}u=0,
$$
for all $u\in\Dom(\widetilde A_k)$ and $0\le r\le s\le2$. From this we get that 
$(\tilde 1_k^{(r)}-\widetilde\chi_k^{(s)})\e^{\i\theta \widetilde A_k}
=(\tilde 1_k^{(r)}-\widetilde\chi_k^{(s)})$. By taking the adjoint we get
\begin{equation}
\e^{\i\theta \widetilde A_k}
(\tilde 1_k^{(r)}-\widetilde\chi_k^{(s)})
=(\tilde 1_k^{(r)}-\widetilde\chi_k^{(s)})
=(\tilde 1_k^{(r)}-\widetilde\chi_k^{(s)})\e^{\i\theta \widetilde A_k},
\label{localAaction}
\end{equation}
and in particular $[\tilde 1_k^{(r)}-\widetilde\chi_k^{(r)},\e^{\i\theta \widetilde A_k}]=0$. 
By combining this identity with the previous result we conclude that 
$[\tilde 1_k^{(r)},\e^{\i\theta \widetilde A_k}]=0$, that is to say that $\tilde 1_k^{(r)}$ reduces 
$\widetilde A_k$.

We denote by $\widetilde A_k^+$ the restriction of the operator $\widetilde A_k$ to the subspace 
$\Ran(\tilde 1_k)$. This operator is self-adjoint on 
$\Dom(\widetilde A_k^+)=\tilde 1_k\Dom(\widetilde A_k)$. We also set $\tilde A_0^+\equiv0$. 
We finish the proof by remarking that
$$
A= U\left(\bigoplus_{k=0}^M\tilde A_k^+\right) U^\ast,\qquad 1_k=U\tilde 1_kU^\ast.
$$
\hfill\qedbox

\bigskip
To prove Theorem \ref{HMourre} we will need several lemmas. 
We shall denote by $R(z)\equiv(H-z)^{-1}$ and $\widetilde R_k(z)\equiv(\widetilde H_k-z)^{-1}$ the 
resolvents of $H$ and of $\widetilde H_k$.

\begin{Lem}[Resolvent equation]\label{ResolvEqu} If Hypotheses \Href{HypReservoir} and 
\Href{HypResDyn} are satisfied, then, for all $r\ge0$ and all $z\in\Res(H)\cap\Res(\widetilde H_k)$, 
we have
$$
M_k^{(r)}\widetilde R_k(z)-R(z)M_k^{(r)}
=R(z)J_k^{(r)}B_k^{(r)}\widetilde R_k(z).
$$ 
\end{Lem}

\begin{demo} Hypotheses \Href{HypResDyn} (i)-(ii) yield 
$J_k^{(r)}B_k^{(r)}u=(H-z)M_k^{(r)}u-M_k^{(r)}(\widetilde H_k-z)u$ for all $u\in\Dom(\widetilde{H}_k)$.
Thus, for all $u,v\in\cH$ 
one has
\begin{align*}
(u,R(z)J_k^{(r)}B_k^{(r)}\widetilde R_k(z)v)&=(R(\bar z)u,(H-z)M_k^{(r)}\widetilde R_k(z)v)
-(R(\bar z)u,M_k^{(r)}(\widetilde H_k-z)\widetilde R_k(z)v)\\
&=(u,M_k^{(r)}\widetilde R_k(z)v)-(R(\bar z)u,M_k^{(r)}v)
=(u,M_k^{(r)}\widetilde R_k(z)v)-(u,R(z)M_k^{(r)}v).
\end{align*}
\hfill\qedbox
\end{demo}

\begin{Coro}\label{Dkcomp} Suppose that Hypotheses \Href{HypReservoir} and \Href{HypResDyn} 
are satisfied. For all $f\in C_0^\infty(\mathbb R)$, define the operator 
$D_k^{(r)}(f)\equiv f(H)M_k^{(r)}-M_k^{(r)}f(\widetilde H_k)$.
\begin{enumerate}[(i)]
\item  $D_k^{(r)}(f)$ is compact.
\item If Hypothesis \Href{HypConjugate} is also satisfied then  
$ D_k^{(r)}(f)[\widetilde A_k,\widetilde H_k]$ is compact.
\end{enumerate}
\end{Coro}

\begin{demo}$(i)$ With the help of Lemma \ref{ResolvEqu} we obtain, by using the Helffer-Sj\"ostrand 
formula,
\begin{equation}
D_k^{(r)}(f)=-\frac{\i}{2\pi}\int_{\mathbb C}\bar\partial \tilde  f(z)\,
R(z)J_k^{(r)}B_k^{(r)}\widetilde R_k(z)\,\d z\wedge\d \bar z,
\label{Dkrform}
\end{equation}
for an appropriate almost-analytic extension $\tilde f$ of $f$. Lemma \ref{ResolvEqu} yields the bound
$$
\|R(z)J_k^{(r)}B_k^{(r)}\widetilde R_k(z)\|\le 2\,|\mathrm{Im}z|^{-1},
$$
which shows that the integral 
(\ref{Dkrform}) converges in norm. The first resolvent equation further gives
$$
R(z)J_k^{(r)}B_k^{(r)}\widetilde R_k(z)
=R(z)J_k^{(r)}B_k^{(r)}\widetilde R_k(\i)(I+(z-\i)\widetilde R_k(z)),
$$
and $B_k^{(r)}\widetilde R_k(\i)$ being compact by Hypothesis \Href{HypResDyn} (iv), we conclude that
$R(z)J_k^{(r)}B_k^{(r)}\widetilde R_k(z)$ is compact for $z\in\mathbb C\setminus\mathbb R$. 
Consequently, $D_k^{(r)}(f)$ is compact.

$(ii)$ First, $[\widetilde A_k,\widetilde H_k]\in\B(\HH_k^1,\HH_k)$ by Hypothesis 
\Href{HypConjugate} (v), implies $[\widetilde H_k,\widetilde A_k]
=[\widetilde A_k,\widetilde H_k]^\ast\in\B(\HH_k,$ $\HH_k^{-1})$ and thus
$$
\widetilde R_k(z)[\widetilde H_k,\widetilde A_k]
=(I+(z-\i)\widetilde R_k(z))\widetilde R_k(\i)[\widetilde H_k,\widetilde A_k],
$$
is bounded on $\HH_k$. Furthermore, there exists a constant $c$ such that 
$$
\|\widetilde R_k(z)[\widetilde H_k,\widetilde A_k]\|\le c\, |\mathrm{Im} z|^{-1},
$$
for $z\in\supp\tilde f$. 

Second,
$$
R(z)J_k^{(r)}B_k^{(r)}=(I+(z+\i)R(z))R(-\i)J_k^{(r)}B_k^{(r)}
=(I+(z+\i)R(z))(B_k^{(r)\ast}J_k^{(r)\ast}R(\i))^\ast,
$$
is compact by Hypothesis \Href{HypResDyn} (v) and there exists a constant $c'$ such that
$$
\|R(z)J_k^{(r)}B_k^{(r)}\|\le c'\,|\mathrm{Im} z|^{-1},
$$
for $z\in\supp\tilde f$. We may conclude that the integral
$$
D_k^{(r)}(f)[\widetilde A_k,\widetilde H_k]=-\frac{1}{2\pi}\int_{\mathbb C}\bar\partial \tilde  f\,
R(z)J_k^{(r)}B_k^{(r)}\widetilde R_k(z)[\widetilde A_k,\widetilde H_k]\,\d z\wedge\d \bar z,
$$
converges in norm and is compact.
\cqfd
\end{demo}

\begin{Lem}\label{Aklocallemma} Under Hypotheses \Href{HypReservoir} and \Href{HypConjugate},
for $r\in[0,2]$ and $k\in\{1,\ldots,M\}$ we have
\begin{equation}
J_k^{(r)}\e^{\i\theta \widetilde A_k}J_k^{(r)\ast}=
M_k^{(r)}\e^{\i\theta \widetilde A_k}M_k^{(r)\ast}+1_k^{(r)}-M_k^{(r)}M_k^{(r)\ast},
\label{JkAkstuf}
\end{equation}
for $\theta\in\mathbb R$. Furthermore, for all $j,l\in\mathbb N$, and $u\in\Dom(\widetilde A_k^l)$, 
\begin{equation}
\tilde1_k^{(r)}\widetilde A_k^l u
=\widetilde\chi_k^{(r)j}\widetilde A_k^lu
=\widetilde A_k^l\widetilde\chi_k^{(r)j}u.
\label{JkAkstuf2}
\end{equation}
\end{Lem}

\begin{demo}
Since $J_k^{(r)}\tilde 1_k^{(r)}=J_k^{(r)}$ and 
$J_k^{(r)}\tilde 1_k^{(r)}J_k^{(r)\ast}=1_k^{(r)}$, the identity (\ref{localAaction}) allows us to write
\begin{align*}
J_k^{(r)}\e^{\i\theta \widetilde A_k}J_k^{(r)\ast}
&=J_k^{(r)}\widetilde\chi_k^{(r)}\e^{\i\theta \widetilde A_k}J_k^{(r)\ast}
+J_k^{(r)}(\tilde 1_k^{(r)}-\chi_k^{(r)})J_k^{(r)\ast}\\
&=J_k^{(r)}\widetilde\chi_k^{(r)}\e^{\i\theta \widetilde A_k}\chi_k^{(r)}J_k^{(r)\ast}+
J_k^{(r)}\widetilde\chi_k^{(r)}(\tilde 1_k^{(r)}-\chi_k^{(r)})J_k^{(r)\ast}
+J_k^{(r)}(\tilde 1_k^{(r)}-\chi_k^{(r)})J_k^{(r)\ast}\\
&=M_k^{(r)}\e^{\i\theta \widetilde A_k}M_k^{(r)\ast}
+J_k^{(r)}\widetilde\chi_k^{(r)}J_k^{(r)\ast}-M_k^{(r)}M_k^{(r)\ast}+\tilde 1_k^{(r)}
-J_k^{(r)}\widetilde\chi_k^{(r)}J_k^{(r)\ast},
\end{align*}
which proves (\ref{JkAkstuf}). If $u\in\Dom(\widetilde A_k^l)$, Hypotheses \Href{HypConjugate} (i) 
and (ii) imply that $\chi_k^{(r)j}u\in\Dom(\widetilde A_k^l)$ and $[A_k^l,\chi_k^{(r)j}]u=0$.
Differentiation of the identity (\ref{localAaction}) w.r.t.\;$\theta$ yields 
$\tilde 1_k^{(r)}\widetilde A_k^lu=\widetilde\chi_k^{(r)}\widetilde A_k^lu
=\widetilde A_k^l\widetilde\chi_k^{(r)}u$ and iterating this identity we conclude that
$$
\tilde 1_k^{(r)}\widetilde A_k^lu=
\widetilde\chi_k^{(r)j}\widetilde A_k^lu=
\widetilde A_k^l\widetilde\chi_k^{(r)j}u.
$$
\cqfd
\end{demo}

\bigskip

{\noindent\bf Proof of Theorem \ref{HMourre}.} $(i)$ By construction we have
$$
A=\bigoplus_{k=0}^M A_k,
$$
with  $A_k=J_k\widetilde A_kJ_k^\ast=1_kA=A1_k$,
$\Dom(A_k)=\{u\in\mathcal H_k\,|\,J_k^\ast u\in\Dom(\widetilde A_k)\}$ and $\widetilde A_0=0$. We may thus write
$$
\e^{\i\theta A}=\bigoplus_{k=0}^M \e^{\i\theta A_k}.
$$
By using the fact that $J_k^\ast J_k=\tilde 1_k$ reduces $\widetilde A_k$ (see the proof of Lemma \ref{ADefOK}) we easily show that
$$
1_k\e^{\i\theta J_k\widetilde A_kJ_k^\ast}1_k=J_k\e^{\i\theta \widetilde A_k}J_k^\ast,
$$
and thus
$$
\e^{\i\theta A}=\bigoplus_{k=0}^M J_k\e^{\i\theta \widetilde A_k}J_k^\ast.
$$
By applying (\ref{JkAkstuf}) and  
$J_0\e^{\i\theta \widetilde A_0}J_0^{\ast}=J_0J_0^{\ast}=1_0$
we obtain,
$$
\e^{\i\theta A_k}=
M_k\e^{\i\theta \widetilde A_k}M_k^\ast
+\left(1_k-M_kM_k^\ast\right).
$$
and after summing over $k$
$$
\e^{\i\theta A}=\sum_{k=1}^M
M_k\e^{\i\theta \widetilde A_k}M_k^\ast
+\left(I-\sum_{k=1}^MM_kM_k^\ast\right).
$$
Since 
$$
M_k^\ast\Dom(H)\subset\Dom(\widetilde H_k),\qquad 
M_k\Dom(\widetilde H_k)\subset\Dom(H),
$$
by Hypothesis \Href{HypResDyn} (i), and 
$\e^{\i\theta \widetilde A_k}\Dom(\widetilde H_k)\subset\Dom(\widetilde H_k)$
by Hypothesis \Href{HypConjugate} (iv), we conclude that $\Dom(H)$  is invariant
under $\e^{\i\theta A_k}$ and $\e^{\i\theta A}$.

$(ii)$ For $u,v\in\Dom(H)\cap\Dom(A^n)$ such that $A^nu,A^nv\in\Dom(H)$ we have
\begin{equation}
(u,\ad_{A_k}^j(H)v)=\i^j\sum_{l=0}^j{j\choose l}(-1)^l
(J_k\widetilde A_k^{j-l}J_k^\ast u,HJ_k\widetilde A_k^lJ_k^\ast v).
\label{Comutexpand}
\end{equation}
For $l>0$ we obtain, by using (\ref{JkAkstuf2})
$$
(J_k\widetilde A_k^{j-l}J_k^\ast u,HJ_k\widetilde A_k^lJ_k^\ast v)
=(\widetilde A_k^{j-l}J_k^\ast u,J_k^\ast HJ_k\widetilde\chi_k^2\widetilde A_k^lJ_k^\ast v).
$$
Hypothesis \Href{HypResDyn} (ii) allows us to continue
$$
(J_k\widetilde A_k^{j-l}J_k^\ast u,HJ_k\widetilde A_k^lJ_k^\ast v)
=(\widetilde A_k^{j-l}J_k^\ast u,\tilde 1_k \widetilde H_k\widetilde\chi_k^2\widetilde A_k^lJ_k^\ast v).
$$
By once again invoking (\ref{JkAkstuf2}) and by remarking that
$$
\tilde 1_k\widetilde A_k^{j-l}J_k^\ast u
=\widetilde A_k^{j-l}\tilde 1_kJ_k^\ast u=\widetilde A_k^{j-l}J_k^\ast u,
$$
we arrive at
$$
(J_k\widetilde A_k^{j-l}J_k^\ast u,HJ_k\widetilde A_k^lJ_k^\ast v)
=(\widetilde A_k^{j-l}J_k^\ast u, \widetilde H_k\widetilde A_k^lJ_k^\ast v).
$$
Reapplying (\ref{JkAkstuf2}) gives 
$$
(J_k\widetilde A_k^{j-l}J_k^\ast u,HJ_k\widetilde A_k^lJ_k^\ast v)
=(\widetilde A_k^{j-l}J_k^\ast u, 
\widetilde H_k\widetilde\chi_k^{(2)}\widetilde A_k^l\widetilde\chi_kJ_k^\ast v),
$$
and Identity (\ref{OldHyp3}) allows us to write
\begin{equation}
(J_k\widetilde A_k^{j-l}J_k^\ast u,HJ_k\widetilde A_k^lJ_k^\ast v)
=(\widetilde A_k^{j-l}J_k^\ast u, 
\widetilde\chi_k\widetilde H_k\widetilde\chi_k^{(2)}\widetilde A_k^l\chi_kJ_k^\ast v)
=(\widetilde A_k^{j-l}\widetilde\chi_kJ_k^\ast u, 
\widetilde H_k\widetilde A_k^l\widetilde\chi_kJ_k^\ast v).
\label{rightcom}
\end{equation}
For $l=0$ we obtain in an analogous way
\begin{align*}
(J_k\widetilde A_k^{j}J_k^\ast u,Hv)&=(HJ_k\widetilde\chi_k^2\widetilde A_k^{j}J_k^\ast u,v)
=(J_k\widetilde H_k\widetilde\chi_k^2\widetilde A_k^{j}J_k^\ast u,v)\\
&=(\widetilde A_k^{j}J_k^\ast u,\widetilde H_kJ_k^\ast v)
=(\widetilde A_k^{j}\widetilde\chi_kJ_k^\ast u,\widetilde\chi_k^{(2)}\widetilde H_kJ_k^\ast v).
\end{align*}
Identity (\ref{OldHyp3}) allows us to write
\begin{equation}
(J_k\widetilde A_k^{j}J_k^\ast u,Hv)
=(\widetilde A_k^{j}\widetilde\chi_kJ_k^\ast u,
\widetilde\chi_k^{(2)}\widetilde H_k\chi_kJ_k^\ast v)
=(\widetilde A_k^{j}\widetilde\chi_kJ_k^\ast u,\widetilde H_k\widetilde\chi_kJ_k^\ast v).
\label{leftcom}
\end{equation}
By gathering (\ref{rightcom}) and (\ref{leftcom}) in (\ref{Comutexpand}) we conclude that
$\ad_{A_k}^j(H)=M_k\ad_{\widetilde A_k}^j(\widetilde H_k)M_k^\ast$ and
\begin{equation}
\ad_{A}^j(H)=\sum_{k=1}^MM_k\ad_{\widetilde A_k}^j(\widetilde H_k)M_k^\ast.
\label{adniceform}
\end{equation}
Hypotheses \Href{HypResDyn} (i) and \Href{HypConjugate} (v) allow us to conclude that
$$
\ad_{A_k}^j(H)\in\B(\mathcal H_H^1,\mathcal H),
$$
for  $j\in\{1,\ldots, n\}$ and by consequence that $H\in\B_{A_k}^n(\mathcal H_H^1,\mathcal H)$.

$(iii)$ Since $A=\oplus_{k=1}^MA_k$, Assertion (ii) implies that 
$H\in\B_{A}^n(\mathcal H_H^1,\mathcal H)$. By Assertion $(i)$ we may invoke 
Theorem~\ref{localchar}~(ii) to conclude that $H\in C^n_\mathrm{loc}(A)$.

$(iv)$ Let $E\in\mathbb R\setminus\cup_k\Sigma_k$. Hypothesis \Href{HypConjugate} stipulates that
for $k\in\{1,\ldots,M\}$ there exists $g_k\in C_0^\infty(\mathbb R)$, $g_k(E)=1$, $0\le g_k\le 1$, 
constants $\theta_k>0$ and compact operators $K_k$ satisfying
\begin{equation}
g_k(\widetilde H_k)
\i[\widetilde H_k,\widetilde A_k]
g_k(\widetilde H_k)
\ge \theta_k g_k(\widetilde H_k)^2+K_k.
\label{hey}
\end{equation}
Since $g_k(E)=1$, there exists $\delta >0$ such that $g_k(x)\ge1/2$ for all $x\in[E-\delta,E+\delta]$ 
and $k\in\{1,\ldots,M\}$. If $g\in C_0^\infty([E-\delta,E+\delta])$ is such that $g(E)=1$ and 
$0\le g\le 1$ then $h_k=g/g_k\ge 0$ and by multiplying both sides of the inequality (\ref{hey}) by 
$h_k(\widetilde H_k)$ we obtain
$$
g(\widetilde H_k)
\i[\widetilde H_k,\widetilde A_k]
g(\widetilde H_k)
\ge \theta_k g(\widetilde H_k)^2+K'_k,
$$
where $K'_k=h_k(\widetilde H_k)K_kh_k(\widetilde H_k)$ is compact.

Formula (\ref{adniceform}) and Corollary \ref{Dkcomp} allow us to write
\begin{align*}
g(H)\i[H,A]g(H)
&=\sum_{k=1}^M
g(H)M_k\i[\widetilde H_k,\widetilde A_k]M_k^\ast g(H)\\
&=\sum_{k=1}^M
(M_kg(\widetilde H_k)+D_k)
\i[\widetilde H_k,\widetilde A_k](g(\widetilde H_k)M_k^\ast+D_k^\ast)\\
&=\sum_{k=1}^M
M_kg(\widetilde H_k)
\i[\widetilde H_k,\widetilde A_k]g(\widetilde H_k)M_k^\ast +K,
\end{align*}
where $D_k\equiv D_k^{(1)}(g)$ and $K$ is compact. We thus have
$$
g(H)\i[H,A]g(H)\ge
\sum_{k=1}^M\theta_k
M_kg(\widetilde H_k)^2M_k^\ast+K',
$$
where $K'=K+\sum_kM_kK'_kM_k^\ast$ is again compact. Since
$$
M_kg(\widetilde H_k)^2M_k^\ast=(g(H)M_k-D_k)
(M_k^\ast g(H)-D_k^\ast),
$$
Corollary \ref{Dkcomp} gives us
$$
g(H)\i[H,A]g(H)\ge\theta\, g(H)
\left(\sum_{k=1}^M
M_kM_k^\ast\right) g(H)+K'',
$$
where $\theta\equiv\min_k\theta_k>0$ and $K''$ is compact. Finally, 
Hypothesis \Href{HypReservoir} (ii) implies
\begin{equation}
\sum_{k=1}^MM_kM_k^\ast=\sum_{k=1}^M\chi_k^{(1)2}\ge\sum_{k=1}^M1_k^{(2)}=I-1_0^{(2)},
\label{summkmk}
\end{equation}
and Hypothesis \Href{HypResDyn} (vi) allows us to conclude
$$
g(H)\i[H,A]g(H)\ge\theta\, g(H)^2+K''',
$$
where $K'''=K''-\theta g(H)1_0^{(2)}g(H)$ is compact. $H$ therefore satisfies a Mourre estimate
with conjugate operator $A$ for all $E\in\mathbb R\setminus\cup_k\Sigma_k$. 

Since each $\Sigma_k$ is closed and countable, so is $\cup_k\Sigma_k$. 
$J\equiv\mathbb R\setminus\cup_k\Sigma_k$ is thus a union of a countable number of open 
intervals $\Delta_j$. Theorem \ref{MourreSingSpec} implies that the singular continuous spectrum 
of $H$ is empty and its eigenvalues can only accumulate at points in $\cup_k\Sigma_k$. This gives 
that
$$
\Sigma_H\equiv\spec_\pp(H)\bigcup\left(\bigcup_{k=1}^M\Sigma_k\right),
$$
is closed and countable.

Let $E\in\mathbb R\setminus\Sigma_H$. The spectrum of $H$ in the neighborhood of $E$ is purely 
absolutely continuous. Let $f_n\in
C_0^\infty(\mathbb R)$ is a sequence such that $0\le f_n\le 1$, $\supp\, f_n\subset[E-1/n,E+1/n]$ 
and $f_n(E)=1$. For $n$ large enough $h_n=f_n/g\in C_0^\infty([E-1/n,E+1/n])$ and $0\le h_n\le2$. 
We conclude that
$$
\slim_{n\to\infty}h_n(H)=0,
$$
and by consequence 
$$
\lim_{n\to\infty}\|h_n(H)K'''h_n(H)\|=0.
$$
If $n_0$ is large enough we have $h_{n_0}(H)K'''h_{n_0}(H)\ge -\theta/4$ and
$$
f_{n_0}(H)\i[H,A]f_{n_0}(H)\ge\theta\, f_{n_0}(H)^2+h_{n_0}(H)K'''h_{n_0}(H)
\ge\theta\, f_{n_0}(H)^2-\theta/4.
$$
Finally, if $f\in C_0^\infty(\mathbb R)$ is such that $0\le f\le 1$, $f(E)=1$ and $f_{n_0}\ge 1/2$ on 
$\supp\, f$ then $\tilde g\equiv ff_{n_0}\ge f/2$ and we have
$$
\tilde g(H)\i[H,A]\tilde g(H)\ge\theta\,\tilde g(H)^2-\theta f(H)^2/4
\ge \theta\tilde g(H)^2/4.
$$
\cqfd
\subsection{Scattering theory}
\label{ScattSect}

In this section we develop some elements of the theory of multi-channel scattering associated with 
the Hamiltonian $H$ and with the decomposition induced by the identification operators $M_k^{(r)}$.

We shall use the ``local smoothness'' approach developed by Lavine \cite{La1,La2} on the basis of 
the theory of ``$H$-smooth'' perturbations due to Kato \cite{Ka1} (see Section XIII.7 of \cite{RS4}). 
This approach has become very effective with the contribution of Mourre theory which allows the 
construction of locally $H$-smooth operators from the Mourre estimate (\cite{M2,M3}, 
{\sl c.f.\;}Corollary \ref{Katosmooth}). We shall make intensive use of the abstract two Hilbert space
scattering theory as exposed in Section XI.3 of \cite{RS3} (see also \cite{DS}).

\subsubsection{Bound states and scattering states}

We remark that under Hypotheses \Href{HypReservoir}, \Href{HypResDyn} and \Href{HypConjugate} 
the Hamiltonians of our system have empty singular continuous spectra 
(by Corollary \ref{NoSingSpec}). We thus have that 
$\HH_{k}=\HH_{k,\pp}(\widetilde H_k)\oplus\HH_{k,\ac}(\widetilde H_k)$ and 
$\mathcal H=\mathcal H_{\pp}(H)\oplus\mathcal H_{\ac}(H)$. To simplify notation, we write 
$$
\HH_{k}=\HH_{k,\pp}\oplus\HH_{k,\ac},\qquad
\mathcal H=\mathcal H_{\pp}\oplus\mathcal H_{\ac},
$$
without explicitly mentioning the Hamiltonians.

Adapting ideas of Ruelle, Amrein and Georgescu \cite{R1,AG}, we can also decompose these spaces 
on the basis of the dynamical properties of the states induced by their elements.
A vector $u\in\mathcal H$ is (improperly) called bound state for $t\to\pm\infty$ if it stays arbitrarily well
localized in a neighborhood of the system $\mathfrak S$ for $\pm t\ge 0$. More precisely the
subspace of bound states for $t\to\pm\infty$ is defined by
$$
\mathcal H_\mathrm{b}^\pm\equiv\left\{u\in\mathcal H\,\bigg|\,\forall \varepsilon>0 
\, \exists R\ge0:
\sup_{\atop{\pm t\ge0}{r>R}}\|(I-1_0^{(r)})\e^{-\i tH}u\|<\epsilon,
\right\}.
$$
The vector $u$ is a scattering state if it escapes any neighborhood of $\mathfrak S$ as $t\to\infty$.
The subspace of scattering states is defined by
$$
\mathcal H_\mathrm{s}^\pm\equiv\left\{u\in\mathcal H\,\bigg|\,\
\forall r\ge0:\lim_{t\to\pm\infty}\|1_0^{(r)}\e^{-\i tH}u\|=0\right\}.
$$

In a similar manner, we define the incoming states of the reservoir $\mathcal R_k$ as the states 
which, when they evolve with the dynamics generated by $\widetilde H_k$, are localized in a 
neighborhood of infinity in the distant past.
$$
\HH_k^\mathrm{in}\equiv\left\{u\in\HH_k\,\bigg|\,\
\forall r\ge0:\lim_{t\to-\infty}\|(I-\tilde 1_k^{(r)})\e^{-\i t\widetilde H_k}u\|=0\right\}.
$$
The outgoing states are also localized in a neighborhood of infinity, but in the distant future
$$
\HH_k^\mathrm{out}\equiv\left\{u\in\HH_k\,\bigg|\,\
\forall r\ge0:\lim_{t\to+\infty}\|(I-\tilde 1_k^{(r)})\e^{-\i t\widetilde H_k}u\|=0\right\}.
$$
The space of all incoming/outgoing states of the full system is
$$
\HH^\mathrm{in/out}\equiv\bigoplus_{k=1}^M\HH_k^\mathrm{in/out}.
$$

We easily verify that $\mathcal H_\mathrm{b}^\pm$ and 
$\mathcal H_\mathrm{s}^\pm$ are closed subspaces of $\mathcal H$ and that 
$\HH_k^\mathrm{in}$ and $\HH_k^\mathrm{out}$ are closed subspaces of $\HH_k$. 
In particular $\HH^\mathrm{in/out}$ are Hilbert spaces. The following result shows the relations 
of these ``dynamical'' subspaces and the spectral subspaces of the corresponding Hamiltonians.
We shall see later in this section that the scattering operator relates the subspaces
$\HH^\mathrm{in}$ and $\HH^\mathrm{out}$.

\begin{Lem}\label{DynSpec} Under the hypotheses \Href{HypReservoir}, \Href{HypResDyn} 
and \Href{HypConjugate} we have
$$
\mathcal H_\mathrm{b}^+=\mathcal H_\mathrm{b}^-=\mathcal H_\pp,\qquad
\mathcal H_\mathrm{s}^+=\mathcal H_\mathrm{s}^-=\mathcal H_\ac,
$$
and
$$
\HH_k^\mathrm{in}\subset\HH_{k,\ac},\qquad
\HH_k^\mathrm{out}\subset\HH_{k,\ac}.
$$
\end{Lem}

Consequently we shall henceforth write
$\mathcal H_\mathrm{b}=\mathcal H_\mathrm{b}^-=\mathcal H_\mathrm{b}^+$ and
$\mathcal H_\mathrm{s}=\mathcal H_\mathrm{s}^-=\mathcal H_\mathrm{s}^+$.

\bigskip
\begin{demo}
If $u$ is an eigenvector of $H$ then
$$
\|(I-1_0^{(r)})\e^{-\i tH}u\|
=\|(I-1_0^{(r)})u\|,
$$
for all $t\in\mathbb R$ and $r\ge0$. We thus have
$$
\|(I-1_0^{(r)})\e^{-\i tH}u\|^2=\sum_{k=1}^M\|1_k^{(r)}u\|^2,
$$
and Hypothesis \Href{HypReservoir} (iv) allows us to conclude that
$$
\lim_{r\to\infty}\,\sup_{t\in\mathbb R}\|(I-1_0^{(r)})\e^{-\i tH}u\|=0,
$$
that is to say that $u\in\mathcal H_\mathrm{b}^-\cap\mathcal H_\mathrm{b}^+$. 
Since $\mathcal H_\mathrm{b}^\pm$ are closed subspaces we get that 
$\mathcal H_\pp\subset\mathcal H_\mathrm{b}^-\cap\mathcal H_\mathrm{b}^+$.

If $u\in\mathcal H_\ac\cap\Dom(H)$ we have
$$
\|1_0^{(r)}\e^{-\i tH}u\|^2=\left(\e^{-\i tH}u,1_0^{(r)}\e^{-\i tH}u\right)
\le \|u\|\,\|1_0^{(r)}(H+\i)^{-1}\e^{-\i tH}(H+\i)u\|,
$$
and since $1_0^{(r)}(H+\i)^{-1}$ is compact by Hypothesis \Href{HypResDyn} (vi) and 
$$
\wlim_{t\to\pm\infty}\e^{-\i tH}(H+\i)u=0,
$$
by the Riemann-Lebesgue lemma (Lemma 2 of Section XI.3 in \cite{RS3}), we may conclude that
$$
\lim_{t\to\pm\infty}\|1_0^{(r)}\e^{-\i tH}u\|=0,
$$
and thus $u\in\mathcal H_\mathrm{s}^-\cap\mathcal H_\mathrm{s}^+$. Since
$\mathcal H_\ac\cap\Dom(H)$ is dense in $\mathcal H_\ac$ and 
$\mathcal H_\mathrm{s}^\pm$ are closed, we conclude that 
$\mathcal H_\ac\subset\mathcal H_\mathrm{s}^-\cap\mathcal H_\mathrm{s}^+$.

We now show that $\mathcal H_\mathrm{s}^\pm$ and $\mathcal H_\mathrm{b}^\pm$ are orthogonal
to each other. If $u\in\mathcal H_\mathrm{s}^+$, $v\in\mathcal H_\mathrm{b}^+$ and $\epsilon>0$ 
there exists $R>0$ such that $\|u\|\,\|(I-1_0^{(R)})\e^{-\i tH}v\|<\epsilon$ for all $t>0$. We deduce that
\begin{align*}
|(u,v)|&=|(\e^{-\i tH}u,\e^{-\i tH}v)|\\
&\le|(\e^{-\i tH}u,(I-1_0^{(R)})\e^{-\i tH}v)|+|(1_0^{(R)}\e^{-\i tH}u,\e^{-\i tH}v)|\\
&\le\|u\|\,\|(I-1_0^{(R)})\e^{-\i tH}v\|+\|1_0^{(R)}\e^{-\i tH}u\|\,\|v\|\\
&\le\epsilon+\|1_0^{(R)}\e^{-\i tH}u\|\,\|v\|,
\end{align*}
for all $t>0$, and as $t\to+\infty$ we obtain $|(u,v)|<\epsilon$. Since $\epsilon>0$ was arbitrary we 
conclude that $(u,v)=0$ and consequently that 
$\mathcal H_\mathrm{s}^+\perp\mathcal H_\mathrm{b}^+$. It is clear that the same argument shows 
that $\mathcal H_\mathrm{s}^-\perp\mathcal H_\mathrm{b}^-$.

We have shown that 
\begin{align*}
\mathcal H&=\mathcal H_\pp\oplus\mathcal H_\ac\subset
\mathcal H_\mathrm{b}^-\oplus\mathcal H_\mathrm{s}^-\subset\mathcal H,\\
\mathcal H&=\mathcal H_\pp\oplus\mathcal H_\ac\subset
\mathcal H_\mathrm{b}^+\oplus\mathcal H_\mathrm{s}^+\subset\mathcal H,
\end{align*}
which immediately gives
$\mathcal H_\ac=\mathcal H_\mathrm{s}^-=\mathcal H_\mathrm{s}^+$ and
$\mathcal H_\pp=\mathcal H_\mathrm{b}^-=\mathcal H_\mathrm{b}^+$.

To show that $\HH_k^\mathrm{out}\subset\HH_{k,\ac}$, it suffices to show that 
$\HH_{k,\pp}\subset\HH_k^{\mathrm{out}\perp}$, that is to say that for each eigenvector 
$u$ of $\widetilde H_k$ and for all $v\in\HH_k^\mathrm{out}$ we have $(u,v)=0$. If 
$\widetilde H_ku=Eu$ then, for all $r>0$,
\begin{align*}
|(u,v)|&=|(\e^{-\i t\widetilde H_k}u,\e^{-\i t\widetilde H_k}v)|=|(u,\e^{-\i t(\widetilde H_k-E)}v)|\\
&\le|(u,(I-\tilde 1_k^{(r)})\e^{-\i t(\widetilde H_k-E)}v)|
+|(\tilde 1_k^{(r)}u,\e^{-\i t(\widetilde H_k-E)}v)|\\
&\le\|u\|\,\|(I-\tilde 1_k^{(r)})\e^{-\i t\widetilde H_k}v\|+\|\tilde 1_k^{(r)}u\|\,\|v\|.
\end{align*}
By Hypothesis \Href{HypReservoir} (iv), for all $\epsilon>0$, there exists $R>1$ such that
$\|\tilde 1_k^{(R)}u\|\,\|v\|<\epsilon$. Since
$$
\lim_{t\to+\infty}\|u\|\,\|(I-\tilde 1_k^{(r)})\e^{-\i t\widetilde H_k}v\|=0,
$$
We conclude that $|(u,v)|<\epsilon$ and since $\epsilon>0$ was arbitrary, we have that $(u,v)=0$.
The last assertion of the lemma is proven in an analogous way.\cqfd
\end{demo}

\subsubsection{The strong topologies of $\B(\mathcal H)$}
\label{strongstuf}
Since the strong and strong-$\ast$ topologies of $\B(\mathcal H)$ play an essential role in scattering 
theory, we start by describing some of their important properties (see Section 2.4.1 of \cite{BR1} 
for a detailed discussion of the various topologies on $\B(\mathcal H)$).

A net $(B_\iota)_{\iota\in I}$ in $\B(\mathcal H)$ is strongly convergent if there exists 
$B\in\B(\mathcal H)$ such that $\lim_\iota B_\iota u=Bu$. We then write $B=\slim_\iota B_\iota$. 
If furthermore the family $(B_\iota^\ast)_{\iota\in I}$ is strongly convergent we say that 
$(B_\iota)_{\iota\in I}$ is strong-$\ast$ convergent. In this case we necessarily have that 
$\slim_\iota B_\iota^\ast=B^\ast$, indeed
$$
(B^\ast u,v)=(u,Bv)=\lim_\iota(u,B_\iota v)=\lim_\iota(B_\iota^\ast u,v)=
((\slim_\iota B_\iota^\ast) u,v),
$$
for all $u,v\in\mathcal H$. We then write $\starslim_\iota B_\iota=B$. Remember however that if 
$\mathcal H$ is infinite dimensional the strong-$\ast$ topology is strictly finer than the strong topology. 
The mapping $B\mapsto B^\ast$ is strong-$\ast$ continuous but not strongly continuous. 
The product $\langle A,B\rangle\mapsto AB$ is not strongly continuous. However its 
restriction to a bounded subset of $\B(\mathcal H)\times\B(\mathcal H)$ is, due to the inequality
$$
\|B_\iota C_\iota u-BCu\|\le \|B_\iota\|\,\|(C_\iota-C)u\|+\|(B_\iota-B)Cu\|.
$$

If $H_1$ and $H_2$ are self-adjoint operators on $\mathcal H_1$ and $\mathcal H_2$ and if 
$B\in\B(\mathcal H_2,\mathcal H_1)$ we denote
$$
\Gamma^\pm(H_1,H_2;B)\equiv\slim_{t\to\pm\infty}\e^{\i tH_1}B\e^{-\i tH_2},
$$
when these limits exist. In the sequel we shall use without explicitly mention 
the following properties of this $\Gamma^\pm$ operation. If $\Gamma^\pm(H_1,H_2;B)$ and 
$\Gamma^\pm(H_1,H_2;C)$ exist, so does $\Gamma^\pm(H_1,H_2;B+C)$ and
$$
\Gamma^\pm(H_1,H_2;B+C)=\Gamma^\pm(H_1,H_2;B)+\Gamma^\pm(H_1,H_2;C).
$$
Similarly, if $\Gamma^\pm(H_1,H_2,B)$ and $\Gamma^\pm(H_2,H_3,C)$ exist, so does
$\Gamma^\pm(H_1,H_3;BC)$ and
\begin{equation}
\Gamma^\pm(H_1,H_3;BC)=\Gamma^\pm(H_1,H_2;B)\Gamma^\pm(H_2,H_3;C).
\label{GammaMult}
\end{equation}
The limits $\Gamma^\pm(H_1,H_2;B)$ and $\Gamma^\pm(H_2,H_1;B^\ast)$ exist simultaneously 
if and only if
$$
\Gamma^\pm(H_1,H_2;B)=\starslim_{t\to\pm\infty}\e^{\i tH_1}B\e^{-\i tH_2},
$$
and in this case
$$
\Gamma^\pm(H_2,H_1;B^\ast)=\Gamma^\pm(H_1,H_2;B)^\ast.
$$
We note that if $\Gamma^\pm(H_1,H_2;B)$ exists then
$$
\slim_{t\to\pm\infty}\e^{\i (s+t)H_1}B\e^{-\i tH_2}
=\slim_{t\to\pm\infty}\e^{\i tH_1}B\e^{-\i (t-s)H_2},
$$
and thus $\e^{\i sH_1}\Gamma^\pm(H_1,H_2;B)=\Gamma^\pm(H_1,H_2;B)\e^{\i sH_2}$. 
We easily conclude that for all measurable functions $f$, 
$\Gamma^\pm(H_1,H_2;B)\Dom(f(H_2))\subset\Dom(f(H_1))$ and 
$$
f(H_1)\Gamma^\pm(H_1,H_2;B)u=\Gamma^\pm(H_1,H_2;B)f(H_2)u,
$$
for all $u\in\Dom(f(H_2))$. In particular,
$$
F(H_1\in I)\Gamma^\pm(H_1,H_2;B)=\Gamma^\pm(H_1,H_2;B)F(H_2\in I),
$$
for all measurable sets $I\subset\mathbb R$. It is then easy to deduce that
\begin{equation}
P_\ac(H_2)\Gamma^\pm(H_1,H_2;B)=\Gamma^\pm(H_1,H_2;B)P_\ac(H_1).
\label{ACpreserv}
\end{equation}
This relation implies an important extension of the identity (\ref{GammaMult}). 
Suppose that the limits $\Gamma^\pm(H_1,H_2;BP_\ac(H_2))$ and $\Gamma^\pm(H_2,H_3;CP_\ac(H_3))$ 
exist. We can therefore decompose
\begin{align*}
\e^{\i tH_1}BCP_\ac(H_3)\e^{-\i tH_3}&=(\e^{\i tH_1}BP_\ac(H_2)\e^{-\i tH_2})
(\e^{\i tH_2}CP_\ac(H_3)\e^{-\i tH_3})\\
&+\e^{\i tH_1}B\e^{-\i tH_2}((I-P_\ac(H_2))\e^{\i tH_2}CP_\ac(H_3)\e^{-\i tH_3}),
\end{align*}
and note that
$\slim_{t\to\pm\infty}(I-P_\ac(H_2))\e^{\i tH_2}CP_\ac(H_3)\e^{-\i tH_3}=0$
by virtue of (\ref{ACpreserv}). We thus obtain
$$
\Gamma^\pm(H_1,H_3;BCP_\ac(H_3))
=\Gamma^\pm(H_1,H_2;BP_\ac(H_2))\Gamma^\pm(H_2,H_3;CP_\ac(H_3)).
$$

The existence of strong limits $\Gamma^\pm(H_1,H_2;J\chi_\Delta(H_2))$ can often be proven 
by combining the propagation estimates with the following result (\cite{Ka1}, \cite{La1,La2}); 
see also Theorem XIII.31 in \cite{RS4} and Chapter 4 of \cite{Y}).

\begin{Prop}\label{Kato} Let $H_1$, $H_2$ be self-adjoint operators on the Hilbert spaces 
$\mathcal H_1$, $\mathcal H_2$. Let $C_1$, $C_2$ be closed operators on $\mathcal H_1$,
$\mathcal H_2$. Finally, let $J,B\in\B(\mathcal H_2,\mathcal H_1)$ and $\Delta\subset\mathbb R$ be
such that
\begin{enumerate}[(i)]
\item $\Dom(H_i)\subset\Dom(C_i)$ and $C_i$ is $H_i$-bounded for $i=1,2$. 
\item For all $u_i\in\Dom(H_i)$,
$$
(H_1u_1,Ju_2)-(u_1,JH_2u_2)=(C_1u_1, BC_2u_2).
$$
\item For all $u_i\in\mathcal H_i$ and for almost all $t\in\mathbb R$,
$\e^{-\i tH_i}\chi_\Delta(H_i)u_i\in\Dom(C_i)$ and there exist constants $c_1$, $c_2$ such that
$$
\int_{-\infty}^\infty\|C_i\e^{-\i tH_i}\chi_\Delta(H_i)u_i\|^2\,\d t\le c_i\,\|u_i\|^2.
$$
\end{enumerate}
Then $\Gamma^\pm(H_1,H_2;J\chi_\Delta(H_2))$ and $\Gamma^\pm(H_2,H_1;J^\ast\chi_\Delta(H_1))$ exist. Furthermore
\begin{enumerate}[(a)]
\item$\Gamma^\pm(H_2,H_1;J^\ast\chi_\Delta(H_1))=\Gamma^\pm(H_1,H_2;J\chi_\Delta(H_2))^\ast$.
\item $\Gamma^\pm(H_1,H_2;J\chi_\Delta(H_2))\Gamma^\pm(H_1,H_2;J\chi_\Delta(H_2))^\ast
=\Gamma^\pm(H_1,H_1;JJ^\ast\chi_\Delta(H_1))$.
\item $\Gamma^\pm(H_2,H_1;J^\ast\chi_\Delta(H_1))\Gamma^\pm(H_2,H_1;J^\ast\chi_\Delta(H_1))^\ast
=\Gamma^\pm(H_2,H_2;J^\ast J\chi_\Delta(H_2))$.
\end{enumerate}
\end{Prop}

{\bf\noindent Remark.} An operator $C_1$ satisfying the condition (iii) is called locally $H_1$-smooth 
on $\Delta$ or simply $H_1$-smooth if $\Delta=\mathbb R$ (see Section XIII.7 of \cite{RS4}). 

\bigskip
\begin{demo} We first show that
$\Gamma^\pm(H_1,H_2;\chi_\Delta(H_1)J\chi_\Delta(H_2))$ exists. With $u_i\in\Dom(H_i)$ we
can write
$$
(u_1,u_2(t'))-(u_1,u_2(t))=\i\int_t^{t'}
(C_1\e^{-\i sH_1}\chi_\Delta(H_1)u_1,BC_2\e^{-\i sH_2}\chi_\Delta(H_2)u_2)\,\d s,
$$
where $u_2(t)\equiv\e^{\i tH_1}\chi_\Delta(H_1)J\chi_\Delta(H_2)\e^{-\i tH_2}u_2$.
We then have
\begin{align*}
|(u_1,u_2(t')-u_2(t))|&\le\|B\|\int_t^{t'}
\|C_1\e^{-\i sH_1}\chi_\Delta(H_1)u_1\|\,\|C_2\e^{-\i sH_2}\chi_\Delta(H_2)u_2\|\,\d s\\
&\le\|B\|\left(\int_t^{t'}\!\!
\|C_1\e^{-\i sH_1}\chi_\Delta(H_1)u_1\|^2\,\d s\right)^{1/2}
\left(\int_t^{t'}\!\!\|C_2\e^{-\i sH_2}\chi_\Delta(H_2)u_2\|^2\,\d s\right)^{1/2}\\
&\le c_1^{1/2}\|B\|\,\|u_1\|
\left(\int_t^{t'}\!\!\|C_2\e^{-\i sH_2}\chi_\Delta(H_2)u_2\|^2\,\d s\right)^{1/2},
\end{align*}
and thus
$$
\|u_2(t')-u_2(t)\|\le
c_1^{1/2}\|B\|
\left(\int_t^{t'}\!\!\|C_2\e^{-\i sH_2}\chi_\Delta(H_2)u_2\|^2\,\d s\right)^{1/2},
$$
which allows us to conclude that $u_2(t)$ converges when $t\to\pm\infty$. This extends by continuity to all
$u_2\in\mathcal{H}_2$.

We now prove that $\Gamma^\pm(H_1,H_2;\chi_{\Delta^c}(H_1))J\chi_\Delta(H_2))=0$. 
To do this it suffices to show that
$$
\Gamma^\pm(H_1,H_2;\chi_{\Delta^c}(H_1)J\chi_{\Delta'}(H_2))=0,
$$
for all compact intervals $\Delta'\subset\Delta$. Let $g\in C_0^\infty(\Delta)$ such that $0\le g\le 1$ 
and $g=1$ on $\Delta'$. For $u_2\in\mathcal H_2$, let
$u_2(t)\equiv\e^{\i tH_1}\chi_{\Delta^c}(H_1)J\chi_{\Delta'}(H_2)\e^{-\i tH_2}u_2$, so that
$$
u_2(t)=\e^{\i tH_1}\chi_{\Delta^c}(H_1)(Jg(H_2)-g(H_1)J)\e^{-\i tH_2}\chi_{\Delta'}(H_2)u_2.
$$
By setting $R_i(z)\equiv(H_i-z)^{-1}$, property (ii) allows us to write, for $u_1\in \mathcal H_1$,
\begin{align*}
&(\e^{-\i tH_1}\chi_{\Delta^c}(H_1)u_1,
(JR_2(z)-R_1(z)J)\e^{-\i tH_2}\chi_{\Delta'}(H_2)u_2)\\
&=(C_1R_1(\bar z)\e^{-\i tH_1}\chi_{\Delta^c}(H_1)u_1,
BC_2R_2(z)\e^{-\i tH_2}\chi_{\Delta'}(H_2)u_2),
\end{align*}
and the Helffer-Sj\"ostrand formula gives
$$
(u_1,u_2(t))=\int_{\mathbb C}\bar\partial\tilde g\,
(C_1R_1(\bar z)\e^{-\i tH_1}\chi_{\Delta^c}(H_1)u_1,
BC_2R_2(z)\e^{-\i tH_2}\chi_{\Delta'}(H_2)u_2)\frac{\d z\wedge\d\bar z}{2\pi},
$$
where $\tilde g$ is an almost-analytic extension of $g$. We therefore obtain the following estimate
\begin{align*}
|(u_1,u_2(t))|\le\|u_1\|\,\|B\|&
\left(
\int_{\mathbb C}|\bar\partial\tilde g|\,\|C_1R_1(\bar z)\|^2\frac{\d z\wedge\d\bar z}{2\pi}
\right)^{1/2}\times\\
&\left(
\int_{\mathbb C}|\bar\partial\tilde g|\,\|C_2\e^{-\i tH_2}R_2(z)\chi_{\Delta'}(H_2)u_2
\|^2\frac{\d z\wedge\d\bar z}{2\pi}
\right)^{1/2}.
\end{align*}
Since $C_1R_1(\bar z)=C_1R_1(\i)(I+(\bar z-\i)R_1(\bar z))$, property (i) allows us to write 
$\|C_1R_1(\bar z)\|\le c\,|\mathrm{Im}z|^{-1}$ for a constant $c$ and $z\in\supp\,\tilde g$. 
Therefore there exists a constant $c'$ such that
$$
\|u_2(t)\|\le c'\,\left(
\int_{\mathbb C}|\bar\partial\tilde g|\,\|C_2\e^{-\i tH_2}R_2(z)\chi_{\Delta'}(H_2)u_2
\|^2\frac{\d z\wedge\d\bar z}{2\pi}\right)^{1/2}.
$$
We denote by $f(z,t)$ the integrand on the right hand side of this inequality. 
Property (i) and an appropriate choice of almost-analytic extension $\tilde g$ (recall 
Estimate~\eqref{ftildeestimate}) show that for a constant 
$c''$ we have
$$
0\le f(z,t)\le c''\,\frac{|\bar\partial \tilde g|}{|\mathrm{Im}z|^2}\in 
L^1(\mathbb C,\d z\wedge\d\bar z).
$$
It is therefore enough for us to show that $\lim_{t\to\pm\infty}f(z,t)=0$ for all 
$z\in\mathbb C\setminus\mathbb R$ to be able to conclude. Property (iii) shows that for such a 
$z$ we have $f\in L^1(\mathbb R,\d t)$. Furthermore, $f$ is differentiable and 
$$
\partial_tf(z,t)=2 |\bar\partial \tilde g|\mathrm{Im}
(C_2\e^{-\i tH_2}R_2(z)\chi_{\Delta'}(H_2)u_2,
C_2\e^{-\i tH_2}R_2(z)H_2\chi_{\Delta'}(H_2)u_2),
$$
shows that $|\partial_tf(z,t)|\in L^1(\mathbb R,\d t)$. The required property follows immediately.

So we have shown that
\begin{align*}
\Gamma^\pm(H_1,H_2;J\chi_\Delta(H_2))&=
\Gamma^\pm(H_1,H_2;\chi_\Delta(H_1)J\chi_\Delta(H_2))
+\Gamma^\pm(H_1,H_2;\chi_{\Delta^c}(H_1)J\chi_\Delta(H_2))\\
&=\Gamma^\pm(H_1,H_2;\chi_\Delta(H_1)J\chi_\Delta(H_2)).
\end{align*}
The existence of the strong limits $\Gamma^\pm(H_2,H_1;J^\ast\chi_\Delta(H_1))$ is therefore a 
consequence of the $1\leftrightarrow2$ symmetry  of our hypotheses.

Assertion (a) is a consequence of the strong-$\ast$ convergence as we explicitly mentioned above. 
Assertion (b) follows from the identity
\begin{align*}
\e^{\i tH_1}JJ^\ast\chi_\Delta(H_1)\e^{-\i tH_1}&=
\e^{\i tH_1}J\chi_\Delta(H_2)\e^{-\i tH_2}\e^{\i tH_2}J^\ast\chi_\Delta(H_1)\e^{-\i tH_1}\\
&+\e^{\i tH_1}J\e^{-\i tH_2}\chi_{\Delta^c}(H_2)
\e^{\i tH_2}J^\ast\chi_\Delta(H_1)\e^{-\i tH_1},
\end{align*}
and the fact that $\chi_{\Delta^c}(H_2)\Gamma^\pm(H_2,H_1;J^\ast\chi_\Delta(H_1))=0$. 
Assertion (c) is proven in the same way.\hfill$\square$
\end{demo}

\bigskip
The following notions shall be useful to us.

\begin{Def} Let $H$ be a self-adjoint operator on the Hilbert space $\mathcal H$.
\begin{enumerate}[(i)]
\item An operator $B\in\B(\mathcal H)$ is called an asymptotic projection for $H$ if the limits
$\Gamma^\pm(H,H;B)$ exist and define orthogonal projections
\item Two operators $B,C\in\B(\mathcal H)$ are said to be 
asymptotically $H$-equivalent if $\Gamma^\pm(H,H;B-C)=0$.
\end{enumerate}
\end{Def}

\subsubsection{M\o ller operators}

To allow us to briefly describe the basic ideas behind multi-channel scattering theory, we start
with a result establishing the existence of asymptotic projection operators for the reference dynamics 
in the reservoirs. We defer its proof to the end of the section.

\begin{Lem}\label{HkProj} Under hypotheses \Href{HypReservoir}, \Href{HypResDyn}, 
and \Href{HypConjugate}, for $k\in\{1,\ldots,M\}$, $\tilde 1_kP_\ac(\widetilde H_k)$ is an asymptotic 
projection for $\widetilde H_k$. Furthermore, for $r\ge1$  the operators 
$M_k^{(r)\ast}M_k^{(r)}P_\ac(\widetilde H_k)$,
$\tilde 1_k^{(r)}P_\ac(\widetilde H_k)$ and $\widetilde\chi_k^{(r)}P_\ac(\widetilde H_k)$
 are $\widetilde H_k$-equivalent to 
$\tilde 1_kP_\ac(\widetilde H_k)$ and
$$
\widetilde P_k^\mathrm{in/out}\equiv
\Gamma^{-/+}(\widetilde H_k,\widetilde H_k;1_kP_\ac(\widetilde H_k)),
$$
are the orthogonal projections onto $\HH_k^\mathrm{in/out}$.
\end{Lem}
In the sequel, we shall sometimes refer to $\widetilde P_k^\mathrm{in/out}$ as $\widetilde P_k^{-/+}$.

\noindent{\bf Working hypothesis:} When it evolves under the dynamics generated by the Hamiltonian 
$H$ a scattering state of the complete system, $u\in\mathcal H_\mathrm{s}$, behaves asymptotically, 
when $t\to-\infty$,  as a state $u^\mathrm{in}\in\HH^\mathrm{in}$  under the dynamics generated by 
$\widetilde H=\oplus_k\widetilde H_k$. Similarly, when $t\to+\infty$, it behaves like a state 
$u^\mathrm{out}\in\HH^\mathrm{out}$. More precisely, we have
\begin{equation}
\e^{-\i tH}u\sim\left\{
\begin{array}{lr}
\sum_{k=1}^M J_k\e^{-\i t\widetilde H_k}u_k^\mathrm{in}& \text{for}\ t\to-\infty,\\[6pt]
\sum_{k=1}^M J_k\e^{-\i t\widetilde H_k}u_k^\mathrm{out}& \text{for}\ t\to+\infty,
\end{array}
\right.
\label{SDef}
\end{equation}
where $u_k^\mathrm{in/out}$ denote the projections of $u^\mathrm{in/out}$ in $\HH_k$ and
$f(t)\sim g(t)$ means that $\lim(f(t)-g(t))=0$.

With this hypothesis, we shall construct the central object of scattering theory: the scattering matrix
\begin{eqnarray*}
S:\HH^\mathrm{in}&\to&\HH^\mathrm{out}\\
u^\mathrm{in}&\mapsto&u^\mathrm{out}.
\end{eqnarray*}
which transforms the incoming asymptote $u^\mathrm{in}\in\HH^\mathrm{in}$ into the outgoing 
asymptote $u^\mathrm{out}\in\HH^\mathrm{out}$. This map can be calculated in two steps. 
We first determine the M\o ller operators which, to an incoming/outgoing asymptote 
$u^\mathrm{in/out}$ associate the scattering state $u$. These transformations are easily obtained 
from (\ref{SDef})
\begin{align*}
\Omega^-:u^\mathrm{in}\mapsto u
&=\lim_{t\to-\infty}\sum_{k=1}^M\e^{\i tH}J_k\e^{-\i t\widetilde H_k}u_k^\mathrm{in},\\
\Omega^+:u^\mathrm{out}\mapsto u
&=\lim_{t\to+\infty}\sum_{k=1}^M\e^{\i tH}J_k\e^{-\i t\widetilde H_k}u_k^\mathrm{out}.
\end{align*}
If the M\o ller operators so defined exist, they are isometric. Indeed Hypothesis \Href{HypReservoir} (iii) 
allows us to write
\begin{align*}
\|\Omega^+ u^\mathrm{out}\|^2&=\lim_{t\to+\infty}\left\|
\sum_{k=1}^MJ_k\e^{-\i t\widetilde H_k}u_k^\mathrm{out}\right\|^2\\
&=\sum_{k=1}^M\lim_{t\to+\infty}(\e^{-\i t\widetilde H_k}u_k^\mathrm{out},
\tilde 1_k\e^{-\i t\widetilde H_k}u_k^\mathrm{out})\\
&=\sum_{k=1}^M\left(\|u_k^\mathrm{out}\|^2-
\lim_{t\to+\infty}\|(I-\tilde 1_k)\e^{-\i t\widetilde H_k}u_k^\mathrm{out}\|^2\right)\\
&=\sum_{k=1}^M\|u_k^\mathrm{out}\|^2=\|u^\mathrm{out}\|^2.
\end{align*}
It is clear that an identical argument shows that $\Omega^-$ is also isometric. In particular the 
M\o ller operators are injective, their images $\Ran\Omega^\pm$ are closed, and
$(\Omega^\pm)^{-1}=\Omega^{\pm\ast}:\Ran\Omega^\pm\to\HH^\mathrm{in/out}$ are isometries. 
It is thus possible to define $S=\Omega^{+\ast}\Omega^-$ if $\Ran\Omega^-=\Ran\Omega^+$.
In this case, we say that the M\o ller operators are {\em weakly asymptotically complete} and we have
\begin{align*}
S^\ast S&=\Omega^{-\ast}\Omega^+\Omega^{+\ast}\Omega^-=\Omega^{-\ast}\Omega^-
=I_{\HH^\mathrm{in}},\\
SS^\ast&=\Omega^{+\ast}\Omega^-\Omega^{-\ast}\Omega^+=\Omega^{+\ast}\Omega^+
=I_{\HH^\mathrm{out}},
\end{align*}
that is to say that the scattering matrix is unitary.

We return now to our working hypothesis. Suppose that the M\o ller operators exist and are weakly 
asymptotically complete. If  $u\in\Ran\Omega^\pm$ then $u=\Omega^\pm u^\mathrm{in/out}$ and 
we have 
$$
0=\|u-\Omega^\pm u^\mathrm{in/out}\|=\lim_{t\to\pm\infty}
\left\|\e^{-\i tH}u-\sum_{j=1}^MJ_k\e^{-\i t\widetilde H_k}u_k^\mathrm{in/out}\right\|.
$$
{\em A posteriori,\,}our working hypothesis is thus verified for all $u\in\Ran\Omega^\pm$. However,
if $\Ran\Omega^\pm\not=\mathcal H_\mathrm{s}$ there exists $u\in\mathcal H_\mathrm{s}$ such 
that $0\not=u\perp\Ran\Omega^\pm$. In this case (\ref{SDef}) and Lemma \ref{HkProj} imply that, 
for all $v^\mathrm{in/out}\in\HH^\mathrm{in/out}$, 
\begin{align*}
0=(u,\Omega^\pm v^\mathrm{in/out})
&=\sum_{k=1}^M\lim_{t\to\pm\infty}(\e^{-\i tH}u,J_k\e^{-\i t\widetilde H_k}v_k^\mathrm{in/out})\\
&=\sum_{k=1}^M\lim_{t\to\pm\infty}
(J_k\e^{-\i t\widetilde H_k}u_k^\mathrm{in/out},
J_k\e^{-\i t\widetilde H_k}v_k^\mathrm{in/out})\\
&=\sum_{k=1}^M\lim_{t\to\pm\infty}
(u_k^\mathrm{in/out},
\e^{\i t\widetilde H_k}\tilde1_k\e^{-\i t\widetilde H_k}v_k^\mathrm{in/out})\\
&=\sum_{k=1}^M
(u_k^\mathrm{in/out},\widetilde P_k^\mathrm{in/out}v_k^\mathrm{in/out})\\
&=(u^\mathrm{in/out},v^\mathrm{in/out}).
\end{align*}
We obtain from this that $u^\mathrm{in/out}=0$ which implies that $u=0$, a contradiction. 
The weak asymptotic completeness is thus not sufficient to assure the validity of (\ref{SDef}). 
It is necessary to also require $\mathcal H_\mathrm{s}=\Ran\Omega^-=\Ran\Omega^+$. 
When this condition is satisfied, we say that the M\o ller operators are {\em asymptotically complete}.

\begin{Rem} In our case, because of Lemma \ref{DynSpec}, asymptotic completeness is equivalent 
to the condition $\Ran\Omega^-=\Ran\Omega^+=\mathcal H_\mathrm{b}^\perp$. However this last 
condition is generally stronger than $\Ran\Omega^-=\Ran\Omega^+=\mathcal H_\mathrm{s}$. 
The M\o ller operators are called complete if 
$\Ran\Omega^-=\Ran\Omega^+=\mathcal H_\mathrm{b}^\perp$.
\end{Rem}

\begin{Prop}\label{Complete} Under Hypotheses \Href{HypReservoir}, \Href{HypResDyn}, 
and \Href{HypConjugate} the partial M\o ller operators
$$
\Omega_k^\pm\equiv\Gamma^\pm(H,\widetilde H_k;J_k^{(r)}P_\ac(\widetilde H_k)),
$$
exist, do not depend on choice of $r>0$, and satisfy
$$
\Omega_k^{\pm\ast}\Omega_l^{\pm}=\delta_{kl}\widetilde P_k^\pm,\qquad
\sum_{k=1}^M\Omega_k^{\pm}\Omega_k^{\pm\ast}=P_\ac(H).
$$
In particular, the M\o ller operators 
$\Omega^\pm=\oplus_{k=1}^M\Omega_k^\pm:\oplus_{k=1}^M\HH_k\to\mathcal H$ are 
complete: 
$$
\Ran\Omega^{-}=\Ran\Omega^{+}=\mathcal H_{\rm s}
=\mathcal H_{\rm b}^\perp=\mathcal H_\ac.
$$
\end{Prop}

\begin{Coro}\label{HProjo}Under the hypotheses  \Href{HypReservoir}, \Href{HypResDyn}, 
and \Href{HypConjugate} the operators $1_k^{(r)}P_\ac(H)$ are asymptotic projections for $H$ and
$$
P_k^\pm\equiv\Gamma^\pm(H,H;1_k^{(r)}P_\ac(H))=\Omega_k^\pm\Omega_k^{\pm\ast}
=\Omega^\pm \widetilde P_k^\pm\Omega^{\pm\ast},
$$
for all $r\ge0$ Furthermore $P_k^\pm P_l^\pm=\delta_{kl}P_k^\pm$,
$\sum_{k=1}^MP_k^\pm=P_\ac(H)$ and
$$
\Ran P_k^\pm=\Ran\Omega_k^\pm=
\{u\in\mathcal H_{\rm s}\,|\,
\lim_{t\to\pm\infty}\|(I-1_k^{(r)})\e^{-\i tH}u\|=0, \text{ for all }r\ge0\}.
$$
\end{Coro}

{\noindent\bf Remarks.} 1. By setting $\widetilde H\equiv\oplus_{k=1}^M\widetilde H_k$ and 
$$
\begin{array}{cccc}
J:&\oplus_{k=1}^M\HH_k&\to&\mathcal H\\
&(u_1,\ldots,u_M)&\mapsto&\sum_{k=1}^MJ_ku_k
\end{array}
$$
we may write
$$
\Omega^\pm=\Gamma^\pm(H,\widetilde H;JP_\ac(\widetilde H)).
$$
We thus have the intertwining relation
$$
\Omega^\pm f(\widetilde H)=f(H)\Omega^\pm.
$$
We also note the identities  $\Omega^\pm\widetilde P_k^\pm=\Omega_k^\pm=P_k^\pm\Omega^\pm$ 
and the intertwining relations for the partial M\o ller operators
$$
\Omega_k^\pm f(\widetilde H_k)=f(H)\Omega_k^\pm.
$$
\noindent 2. The scattering matrix $S=\Omega^{+\ast}\Omega^-:\HH^\mathrm{in}\to\HH^\mathrm{out}$ is unitary. The decompositions
$\HH^\mathrm{in}=\oplus_{k=1}^M\HH^\mathrm{in}_k$ and 
$\HH^\mathrm{out}=\oplus_{k=1}^M\HH^\mathrm{out}_k$ allow us to write $S=(S_{kj})$ and $S^\ast=(S^\ast_{kj})$ with
\begin{align*}
S_{kj}=\Omega^{+\ast}_k\Omega^-_j&:\HH^\mathrm{in}_j\to\HH^\mathrm{out}_k,\\
S^\ast_{kj}=(S_{jk})^\ast=\Omega^{-\ast}_k\Omega^+_j&:\HH^\mathrm{out}_j\to\HH^\mathrm{in}_k.
\end{align*}
The unitarity of $S$ is thus written as
$$
\sum_{j=1}^MS^\ast_{kj}S_{jl}=\delta_{kl}\,I_{\HH_k^\mathrm{in}},
\qquad \sum_{j=1}^MS_{kj}S^\ast_{jl}=\delta_{kl}\,I_{\HH_k^\mathrm{out}}.
$$

\bigskip

For the proof of these results we shall use the following lemma.

\begin{Lem}\label{CompLemmA} Under Hypotheses \Href{HypReservoir}, \Href{HypResDyn} 
and \Href{HypConjugate} the following assertions hold.
\begin{enumerate}[(i)]
\item For all $k\in\{1,\ldots,M\}$ and $g\in C_0^\infty(\mathbb R)$, the quadratic form
$$
\widetilde\xi_{k,g}(u,v)=(g(\widetilde H_k)(\widetilde A_k^2+1)u,
\i[\widetilde H_k,\widetilde\chi_k]g(\widetilde H_k)
(\widetilde A_k^2+1)v),
$$
defined on $\Dom(\widetilde A_k^2)\times\Dom(\widetilde A_k^2)$ is bounded. We denote by 
$\widetilde\Xi_{k,g}\in\B(\mathcal H)$ the operator it defines.
\item For all $k\in\{1,\ldots,M\}$ and $g\in C_0^\infty(\mathbb R)$, the quadratic form
$$
\xi_{k,g}(u,v)=(g(H)(A^2+1)u,J_k
\,\i[\widetilde H_k,\widetilde\chi_k]g(\widetilde H_k)
(\widetilde A_k^2+1)v),
$$
defined on $\Dom(A^2)\times\Dom(\widetilde A_k^2)$ is bounded and
$\Xi_{k,g}\in\B(\mathcal H)$ denotes the associated operator.
\end{enumerate}
\end{Lem} 

\begin{demo} $(i)$ By invoking Theorem \ref{localchar} and Definition \ref{Cnlocdef}, we remark 
that Hypothesis \Href{HypConjugate} (v) implies that $[\widetilde A_k,g(\widetilde H_k)]$ is bounded. 
Furthermore, Lemma \ref{BUcomm} ensures that
$g(\widetilde H_k)\Dom(\widetilde A_k)\subset\Dom(\widetilde A_k)$. We may thus write, for 
all $u,v\in\Dom(\widetilde A_k)$,
$$
\widetilde\xi_{k,g}(u,v)=
((g(\widetilde H_k)+\widetilde A_kg(\widetilde H_k)\widetilde A_k
-[\widetilde A_k,g(\widetilde H_k)]\widetilde A_k)
u,\i[\widetilde H_k,\widetilde\chi_k]g(\widetilde H_k)(\widetilde A_k^2+1)v).
$$
By a repeated use of the identities (\ref{OldHyp3}) and (\ref{JkAkstuf2}), we obtain
\begin{align}
(\widetilde A_ku,[\widetilde H_k,\widetilde\chi_k]v)&=
(\widetilde A_ku,\widetilde H_k\widetilde\chi_k^{(1)}v
-\widetilde\chi_k^{(1)}\widetilde H_kv)\nonumber\\
&=(\widetilde A_ku,\widetilde\chi_k^{(0)}\widetilde H_k\widetilde\chi_k^{(1)}v
-\widetilde\chi_k^{(1)}\widetilde H_kv)\nonumber\\
&=(\widetilde\chi_k^{(0)}\widetilde A_ku,\widetilde H_k\widetilde\chi_k^{(1)}v)
-(\widetilde A_ku,\widetilde\chi_k^{(1)}\widetilde H_kv)\nonumber\\
&=(\widetilde\chi_k^{(2)}\widetilde A_ku,\widetilde H_k\widetilde\chi_k^{(1)}v)
-(\widetilde A_ku,\widetilde\chi_k^{(1)}\widetilde H_kv)\label{magiczero}\\
&=(\widetilde A_ku,\widetilde\chi_k^{(2)}\widetilde H_k\widetilde\chi_k^{(1)}v
-\widetilde\chi_k^{(1)}\widetilde H_kv)\nonumber\\
&=(\widetilde A_ku,\widetilde\chi_k^{(2)}\widetilde H_kv
-\widetilde\chi_k^{(1)}\widetilde H_kv)\nonumber\\
&=((\widetilde\chi_k^{(2)}-\widetilde\chi_k^{(1)})\widetilde A_ku,\widetilde H_kv)\nonumber\\
&=0,\nonumber
\end{align}
for all $u\in\Dom(\widetilde A_k)$ and $v\in\Dom(\widetilde H_k)$. We may conclude that
$$
\widetilde\xi_{k,g}(u,v)=
((g(\widetilde H_k)-[\widetilde A_k,g(\widetilde H_k)]\widetilde A_k)
u,\i[\widetilde H_k,\widetilde\chi_k]g(\widetilde H_k)(\widetilde A_k^2+1)v).
$$
The same reasoning shows us that Hypothesis \Href{HypConjugate} (v) also implies that 
$[\widetilde A_k,[\widetilde A_k,g(\widetilde H_k)]]$ is bounded and that 
$[\widetilde A_k,g(\widetilde H_k)]\Dom(\widetilde A_k)\subset\Dom(\widetilde A_k)$. 
From this we get, with the help of the identity (\ref{magiczero}),
$$
\widetilde\xi_{k,g}(u,v)=
((g(\widetilde H_k)+
[\widetilde A_k,[\widetilde A_k,g(\widetilde H_k)]])u,
\i[\widetilde H_k,\widetilde\chi_k]g(\widetilde H_k)(\widetilde A_k^2+1)v).
$$
By using the expansion
\begin{align*}
[\widetilde A_k,[\widetilde A_k, \widetilde R_k(z)]]=
2\widetilde R_k(z)[\widetilde A_k, \widetilde H_k]
\widetilde R_k(z)[\widetilde A_k, \widetilde H_k]
\widetilde R_k(z)
-\widetilde R_k(z)[\widetilde A_k,[\widetilde A_k, \widetilde H_k]]
\widetilde R_k(z),
\end{align*}
Hypothesis \Href{HypConjugate} (v) and the Helffer-Sj\"ostrand formula, we easily show that 
$[\widetilde A_k[\widetilde A_k, g(\widetilde H_k)]]\in
\B(\mathcal H_k,\mathcal H_{k}^1)$. Since
$[\widetilde H_k,\widetilde\chi_k]\in\B(\mathcal H_k^1,\mathcal H_{k})$ we may write
$$
\widetilde\xi_{k,g}(u,v)=
(\i[\widetilde H_k,\widetilde\chi_k](g(\widetilde H_k)+
[\widetilde A_k,[\widetilde A_k,g(\widetilde H_k)]]
)u,g(\widetilde H_k)(\widetilde A_k^2+1)v).
$$
By repeating our argument on the second factor of the scalar product on the right hand side of this identity we show that
$$
\widetilde\xi_{k,g}(u,v)\!=\!
(\i[\widetilde H_k,\widetilde\chi_k]
(g(\widetilde H_k)+
[\widetilde A_k,[\widetilde A_k,g(\widetilde H_k)]])u,
(g(\widetilde H_k)+
[\widetilde A_k,[\widetilde A_k,g(\widetilde H_k)]])v),
$$
which allows us to conclude that
$$
|\widetilde\xi_{k,g}(u,v)|\le\|[\widetilde H_k,\widetilde\chi_k]\|_{\B(\mathcal H_k^1,\mathcal H_{k})}\,
\|g(\widetilde H_k)
+[\widetilde A_k[\widetilde A_k,g(\widetilde H_k)]]\|^2_{\B(\mathcal H_k,\mathcal H_{k}^1)}\,
\|u\|\,\|v\|.
$$

$(ii)$ The second assertion is proven in a very similar way. It suffices to note that, for all $u\in\Dom(A)$ and $v\in\Dom(\widetilde H_k)$, we have $J_k^\ast u\in\Dom(\widetilde A_k)$ (see Lemma \ref{ADefOK}) and thus, by the identity (\ref{magiczero}),
\begin{align}
(Au,J_k[\widetilde H_k,\widetilde\chi_k]v)&=
(Au,1_kJ_k[\widetilde H_k,\widetilde\chi_k]v)=(1_kAu,J_k[\widetilde H_k,\widetilde\chi_k]v)
\nonumber\\
&=(J_k\widetilde A_kJ_k^\ast u,J_k[\widetilde H_k,\widetilde\chi_k]v)\label{magiczero2}\\
&=(\widetilde A_kJ_k^\ast u,[\widetilde H_k,\widetilde\chi_k]v)\nonumber\\
&=0.\nonumber
\end{align}
By invoking Theorem \ref{HMourre} (ii) we obtain, for all $u\in\Dom(A^2)$ and for all $v\in\Dom(\widetilde A_k)$,
\begin{align*}
\xi_{k,g}(u,v)\!=\!(
(g(H)+[A,[A,g(H)]])u,J_k\i[\widetilde H_k,\widetilde\chi_k]
(g(\widetilde H_k)+
[\widetilde A_k[\widetilde A_k,g(\widetilde H_k)]])v),
\end{align*}
an identity which allows us to conclude our proof easily. \cqfd
\end{demo}

{\bigskip\noindent\bf Proof of Lemma \ref{HkProj}.} If $K$ is  $\widetilde H_k$-compact, for all $u\in\Dom(\widetilde H_k)$ we have
$$
\lim_{t\to\pm\infty}K(\widetilde H_k+\i)^{-1}\e^{-\i t\widetilde H_k}P_\ac(\widetilde H_k)
(\widetilde H_k+\i)u=0,
$$
by the Riemann-Lebesgue lemma. Since $\Dom(\widetilde H_k)$ is dense we may conclude that
$$
\slim_{t\to\pm\infty}K\e^{-\i t\widetilde H_k}P_\ac(\widetilde H_k)=0.
$$
We note that for all $r,s>0$, $\tilde1_k^{(r)}-\tilde1_k^{(s)}$ is 
$\widetilde H_k$-compact by Hypothesis \Href{HypResDyn} (vii).
Hypothesis \Href{HypReservoir} (ii) and the inequalities (\ref{chidomin}) imply
$$
0\le\tilde1_k^{(r)}-\widetilde\chi_k^{(r)}\le\tilde1_k^{(r)}-\widetilde\chi_k^{(r)2}\le
\tilde1_k^{(r)}-\tilde1_k^{(r+1)},
$$
and consequently $\tilde1_k^{(r)}-\widetilde\chi_k^{(r)}$ and 
$\tilde1_k^{(r)}-\widetilde\chi_k^{(r)2}$ are $\widetilde H_k$-compact. Thus, the operators 
$\widetilde\chi_k^{(r)2}P_\ac(\widetilde H_k)=
M_k^{(r)\ast}M_k^{(r)}P_\ac(\widetilde H_k)$,
$\widetilde\chi_k^{(r)}P_\ac(\widetilde H_k)$, $\tilde 1_k^{(r)}P_\ac(\widetilde H_k)$, and 
$\tilde 1_kP_\ac(\widetilde H_k)$ are asymptotically $\widetilde H_k$-equivalent and it suffices 
for us to consider the strong limits 
$\Gamma^\pm(\widetilde H_k,\widetilde H_k;\widetilde\chi_kP_\ac(\widetilde H_k))$. 
Furthermore, Hypothesis \Href{HypConjugate} ensures that the spectrum of $\widetilde H_k$ is 
purely absolutely continuous on $\mathbb R\setminus\Sigma_k$. The Lebesgue measure of 
$\Sigma_k$ being zero thus allows us to restrict to 
$\Gamma^\pm(\widetilde H_k,\widetilde H_k;\widetilde\chi_kg(\widetilde H_k)^2)$ where 
$g\in C_0^\infty(\mathbb R\setminus\Sigma_k)$ and $\widetilde H_k$ satisfy a strict Mourre 
estimate on $\Delta=\supp\,g$. A simple variation of Corollary \ref{Dkcomp} shows that 
$[\widetilde \chi_k,g(\widetilde H_k)]$ is compact. It thus suffices to consider the case of
$\Gamma^\pm(\widetilde H_k,\widetilde H_k;g(\widetilde H_k)\widetilde\chi_kg(\widetilde H_k))$. 
We may now write, as a quadratic form,
$$
[\widetilde H_k,g(\widetilde H_k)\widetilde\chi_kg(\widetilde H_k)]
=g(\widetilde H_k)[\widetilde H_k,\widetilde\chi_k]g(\widetilde H_k)
=C_k^\ast \,\widetilde\Xi_{k,g} C_k,
$$
with $C_k=(\widetilde A_k^2+1)^{-1}$. The operator $\widetilde\Xi_{k,g}$ is bounded by 
Lemma \ref{CompLemmA}. The existence of the strong limit 
$\Gamma^\pm(\widetilde H_k,\widetilde H_k;g(\widetilde H_k)\widetilde\chi_kg(\widetilde H_k))$ 
now follows from Corollary \ref{Katosmooth} and from Proposition \ref{Kato}.

Having established the existence of  
$\widetilde P_k^\pm=\Gamma^\pm(\widetilde H_k,\widetilde H_k;\tilde1_kP_\ac(\widetilde H_k))$,
it follows from the general remarks in Section \ref{strongstuf} that 
$$
\Gamma^\pm(\widetilde H_k,\widetilde H_k;\tilde1_kP_\ac(\widetilde H_k))=
\Gamma^\pm(\widetilde H_k,\widetilde H_k;P_\ac(\widetilde H_k)\tilde1_kP_\ac(\widetilde H_k)),
$$
which implies that $\widetilde P_k^\pm$ is self-adjoint and that
$$
\Gamma^\pm(\widetilde H_k,\widetilde H_k;\tilde1_kP_\ac(\widetilde H_k))^2
=\Gamma^\pm(\widetilde H_k,\widetilde H_k;\tilde1_kP_\ac(\widetilde H_k)).
$$
This confirms that $\widetilde P_k^\pm$ is an orthogonal projection. 

Finally, we note that $u\in\Ran\widetilde P_k^\pm$ if and only if, for all $r>0$,
$$
0=\|u-\widetilde P_k^\pm u\|=\lim_{t\to\pm\infty}\|
\e^{-\i t\widetilde H_k}u-\tilde 1_k^{(r)} \e^{-\i t\widetilde H_k}u
\|=\lim_{t\to\pm\infty}\|(I-\tilde 1_k^{(r)}) \e^{-\i t\widetilde H_k}u\|.
$$
The range of the projection $\widetilde P_k^\mp$ is thus $\HH_k^\mathrm{in/out}$.
\cqfd

{\bigskip\noindent\bf Proof of Proposition \ref{Complete}.} We use the same approach as in the 
previous proof. We remark first that if $0\le r\le s$, Hypothesis \Href{HypReservoir} (vi) implies that 
$$
J_k^{(r)}-J_k^{(s)}=J_k^{(r)}(\tilde1_k^{(r)}-\tilde1_k^{(s)}),
$$
and Hypothesis \Href{HypResDyn} (vii)  allows us to conclude that
$J_k^{(r)}-J_k^{(s)}$ is $\widetilde H_k$-compact. $J_k^{(r)}P_\ac(\widetilde H_k)$ and 
$J_k^{(s)}P_\ac(\widetilde H_k)$ are thus asymptotically $\widetilde H_k$-equivalent, 
which shows that if the partial M\o ller operators exist, they are independent of choice of $r>0$.

We note again that for all $r\ge0$
\begin{equation}
(\tilde1_k^{(r)}-\widetilde\chi_k^{(r)})\tilde1_k^{(r)}=
(\tilde1_k^{(r)}-\widetilde\chi_k^{(r)}),
\label{mangle1}
\end{equation}
while Hypothesis \Href{HypReservoir} (iv) implies 
\begin{equation}
\tilde1_k^{(r)}\tilde1_k^{(r+1)}=\tilde1_k^{(r+1)}.
\label{mangle2}
\end{equation}
and identity (\ref{chidomin2}) implies
\begin{equation}
\widetilde\chi_k^{(r)}\tilde1_k^{(r+1)}=\tilde1_k^{(r+1)}.
\label{mangle3}
\end{equation}
The relations (\ref{mangle1}),(\ref{mangle2}) and (\ref{mangle3}) allow us to conclude that
$$
J_k^{(r)}-M_k^{(r)}=J_k^{(r)}(\tilde1_k^{(r)}-\widetilde\chi_k^{(r)})
=J_k^{(r)}(\tilde1_k^{(r)}-\widetilde\chi_k^{(r)})(\tilde1_k^{(r)}-\tilde1_k^{(r+1)}),
$$
is $\widetilde H_k$-compact by Hypothesis \Href{HypResDyn} (vii).
$J_k^{(r)}$ and $M_k^{(r)}$  are thus asymptotically $\widetilde H_k$-equivalent. To prove the 
existence of the partial M\o ller operators it is  thus sufficient to consider the limits
$\Gamma^\pm(H,\widetilde H_k;M_kP_\ac(\widetilde H_k))$. As in the proof of 
Lemma \ref{HkProj}, we may restrict ourselves to 
$\Gamma^\pm(H,\widetilde H_k;M_kg(\widetilde H_k)^2)$ with 
$g\in C_0^\infty(\mathbb R\setminus S_k)$ where $S_k=\Sigma_k\cup\spec_\pp(H)$ is 
discrete by Corollary \ref{NoSingSpec}  and $H$ and $\widetilde H_k$ both satisfy a strict 
Mourre estimate on $\Delta=\supp\,g$. By invoking Corollary \ref{Dkcomp} it suffices for 
us to consider $\Gamma^\pm(H,\widetilde H_k;g(H)M_kg(\widetilde H_k))$. As a quadratic 
form we have
$$
Hg(H)M_kg(\widetilde H_k)-g(H)M_kg(\widetilde H_k)\widetilde H_k=
C^\ast\,\Xi_{k,g}C_k,
$$
with $C=(A^2+1)$, $C_k=(\widetilde A_k^2+1)$ and $\Xi_{k,g}$ bounded by 
Lemma \ref{CompLemmA}. Invoking Theorem \ref{HMourre}, we obtain the existence of 
$\Gamma^\pm(H,\widetilde H_k;g(H)M_kg(\widetilde H_k))$ and of 
$\Gamma^\pm(\widetilde H_k,H;g(\widetilde H_k)M_k^\ast g(H))$
as in the proof of Lemma \ref{HkProj}. It follows that
$\Omega_k^\pm=\Gamma^\pm(H,\widetilde H_k;M_kP_\ac(\widetilde H_k))$ and
$\Omega_k^{\pm\ast}=\Gamma^\pm(\widetilde H_k,H;M_k^\ast P_\ac(H))$ exist.
Hypothesis \Href{HypReservoir} (iii) and Lemma \ref{HkProj} imply
$$
\Omega_k^{\pm\ast}\Omega_l^{\pm}
=\Gamma^\pm(\widetilde H_k,\widetilde H_k;M_k^\ast M_l P_\ac(H))
=\delta_{kl}\widetilde P_k^\pm,
$$
and thus $\Ran\Omega_k^{\pm\ast}=\HH_k^\mathrm{in/out}$. The inequality (\ref{summkmk}) and 
Hypothesis \Href{HypResDyn} (vi) show that $I-\sum_{k=1}^MM_kM_k^\ast$ is $H$-compact, from 
which we get that
\begin{align*}
\sum_{k=1}^M\Omega_k^\pm\Omega_k^{\pm\ast}
&=\sum_{k=1}^M\Gamma^\pm(H,H;M_kM_k^\ast P_\ac(H))\\
&=\Gamma^\pm(H,H;P_\ac(H))=P_\ac(H).
\end{align*}
\cqfd

{\noindent\bf Proof of Corollary \ref{HProjo}.} Hypothesis \Href{HypReservoir} (iii) (v) implies 
that, for $0\le r\le s$,
$$
1_k^{(r)}-1_k^{(s)}=1_k^{(r)}(1_k^{(r)}-1_k^{(s)})=1_k^{(r)}(1_0^{(s)}-1_0^{(r)}).
$$
Hypothesis \Href{HypResDyn} (vi) allows us to conclude that
$1_k^{(r)}-1_k^{(s)}$ is $H$-compact. $1_k^{(r)}P_\ac(H)$ and $1_k^{(s)}P_\ac(H)$  
are thus asymptotically $H$-equivalent. Hypothesis \Href{HypReservoir} (ii) (vi) implies
$$
0\le1_k-M_kM_k^\ast=J_k(\tilde1_k-\widetilde\chi_k^2)J_k^\ast
\le J_k(\tilde1_k-\tilde1_k^{(2)})J_k^\ast=1_k^{(1)}-1_k^{(2)},
$$
$1_kP_\ac(H)$ and $M_kM_k^\ast P_\ac(H)$ are thus asymptotically $H$-equivalent. We thus have
\begin{align*}
\Omega^\pm\widetilde P_k^\pm\Omega^{\pm\ast}=
\Omega_k^\pm\Omega_k^{\pm\ast}
&=\Gamma^\pm(H,H;M_kM_k^\ast P_\ac(H))\\
&=\Gamma^\pm(H,H;1_kP_\ac(H))\\
&=\Gamma^\pm(H,H;1_k^{(r)}P_\ac(H))\\
&=P_k^\pm,
\end{align*}
for all $r\ge0$. Since
$$
P_k^{\pm\ast}=(\Omega_k^\pm\Omega_k^{\pm\ast})^\ast
=\Omega_k^\pm\Omega_k^{\pm\ast}=P_k^\pm,
$$
and
$$
P_k^{\pm}P_l^{\pm}
=(\Omega_k^\pm\Omega_k^{\pm\ast})(\Omega_l^\pm\Omega_l^{\pm\ast})=
\Omega_k^\pm(\Omega_k^{\pm\ast}\Omega_l^\pm)\Omega_l^{\pm\ast}=\delta_{kl}
\Omega_k^\pm\widetilde P_k^\pm\Omega_k^{\pm\ast}=\delta_{kl}P_k^\pm,
$$
$P_k^\pm$ are the disjoint orthogonal projections. Furthermore
$$
\sum_{k=1}^MP_k^\pm=\sum_{k=1}^M\Omega_k^\pm\Omega_k^{\pm\ast}
=\Omega^\pm\Omega^{\pm\ast}=P_\ac(H).
$$

Finally, $u\in\Ran P_k^\pm$ if and only if
$$
0=\|u-P_k^\pm u\|=\lim_{t\to\pm\infty}\|\e^{-\i tH}u-1_k^{(r)}\e^{-\i tH}u\|,
$$
for all $r\ge0$.
\cqfd

\subsection{Non-equilibrium steady states (NESS)}
\label{Sect-NESS}

We set $\mathcal O_k=\mathrm{CAR}(\HH_k)$  and we denote by $\tau_k^t$  the group of 
Bogoliubov automorphisms on $\mathcal O_k$ generated by $\widetilde H_k$. The following result 
is a slightly adapted version of Theorem 3.2 of \cite{AJPP2}. This is the fundamental result 
which ensures the existence of a rich enough family of nonequilibrium steady states.

\begin{Prop}\label{NESSexist} We suppose Hypotheses \Href{HypReservoir} and \Href{HypResDyn} 
hold. For all $k\in\{1,\ldots,M\}$ let $\widetilde T_k\in\B(\HH_k)$  be the generator of a gauge invariant 
quasi-free state which is also $\tau_k$-invariant on $\mathcal O_k$. Then, for all
$r\ge 0$,
$$
T=\sum_{k=1}^MJ_k^{(r)}\widetilde T_k J_k^{(r)\ast},
$$
generates a gauge invariant quasi-free state $\omega_T$ on $\mathcal O$. If 
Hypothesis \Href{HypConjugate} also holds, the NESS
\begin{equation}
\omega_T^+=\wstarlim_{t\to\infty}\frac{1}{t}\int_0^t\omega_T\circ\tau^s\,\d s,
\label{NESSconv}
\end{equation}
exists. Furthermore, the following hold.
\begin{enumerate}[(i)]
\item The restriction $\omega_T^+|_{\mathrm{CAR}(\mathcal H_\ac)}$ is the gauge invariant 
quasi-free state generated by
$$
T^+=\sum_{k=1}^M\Omega_k^-\widetilde T_k\Omega_k^{-\ast}.
$$
In particular, this state does not depend on $r\ge0$.
\item For any gauge invariant and $\omega_T$-normal state $\eta$
  on $\mathcal O$ and for all $A\in\mathrm{CAR}(\mathcal H_\ac)$,
$$
\lim_{t\to\infty}\eta\circ\tau^t(A)=\omega_{T^+}(A).
$$
\item For each trace-class operator $c$ on $\mathcal H$,
\begin{equation}
\omega_T^+(\d\Gamma(c))=\tr(T^+c)
+\sum_{\varepsilon\in\spec_\pp(H)}\tr(P_\varepsilon TP_\varepsilon c),
\label{OneBodyplus}
\end{equation}
where $P_\varepsilon$  denotes the orthogonal projection onto the eigenspace of  $H$ 
associated to the eigenvalue $\varepsilon$.
\end{enumerate}
\end{Prop}

\begin{demo} By hypothesis we have $0\le\widetilde T_k\le I$ for all $k\in\{1,\dots,M\}$. By invoking 
Hypothesis \Href{HypReservoir}  we get that
\begin{align*}
(u,Tu)&=\sum_{k=1}^M(J_k^{(r)\ast}u,\widetilde T_kJ_k^{(r)\ast}u)\le
\sum_{k=1}^M(J_k^{(r)\ast}u,J_k^{(r)\ast}u)\\
&=\sum_{k=1}^M(u,1_k^{(r)\ast}u)=(u,(I-1_0^{(r)})u)\le(u,u),
\end{align*}
and thus $0\le T\le I$ which shows that $T$ is the generator of a gauge invariant quasi-free state 
on $\mathcal O$.

The rest of the proposition is proven like Theorem 3.2 of \cite{AJPP2}, by remarking that
\begin{align*}
\lim_{t\to\infty}
(\e^{\i tH}P_\ac(H)g,T\e^{\i tH}P_\ac(H)f)&=
\sum_{k=1}^M\lim_{t\to\infty}(\e^{-\i t\widetilde H_k}J_k^{(r)\ast}\e^{\i tH}P_\ac(H)g,
\widetilde T_k\e^{-\i t\widetilde H_k}J_k^{(r)\ast}\e^{\i tH}P_\ac(H)f)\\
&=(g,T^+f).
\end{align*}
\cqfd
\end{demo}

\bigskip It is clear that the $\tau_k$-invariance of
$\omega_{\widetilde T_k}$, that is to say the fact that, for all 
$t\in\mathbb R$, $\e^{\i t\widetilde H_k}\widetilde T_k\e^{-\i
  t\widetilde H_k}=\widetilde T_k$  is crucial in the preceding proposition. We may choose for 
example $\widetilde T_k=f_k(\widetilde H_k)$ where $f_k:\spec(\widetilde H_k)\to[0,1]$ is a 
measurable function. We therefore have, since $f(H)$ and $P_k^-$ commute,
\begin{equation}
T^+=\sum_{k=1}^M\Omega_k^-f_k(\widetilde H_k)\Omega_k^{-\ast}=
\sum_{k=1}^Mf_k(H)\Omega_k^-\Omega_k^{-\ast}=
\oplus_{k=1}^MP_k^-f_k(H)P_k^-.
\label{Tplusform}
\end{equation}
The particular case 
$$
f_k(\varepsilon)=\frac{1}{1+\e^{\beta_k(\varepsilon-\mu_k)}},
$$
of course plays a crucial role in applications to statistical mechanics 
(recall Section~\ref{QFKMSsubsect}).

The following result shows that we can, in an equivalent fashion, choose $\omega_T$ as 
a ``superposition'' of $\tau$-invariant states.

\begin{Theo}\label{NESSalt}We suppose that Hypotheses \Href{HypReservoir} 
and \Href{HypResDyn} hold. For all $k\in\{1,\ldots,M\}$ let $T_k\in\B(\mathcal H)$ be the 
generator of a gauge invariant, quasi-free state which is also $\tau$-invariant on $\mathcal O$. 
Then, for all $r\ge 0$,
$$
T=\sum_{k=1}^M1_k^{(r)}T_k 1_k^{(r)},
$$
generates a gauge invariant quasi-free state on $\mathcal O$. If Hypothesis \Href{HypConjugate} 
also holds, the NESS (\ref{NESSconv}) exists and Assertions (i)-(iii) of Proposition \ref{NESSexist} 
hold true with
$$
T^+=\sum_{k=1}^MP_k^-T_kP_k^{-}=\sum_{k=1}^M\Omega_k^{-}\widetilde T_k\Omega_k^{-\ast},
$$
where $\widetilde T_k=\Omega_k^{-\ast}T_k\Omega_k^{-}$. In particular, if $T_k=f_k(H)$, 
we obtain the formula
$$
T^+=\sum_{k=1}^MP_k^-f_k(H)P_k^{-},
$$
which is identical to (\ref{Tplusform}).
\end{Theo}

\begin{demo} We argue in the same manner as in the proof of Proposition \ref{NESSexist}, by
remarking that this time
\begin{align*}
\lim_{t\to\infty}
(\e^{\i tH}P_\ac(H)g,&T\e^{\i tH}P_\ac(H)f)=\\
\sum_{k=1}^M\lim_{t\to\infty}(\e^{-\i H}1_k^{(r)\ast}\e^{\i tH}P_\ac(H)g,
&T_k\e^{-\i tH}1_k^{(r)\ast}\e^{\i tH}P_\ac(H)f)\\
=(g&,T^+f).
\end{align*}
\cqfd
\end{demo}

\section{The geometric Landauer-B\"uttiker formula}
\label{LBForm}

In this section we derive a Landauer-B\"uttiker formula for currents associated
to a general class of conserved charges. As opposed to previous derivations of this formula
\cite{AJPP2,N} or of its linearized version \cite{CJM}, which exploit the stationary formalism of
scattering theory we shall continue to use the time dependent framework.

\subsection{Hypotheses}

In the remaining parts of these notes, and unless otherwise stated,
 we shall assume that Hypotheses \Href{HypReservoir},
\Href{HypResDyn}, and \Href{HypConjugate} of Section~\ref{NESSChapter} hold. To establish the 
existence of current observable and to study their properties we must however make a few additional 
hypotheses.

\begin{hyp}
\item There exists an integer $m>2$ such that, for $k\in\{1,\ldots,M\}$, 
\begin{enumerate}[(i)]
 \item $(\widetilde A_k+\i)^{-m}g(\widetilde H_k)$ is trace-class for all 
$g\in C_0^\infty(\mathbb R\setminus\Sigma_k)$.
\item $\widetilde H_k\in\B_{\widetilde A_k}^{m+2}(\HH_k^1,\HH_k)$.
\item  $1_0^{(r)}g(H)$ is trace-class for all $g\in C_0^\infty(\mathbb R)$ and $0\le r\le 2$.
\end{enumerate}
\label{HypAHTrace}
\item There exists an integer $\nu\ge1$ such that, for $k\in\{1,\ldots,M\}$, 
\begin{enumerate}[(i)]
\item $\ad_{\widetilde H_k}^j(\widetilde\chi_k^{(r)})\in\B(\HH_k^{j/2},\HH_k)$ for $j=1,\ldots,4\nu$.
\item $(I-\tilde1_k^{(r)})\ad_{\widetilde H_k}^j(\widetilde\chi_k^{(r)})=0$ for $j=1,\ldots,2\nu$.
\item $\tilde1_k^{(r+1)}\ad_{\widetilde H_k}^j(\widetilde\chi_k^{(r)})=0$ for $j=1,\ldots,2\nu$.
\item $(\widetilde\chi_k^{(r)}-\widetilde\chi_k^{(s)})(\widetilde H_k+\i)^{-\nu}$ is trace-class for $r,s\ge0$.
\end{enumerate}
\label{HypChi}\label{HypTrace0}
\end{hyp}

\bigskip
We insist on the fact that only Hypothesis \Href{HypAHTrace} (iii) concerns the sample $\mathfrak S$ 
through the Hamiltonian $H$. Moreover, this hypothesis, which quantifies the confinement of 
$\mathfrak S$ is very weak. All the other hypotheses concern only the
extended reservoirs $\widetilde{\mathfrak{R}}_k$.

We begin by deducing several important consequences of these hypotheses which will be useful 
to us later on.

\begin{Lem}\label{fHMLemma} Under hypotheses \Href{HypReservoir},
\Href{HypResDyn}, and \Href{HypChi} the operators
$$
(H-z)^{-\ell}M_k^{(r)}-M_k^{(r)}(\widetilde H_k-z)^{-\ell},
$$
and
$$
D_k^{(r)}(f)=f(H)M_k^{(r)}-M_k^{(r)}f(\widetilde H_k),
$$
are trace-class for all $f\in C_0^\infty(\mathbb R)$, $k\in\{1,\ldots,M\}$, $r\ge 1$, 
$z\in\Res(H)\cap\Res(\widetilde H_k)$ and $\ell\ge\nu$.
\end{Lem}

The proof of this lemma being quite long and technical, we have chosen to defer it to
Appendix \ref{HTraceProof}.

\begin{Lem}\label{HchiLemma} Under the hypotheses of Lemma \ref{fHMLemma} the operators 
\begin{equation*}
\begin{array}{llll}
[f(\widetilde H_k),\widetilde\chi_k^{(r)}],\quad &[f(\widetilde H_k),\tilde1_k^{(r)}],\quad
&f(\widetilde H_k)(\widetilde\chi_k^{(r)}-\widetilde\chi_k^{(s)})\quad
&f(\widetilde H_k)(\tilde 1_k^{(r)}-\tilde 1_k^{(s)})\cr
[f(H),\chi_k^{(r)}], &[f(H),1_k^{(r)}], &f(H)(\chi_k^{(r)}-\chi_k^{(s)})\quad&f(H)(1_k^{(r)}-1_k^{(s)}),
\end{array}
\end{equation*}
are trace-class for all $f\in C_0^\infty(\mathbb R)$, $k\in\{1,\ldots,M\}$ and $r,s\ge1$.
\end{Lem}

\begin{demo} Hypothesis \Href{HypChi} (iv) implies that for $r,s\ge0$
$$
f(\widetilde H_k)(\widetilde\chi_k^{(r)}-\widetilde\chi_k^{(s)})=
f(\widetilde H_k)(\widetilde H_k+\i)^\nu
(\widetilde H_k+\i)^{-\nu}(\widetilde\chi_k^{(r)}-\widetilde\chi_k^{(s)}),
$$
is trace-class. We set $D\equiv D_k^{(r)}(f)$ and remark that Lemma \ref{fHMLemma} 
implies that, for $r\ge1$,
$$
[f(\widetilde H_k),\widetilde\chi_k^{(r)2}]
=f(\widetilde H_k)M_k^{(r)\ast}M_k^{(r)}-M_k^{(r)\ast}M_k^{(r)}f(\widetilde H_k)
=M_k^{(r)\ast}D-D^\ast M_k^{(r)},
$$
is trace-class. Hypothesis  \Href{HypReservoir} (iv) and  Identity (\ref{chidomin2}) imply
$$
(\tilde1_k^{(r)}-\widetilde\chi_k^{(r)2})\widetilde\chi_k^{(r+1)}
=(\tilde1_k^{(r)}-\widetilde\chi_k^{(r)2})\tilde 1_k^{(r+1)}\widetilde\chi_k^{(r+1)}
=(\tilde1_k^{(r+1)}-\tilde1_k^{(r+1)})\widetilde\chi_k^{(r+1)}=0,
$$
$$
(\tilde1_k^{(r)}-\widetilde\chi_k^{(r)2})\widetilde\chi_k^{(r-1)}
=(\tilde1_k^{(r)}-\widetilde\chi_k^{(r)2})\tilde 1_k^{(r)}\widetilde\chi_k^{(r-1)}
=(\tilde1_k^{(r)}-\widetilde\chi_k^{(r)2})\tilde 1_k^{(r)}
=(\tilde1_k^{(r)}-\widetilde\chi_k^{(r)2}),
$$
from which we conclude that
\beq
\tilde1_k^{(r)}-\widetilde\chi_k^{(r)2}=(\tilde1_k^{(r)}-\widetilde\chi_k^{(r)2})
(\widetilde\chi_k^{(r-1)}-\widetilde\chi_k^{(r+1)})
=(\widetilde\chi_k^{(r-1)}-\widetilde\chi_k^{(r+1)})(\tilde1_k^{(r)}-\widetilde\chi_k^{(r)2}).
\label{fanny}
\eeq
In the same way, one shows that
\beq
\tilde1_k^{(r)}-\widetilde\chi_k^{(r)}=(\tilde1_k^{(r)}-\widetilde\chi_k^{(r)})
(\widetilde\chi_k^{(r-1)}-\widetilde\chi_k^{(r+1)})
=(\widetilde\chi_k^{(r-1)}-\widetilde\chi_k^{(r+1)})(\tilde1_k^{(r)}-\widetilde\chi_k^{(r)}).
\label{fanny1}
\eeq
From \eqref{fanny} we get that
\begin{align*}
[f(\widetilde H_k),\tilde1_k^{(r)}]=[f(\widetilde H_k),\tilde\chi_k^{(r)2}]&+f(\widetilde H_k)
(\widetilde\chi_k^{(r-1)}-\widetilde\chi_k^{(r+1)})(\tilde1_k^{(r)}-\widetilde\chi_k^{(r)2})\\
&-(\tilde1_k^{(r)}-\widetilde\chi_k^{(r)2})(\widetilde\chi_k^{(r-1)}-\widetilde\chi_k^{(r+1)})
f(\widetilde H_k),
\end{align*}
and Hypothesis \Href{HypChi} (iv) allows us to conclude that $[f(\widetilde H_k),\tilde1_k^{(r)}]$ 
is trace-class.
In an analogous manner we deduce from \eqref{fanny1} that $[f(\widetilde H_k),\widetilde\chi_k^{(r)}]$ 
is trace-class. Since
\begin{align*}
[f(H),\chi_k^{(r)}]&=f(H)M_k^{(r)}J_k^{(r)\ast}-J_k^{(r)}M_k^{(r)\ast}f(H)\\
&=M_k^{(r)}f(\widetilde H_k)J_k^{(r)\ast}+DJ_k^{(r)\ast}
-J_k^{(r)}f(\widetilde H_k)M_k^{(r)\ast}-J_k^{(r)}D^\ast\\
&=-J_k^{(r)}[f(\widetilde H_k),\widetilde\chi_k^{(r)}]J_k^{(r)\ast}
+DJ_k^{(r)\ast}-J_k^{(r)}D^\ast,
\end{align*}
we may also conclude that $[f(H),\chi_k^{(r)}]$ is trace-class. The identity
\begin{align*}
f(H)(\chi_k^{(r)}-\chi_k^{(s)})&=f(H)(M_k^{(r)}J_k^{(r)\ast}-M_k^{(s)}J_k^{(s)\ast})\\
&=(f(H)M_k^{(r)}-M_k^{(r)}f(\widetilde H_k))J_k^{(r)\ast}\\
&-(f(H)M_k^{(s)}-M_k^{(s)}f(\widetilde H_k))J_k^{(s)\ast}\\
&+J_k^{(s)}[f(\widetilde H_k),\widetilde\chi_k^{(s)}]J_k^{(s)\ast}
-J_k^{(r)}[f(\widetilde H_k),\widetilde\chi_k^{(r)}]J_k^{(r)\ast}\\
&+J_k^{(r)}f(\widetilde H_k)(\widetilde \chi_k^{(r)}-\widetilde \chi_k^{(s)})J_k^{(r)\ast},
\end{align*}
implies that $f(H)(\chi_k^{(r)}-\chi_k^{(s)})$ is trace-class for $r,s\ge1$. The identity
 \beq
1_k^{(r)}-\chi_k^{(r)}=(1_k^{(r)}-\chi_k^{(r)})
(\chi_k^{(r-1)}-\chi_k^{(r+1)})
=(\chi_k^{(r-1)}-\chi_k^{(r+1)})(1_k^{(r)}-\chi_k^{(r)}),
\label{fanny2}
\eeq
(a version of \eqref{fanny1} without tilde) allows us to show that $[f(H),1_k^{(r)}]$ is 
trace-class. Finally, one deduces from Identities \eqref{fanny1}, \eqref{fanny2} and
the previous results that  $f(\widetilde H_k)(\tilde 1_k^{(r)}-\tilde 1_k^{(s)})$
and $f(H)(1_k^{(r)}-1_k^{(s)})$ are trace class. \cqfd
\end{demo}

\bigskip

\begin{Lem}\label{AHTraceLemma} 
If Hypotheses \Href{HypReservoir}--\Href{HypTrace0} are 
satisfied, then $f(H)\chi_k^{(r)}g(A_k)$ and $f(H)1_k^{(r)}g(A_k)$
are trace-class for all $r\ge1$,
$f\in C_0^\infty(\mathbb R\setminus\Sigma_H)$ and $g\in C_0^\infty(\mathbb R)$.
\end{Lem}

\begin{demo}
Since $A_k=J_k\widetilde A_kJ_k^\ast$, we have
$$
(A_k-z)^{-1}=J_k(\widetilde A_k-z)^{-1}J_k^\ast-z^{-1}(I-1_k),
$$
and the Helffer-Sj\"ostrand formula yields
\begin{equation}
g(A_k)=J_kg(\widetilde A_k)J_k^\ast-g(0)(I-1_k),
\label{gAkfirst}
\end{equation}
for all $g\in C_0^\infty(\mathbb R)$. We thus have
\begin{align*}
f(H)\chi_k^{(r)}g(A_k)&=f(H)M_k^{(r)}g(\widetilde A_k)J_k^\ast\\
&=(f(H)M_k^{(r)}-M_k^{(r)}f(\widetilde H_k))g(\widetilde A_k)J_k^\ast\\
&+M_k^{(r)}f(\widetilde H_k)(\widetilde A_k+\i)^{-m}(\widetilde A_k+\i)^{m}
g(\widetilde A_k)J_k^\ast,
\end{align*}
and Hypotheses \Href{HypAHTrace} (i) and Lemma \ref{fHMLemma} imply the first
assertion. Writing 
$$
f(H)1_k^{(r)}g(A_k)=f(H)\chi_k^{(r)}g(A_k)+f(H)(1_k^{(r)}-\chi_k^{(r)})g(A_k),
$$
the second assertion is a direct consequence of the first one, Identity~\eqref{fanny1} and
Lemma~\ref{HchiLemma}.\cqfd
\end{demo}

\bigskip
We note for later reference that Identity (\ref{JkAkstuf}) implies
\begin{equation}
J_k^{(r)}(\widetilde A_k-z)^{-j}J_k^{(r)\ast}=
M_k^{(r)}(\widetilde A_k-z)^{-j}M_k^{(r)\ast}+(-z)^{-j}(1_k^{(r)}-M_k^{(r)}M_k^{(r)\ast}),
\label{SmoothResolv}
\end{equation}
for $z\in\mathbb C\setminus\mathbb R$ and $j\in\mathbb N$, and hence
$$
J_k^{(r)}g(\widetilde A_k)J_k^{(r)\ast}=
M_k^{(r)}g(\widetilde A_k)M_k^{(r)\ast}-g(0)(1_k^{(r)}-M_k^{(r)}M_k^{(r)\ast}).
$$
Together with Eq.~(\ref{gAkfirst}), we thus obtain
\begin{equation}
g(A_k)=
M_k^{(r)}g(\widetilde A_k)M_k^{(r)\ast}-g(0)(I-M_k^{(r)}M_k^{(r)\ast}).
\label{MgMstarId}
\end{equation}

\subsection{A simple model (continued)}

We come back to the simple example of Section \ref{cylindricends}.
As already remarked there, Hypotheses~\Href{HypAHTrace}~(ii) holds
for any integer $m$. For any $\nu\ge1$, the verification of Hypotheses \Href{HypChi}~(ii)--(iii) 
reduces to straightforward calculations while \Href{HypChi}~(i) follows from the easily established fact
that
$$
\ad_{\widetilde H_\mp}^n(\widetilde\chi_\mp^{(r)})=\sum_{j=0}^nc_{nj}(x)\partial_x^j,
$$
with $c_{nj}\in C_0^\infty(\rr)$. 
To check the remaining Hypotheses~\Href{HypAHTrace}~(i)+(iii) and~\Href{HypChi}~(iv) we 
need some trace-class estimates.

For any $r>0$ there exists a compact subset  $K\subset\mathfrak{M}$ such that
$1_0^{(r)}g(H)=1_0^{(r)}1_K(H+1)^{-s}(H+1)^sg(H)$ holds for any $s>0$
($1_K$ denotes the operator of multiplication by the  characteristic function of $K$).
Reciprocally, for any compact subset $K\subset\mathfrak{M}$ and $s\ge1$ one has
$1_K(H+1)^{-s}=1_K1_0^{(r)}(H+1)^{-1}(H+1)^{-(s-1)}$ for large enough $r>0$.
Thus, to prove Property~\Href{HypAHTrace}~(iii) it suffices to show that for sufficiently 
large $s>0$ and any compact subset $K\subset\mathfrak{M}$ the operator $1_K(H+1)^{-s}$
is trace-class. Moreover, we already know, from Hypothesis~\Href{HypResDyn}~(vi), that this 
operator is compact. Set $C=(H+1)^{-s}1_K$ and denote by $\{\mu_j\}$ the decreasing
sequence of repeated eigenvalues of $C^\ast C=1_K(H+1)^{-2s}1_K\ge0$. We have to
show that
\beq
\|C\|_1=\tr|C|=\sum_j\mu_j^{1/2}<\infty.
\label{tracC}
\eeq
Since $\spec(AB)\setminus\{0\}=\spec(BA)\setminus\{0\}$ holds for any bounded operators 
$A$ and $B$, the sequence $\{\mu_j\}$ coincide with the sequence of non-zero eigenvalues
of $\widetilde C=1_K(H+1)^{-2s}$. By Proposition 1 and Theorem 3 in \cite{Sk}, we have
$$
\mu_j=\mathcal{O}(j^{-2s}),
$$
as $j\to\infty$ so that \eqref{tracC} holds provided $s>1$. We can use the same argument
to show that Hypothesis~\Href{HypChi}~(iv) holds provided $\nu>1$.

The case of Hypothesis~\Href{HypAHTrace}~(i) is more delicate. We first claim that we 
can replace $\widetilde{A}_\mp$ with the generator $A_0=\frac1{2\i}(x\partial_x+\partial_xx)$
of the dilation group $(U_0^tf)(x,\varphi)=\e^{t/2}f(\e^t x,\varphi)$. Indeed, with $\chi\in C_0^\infty(\rr)$
such that $0\le\chi(x)\le1$ for all $x\in\rr$ and $\chi(x)=1$ for $|x|\le20$ we can write
$$
(\widetilde{A}_\mp+\i)^{-m}g(\widetilde{H}_\mp)
=(\widetilde{A}_\mp+\i)^{-m}\chi(x) g(\widetilde{H}_\mp)
+(\widetilde{A}_\mp+\i)^{-m}(I-\chi(x))(A_0+\i)^m(A_0+\i)^{-m}g(\widetilde{H}_\mp).
$$
The first term on the right hand side of this identity is trace class by the argument previously used
to prove  Hypothesis~\Href{HypChi}~(iv). To deal with the second term, we invoke 
Lemma~\ref{ExactCummutExpansion} and use the fact that $v(x)=x$ on the support of $1-\chi$
to write
\begin{align*}
(\widetilde{A}_\mp+\i)^{-m}(I-\chi(x))(A_0+\i)^m
&=(\widetilde{A}_\mp+\i)^{-m}(I-\chi(x))(A_\mp+\i)^m\\
&=I-\chi(x)+[(\widetilde{A}_\mp+\i)^{-m},(I-\chi(x))](A_\mp+\i)^m\\
&=I-\chi(x)-\sum_{j=1}^m{m\choose j}\i^j(\widetilde{A}_\mp+\i)^{-j}{\rm ad}_{\widetilde{A}_\mp}^j(\chi).
\end{align*}
Since ${\rm ad}_{\widetilde{A}_\mp}^j(\chi)=(v(x)\partial_x)^j\chi\in C_0^\infty(\rr)$, we conclude
that $(\widetilde{A}_\mp+\i)^{-m}(I-\chi(x))(A_0+\i)^m$ is bounded. Thus, as claimed, it is sufficient
to show that $(A_0+\i)^{-m}g(\widetilde{H}_\mp)$ is trace class.

We shall derive a more convenient representation of this operator using the following unitary maps:
\begin{enumerate}
\item The Fourier transform $\mathcal{F}:\HH_\mp=L^2(\rr\times\gamma_\mp,\d x\,R\d\varphi)\to
L^2(\rr,\d p)\otimes\ell^2(\zz)$,
$$
(\mathcal{F}f)(p,k)=\frac{R^{1/2}}{2\pi}\int_\rr\d x\int_0^{2\pi}\d\varphi
f(x,\varphi)\e^{-\i(px+k\varphi)},
$$
maps the Hamiltonian $\widetilde{H}_\mp$ to the multiplication operator
$$
(\mathcal{F}\widetilde{H}_\mp f)(p,k)=(p^2+\lambda_k^2)(\mathcal{F}f)(p,k),
$$
and the dilation group
$U_0^t$ to its inverse, $(\mathcal{F}U_0^t f)(p,k)=\e^{-t/2}(\mathcal{F}f)(\e^{-t}p,k)$.
\item The map $\mathcal{P}:L^2(\rr,\d p)\otimes\ell^2(\zz)
\to L^2(\rr_+,\d p)\otimes\cc^2\otimes\ell^2(\zz)$
defined by
$$
(\mathcal{P}f)(p,k)=\frac1{\sqrt2}\left(
\begin{array}{ll}
f(-p,k)-f(p,k)\\f(-p,k)+f(p,k)
\end{array}
\right)=\left(
\begin{array}{ll}
(\mathcal{P}f)_-(p,k)\\(\mathcal{P}f)_+(p,k)
\end{array}
\right),
$$
decomposes $f$ into its odd/even parts w.r.t.\;the $p$-variable. It clearly commutes with
the actions of $\widetilde{H}_\mp$ and $U_0^t$.
\item The map $\mathcal{V}:L^2(\rr_+,\d p)\otimes\cc^2\otimes\ell^2(\zz)\to
L^2(\rr,\d s)\otimes\cc^2\otimes\ell^2(\zz)$ defined by
$$
(\mathcal{V}f)(s,k)=\e^{s/2}f(\e^s,k),
$$
implements the change of variable $s=\log p$.
\item The Fourier transform in the $s$-variable
$\mathcal{S}:L^2(\rr,\d s)\otimes\cc^2\otimes\ell^2(\zz)\to
L^2(\rr,\d a)\otimes\cc^2\otimes\ell^2(\zz)$,
$$
(\mathcal{S}f)(a,k)=\frac1{\sqrt{2\pi}}\int_\rr\e^{\i a s}f(s,k)\d s.
$$
\item  The Mellin transform $\mathcal{M}=\mathcal{S}\mathcal{V}:
L^2(\rr_+,\d p)\otimes\cc^2\otimes\ell^2(\zz)\to
L^2(\rr,\d a)\otimes\cc^2\otimes\ell^2(\zz)$,
$$
(\mathcal{M}f)(a,k)=\frac1{\sqrt{2\pi}}\int_0^\infty p^{\i a-1/2}f(p,k)\d p,
$$
satisfies $(\mathcal{M}\mathcal{P}\mathcal{F}U_0^tf)(a,k)
=\e^{\i ta}(\mathcal{M}\mathcal{P}\mathcal{F}f)(a,k)$. Thus, $\mathcal{M}\mathcal{P}\mathcal{F}$
maps the operator $A_0$ to a multiplication operator
$(\mathcal{M}\mathcal{P}\mathcal{F}A_0f)(a,k)=a(\mathcal{M}\mathcal{P}\mathcal{F}f)(a,k)$.
\end{enumerate}

It follows that $(A_0+\i)^{-m}g(\widetilde{H}_\mp)
=\mathcal{F}^\ast\mathcal{P}^\ast\mathcal{M}^\ast C\mathcal{V}\mathcal{P}\mathcal{F}$
where $C$ is the operator acting on $L^2(\rr)\otimes\cc^2\otimes\ell^2(\zz)$ as
$$
(Cf)_\pm(a,k)=F(a)(\mathcal{S}G_kf_\pm(\,\cdot\,,k))(a),
$$
where $F(a)=(a+\i)^{-m}$ and $G_k$ is the operator of multiplication by the function
$G_k(s)=g(\e^{2s}+\lambda_k^2)$. Writing our Hilbert space as a direct sum
$$
L^2(\rr)\otimes\cc^2\otimes\ell^2(\zz)
=\bigoplus_{\langle k,\sigma\rangle\in\zz\times\{\pm\}}L^2(\rr),
$$
and denoting by $C_{k\pm}$ the operator defined on $L^2(\rr)$
by $(C_{k\pm} f)(a)=F(a)(\mathcal{S}G_kf)(a)$ we get
$$
C=\bigoplus_{\langle k,\sigma\rangle\in\zz\times\{\pm\}}C_{k\sigma}.
$$

Let $\delta={\rm dist}(\supp(g),\Sigma_\pm)>0$ and $\rho=\sup\supp(g)$.
For $s\in\supp(G_k)$ one has
$$
\inf_{j\in\zz}|\e^{2s}+\lambda_k^2-\lambda_j^2|\ge\delta,\qquad \e^{2s}+\lambda_k^2\le\rho.
$$
It follows that $G_k\in C_0^\infty(\rr)$ with
$$
\left\{
\begin{array}{ll}
G_k=0&\text{ if }|k|>R\rho^{1/2};\\[6pt]
\supp(G_k)\subset[\log(\delta)/2, \log(\rho)/2]&\text{ otherwise}.
\end{array}
\right.
$$
Thus, it suffices to show that $C_{k\pm}$ is trace class for any
$\langle k,\sigma\rangle\in\zz\times\{\pm\}$. 

The (quasi-)Banach space $\ell^p(\zz,L^2([0,1]))$ is the space of measurable functions $f:\rr\to\cc$
such that
$$
\|f\|_{2,p}=
\left(\sum_{j\in\zz}\left(\int_0^1|f(j+s)|^2\d s\right)^{p/2}\right)^{1/p}<\infty.
$$
One easily checks that $\|F\|_{2,p}<\infty$ for $p>m^{-1}$ and $\|G_k\|_{2,p}<\infty$ for
any $p>0$. It follows from Paragraph~5.7 in \cite{BKS} (see also Theorem 4.5 in \cite{S})
that the singular values of $C_{k\pm}$ satisfy
$$
\kappa_j(C_{k\pm})=\mathcal{O}(j^{-1/p}),
$$
as $j\to\infty$ for any $p>m^{-1}$. In particular, $C_{k\pm}$ is trace-class for $m>1$.

The arguments presented in this section can easily be adapted to the various extensions discussed
at the end of Section \ref{cylindricends}. We note  however that Hypothesis~\Href{HypAHTrace}~(i)
fails for super-reservoirs of the type $\widetilde{\mathfrak{R}}_j=]1,\infty[\times\gamma_j$ equipped 
with a metric $g_{\widetilde{\mathfrak{R}}_j}=\d x^2+r(x)^2g_{\gamma_j}$, such that $r(x)\to\infty$ as
$x\to\infty$. In fact, it follows from Hypotheses \Href{HypReservoir}--\Href{HypTrace0} that the
Hamiltonian $H$ has locally finite spectral multiplicity (see Proposition \ref{SpectralMult}).
In physical terms, our hypotheses only allow for a finite number of open scattering channels 
at any given finite energy $E$.

\subsection{Charges and conserved currents}

\subsubsection{Charges}

\begin{Def}\label{chargedef} A charge of the  of one-particle system is an observable, described by 
the self-adjoint operator $Q$ on $\mathcal H$, such that
\begin{enumerate}[(i)]
\item $\e^{\i tH}Q\e^{-\i tH}=Q$ for all $t\in\mathbb R$.
\item For each $k\in\{1,\ldots,M\}$ there exists a self-adjoint operator $\widetilde Q_k$ on $\HH_k$ 
such that $$\e^{\i t\widetilde H_k}\widetilde Q_k\e^{-\i t\widetilde H_k}=\widetilde Q_k,$$ for all 
$t\in\mathbb R$.
\item For each scattering state $u\in\mathcal H_\mathrm s$ and for all $r\ge 0$ we have
$$
\lim_{t\to\pm\infty}\sum_{k=1}^M(M_k^{(r)\ast}\e^{-\i tH}u,\widetilde Q_kM_k^{(r)\ast}\e^{-\i tH}u)
=(u,Qu).
$$
\end{enumerate} 
\end{Def}

Condition (i) expresses charge conservation in the system. Conditions (ii) and (iii) ensure that it is 
possible to determine the total charge of a scattering state by performing a measurement in the 
reservoirs (and waiting long enough). They reflect the fact that charge transport
across the sample can be determined by measuring the charges in the reservoirs at two
well separated times, as in the full counting scheme described in Section~\ref{Sect-FCS}.

\bigskip
Since Condition (ii) implies
$$
(M_k^{(r)\ast}\e^{-\i tH}u,\widetilde Q_kM_k^{(r)\ast}\e^{-\i tH}u)=
(\e^{\i t\widetilde H_k}
M_k^{(r)\ast}\e^{-\i tH}u,\widetilde Q_k\e^{\i t\widetilde H_k}M_k^{(r)\ast}\e^{-\i tH}u),
$$
Condition (iii) is equivalent to
\begin{align*}
\sum_{k=1}^M
(u^\mathrm{in/out}_k,\widetilde Q_ku_k^\mathrm{in/out})=
\sum_{k=1}^M
(\Omega_k^{\mp\ast}u,\widetilde Q_k\Omega_k^{\mp\ast}u)
=(u,Qu),
\end{align*}
that is to say that $\mathcal H_\mathrm{s}=\mathcal H_\ac$ and
$$
\left. Q\right|_{\mathcal H_\ac}=\sum_{k=1}^M\Omega_k^{+}\widetilde Q_k\Omega_k^{+\ast}
=\sum_{k=1}^M\Omega_k^{-}\widetilde Q_k\Omega_k^{-\ast}.
$$
We have in particular
\begin{equation}
\Omega_k^{\pm\ast}Q\Omega_k^\pm=\widetilde P_k^\pm\widetilde Q_k\widetilde P_k^\pm,
\label{QConjug}
\end{equation}
and, by setting $\widetilde Q=\oplus_{k=1}^M\widetilde Q_k$,
\begin{equation}
\left. Q\right|_{\mathcal H_\ac}
=\Omega^{+}\widetilde Q\Omega^{+\ast}
=\Omega^{-}\widetilde Q\Omega^{-\ast}.
\label{Cond3}
\end{equation}
We note that this condition implies
\begin{equation}
S\left(\widetilde Q\left.\right|_{\HH^\mathrm{in}} \right)S^\ast
=\widetilde Q\left.\right|_{\HH^\mathrm{out}},
\label{QkConjug}
\end{equation}
which expresses the fact that the total charge of the reservoirs is conserved by the scattering process.

The two canonical examples of charge are the following.
\begin{enumerate}[1.]
\item The electric charge: by supposing that each fermion carries a unit charge, the observable of 
electric charge is simply $Q=I$. The corresponding observables in the reservoirs are 
$\widetilde Q_k=I$. Conditions (i) and (ii) are thus trivially satisfied. 
Condition (iii), or more precisely the equivalent condition (\ref{Cond3}) reduces to
$$
P_\ac(H)=\Omega^\pm\Omega^{\pm\ast},
$$
that is to say to asymptotic completeness.
\item The energy: corresponds to $Q=H$ and $\widetilde Q_k=\widetilde H_k$. Conditions (i) and (ii) 
are again trivially satisfied, while (\ref{Cond3}) translates into 
$$
H_\ac=\Omega^\pm\left(\oplus_{k=1}^M\widetilde H_k\right)\Omega^{\pm\ast}.
$$
\end{enumerate}

\subsubsection{Currents and regularized currents}
\label{CurSubSct}
Let $Q$ be a charge of the one-particle system. The total charge in reservoir $\mathfrak R_k$
at a ``distance'' greater than $r$ of the sample $\mathfrak S$  is described by the observable 
$\d\Gamma(1_k^{(r)}Q1_k^{(r)})$. The observable of the corresponding current is 
$$
\left.\frac{\d\ }{\d t}\tau^t(\d\Gamma(1_k^{(r)}Q1_k^{(r)}))\right|_{t=0}
=\d\Gamma(\i[H,1_k^{(r)}Q1_k^{(r)}]).
$$
Our goal in this section is to give a meaning to the notion of steady current
\begin{equation}
\omega_T^+(\d\Gamma(\i[H,1_k^{(r)}Q1_k^{(r)}])),
\label{JunkCurrentId}
\end{equation}
expectation value of the current observable in the NESS which we constructed in 
Section \ref{NESSChapter}.

Several problems arise:
\begin{enumerate}[1.]
\item If the charge $Q$ is an unbounded operator, the product $1_k^{(r)}Q1_k^{(r)}$ does not make 
sense in general. This is the case, for example, for the energy current.
\item The commutator $[H,1_k^{(r)}Q1_k^{(r)}]$ is not well defined, even if $Q$ is bounded.
\item To be able to use Assertion (iii) of Proposition \ref{NESSexist} for the calculation of the 
expectation value of the current, the operator $[H,1_k^{(r)}Q1_k^{(r)}]$  must be trace-class
\end{enumerate}

\bigskip
The appearance of these problems is not really a surprise. In fact, if the operator 
$\d\Gamma(1_k^{(r)}Q1_k^{(r)})$ was a {\sl bona fide} observable then its expectation
$$
\omega_T^+\circ\tau^t(\d\Gamma(1_k^{(r)}Q1_k^{(r)}))
=\tr(T^+\e^{\i tH}1_k^{(r)}Q1_k^{(r)}\e^{-\i tH}),
$$
would be independent of $t$ and consequently the expectation of the current (\ref{JunkCurrentId}) 
would be zero. The problem resides in the fact that with the density of fermions being non-zero in 
the state $\omega_T^+$, the total charge in the reservoir  $\mathfrak R_k$,
$\omega_T^+(\d\Gamma(1_k^{(r)}Q1_k^{(r)}))$, is actually infinite. Alternatively stated, the 
operator $T^+(1_k^{(r)}Q1_k^{(r)})$ is not trace-class.

We can easily resolve the first problem by regularizing the charge $Q$. For example, by replacing 
$Q$ with $Q_q\equiv QF(|Q|\le q)$ for a $q>0$. We shall only consider charges which are temperate
according to the following definition.

\begin{Def}\label{TempQ}A charge $Q$ is called temperate if $\Dom(|H|^\alpha)\subset\Dom(Q)$ 
for some $\alpha>0$.
\end{Def}

We may regularize a temperate charge by a localization in energy. In fact, if $Q$ is a temperate 
charge, then $Q_\epsilon\equiv Q(1+\epsilon H^2)^{-\alpha/2}$  is a bounded charge. To resolve 
the second problem, we regularize the commutator by localizing it with the help of a function 
$g\in C_0^\infty(\mathbb R\setminus\Sigma_H)$. We remark that if $f\in C_0^\infty(\mathbb R)$
is such that $g(x)f(x)=xg(x)$ for all $x\in\mathbb R$, then the expression
$$
\Phi_{Q_\epsilon,g,k}^{(r)}\equiv 
g(H)\i[H,1_k^{(r)}Q_\epsilon 1_k^{(r)}]g(H)=g(H)\i[f(H),1_k^{(r)}Q_\epsilon 1_k^{(r)}]g(H),
$$
is well-defined: it only involves bounded operators. Furthermore, since
$f(H)$ commutes with $Q_\epsilon$, we have
$$
\Phi_{Q_\epsilon,g,k}^{(r)}=
g(H)\i[f(H),1_k^{(r)}]Q_\epsilon1_k^{(r)}g(H)+g(H)1_k^{(r)}Q_\epsilon\i[f(H),1_k^{(r)}]g(H),
$$
and Lemma \ref{HchiLemma} shows that $\Phi_{Q_\epsilon,g,k}^{(r)}$ is trace-class. 
The third problem is thus resolved too.

\begin{Lem}\label{Phiistraceclass}Under the hypotheses of Lemma \ref{HchiLemma} the 
regularized current operator of a temperate charge $Q$
$$
\Phi_{Q_\epsilon,g,k}^{(r)}\equiv g(H)\i[f(H),1_k^{(r)}Q_\epsilon1_k^{(r)}]g(H),
$$
is trace-class for all $g\in C_0^\infty(\mathbb R\setminus\Sigma_H)$, 
$f\in C_0^\infty(\mathbb R)$, $\epsilon>0$ and $r\ge1$.
\end{Lem}

To formulate the main result of this section, Theorem \ref{LBTheo} below, we need to introduce the 
spectral representations of the Hamiltonians $\widetilde H_k$. For $k\in\{1,\ldots,M\}$ there exists a 
measurable family 
$(\mathfrak h_k(\varepsilon))_{\varepsilon\in\mathbb R}$ of Hilbert spaces and a unitary operator
\begin{equation}
U_k:\HH_{k,\ac}\to\int^\oplus\mathfrak h_k(\varepsilon)\,\d\varepsilon,
\label{UkRep}
\end{equation}
such that $(U_k\widetilde H_ku)(\varepsilon)=\varepsilon
(U_ku)(\varepsilon)$ for all $u\in\HH_{k,\ac}$. 
To each operator $B\in\B(\HH_k,\HH_j)$ such that $f(\widetilde H_j) B=Bf(\widetilde H_k)$ for all
bounded measurable function $f$, the spectral representation \eqref{UkRep} associates a 
measurable family $b(\varepsilon)\in\B(\mathfrak h_k(\varepsilon),\mathfrak h_j(\varepsilon))$ 
such that $(U_jBu)(\varepsilon)=b(\varepsilon)(U_ku)(\varepsilon)$ for all $u\in\HH_{k,\ac}$. 
We have in particular the following correspondences
\begin{align*}
\widetilde T_k&\longrightarrow t_k(\varepsilon),\\
\widetilde Q_k&\longrightarrow q_k(\varepsilon),\\
\widetilde P_k^\pm&\longrightarrow p_k^\pm(\varepsilon),\\
S_{kj}&\longrightarrow s_{kj}(\varepsilon).
\end{align*}
\begin{Theo}\label{LBTheo}We suppose that Hypotheses \Href{HypReservoir}--\Href{HypChi} are
satisfied. Let $\omega_T^+$ be the NESS described by Proposition \ref{NESSexist} and $Q$ a 
temperate charge such that
$$
\Ran\widetilde T_j\subset\Dom |\widetilde H_j|^{\nu+\alpha+1},\qquad
\Dom |H|^\alpha\subset\Dom Q.
$$
Then, for any sequence $g_n\in C_0^\infty(\mathbb R\setminus\Sigma_H)$  such that $0\le g_n\le 1$ 
and $\lim_ng_n(x)=1$ almost everywhere the limit
$$
\omega_T^+(\d\Gamma(\Phi_{Q,k}))\equiv\lim_{n}
\lim_{\epsilon\to0}\omega_T^+(\d\Gamma(\Phi_{Q_\epsilon,g_n,k}^{(r)})),
$$
exists, is independent of $r\ge0$, and is given by the Landauer-B\"uttiker  formula
$$
\omega_T^+(\d\Gamma(\Phi_{Q,k}))
=\sum_{j=1}^M\int\tr_{\mathfrak h_j(\varepsilon)}
\left\{t_j(\varepsilon)\left(s_{jk}^\ast(\varepsilon)q_k(\varepsilon)s_{kj}(\varepsilon)
-\delta_{kj}p_j^-(\varepsilon)q_j(\varepsilon)p_j^-(\varepsilon)\right)
\right\}\frac{\d\varepsilon}{2\pi}.
$$
\end{Theo}

We finish this section with an important property of current observables.

\begin{Lem}\label{CurrentExpect}Let $T,Q$ be bounded operators and $H$ a self-adjoint operator 
on the Hilbert space $\mathcal H$. Let $f,g$ be continuous and bounded functions on $\mathbb R$ 
such that $g(H)[f(H),Q]g(H)$ is trace-class. Finally, let $P_\varepsilon$ be the orthogonal projection 
onto the eigenspace of $H$ associated with the eigenvalue $\varepsilon$. Then
$$
\tr\left(P_\varepsilon TP_\varepsilon g(H)[f(H),Q]g(H) \right)=0.
$$
\end{Lem}

\begin{demo} By using the cyclic property of the trace, we may write
\begin{align*}
\tr\left(P_\varepsilon TP_\varepsilon g(H)[f(H),Q]g(H) \right)
&=\tr\left(TP_\varepsilon g(H)[f(H),Q]g(H)P_\varepsilon\right)\\
&=g(\varepsilon)^2\tr\left(TP_\varepsilon[f(H),Q]P_\varepsilon\right),
\end{align*}
and the result is a consequence of the identity
$$
P_\varepsilon[f(H),Q]P_\varepsilon=P_\varepsilon[f(\varepsilon),Q]P_\varepsilon=0.
$$
\cqfd
\end{demo}

\bigskip
Current observables thus have the property of being insensitive to the contributions of the point 
spectrum of the Hamiltonian $H$ to the NESS (the second term on the right hand side of 
(\ref{OneBodyplus})). If $\omega_T^+$ is a NESS described by Proposition \ref{NESSexist} or 
Theorem \ref{NESSalt} we have
$$
\omega_T^+(\d\Gamma(\Phi_{Q_\epsilon,g,k}^{(r)}))
=\omega_{T^+}(\d\Gamma(\Phi_{Q_\epsilon,g,k}^{(r)}))=
\tr(T^+\Phi_{Q_\epsilon,g,k}^{(r)}).
$$
This last formula is our starting point for the calculation of steady currents. The rest of this 
section is dedicated to the calculation of the limit
$$
\omega_T^+(\d\Gamma(\Phi_{Q,k}))\equiv\lim_{g\to 1}
\lim_{\epsilon\to0}\tr(T^+\Phi_{Q_\epsilon,g,k}^{(r)}),
$$
under the hypotheses of Theorem \ref{LBTheo}.

To simplify notation we shall suppose, up to Section \ref{LBFormSect}, that the charge $Q$ 
is bounded and does not require regularization, i.e., we write $Q$ instead of $Q_\epsilon$.

\subsection{Equivalence of currents}

In the preceding section, we introduced the observables of current. To do this we have localized the 
charge in the reservoir with the help of operators $1_k^{(r)}$. In this section, we show that it is 
possible to achieve the same effect by localizing the charge with an appropriate functions of the 
conjugate operator $A_k$ by exploiting Remark \ref{AlocalRemark}. Since the propagation of
$A_k$ is controlled by the Mourre estimate, this localization method, introduced in \cite{AEGSS} is 
best adapted to the calculation of steady current by the time-dependent approach which we have 
adopted.

The following result expresses the fact (obvious from a physical point of view) that in a stationary 
regime, the total current in a reservoir does not depend on the depth at which we measure it.

\begin{Theo}\label{Phirindep} If the hypotheses of Lemma \ref{HchiLemma} hold, and if $T$ is the 
generator of a gauge invariant quasi-free state on $\mathcal O$ which is also $\tau$-invariant  then
$$
\tr(T\Phi_{Q,g,k}^{(r)}),
$$
is independent of $r\ge1$.
\end{Theo}

\begin{demo}For $r,s\ge1$ we have $\Phi_k^{(r)}-\Phi_k^{(s)}=\Phi_1+\Phi_2$ with
\begin{align*}
\Phi_1&\equiv g(H)\i[f(H),(1_k^{(r)}-1_k^{(s)})Q1_k^{(r)}]g(H),\\
\Phi_2&\equiv g(H)\i[f(H),1_k^{(s)}Q(1_k^{(r)}-1_k^{(s)})]g(H).
\end{align*}
Lemma \ref{HchiLemma} implies that
$$
g(H)f(H)(1_k^{(r)}-1_k^{(s)})Q1_k^{(r)}g(H),\qquad
g(H)(1_k^{(r)}-1_k^{(s)})Q1_k^{(r)}f(H)g(H),
$$
are both trace-class. $\Phi_1$ is thus trace-class and since $T$ and $f(H)$ commute we may write, 
using the cyclicity of the trace,
\begin{align*}
\tr(T\Phi_1)
&=\i\tr(Tg(H)f(H)(1_k^{(r)}-1_k^{(s)})Q1_k^{(r)}
-Tg(H)(1_k^{(r)}-1_k^{(s)})Q1_k^{(r)}f(H))\\
&=\i\tr(Tg(H)f(H)(1_k^{(r)}-1_k^{(s)})Q1_k^{(r)}
-Tg(H)f(H)(1_k^{(r)}-1_k^{(s)})Q1_k^{(r)})\\
&=0.
\end{align*}
In the same way, we show that $\Phi_2$ is trace class and that $\tr(T\Phi_2)=0$.\cqfd
\end{demo}

\bigskip

Let $h\in C^{\infty}(\rr)$ be such that $0\le h\le 1$ and
$$
h(x)=\left\{\begin{array}{ccl}  
0 & \text{if} &x< -1;  \\  
1 & \text{if} &x> 1.
\end{array}\right.
$$
For $a\ge1$ we set $h_\pm^{(a)}(x)\equiv h(\pm x-a)$ and $h^{(a)}\equiv h_-^{(a)}+h_+^{(a)}$. 
We note that $g^{(a)}=1-h^{(a)}\in C_0^\infty(\mathbb R)$ with $\supp\,g^{(a)}\subset[-a-1,a+1]$ 
while $\supp\,h^{(a)}\subset\mathbb R\setminus]-a+1,a-1[$.

\begin{Theo}\label{Psiaindep}If Hypotheses \Href{HypAHTrace} and \Href{HypTrace0} are satisfied 
and $a\ge1$ then the operator
\begin{equation}
\label{courant}
\Psi^{(a)}_{Q,g,k}
\equiv g(H)i[f(H),h^{(a)}(A_k)Qh^{(a)}(A_k)]g(H).
\end{equation}
is trace-class. Furthermore, if $T$ is the generator of a $\tau$-invariant, gauge invariant, 
quasi-free state on $\mathcal O$ then
$$
\tr(T\Psi_{Q,g,k}^{(a)})=\tr(T\Phi_{Q,g,k}^{(1)}).
$$
\end{Theo}

\begin{demo} We write
$$
\Phi_{Q,g,k}^{(1)}\!=\!\i[f(H),g(H)1_k(h^{(a)}(A_k)+g^{(a)}(A_k))Q(h^{(a)}(A_k)+g^{(a)}(A_k))1_kg(H)].
$$
By Lemma \ref{ADefOK}, $\Ran(I-1_k)\subset\Ker A_k$ and since $h^{(a)}(0)=0$  for
$a\ge 1$ one has $h^{(a)}(A_k)(I-1_k)=(I-1_k)h^{(a)}(A_k)=0$.
By Lemma \ref{AHTraceLemma}, $g(H)1_kg^{(a)}(A_k)$ is trace-class and we get
$$
\Phi_{Q,g,k}^{(1)}=\Psi^{(a)}_{Q,g,k}+[f(H),C],
$$
where $C$ is trace-class. Lemma \ref{Phiistraceclass} thus implies that $\Psi^{(a)}_{Q,g,k}$ is 
trace-class.

If $T$ commutes with $f(H)$, the cyclic property of the trace implies that
$$
\tr(T\Phi_{Q,g,k}^{(1)})=\tr(T\Psi_{Q,g,k}^{(a)}).
$$
\cqfd
\end{demo}

If the hypotheses of the previous theorem hold, we have
\begin{equation}
\omega_T^+(\d\Gamma(\Phi_{Q,k}^{(1)}))
=\lim_{g\to 1}\lim_{\epsilon\to0}\tr(T^+\Psi_{Q_\epsilon,g,k}^{(a)}),
\label{CourantFormzero}
\end{equation}
for all $a\ge 1$. The evaluation of the trace on the right hand side of this relation is a difficult 
problem which is the aim of the two following sections.

\subsection{Calculation of steady current I}
\label{CurrentSectI}

The first step in the evaluation of the formula (\ref{CourantFormzero}) is largely  inspired 
by the article of Avron {\sl et al.\,}\cite{AEGSS}. The idea is to develop the current operator
\begin{equation}
\Psi^{(a)}_{Q,g,k}=\sum_{\sigma,\sigma'\in\{\pm\}}\Psi^{(a,\sigma,\sigma')}_{Q,g,k},
\label{PsiDecompsigmasigma}
\end{equation}
where
\begin{equation}
\label{courantpm}
\Psi^{(a,\sigma,\sigma')}_{Q,g,k}
\equiv g(H)i[f(H),h_\sigma^{(a)}(A_k)Qh_{\sigma'}^{(a)}(A_k)]g(H),
\end{equation}
and to exploit the property of the commutator $[H,h_\sigma^{(a)}(A_k)]$ which appears in this last 
expression (recall that $f(H)g(H)=Hg(H)$). By expanding this commutator
$$
[H,h_\sigma^{(a)}(A_k)]\sim h_\sigma^{(a)'}(A_k)[H,A_k],
$$
we note that it is localized in a spectral neighborhood of $A_k=\sigma a$  (see 
Lemma~\ref{TraceOmega} below for a precise statement). The Mourre estimate tells us how 
states in the range of this localized operator propagates,
$$
\e^{\i\sigma tH}h_\sigma^{(a)'}(A_k)\sim h_\sigma^{(a)'}(A_k+\theta t)\e^{\i\sigma tH}.
$$
The sample $\mathfrak S$ being confined to the the subspace $A_k=0$ by 
Hypothesis~\Href{HypConjugate}, we conclude that these states do not undergo scattering when 
$t\to\sigma\infty$. The M\o ller operator $\Omega^{\sigma\ast}$ thus acts trivially on such states,
while $\Omega^{-\sigma\ast}=\Omega^{-\sigma\ast}\Omega^{\sigma}\Omega^{\sigma\ast}$ acts 
like the scattering matrix $\Omega^{-\sigma\ast}\Omega^{\sigma}$.

The result of this first reduction of the problem is the following.

\begin{Theo}\label{currentpropone} We suppose that Hypotheses \Href{HypReservoir}, 
\Href{HypResDyn}, \Href{HypConjugate}, and \Href{HypAHTrace} are satisfied. Let $\widetilde T_j$ 
be the generator of a $\tau_j$-invariant, gauge invariant, quasi-free state on $\mathcal O_j$ such that
$\Ran\widetilde T_j^{1/2}\subset\Dom\widetilde H_j$. For all
$g\in C_0^\infty(\mathbb R\setminus\Sigma_H)$ we have
\begin{align}
\lim_{a\to\infty}&\left[\tr(\Omega_j^{-}\widetilde T_j\Omega_j^{-\ast}\Psi^{(a)}_{Q,g,k})
-\tr\left(\widetilde T_j\left\{S_{jk}^\ast\widetilde\Psi^{(a,+,+)}_{Q,g,k}S_{kj}
\right.\right.\right.\nonumber\\
+\delta_{jk}&\left.\left.\left.\left(
\widetilde P_k^-
\widetilde\Psi^{(a,-,-)}_{Q,g,k}\widetilde P_k^-
+S_{kk}^\ast\widetilde\Psi^{(a,+,-)}_{Q,g,k}\widetilde P_k^-
+\widetilde P_k^-\widetilde\Psi^{(a,-,+)}_{Q,g,k}S_{kk}  
\right)
\right\}\right)\right]=0.\label{FluxOne}
\end{align}
where $\widetilde\Psi^{(a,\sigma,\sigma')}_{Q,g,k}
\equiv M_k^\ast\Psi^{(a,\sigma,\sigma')}_{Q,g,k}M_k$.
\end{Theo}

The main technical tool necessary for the proof of this result is the following localization lemma due 
to \cite{AEGSS}. Since its proof is quite long, we defer it to Appendix \ref{LocalisationProof}.

\begin{Lem}\label{TraceOmega} If Hypotheses  \Href{HypReservoir}, \Href{HypResDyn}, 
\Href{HypConjugate}, and \Href{HypAHTrace} are satisfied then, for all $f\in C_0^\infty(\mathbb R)$ 
and all $g\in C_0^\infty(\mathbb R\setminus\Sigma_H)$:
\begin{enumerate}[(i)]
\item $\sup_{a\ge1}\|[f(H),h^{(a)}_\pm(A_k)]g(H)\|_1<\infty$.
\item There exist constants $s>1$ and $C$ such that, for all $a,\alpha\ge1$,
$$
\|F(\pm A<a-\alpha)[f(H),h_\pm^{(a)}(A_k)]g(H)\|_1<C\langle\alpha\rangle^{-s}.
$$
\end{enumerate}
\end{Lem}

\noindent{\bf Remark.} It follows from (i) that the components $\Psi^{(a,\sigma,\sigma')}_{Q,g,k}$ of 
the current are trace-class operators.

\bigskip
To prepare for the proof of Theorem \ref{currentpropone} we begin by proving the following lemma.

\begin{Lem}\label{TracePropagation} We suppose that the hypotheses of Theorem \ref{currentpropone}
are satisfied. For $\alpha\in\{j,l\}\subset\{1,\ldots,M\}$ let $\widetilde T_\alpha$ be the generator of a 
$\tau_\alpha$-invariant, gauge invariant, quasi-free state on $\mathcal O_\alpha$ such that 
$\Ran\widetilde T_\alpha^{1/2}\subset\Dom\widetilde H_\alpha$. For all 
$g\in C_0^\infty(\mathbb R\setminus\Sigma_H)$ we have:
\begin{equation}
\lim_{a\to\infty}\|\widetilde T_j^{1/2}(
\Omega_j^{\sigma\ast}\Psi^{(a,\sigma,\sigma')}_{Q,g,k}\Omega_l^{\sigma'}-\delta_{jk}
M_k^{\ast}\Psi^{(a,\sigma,\sigma')}_{Q,g,k}M_k\delta_{kl})\widetilde T_l^{1/2}\|_1=0.\label{OmegaPsi}
\end{equation}
\end{Lem}

\begin{demo} 
Since $Q$ commutes with $f(H)$, we may write
\begin{align}
\Psi^{(a,\sigma,\sigma')}_{Q,g,k}
&=g(H)\i[f(H),h^{(a)}_\sigma(A_k)]Qh^{(a)}_{\sigma'}(A_k)g(H)\nonumber\\
&+g(H)h^{(a)}_\sigma(A_k)Q\i[f(H),h^{(a)}_{\sigma'}(A_k)]g(H).
\label{PsiDecompos}
\end{align}
$Q$ being bounded, this decomposition, the inequality $\|BC\|_1\le\|B\|\,\|C\|_1$, and a telescopic 
expansion allow us to reduce the proof to the following three assertions
\begin{align}
&\sup_{a\ge1}\|g(H)[f(H),h^{(a)}_\sigma(A_k)]\|_1<\infty,\label{estone}\\
&\lim_{a\to\infty}\|\widetilde T_j^{1/2}(\Omega_j^{\sigma\ast}-\delta_{jk}M_j^\ast)
g(H)h^{(a)}_\sigma(A_k)\|=0,\label{esttwo}\\
&\lim_{a\to\infty}\|\widetilde T_j^{1/2}(\Omega_j^{\sigma\ast}-\delta_{jk}M_j^\ast)
g(H)[f(H),h^{(a)}_\sigma(A_k)]\|_1=0.\label{estthree}
\end{align}
We immediately remark that \eqref{estone} follows from Assertion (i) of Lemma \ref{TraceOmega} by 
taking the adjoint.

Hypothesis \Href{HypResDyn} (ii) (iv) allows us to write
$$
(HM_j-M_j\widetilde H_j)\widetilde T_j^{1/2}
=J_j[\widetilde H_j,\widetilde\chi_j]\widetilde T_j^{1/2}=J_jB_j\widetilde T_j^{1/2},
$$
and to conclude that $(HM_j-M_j\widetilde H_j)\widetilde T_j^{1/2}$ is bounded. 
We now remark that first, the identity (\ref{magiczero2}) implies
$$
\Ran(HM_j-M_j\widetilde H_j)\widetilde T_j^{1/2}\subset\Ker A_j,
$$
and second, Hypothesis \Href{HypReservoir} (iii) shows that, for $k\not=j$,
$$
1_k(HM_j-M_j\widetilde H_j)\widetilde T_j=1_kJ_j[\widetilde H_j,\widetilde\chi_j]
\widetilde T_j^{1/2}=0.
$$
Since $A=\sum_{k=1}^M A_k\,1_k$ we get that
$$
\Ran(HM_j-M_j\widetilde H_j)\widetilde T_j^{1/2}\subset\Ker A,
$$
and that, consequently,
\begin{equation}
(HM_j-M_j\widetilde H_j)\widetilde T_j^{1/2}=F(\sigma A\le \vartheta t+a/2-1)
(HM_j-M_j\widetilde H_j)\widetilde T_j^{1/2},
\label{KerAtrick}
\end{equation}
for all $\vartheta t\ge0$ and $a\ge2$.

The adjoint of this last relation and the fact that $\widetilde T_j$ and $\e^{\i t\widetilde H_j}$ commute 
allow us to write the Cook representation
\begin{align}
\widetilde T_j^{1/2}(\Omega_j^{\sigma\ast}-M_j^\ast)g(H)h^{(a)}_\sigma(A_k)&=
\label{CookOne}\\
\i\sigma\!\!\int_0^\infty\!\!\e^{\i\sigma t\widetilde H_j}
\widetilde T_j^{1/2}(\widetilde H_jM_j^\ast&-M_j^\ast H)
F(\sigma A\le \vartheta t+a/2-1)\e^{-\i\sigma tH}g(H)h_\sigma^{(a)}(A_k)\,\d t,\nonumber
\end{align}
valid for all $\vartheta>0$. By invoking the fact that $A=\oplus_{k=1}^MA_k$, it follows from the 
definition of the functions $h_\pm^{(a)}$ that
$$
h_\sigma^{(a)}(A_k)=F(\sigma A\ge a-1)h_\sigma^{(a)}(A_k),
$$
which allows us to write, with $a'=a-1$,
$$
\|F(\sigma A\le\vartheta t+a/2-1)\e^{-\i\sigma tH}g(H)h_\sigma^{(a)}(A_k)\|
\le\|F(\sigma A\le a'-a/2+\vartheta t)\e^{-\i\sigma tH}g(H)F(\sigma A\ge a')\|.
$$
By Hypothesis \Href{HypAHTrace} (ii), Assertion (ii) of Theorem \ref{HMourre} and 
Proposition \ref{MainPropEst} it is possible to choose  $\vartheta>0$ in such a way that there exists 
a constant $C_1$ such that
\begin{equation}
\|F(\sigma A\le\vartheta t+a/2-1)\e^{-\i\sigma tH}g(H)h_\sigma^{(a)}(A_k)\|
\le C_1\langle a/2+\vartheta t\rangle^{-s},
\label{FirstHalfOfIt}
\end{equation}
for some $s>1$ and for all $a\ge2$, $t\ge0$. With the representation (\ref{CookOne}) we obtain
$$
\|\widetilde T_j^{1/2}(\Omega_j^{\sigma\ast}-M_j^\ast)g(H)h^{(a)}_\sigma(A_k)\|\le
\|\widetilde T_j^{1/2}(\widetilde H_jM_j^\ast-M_j^\ast H)\|\!\!
\int_0^\infty \!\!\!C_1\langle a/2+\vartheta t\rangle^{-s}\,\d s,
$$
and if $j=k$, (\ref{esttwo}) follows from an dominated convergence argument. To finish the proof 
of (\ref{esttwo}), it suffices to show that
$$
\lim_{a\to\infty}\|M_j^\ast g(H)h^{(a)}_\sigma(A_k)\|=0,
$$
when $j\not=k$. To do this we begin by remarking that $A_k1_j=0$ and thus 
$\Ran M_j\subset\Ker A_k$. We get that
\begin{align}
\|M_j^\ast g(H)h^{(a)}_\sigma(A_k)\|&= \|M_j^\ast F(A_k=0)g(H)h^{(a)}_\sigma(A_k)\|\nonumber\\
&\le \|F(A_k=0)g(H)h^{(a)}_\sigma(A_k)\|\nonumber\\
&\le \|1_kF(A=0)g(H)h^{(a)}_\sigma(A_k)\|\nonumber\\
&\le \|F(\sigma A\le a/2-1)g(H)h^{(a)}_\sigma(A_k)\|,\label{ResidueOne}
\end{align}
for $a\ge2$. The estimate (\ref{FirstHalfOfIt}), with $t=0$ allows us to conclude.

By proceeding in a similar way we obtain the Cook representation
\begin{align}
\widetilde T_j^{1/2}(&\Omega_j^{\sigma\ast}-M_j^\ast)g(H)[f(H),h^{(a)}_\sigma(A_k)]=
\i\sigma\int_0^\infty\e^{\i\sigma t\widetilde H_j}\widetilde T_j^{1/2}
(\widetilde H_jM_j^\ast-M_j^\ast H)\nonumber\\
&\times F(\sigma A\le\vartheta t/2)\e^{-\i\sigma tH}g(H)[f(H),h_\sigma^{(a)}(A_k)]g(H)\,\d t,
\label{cooktwo}
\end{align}
valid for all $\vartheta>0$. To estimate the integral of the right hand side of this identity, 
we decompose
\begin{align*}
F(\sigma A\le&\vartheta t/2)\e^{-\i\sigma tH}g(H)[f(H),h_\sigma^{(a)}(A_k)]g(H)=\\
&F(\sigma A\le\vartheta t/2)\e^{-\i\sigma tH}g(H)F(\sigma A\ge(a-\vartheta t)/2)[f(H),h_\sigma^{(a)}(A_k)]g(H)\\
+&F(\sigma A\le\vartheta t/2)\e^{-\i\sigma tH}g(H)F(\sigma A<(a-\vartheta t)/2)[f(H),h_\sigma^{(a)}(A_k)]g(H),
\end{align*}
and thus obtain
\begin{align}
\|F(&\sigma A\le\vartheta t/2)\e^{-\i\sigma tH}g(H)[f(H),h_\sigma^{(a)}(A_k)]g(H)\|_1\le
\nonumber\\
&\|F(\sigma A\le\vartheta t/2)\e^{-\i\sigma tH}g(H)F(\sigma A\ge(a-\vartheta t)/2)\|\,
\|[f(H),h_\sigma^{(a)}(A_k)]g(H)\|_1\nonumber\\
+&\|F(\sigma A<(a-\vartheta t)/2)[f(H),h_\sigma^{(a)}(A_k)]g(H)\|_1,\label{bigmess}
\end{align}
We consider the first term on the right hand side of this inequality. Its second factor is uniformly 
bounded by (i) of Lemma \ref{TraceOmega}. By setting $a'=(a-\vartheta t)/2$ its first factor 
can be written as
$$
\|F(\sigma A\le a'-a/2+\vartheta t)\e^{-\i\sigma tH}g(H)F(\sigma A\ge a')\|.
$$
As before, we invoke Assertion (ii) of Theorem \ref{HMourre} and Proposition \ref{MainPropEst} 
to choose $\vartheta>0$ so that this factor is bounded by $C_1\langle a/2+\vartheta t\rangle^{-s}$.

Writing the second term on the right hand side  of (\ref{bigmess}) as
$$
\|F(\sigma A<a-(a+\vartheta t)/2)[f(H),h_\sigma^{(a)}(A_k)]g(H)\|_1,
$$
Assertion (ii) of Lemma \ref{TraceOmega} allows us to conclude that it is bounded by 
$C\langle a/2+\vartheta t/2\rangle^{-s}$ for a constant $C$ and some $s>1$. 
We have thus shown that
$$
\|F(\sigma A\le\vartheta t/2)\e^{-\i\sigma tH}g(H)[f(H),h_\sigma^{(a)}(A_k)]g(H)\|_1\le
C_2\langle a+\vartheta t\rangle^{-s},
$$
for a constant $C_2$, some $s>1$, and for all $a\ge2$. For $j=k$, Assertion (\ref{estthree}) is an 
immediate consequence of this estimate and the representation (\ref{cooktwo}). To finish the proof 
of (\ref{estthree}) it suffices to show that
$$
\lim_{a\to\infty}\|M_j^\ast g(H)[f(H),h^{(a)}_\sigma(A_k)]\|_1=0,
$$
when $j\not=k$. We proceed as in \eqref{ResidueOne} to obtain
\begin{align*}
\|M_j^\ast g(H)[f(H),h^{(a)}_\sigma(A_k)]\|_1
&= \|M_j^\ast F(A_k=0)g(H)[f(H),h^{(a)}_\sigma(A_k)]\|_1\\
&\le \|F(A_k=0)g(H)[f(H),h^{(a)}_\sigma(A_k)]\|_1\\
&\le \|1_kF(A=0)g(H)[f(H),h^{(a)}_\sigma(A_k)]\|_1\\
&\le \|F(\sigma A< a/2)g(H)[f(H),h^{(a)}_\sigma(A_k)]\|_1,
\end{align*}
and we conclude by once again invoking (ii) of Lemma \ref{TraceOmega}.\cqfd
\end{demo}

\bigskip
{\bf\noindent Proof of Theorem \ref{currentpropone}.} We write $C\sim D$ whenever
$C$ and $D$ are operators depending on $a$ and such that
$\lim_{a\to\infty}\|\widetilde T_j^{1/2}(C-D)\widetilde T_j^{1/2}\|_1=0$ holds.

By invoking Corollary \ref{HProjo}, we may write
$$
\Omega_j^{-\ast}\Psi^{(a,\sigma,\sigma')}_{Q,g,k}\Omega_j^{-}
=\sum_{m,n=1}^M\Omega_j^{-\ast}P_m^\sigma
\Psi^{(a,\sigma,\sigma')}_{Q,g,k}P_n^{\sigma'}\Omega_j^{-}
=\sum_{m,n=1}^M\Omega_j^{-\ast}
\Omega_m^{\sigma}\Omega_m^{\sigma\ast}
\Psi^{(a,\sigma,\sigma')}_{Q,g,k}
\Omega_n^{\sigma'}\Omega_n^{\sigma'\ast}\Omega_j^{-}.
$$
The properties of the M\o ller operators show that
$$
\Omega_\alpha^{\sigma\ast}\Omega_j^{-}\Dom\widetilde H_j\subset\Dom\widetilde H_\alpha.
$$
We may thus invoke Lemma \ref{TracePropagation} to continue with
$$
\Omega_j^{-\ast}\Psi^{(a,\sigma,\sigma')}_{Q,g,k}\Omega_j^{-}
\sim \Omega_j^{-\ast}
\Omega_k^{\sigma}
M_k^{\ast}
\Psi^{(a,\sigma,\sigma')}_{Q,g,k}
M_k
\Omega_k^{\sigma'\ast}
\Omega_j^{-}.
$$
Since
$$
\Omega_k^{\sigma\ast}\Omega_j^{\sigma'}=\left\{
\begin{array}{ccc}
S_{kj}&\text{if}&\sigma=+\ \text{and}\ \sigma'=-;\\[4pt]
\delta_{kj}\widetilde P_j^\sigma&\text{if}&\sigma=\sigma';\\[4pt]
S_{kj}^\ast&\text{if}&\sigma=-\ \text{and}\ \sigma'=+;
\end{array}
\right.
$$
we obtain
\begin{align*}
\Omega_j^{-\ast}
\Psi^{(a)}_{Q,g,k}
\Omega_j^{-}&=
\sum_{\sigma,\sigma'\in\{\pm\}}
\Omega_j^{-\ast}
\Psi^{(a,\sigma,\sigma')}_{Q,g,k}
\Omega_j^{-}\\
&\sim
S_{jk}^\ast M_k^\ast\Psi^{(a,+,+)}_{Q,g,k}M_kS_{kj}\\
&+
\delta_{kj}\left(
S_{kk}^\ast M_k^\ast\Psi^{(a,+,-)}_{Q,g,k}M_k\widetilde P_k^-\right.
\left.+\widetilde P_k^-M_k^\ast\Psi^{(a,-,+)}_{Q,g,k}M_kS_{kk}
+\widetilde P_k^-
M_k^\ast\Psi^{(a,-,-)}_{Q,g,k}M_k\widetilde P_k^-
\right).
\end{align*}
\cqfd

\subsection{Calculation of steady current II}
\label{CurrentSectII}

In this section we finish the calculation of the steady current starting from the 
formula (\ref{FluxOne}). The method which we use here differs from that 
of \cite{AEGSS}. In fact, in this work the reservoirs are straight, one-dimensional wires without 
internal structure. In this case we may choose the conjugate operator in such a  way that 
$\i[\widetilde H_k,\widetilde A_k]=2\widetilde H_k$ and it is easy to explicitly construct the spectral 
representations of $\widetilde H_k$ and $\widetilde A_k$ and to compute the integral kernel of
operators of the form $f(\widetilde H_k)g(\widetilde A_k)$. This reduces the calculation of their
trace to an integral over the diagonal of this integral kernel.
This approach is inapplicable at the level of generality where we have placed ourselves. We shall use 
a more systematic approach based uniquely on the propagation estimates and on the abstract 
spectral representation of the operator $\widetilde H_k$.

\subsubsection{Preliminaries}
We begin by showing that it is possible to replace $H$ with $\widetilde H_k$ and $A_k$ by 
$\widetilde A_k$ in the definition of $\widetilde\Psi_{Q,g,k}^{(a,\sigma,\sigma')}$ without 
altering its NESS expectation.

\begin{Lem}\label{PsiReduct} Under Hypothesis \Href{HypTrace0} we have
$$
\lim_{a\to\infty}\left\|\widetilde\Psi_{Q,g,k}^{(a,\sigma,\sigma')}-
\widetilde\Psi^{\#(a,\sigma,\sigma')}_{Q,g,k}\right\|_1=0.
$$
where $\widetilde\Psi^{\#(a,\sigma,\sigma')}_{Q,g,k}
\equiv g(\widetilde H_k)\i[f(\widetilde H_k),
h_\sigma^{(a)}(\widetilde A_k)M_k^\ast QM_kh_{\sigma'}^{(a)}(\widetilde A_k)]
g(\widetilde H_k)$.
\end{Lem}

\begin{demo} We shall write $C\sim D$ when $\lim_{a\to\infty}\|C-D\|_1=0$. By 
Lemma \ref{fHMLemma}, $M_k^\ast g(H)-g(\widetilde H_k)M_k^\ast$ and 
$g(H)M_k-M_kg(\widetilde H_k)$ are trace-class. Since
$$
\starslim_{a\to\infty}h_\sigma^{(a)}(A_k)Qh_{\sigma'}^{(a)}(A_k)=0,
$$
we have
$$
M_k^\ast g(H)\i[f(H),h_\sigma^{(a)}(A_k)Qh_{\sigma'}^{(a)}(A_k)]g(H)M_k
\sim
g(\widetilde H_k)M_k^\ast \i[f(H),h_\sigma^{(a)}(A_k)Qh_{\sigma'}^{(a)}(A_k)]M_k
g(\widetilde H_k).
$$
We prove in the same way that
$$
g(\widetilde H_k)M_k^\ast \i[f(H),h_\sigma^{(a)}(A_k)Qh_{\sigma'}^{(a)}(A_k)]M_k
g(\widetilde H_k)
\sim
g(\widetilde H_k)\i[f(\widetilde H_k),M_k^\ast h_\sigma^{(a)}(A_k)Qh_{\sigma'}^{(a)}(A_k)M_k]
g(\widetilde H_k).
$$
Finally, it follows from the identity (\ref{MgMstarId}) that
$$
M_k^\ast h_\sigma^{(a)}(A_k)=
M_k^\ast M_kh_\sigma^{(a)}(\widetilde A_k)M_k^{\ast}=
1_kh_\sigma^{(a)}(\widetilde A_k)M_k^{\ast}=
h_\sigma^{(a)}(\widetilde A_k)M_k^{\ast},
$$
from which it follows that
$$
g(\widetilde H_k)\i[f(\widetilde H_k),M_k^\ast h_\sigma^{(a)}(A_k)Qh_{\sigma'}^{(a)}(A_k)M_k]
g(\widetilde H_k)
=
g(\widetilde H_k)\i[f(\widetilde H_k),h_\sigma^{(a)}(\widetilde A_k)M_k^\ast QM_k
h_{\sigma'}^{(a)}(\widetilde A_k)]
g(\widetilde H_k).
$$
\cqfd
\end{demo}

\begin{Coro}\label{currentproptwo} Under the hypotheses of Theorem \ref{currentpropone} and 
Hypothesis \Href{HypTrace0} we have
\begin{align}
&\tr(\Omega_j^{-}\widetilde T_j\Omega_j^{-\ast}\Psi^{(a)}_{Q,g,k})
=\tr\left(\widetilde T_j\left\{S_{jk}^\ast\widetilde\Psi^{\#(a,+,+)}_{Q,g,k}S_{kj}
\right.\right.\nonumber\\
+\delta_{jk}&\left.\left.\left(
\widetilde P_k^-
\widetilde\Psi^{\#(a,-,-)}_{Q,g,k}\widetilde P_k^-
+S_{kk}^\ast\widetilde\Psi^{\#(a,+,-)}_{Q,g,k}\widetilde P_k^-
+\widetilde P_k^-\widetilde\Psi^{\#(a,-,+)}_{Q,g,k}S_{kk}  
\right)
\right\}\right).
\label{FluxTwo}
\end{align}
for all $a\ge 1$.
\end{Coro}

\begin{demo} Since $\Omega_j^{-}\widetilde T_j\Omega_j^{-\ast}$ commutes with $H$, 
Theorem \ref{Psiaindep} shows that the left hand side of (\ref{FluxTwo}) is independent of 
$a\ge 1$. By taking into account Proposition \ref{currentpropone} and Lemma \ref{PsiReduct}, 
it suffices to show that that the right hand side is also independent of $a$.

For $a,b\ge1$, we have $g_\sigma\equiv h_\sigma^{(a)}-h_\sigma^{(b)}\in C_0^\infty(\mathbb R)$. 
It follows from Hypothesis \Href{HypAHTrace} (i) that 
$C_\sigma\equiv g(\widetilde H_k)g_\sigma(\widetilde A_k)=g(\widetilde H_k)(\widetilde A_k+\i)^{-m}
(\widetilde A_k+\i)^mg_\sigma(\widetilde A_k)$ is trace-class. Since
\begin{align*}
\widetilde\Psi^{\#(a,\sigma,\sigma')}_{Q,g,k}-\widetilde\Psi^{\#(b,\sigma,\sigma')}_{Q,g,k}
&= \i[f(\widetilde H_k),C_\sigma M_k^\ast QM_kh_{\sigma'}^{(a)}(\widetilde A_k)g(\widetilde H_k)]\\
&+\i[f(\widetilde H_k),g(\widetilde H_k)h_{\sigma}^{(b)}(\widetilde A_k)
M_k^\ast QM_kC^\ast_{\sigma'}],
\end{align*}
and $S_{kj}\widetilde T_jS_{jk}^\ast$ commutes with $\widetilde H_k$ the cyclic property of the trace allows us to conclude that
$$
\tr\left(S_{kj}\widetilde T_jS_{jk}^\ast\left(\widetilde\Psi^{\#(a,\sigma,\sigma')}_{Q,g,k}
-\widetilde\Psi^{\#(b,\sigma,\sigma')}_{Q,g,k}\right)\right)=0.
$$
The 3 other terms of the right hand side of (\ref{FluxTwo}) are treated in a similar manner. \cqfd
\end{demo}

\subsubsection{Spectral representation of the current}

We are now in position to pass to the spectral representation (\ref{UkRep}).

The main result of this section is the following.
\begin{Theo}\label{Kernelpsi} We suppose that Hypotheses \Href{HypReservoir}, \Href{HypResDyn} 
and \Href{HypConjugate} are satisfied. If $g\in C_0^\infty(\mathbb R\setminus\Sigma_H)$  and if 
the operator
$$
\widetilde\Psi^{\#(a,\sigma,\sigma')}_{Q,g,k}=g(\widetilde H_k)\i[f(\widetilde H_k),
h_\sigma^{(a)}(\widetilde A_k)M_k^\ast QM_kh_{\sigma'}^{(a)}(\widetilde A_k)]
g(\widetilde H_k),
$$
is trace-class, then:
\begin{enumerate}[(i)]
\item $\widetilde\Psi^{\#(a,\sigma,\sigma')}_{Q,g,k}$ reduces to its part in $\HH_{k,\ac}$, that is to
say that
$$
\widetilde\Psi^{\#(a,\sigma,\sigma')}_{Q,g,k}=
P_\ac(\widetilde H_k)\widetilde\Psi^{\#(a,\sigma,\sigma')}_{Q,g,k}P_\ac(\widetilde H_k).
$$
\item There exists a measurable set $\Delta\subset\mathbb R$, with
  $\mathbb R\setminus\Delta$ having Lebesgue measure zero, and a mapping
$$
\Delta\times\Delta\ni\langle\varepsilon,\varepsilon'\rangle\mapsto
\psi^{\#(a,\sigma,\sigma')}_{Q,g,k}(\varepsilon',\varepsilon)\in
\mathcal L^1(\mathfrak h_k(\varepsilon),\mathfrak h_k(\varepsilon')),
$$
such that, for all $u,v\in\HH_{k,\ac}$,
$$
(u,\widetilde\Psi^{\#(a,\sigma,\sigma')}_{Q,g,k}v)=\int((U_ku)(\varepsilon'),
\psi^{\#(a,\sigma,\sigma')}_{Q,g,k}(\varepsilon',\varepsilon)
(U_kv)(\varepsilon))_{\mathfrak h_k(\varepsilon')}\,\d\varepsilon\d\varepsilon',
$$
\item For all $\varepsilon\in\Delta$, 
\begin{equation}
\psi^{\#(a,\sigma,\sigma')}_{Q,g,k}
(\varepsilon,\varepsilon)=\frac{\sigma}{2\pi}\,g(\varepsilon)^2\,p_k^\sigma(\varepsilon)
q_k(\varepsilon)p_k^\sigma(\varepsilon)\delta_{\sigma\sigma'}.
\label{psikernelId}
\end{equation}
\end{enumerate}
\end{Theo}

We shall show in Section \ref{LBFormSect} how Formula (\ref{psikernelId}) can be used to complete 
the calculation of the steady current. The rest of this section is dedicated to the proof of 
Theorem \ref{Kernelpsi} which is organized as follows. In Section \ref{TraceOps} we prove 
a theorem about the general structure of trace-class operators on a direct integral of Hilbert spaces 
and in particular the existence of the integral kernel of such an operator. 
In Section \ref{DiagSubSect} we show how to compute the diagonal of this integral kernel. 
These results allow us finally to prove Theorem \ref{Kernelpsi} in Section \ref{Kernelpsidemo}.

\subsubsection{Trace-class operators on $\int^\oplus\mathfrak h_\varepsilon\d\mu(\varepsilon)$}
\label{TraceOps}

\begin{Theo} \label{TraceKernel} Let $\Delta\subset\mathbb R$ be a measurable set, 
$\mu$ a $\sigma$-finite measure on $\Delta$ and $(\mathfrak h_\varepsilon)_{\varepsilon\in\Delta}$  
a family of $\mu$-measurable, separable Hilbert spaces. If $C$ is a trace-class operator on 
$\mathfrak H\equiv\int_\Delta^\oplus\mathfrak h_\varepsilon\,\d\mu(\varepsilon)$ then:
\begin{enumerate}[(i)]
\item There exists a measurable set $\Delta_0\subset\Delta$ such that
$\Delta\setminus\Delta_0$ is $\mu$-negligible and, for all $\langle \varepsilon',\varepsilon\rangle\in\Delta_0\times\Delta_0$, a trace-class operator $c(\varepsilon',\varepsilon):\mathfrak h_{\varepsilon}
\to\mathfrak h_{\varepsilon'}$ such that $\langle \varepsilon',\varepsilon\rangle\mapsto (u(\varepsilon'),c(\varepsilon',\varepsilon)v(\varepsilon))_{\mathfrak h_{\varepsilon'}}$ is measurable for all $u,v\in\mathfrak H$.
\item For all $u,v\in\mathfrak H$, $(u,Cv)=\int_\Delta(u(\varepsilon'), c(\varepsilon',\varepsilon)v(\varepsilon))_{\mathfrak h_{\varepsilon'}}\,
\d\mu(\varepsilon)\d\mu(\varepsilon')$.
\item $\int_\Delta\|c(\varepsilon,\varepsilon)\|_1\,\d\mu(\varepsilon)\le \|C\|_1$.
\item $\int_\Delta\tr_{\mathfrak h_\varepsilon}(c(\varepsilon,\varepsilon))\,\d \mu(\varepsilon)
=\tr(C)$.
\end{enumerate}
\end{Theo}

Since this result does not seem to be widely known, we give a proof by following \cite{Y}.

\begin{demo} $C$ being compact, it admits a canonical representation
$$
C=\sum_{n\in N}\kappa_n u_n^+(u_n^-,\,\cdot\,),
$$
where $N$ is a set which is at most countable, $(u_n^\pm)_{n\in N}$ are orthonormal families in
$\mathfrak H$, and $(\kappa_n)_{n\in N}$ is the family of singular values of $C$.
In particular we have $\kappa_n>0$ and  $\sum_{n\in N}\kappa_n=\|C\|_1<\infty$. Since
$$
\|C\|_1=\sum_{n\in N}\kappa_n\int_\Delta\|u_n^\pm(\varepsilon)\|_{\mathfrak h_\varepsilon}^2
\,\d \mu(\varepsilon)<\infty,
$$
Fubini's theorem implies the existence of a measurable set $\Delta^\pm\subset\Delta$, such that 
$\Delta\setminus\Delta^\pm$  is $\mu$-negligible and 
$\sum_{n\in N}\kappa_n\|u^\pm_n(\varepsilon)\|_{\mathfrak h_\varepsilon}^2<\infty$ for all 
$\varepsilon\in\Delta^\pm$. We set $\Delta_0\equiv\Delta^+\cap\Delta^-$. $\Delta\setminus\Delta_0$
is $\mu$-negligible and, for all $\varepsilon,\varepsilon'\in\Delta_0$,
$$
c(\varepsilon',\varepsilon)\equiv\sum_{n\in N}\kappa_n u_n^+(\varepsilon')(u_n^-(\varepsilon),
\,\cdot\,)_{\mathfrak h_{\varepsilon}},
$$
converges in norm in $\B(\mathfrak h_{\varepsilon},\mathfrak h_{\varepsilon'})$. In fact, the 
Cauchy-Schwarz inequality implies
$$
\sum_{n\in N}\kappa_n \|u_n^+(\varepsilon')\|_{\mathfrak h_{\varepsilon'}}\,
\|u_n^-(\varepsilon)\|_{\mathfrak h_{\varepsilon}}\!\le\!
\left(\sum_{n\in N}\kappa_n \|u_n^+(\varepsilon')\|_{\mathfrak h_{\varepsilon'}}^2
\sum_{m\in N}\kappa_m \|u_m^-(\varepsilon)\|_{\mathfrak h_{\varepsilon}}^2\right)^{1/2}
\!\!\!\!<\infty.
$$
For $u,v\in\mathfrak H$ we thus have
\begin{align*}
\int_\Delta
(u(\varepsilon'),c(\varepsilon',\varepsilon)v(\varepsilon))_{\mathfrak h_{\varepsilon'}}\,
\d\mu(\varepsilon)\d\mu(\varepsilon')
&=\sum_{n\in N}\kappa_n \int_\Delta(u(\varepsilon'),
u_n^+(\varepsilon'))_{\mathfrak h_{\varepsilon'}}\,\d\mu(\varepsilon')\int_\Delta
(u_n^-(\varepsilon),v(\varepsilon))_{\mathfrak h_{\varepsilon}}\,\d \mu(\varepsilon)\\
&=\sum_{n\in N}\kappa_n (u,u_n^+)(u_n^-,v)
=(u,Cv).
\end{align*}
Since $\|u_n^+(\varepsilon)(u_n^-(\varepsilon),\,\cdot\,)_{\mathfrak h_\varepsilon}\|_1
=\|u_n^+(\varepsilon)\|_{\mathfrak h_\varepsilon}\,
\|u_n^-(\varepsilon)\|_{\mathfrak h_\varepsilon}$ we have
\begin{align*}
\int_\Delta\|c(\varepsilon,\varepsilon)\|_1\,\d\mu(\varepsilon)
&\le\sum_{n\in N}\kappa_n\int_\Delta
\|u_n^+(\varepsilon)\|_{\mathfrak h_\varepsilon}\,
\|u_n^-(\varepsilon)\|_{\mathfrak h_\varepsilon}\,\d \mu(\varepsilon)\\
&\le\sum_{n\in N}\kappa_n\left(\int_\Delta
\|u_n^+(\varepsilon)\|_{\mathfrak h_\varepsilon}^2\,\d\mu(\varepsilon)
\int_\Delta\|u_n^-(\varepsilon')\|_{\mathfrak h_{\varepsilon'}}^2\,
\d \mu(\varepsilon')\right)^{1/2}\\
&=\sum_{n\in N}\kappa_n\|u_n^+\|\,\|u_n^-\|\\
&= \sum_{n\in N}\kappa_n=\|C\|_1.
\end{align*}
Similarly, since $\tr(u_n^+(\varepsilon)(u_n^-(\varepsilon),\,\cdot\,))_{\mathfrak h_\varepsilon}
=(u_n^-(\varepsilon),u_n^+(\varepsilon))_{\mathfrak h_\varepsilon}$,
$$
\int_\Delta\tr(c(\varepsilon,\varepsilon))\,\d \mu(\varepsilon)
=\sum_{n\in N}\kappa_n\int_\Delta(u_n^-(\varepsilon),
u_n^+(\varepsilon))_{\mathfrak h_\varepsilon}\,
\d \mu(\varepsilon)
=\sum_{n\in N}\kappa_n(u_n^-,u_n^+)=\tr(C).
$$
\cqfd
\end{demo}

\bigskip 
To compare the operators in a direct sum of Hilbert spaces, the following result is also useful.

\begin{Lem} \label{CompareKernel} Let $\Delta\subset\mathbb R$ be a measurable set equipped
with a $\sigma$-finite measure $\mu$, $(\mathfrak h_\varepsilon)_{\varepsilon\in\Delta}$ 
a family of $\mu$-measurable, separable Hilbert spaces, and 
$\mathcal D\subset\mathfrak H\equiv\int_\Delta^\oplus\mathfrak h_\varepsilon\,\d\mu(\varepsilon)$ 
a dense subspace. If  $\Delta\ni\varepsilon\mapsto a(\varepsilon)\in\B(\mathfrak h_\varepsilon)$ is a 
mapping such that, for all $u,v\in\mathcal D$, there exists a $\mu$-negligible set 
$\Delta_{uv}\subset\Delta$ with the property that
$$
(u(\varepsilon),a(\varepsilon)v(\varepsilon))_{\mathfrak h_\varepsilon}=0,
$$
for all $\varepsilon\in\Delta\setminus\Delta_{uv}$, then $a(\varepsilon)=0$ for $\mu$-almost 
every $\varepsilon\in\Delta$.
\end{Lem}

\begin{demo} Since $\mathfrak H$ is separable it is possible to extract a countable family 
$(u_n)_{n\in N}\subset\mathcal D$ which is dense. For each $n\in N$, let $\Delta_n$ be the set of 
$\varepsilon\in\Delta$ for which $u_n(\varepsilon)\in\mathfrak h_\varepsilon$ is defined.
Then $\Delta\setminus\Delta_n$ is $\mu$-negligible and the same is true of 
$\Delta\setminus\tilde\Delta$ where
$\tilde\Delta\equiv\cap_{n\in N}\Delta_n$.

For all $\varepsilon\in\tilde\Delta$, we may apply to the family 
$(u_n(\varepsilon))_{n\in N}$ the Gramm-Schmidt procedure to obtain an orthonormal basis
$(g_m(\varepsilon))_{m\in M}$ of the closed subspace 
$\mathfrak u_\varepsilon\subset\mathfrak h_\varepsilon$ generated by 
$(u_n(\varepsilon))_{n\in N}$. For all $m\in M$ we have 
$g_m(\varepsilon)=\sum_{n\in N}\alpha_{mn}(\varepsilon)u_n(\varepsilon)$, this sum being finite
 (i.e.,\;$\{n\in N\,|\,\alpha_{mn}(\varepsilon)\not=0\}$ is finite for all $m\in M$). 
Furthermore each coefficient $\alpha_{mn}(\varepsilon)$ is a measurable function of a finite number 
of scalar products $(u_i(\varepsilon),u_j(\varepsilon))_{\mathfrak h_\varepsilon}$  which are 
measurable functions of $\varepsilon$. We conclude that for all $u\in\mathfrak H$ the functions
$\tilde\Delta\ni\varepsilon\mapsto(g_m(\varepsilon), u(\varepsilon))_{\mathfrak h_\varepsilon}$ are 
measurable. Let
$p(\varepsilon)$ be the orthogonal projection onto $\mathfrak u_\varepsilon$. The Cauchy-Schwarz 
and Bessel inequalities show that for all $u,v\in\mathfrak H$ and for all $\varepsilon\in\tilde\Delta$ 
for which $u(\varepsilon)$ and $v(\varepsilon)$ are defined, the series
$$
(u(\varepsilon),p(\varepsilon)v(\varepsilon))_{\mathfrak h_\varepsilon}=
\sum_{m\in M}(u(\varepsilon),g_m(\varepsilon))_{\mathfrak h_\varepsilon}
(g_m(\varepsilon),v(\varepsilon))_{\mathfrak h_\varepsilon},
$$
converges absolutely. Its sum is thus a measurable function defined $\mu$-almost everywhere 
which satisfies
$$
\left|\sum_{m\in M}(u(\varepsilon),g_m(\varepsilon))_{\mathfrak h_\varepsilon}
(g_m(\varepsilon),v(\varepsilon))_{\mathfrak h_\varepsilon}\right|\le
\|u(\varepsilon)\|_{\mathfrak h_\varepsilon}\|v(\varepsilon)\|_{\mathfrak h_\varepsilon}.
$$
The Cauchy-Schwarz inequality and Fubini's theorem allow us to conclude that
$$
(u,Pv)\equiv
\int(u(\varepsilon),p(\varepsilon)v(\varepsilon))_{\mathfrak h_\varepsilon}\,\d\varepsilon,
$$
defines a bounded operator on $\mathfrak H$. By construction, $Pu_n=u_n$ for all $n\in N$.
Since  $(u_n)_{n\in N}$ is dense in $\mathfrak H$, it follows that $P=I$ and consequently,
$$
0=(u,(I-P)u)=
\int(u(\varepsilon),(I-p(\varepsilon))u(\varepsilon))_{\mathfrak h_\varepsilon}\,\d\varepsilon,
=\int\|(I-p(\varepsilon))u(\varepsilon)\|^2_{\mathfrak h_\varepsilon}\,\d\varepsilon,
$$
for all $u\in\mathfrak H$. This implies that 
$\|(I-p(\varepsilon))u(\varepsilon)\|_{\mathfrak h_\varepsilon}=0$,
$\mu$-almost everywhere, that is to say that there exists a measurable set 
$\tilde\Delta_0\subset\tilde\Delta$ such that $\Delta\setminus\tilde\Delta_0$ is $\mu$-negligible and 
$(u_n(\varepsilon))_{n\in N}$ is total in $\mathfrak h_\varepsilon$ for all $\varepsilon\in\tilde\Delta_0$.

Let 
$$
\Delta_0\equiv\left(\bigcap_{n,m\in N}\Delta\setminus\Delta_{u_nu_m}\right)
\bigcap\tilde\Delta_0.
$$
We may conclude the proof of the lemma by remarking that $\Delta\setminus\Delta_0$ is 
$\mu$-negligible and that for $\varepsilon\in\Delta_0$ we have 
$(u_n(\varepsilon),a(\varepsilon),u_m(\varepsilon))_{\mathfrak h_\varepsilon}=0$ for all $n,m\in N$ 
and thus $a(\varepsilon)=0$.\cqfd
\end{demo}

\subsubsection{The diagonal}
\label{DiagSubSect}
The diagonal $c(\varepsilon,\varepsilon)$ of the integral kernel of a trace-class operator $C$ 
on the Lebesgue direct integral $\int^\oplus\mathfrak h_\varepsilon\,\d\varepsilon$ is defined 
almost everywhere. To calculate this diagonal, the following result is often useful.

\begin{Lem}\label{DiagKernel} Let $\Delta\subset\mathbb R$ be a measurable set and 
$(\mathfrak h_\varepsilon)_{\varepsilon\in\Delta}$ a family of Lebesgue-measurable Hilbert spaces. 
Let $E$ be the self-adjoint operator on 
$\mathfrak H\equiv\int^\oplus_\Delta\mathfrak h_\varepsilon\,\d \varepsilon$ defined by 
$(Eu)(\varepsilon)=\varepsilon u(\varepsilon)$. If $C\in\mathcal L^1(\mathfrak H)$ and 
$c(\varepsilon',\varepsilon)$ denotes its integral kernel, then there exists a dense subspace 
$\mathfrak F\subset\mathfrak H$ such that
$$
\int_\Delta(u(\varepsilon),c(\varepsilon,\varepsilon)v(\varepsilon))\,\d\varepsilon=
\lim_{\eta\downarrow0}\frac{1}{2\pi}\int_{-\infty}^\infty\e^{-\eta|t|}(\e^{\i tE}u,C\e^{\i tE}v)\,\d t,
$$
for all $u,v\in\mathfrak F$.
\end{Lem}

\begin{demo}
Theorem \ref{TraceKernel} and its proof show that there exists a measurable set 
$\Delta_0\subset\Delta$ such that $\Delta\setminus\Delta_0$ has Lebesgue measure zero and, 
for all $\varepsilon,\varepsilon'\in\Delta_0$,
$$
c(\varepsilon',\varepsilon)=\sum_{n\in N}\kappa_n\,
u_n^+(\varepsilon')(u_n^-(\varepsilon),\,\cdot\,)_{\mathfrak h_\varepsilon},
$$
where $N$  is a set which is at most countable, $\kappa_n>0$, $\sum_{n\in N}\kappa_n=\|C\|_1$  and $(u_n^\pm)_{n\in N}$  are orthonormal families of $\mathfrak H$ such that
$$
\sum_{n\in N}\kappa_n\,\|u_n^\pm(\varepsilon)\|_{\mathfrak h_\varepsilon}^2<\infty,
$$
for all $\varepsilon\in\Delta_0$.

Let $u,v\in\mathfrak H$, for all $t\in\mathbb R$ we have
\begin{align*}
(\e^{\i tE}u,&C\e^{\i tE}v)=
\int(u(\varepsilon'),c(\varepsilon',\varepsilon)
v(\varepsilon))_{\mathfrak h_{\varepsilon'}}\,
\e^{\i t(\varepsilon-\varepsilon')}\,\d\varepsilon\d\varepsilon'\\
&=\int\left(
\sum_{n\in N}\kappa_n(u(\varepsilon'),u_n^+(\varepsilon'))_{\mathfrak h_{\varepsilon'}}
(u_n^-(\varepsilon),v(\varepsilon))_{\mathfrak h_\varepsilon}\e^{\i t(\varepsilon-\varepsilon')}
\right)\d\varepsilon\d\varepsilon'.
\end{align*}
Since $\|u_n^\pm\|=1$, we have
\begin{align*}
\int|(u(\varepsilon'),u_n^+(\varepsilon'))_{\mathfrak h_{\varepsilon'}}\,
(u_n^-(\varepsilon),v(\varepsilon))_{\mathfrak h_k(\varepsilon)}|
\,\d\varepsilon\d\varepsilon' \le\|u\|\,\|v\|,
\end{align*}
and thus
\begin{align*}
\sum_{n\in N}\kappa_n\int|(u(\varepsilon'),u_n^+(\varepsilon'))_{\mathfrak h_{\varepsilon'}}
(u_n^-(\varepsilon),v(\varepsilon))_{\mathfrak h_\varepsilon}|
\,\d\varepsilon\d\varepsilon'\le\sum_{n\in N}\kappa_n\|u\|\,\|v\|<\infty.
\end{align*}
Fubini's theorem allows us to conclude that for $\eta>0$,
\begin{align}
&\int_{-\infty}^{\infty}\e^{-\eta|t|}(\e^{\i tE}u,C\e^{\i tE}v)\,\d t\nonumber\\
&=\sum_{n\in N}\kappa_n\int\left(\int_{-\infty}^\infty\e^{-\eta|t|}
(u(\varepsilon'),u_n^+(\varepsilon'))_{\mathfrak h_{\varepsilon'}}
(u_n^-(\varepsilon),v(\varepsilon))_{\mathfrak h_\varepsilon}
\e^{\i t(\varepsilon-\varepsilon')}\,\d t\right)
\d\varepsilon\d\varepsilon'\nonumber\\
&=2\pi\sum_{n\in N}\kappa_n\int
(u(\varepsilon'),u_n^+(\varepsilon'))_{\mathfrak h_{\varepsilon'}}
(u_n^-(\varepsilon),v(\varepsilon))_{\mathfrak h_\varepsilon}
\,\delta_\eta(\varepsilon-\varepsilon')\,
\d\varepsilon\d\varepsilon'\nonumber\\
&=2\pi\sum_{n\in N}\kappa_n\int\overline{F_n^+(\varepsilon)}(\delta_\eta\star F_n^-)(\varepsilon)
\,\d\varepsilon,\label{etaintegral}
\end{align}
where $F_n^+(\varepsilon)=(u_n^+(\varepsilon),u(\varepsilon))_{\mathfrak h_\varepsilon}$,
$F_n^-(\varepsilon)=(u_n^-(\varepsilon),v(\varepsilon))_{\mathfrak h_\varepsilon}$
and
$$
\delta_\eta(\varepsilon)=\frac{1}{2\pi}\int_{-\infty}^\infty
\e^{\i t\varepsilon-\eta|t|}\,\d t
=\frac{1}{\pi}\,\frac{\eta}{\varepsilon^2+\eta^2}.
$$
Recall that the set $\mathfrak F\equiv\{u\in\mathfrak H\,|\,
\VERT u\VERT\equiv\sup_{\varepsilon\in\Delta} \|u(\varepsilon)\|_{\mathfrak h_\varepsilon}<\infty\}$  
is dense in $\mathfrak H$.
For $u,v\in\mathfrak F$ the Cauchy-Schwarz inequality in $\mathfrak h_\varepsilon$ implies that 
$F_n^\pm\in L^2(\Delta,\d\varepsilon)$. Since
$\slim_{\eta\downarrow0}\delta_\eta\star=I$ in $L^2(\mathbb R)$ we have
$$
\lim_{\eta\downarrow0}\int\overline{F_n^+(\varepsilon)}(\delta_\eta\star F_n^-)(\varepsilon)
\,\d\varepsilon=\int\overline{F_n^+(\varepsilon)}F_n^-(\varepsilon)
\,\d\varepsilon
=\int(u(\varepsilon),u_n^+(\varepsilon))_{\mathfrak h_\varepsilon}
(u_n^-(\varepsilon),v(\varepsilon))_{\mathfrak h_\varepsilon}
\,\d\varepsilon,
$$
for all $n\in N$ and all $u,v\in\mathfrak F$. Since $\delta_\eta\in L^1(\mathbb R)$ and 
$\|\delta_\eta\|_1=\int\delta_\eta(\varepsilon)\,\d\varepsilon=1$, the Cauchy-Schwarz and Young 
inequalities imply
\begin{align*}
&\left|\int\overline{F_n^+(\varepsilon)}(\delta_\eta\star F_n^-)(\varepsilon)
\,\d\varepsilon\right|
\le\left(\int|F_n^+(\varepsilon)|^2\,\d\varepsilon\right)^{1/2}
\left(\int|F_n^-(\varepsilon)|^2\,\d\varepsilon\right)^{1/2}\\
&\le\left(\int\|u_n^+(\varepsilon)\|^2_{\mathfrak h_\varepsilon}\,
\|u(\varepsilon)\|^2_{\mathfrak h_\varepsilon}
\,\d\varepsilon\right)^{1/2}
\left(\int\|u_n^-(\varepsilon)\|^2_{\mathfrak h_\varepsilon}\,
\|v(\varepsilon)\|^2_{\mathfrak h_\varepsilon}
\,\d\varepsilon\right)^{1/2}\\
&\le\left(\VERT u\VERT^2\int\|u_n^+(\varepsilon)\|^2_{\mathfrak h_\varepsilon}
\,\d\varepsilon\right)^{1/2}
\left(\VERT v\VERT^2\int\|u_n^-(\varepsilon)\|^2_{\mathfrak h_\varepsilon}
\,\d\varepsilon\right)^{1/2}\!\!\!=\VERT u\VERT\,\VERT v\VERT,
\end{align*}
and the dominated convergence theorem applies to the right hand side of the identity 
(\ref{etaintegral})
\begin{align*}
\lim_{\eta\downarrow0}&
\int_{-\infty}^{\infty}\e^{-\eta|t|}(\e^{\i tE}u,C\e^{\i tE}v)\,\d t
=2\pi\sum_{n\in N}\kappa_n\int\overline{F_n^+(\varepsilon)}F_n^-(\varepsilon)
\,\d\varepsilon\\
&=2\pi\sum_{n\in N}\kappa_n\int
(u(\varepsilon),u_n^+(\varepsilon))_{\mathfrak h_\varepsilon}
(u_n^-(\varepsilon),v(\varepsilon))_{\mathfrak h_\varepsilon}\,\d\varepsilon.
\end{align*}
Finally, Fubini's theorem allows us to conclude that
\begin{align*}
\lim_{\eta\downarrow0}
\int_{-\infty}^{\infty}\e^{-\eta|t|}&(\e^{\i tE}u,C\e^{-\i tE}v)\,\d t\nonumber\\
&=2\pi\int\sum_{n\in N}\kappa_n
(u(\varepsilon),u_n^+(\varepsilon))_{\mathfrak h_\varepsilon}
(u_n^-(\varepsilon),v(\varepsilon))_{\mathfrak h_\varepsilon}\,\d\varepsilon\nonumber\\
&=2\pi\int(u(\varepsilon),c(\varepsilon,\varepsilon)
v(\varepsilon))_{\mathfrak h_\varepsilon}\,\d\varepsilon,
\end{align*}
for all $u,v\in\mathfrak F$.
\cqfd
\end{demo}

\bigskip

\begin{Rem} Theorem \ref{TraceKernel} and Lemmas \ref{CompareKernel} and \ref{DiagKernel} 
are generalized without difficulty to operators
$$
C:\int^\oplus\mathfrak h_1(\varepsilon)\,\d\varepsilon
\to\int^\oplus\mathfrak h_2(\varepsilon)\,\d\varepsilon,
$$
it suffices in fact to identify them with operators
$$
C':\int^\oplus\mathfrak h_1(\varepsilon)\oplus \mathfrak h_2(\varepsilon)\,\d\varepsilon
\to\int^\oplus\mathfrak h_2(\varepsilon)\oplus \mathfrak h_2(\varepsilon)\,\d\varepsilon.
$$

\end{Rem}

To apply the previous lemma to the calculation of the current, we shall use the following result.

\begin{Prop}\label{AbelStuff} Under Hypotheses\Href{HypReservoir},
\Href{HypResDyn}, and \Href{HypConjugate}, for all $u,v\in\HH_k$ we have
\begin{equation*}
\lim_{\eta\downarrow0}
\int_{-\infty}^{\infty}\e^{-\eta|t|}(u,
\e^{\i t\widetilde H_k}\widetilde\Psi^{\#(a,\sigma,\sigma')}_{Q,g,k}\e^{-\i t\widetilde H_k}v)
\,\d t
=\sigma\delta_{\sigma\sigma'}(g(\widetilde H_k)u,\widetilde P_k^\sigma
\widetilde Q_k \widetilde P_k^\sigma
g(\widetilde H_k)v).
\end{equation*}
\end{Prop}

\begin{demo}
By remarking that
$$
(u,\e^{\i t\widetilde H_k}\widetilde\Psi^{\#(a,\sigma,\sigma')}_{Q,g,k}\e^{-\i t\widetilde H_k}v)
=\frac{\d\ }{\d t}\, G^{(\sigma,\sigma')}(t),
$$
with
$$
G^{(\sigma,\sigma')}(t)=(h_\sigma^{(a)}(\widetilde A_k)\e^{\i t\widetilde H_k}g(\widetilde H_k)u,
M_k^\ast QM_kh_{\sigma'}^{(a)}(\widetilde A_k)
\e^{\i t\widetilde H_k}g(\widetilde H_k)v),
$$
we obtain, after an integration by parts,
$$
\int_{-\infty}^{\infty}\e^{-\eta|t|}(u,
\e^{\i t\widetilde H_k}\widetilde\Psi^{\#(a,\sigma,\sigma')}_{Q,g,k}\e^{-\i t\widetilde H_k}v)
\,\d t
=\eta\int_0^\infty\e^{-\eta t}(G^{(\sigma,\sigma')}(t)-G^{(\sigma,\sigma')}(-t))\,\d t,
$$
and consequently
\begin{align*}
\lim_{\eta\downarrow0}\int_{-\infty}^{\infty}\e^{-\eta|t|}(u,
\e^{\i t\widetilde H_k}&\widetilde\Psi^{\#(a,\sigma,\sigma')}_{Q,g,k}\e^{-\i t\widetilde H_k}v)
\,\d t\\
&=\alim_{t\to\infty}G^{(\sigma,\sigma')}(t)-\alim_{t\to-\infty}G^{(\sigma,\sigma')}(t).
\end{align*}
To evaluate these Abelian limits, we invoke the propagation estimates. For $t>0$, we have
\begin{align*}
\|h_\sigma^{(a)}(\widetilde A_k)\e^{\i\sigma t\widetilde H_k}&g(\widetilde H_k)u\|
\le\|F(\sigma\widetilde A_k\ge a-1)
\e^{\i\sigma t\widetilde H_k}g(\widetilde H_k)u\|\\
&\le\|F(\sigma\widetilde A_k\ge a-1)
\e^{\i\sigma t\widetilde H_k}g(\widetilde H_k)F(\sigma\widetilde A_k>\vartheta t)u\|\\
&+\|F(\sigma\widetilde A_k\ge a-1)
\e^{\i\sigma t\widetilde H_k}g(\widetilde H_k)F(\sigma\widetilde A_k\le\vartheta t)u\|\\
&\le\|F(\sigma\widetilde A_k>\vartheta t)u\|\\
&+\|F((-\sigma)\widetilde A_k\le1-a)
\e^{-\i(-\sigma) t\widetilde H_k}g(\widetilde H_k)F((-\sigma)\widetilde A_k\ge-\vartheta t)u\|.
\end{align*}
When $t\to+\infty$ the first term on the right hand side of this inequality tends clearly to $0$ for all $\vartheta>0$. By applying Proposition \ref{MainPropEst} to the second term it is possible to choose $\vartheta>0$ such that
$$
\|F((-\sigma)\widetilde A_k\le1-a)
\e^{-\i(-\sigma) t\widetilde H_k}g(\widetilde H_k)F((-\sigma)\widetilde A_k\ge-\vartheta t)u\|
\le c\,\langle a-1+\vartheta t\rangle^{-s},
$$
for constants $c$ and $s>0$. We thus have
$$
\lim_{t\to+\infty}\|h_\sigma^{(a)}(\widetilde A_k)
\e^{\i\sigma t\widetilde H_k}g(\widetilde H_k)u\|=0,
$$
and it follows that
$$
\alim_{t\to+\infty}G^{(\sigma,\sigma')}(t)=
\lim_{t\to+\infty}G^{(\sigma,\sigma')}(t)=0,
$$
if $\langle\sigma,\sigma'\rangle\not=\langle+,+\rangle$ and that
$$
\alim_{t\to-\infty}G^{(\sigma,\sigma')}(t)=
\lim_{t\to-\infty}G^{(\sigma,\sigma')}(t)=0,
$$
if $\langle\sigma,\sigma'\rangle\not=\langle-,-\rangle$. It remains to consider 
$G^{(\sigma,\sigma)}(\sigma t)$ for $t\to+\infty$. By writing
\begin{align}
&G^{(\sigma,\sigma)}(\sigma t)=
(h_\sigma^{(a)}(\widetilde A_k)\e^{-\i\sigma t\widetilde H_k}g(\widetilde H_k)u,
M_k^\ast QM_kh_{\sigma}^{(a)}(\widetilde A_k)
\e^{-\i\sigma t\widetilde H_k}g(\widetilde H_k)v)\nonumber\\
&=(\e^{-\i\sigma t\widetilde H_k}g(\widetilde H_k)u,
M_k^\ast QM_k\e^{-\i\sigma t\widetilde H_k}g(\widetilde H_k)v)\nonumber\\
&-(\e^{-\i\sigma t\widetilde H_k}g(\widetilde H_k)u,
M_k^\ast QM_k(1-h_{\sigma}^{(a)}(\widetilde A_k))
\e^{-\i\sigma t\widetilde H_k}g(\widetilde H_k)v)\label{Gss}\\
&-((1-h_\sigma^{(a)}(\widetilde A_k))\e^{-\i\sigma t\widetilde H_k}g(\widetilde H_k)u,
M_k^\ast QM_k
\e^{-\i\sigma t\widetilde H_k}g(\widetilde H_k)v)\nonumber\\
&+((1-h_\sigma^{(a)}(\widetilde A_k))\e^{-\i\sigma t\widetilde H_k}g(\widetilde H_k)u,
M_k^\ast QM_k(1-h_{\sigma}^{(a)}(\widetilde A_k))
\e^{-\i\sigma t\widetilde H_k}g(\widetilde H_k)v),\nonumber
\end{align}
we remark that
\begin{align*}
\|(1-h_\sigma^{(a)}(\widetilde A_k))&\e^{-\i\sigma t\widetilde H_k}g(\widetilde H_k)u\|
\le\|F(\sigma\widetilde A_k\le a+1)\e^{-\i\sigma t\widetilde H_k}g(\widetilde H_k)u\|\\
&\le\|F(\sigma\widetilde A_k\le a+1)\e^{-\i\sigma t\widetilde H_k}
g(\widetilde H_k)F(\sigma\widetilde A_k<a+1-\vartheta t)u\|\\
&+\|F(\sigma\widetilde A_k\le a+1)\e^{-\i\sigma t\widetilde H_k}
g(\widetilde H_k)F(\sigma\widetilde A_k\ge a+1-\vartheta t)u\|,
\end{align*}
and we conclude as before that
$$
\lim_{t\to+\infty}\|(1-h_\sigma^{(a)}(\widetilde A_k))
\e^{-\i\sigma t\widetilde H_k}g(\widetilde H_k)u\|=0,
$$
which shows that the three last terms of the right hand side of the identity (\ref{Gss}) vanish in this 
limit. The first term in turn is calculated as follows, by using the fact that $Q$ commutes with $H$ 
and the identity (\ref{QConjug}),
\begin{align*}
\lim_{t\to+\infty}&(\e^{-\i\sigma t\widetilde H_k}g(\widetilde H_k)u,
M_k^\ast QM_k\e^{-\i\sigma t\widetilde H_k}g(\widetilde H_k)v)\\
&=\lim_{t\to+\infty}
(g(\widetilde H_k)u,\e^{\i\sigma t\widetilde H_k}M_k^\ast\e^{-\i\sigma tH} Q
\e^{\i\sigma tH}M_k\e^{-\i\sigma t\widetilde H_k}g(\widetilde H_k)v)\\
&=(g(\widetilde H_k)u,\Omega_k^{\sigma\ast} Q\Omega_k^\sigma
g(\widetilde H_k)v)\\
&=(g(\widetilde H_k)u,\widetilde P_k^\sigma
\widetilde Q_k \widetilde P_k^\sigma
g(\widetilde H_k)v).
\end{align*}
\cqfd
\end{demo}

\subsubsection{Proof of Theorem \ref{Kernelpsi}} 
\label{Kernelpsidemo}

\noindent (i) We note that since 
$g\in C_0^\infty(\mathbb R\setminus\Sigma_H)\subset  C_0^\infty(\mathbb R\setminus\Sigma_k)$,
the spectrum of $\widetilde H_k$ is purely absolutely continuous on $\supp\,g$ by 
Corollary \ref{NoSingSpec}. We thus have
$\widetilde\Psi^{\#(a,\sigma,\sigma')}_{Q,g,k}=P_\ac(\widetilde
H_k)\widetilde\Psi^{\#(a,\sigma,\sigma')}_{Q,g,k}P_\ac(\widetilde
H_k)$.

\noindent (ii) The existence of the integral kernel $\psi^{\#(a,\sigma,\sigma')}_{Q,g,k}$ is a direct 
consequence of the fact that $\widetilde\Psi^{\#(a,\sigma,\sigma')}_{Q,g,k}$ is trace-class and of
Theorem~\ref{TraceKernel}.

\noindent (iii) Lemma \ref{DiagKernel} and Proposition \ref{AbelStuff} give us
\begin{align*}
\int (u(\varepsilon),&\psi^{\#(a,\sigma,\sigma')}_{Q,g,k}
(\varepsilon,\varepsilon)v(\varepsilon))_{\mathfrak h_k(\varepsilon)}\,\d\varepsilon
=\frac{1}{2\pi}\sigma\delta_{\sigma\sigma'}(g(\widetilde H_k)u,\widetilde P_k^\sigma
\widetilde Q_k \widetilde P_k^\sigma
g(\widetilde H_k)v)\\
&=\frac{1}{2\pi}\sigma\delta_{\sigma\sigma'}\int (u(\varepsilon),g(\varepsilon)
p_k^\sigma(\varepsilon)q_k(\varepsilon)p_k^\sigma(\varepsilon)
g(\varepsilon)
v(\varepsilon))_{\mathfrak h_k(\varepsilon)}\,\d\varepsilon,
\end{align*}
for all $u,v\in\mathfrak F_k$, a dense subspace of $\HH_{k,\ac}$. Lemma \ref{CompareKernel} allows us to 
conclude that
$$
\psi^{\#(a,\sigma,\sigma')}_{Q,g,k}(\varepsilon,\varepsilon)=
\frac{\sigma}{2\pi}\delta_{\sigma\sigma'}g(\varepsilon)p_k^\sigma(\varepsilon)
q_k(\varepsilon) p_k^\sigma(\varepsilon)
g(\varepsilon),
$$
for almost all $\varepsilon\in\mathbb R$.\cqfd

\subsection{The Landauer-B\"uttiker formula}
\label{LBFormSect}
We are now prepared to prove Theorem \ref{LBTheo}.

Starting from (\ref{FluxTwo}) we obtain the following representation of the steady current
\begin{align*}
\tr(&\Omega_j^{-}\widetilde T_j\Omega_j^{-\ast}\Psi^{(a)}_{Q,g,k})
=\int
\tr_{\mathfrak h_j(\varepsilon)}\left(t_j(\varepsilon)\left\{s_{jk}^\ast(\varepsilon)\psi^{\#(a,+,+)}_{Q,g,k}
(\varepsilon,\varepsilon)
s_{kj}(\varepsilon)
\right.\right.\nonumber\\
&+\delta_{jk}\left(
s_{kk}^\ast(\varepsilon)\psi^{\#(a,+,-)}_{Q,g,k}(\varepsilon,\varepsilon)p_k^-(\varepsilon)
+p_k^-(\varepsilon)\psi^{\#(a,-,+)}_{Q,g,k}(\varepsilon,\varepsilon)s_{kk}(\varepsilon)
\left.\left.
+p_k^-(\varepsilon)\psi^{\#(a,-,-)}_{Q,g,k}(\varepsilon,\varepsilon)p_k^-(\varepsilon)
\right)\right\}\right)\,\d\varepsilon.
\end{align*}
By inserting Expression (\ref{psikernelId}) we find 
$$
\tr(\Omega_j^{-}\widetilde T_j\Omega_j^{-\ast}\Psi^{(a)}_{Q,g,k})
=\int g(\varepsilon)^2 \,
\tr_{\mathfrak h_j(\varepsilon)}\left(t_j(\varepsilon)\left\{s_{jk}^\ast(\varepsilon)
q_k^+(\varepsilon)s_{kj}(\varepsilon) 
-\delta_{kj}q_j^-(\varepsilon)\right\}\right)\,\frac{\d\varepsilon}{2\pi},
$$
where $q_k^\pm(\varepsilon)\equiv p_k^\pm(\varepsilon)q_k(\varepsilon)p_k^\pm(\varepsilon)$.

To control the limit (\ref{CourantFormzero}) and thus obtain the steady current of an eventually 
unbounded charge we shall use the following result, adapted from Theorem 6 of Section 7.6 
in \cite{Y} to our situation.

\begin{Lem}\label{Sbound} Under Hypotheses \Href{HypReservoir}, \Href{HypResDyn}, 
\Href{HypConjugate}, and \Href{HypChi}, 
$s_{jk}(\varepsilon)-\delta_{jk}p_k^\pm(\varepsilon)$ is trace-class for almost all 
$\varepsilon\in\mathbb R$. Furthermore, there exists a constant $c$ such that, for any measurable 
bounded function $f:\mathbb R\to\mathbb R$ and any interval $\Delta\subset\mathbb R$
$$
\int_\Delta f(\varepsilon)^2\,\|s_{jk}(\varepsilon)-\delta_{jk}p_k^\pm(\varepsilon)
\|_1\,\d\varepsilon\le
c\,\left(\esup_{\varepsilon\in\Delta}\,\langle\varepsilon\rangle^{\nu+1}|f(\varepsilon)|\right)^2.
$$
\end{Lem}

\begin{demo} Without loss of generality we may suppose that $f\ge0$. Let  $K\subset\Delta$ be 
compact, set $g\equiv f1_K$ and note that
\begin{align}
(g(\widetilde H_j)u,(S_{jk}-\delta_{jk}\widetilde P_k^+)g(\widetilde H_k)v)
&=-(\Omega_j^+g(\widetilde H_j)u,(\Omega_k^+-\Omega_k^-)g(\widetilde H_k)v),\label{horiz1}\\
(g(\widetilde H_j)u,(S_{jk}-\delta_{jk}\widetilde P_k^-)g(\widetilde H_k)v)
&=-((\Omega_j^+-\Omega_j^-)g(\widetilde H_j)u, \Omega_k^-g(\widetilde H_k)v).\label{horiz2}
\end{align}
For all $u\in\HH_{j,\ac}$, $v\in\HH_{k,\ac}$, Relation  \eqref{horiz1} yields, after passing to the Abelian  
limit and integrating by parts,
\begin{align*}
(g(\widetilde H_j)u,&(S_{jk}-\delta_{jk}\widetilde P_k^+)g(\widetilde H_k)v)
=-\lim_{t\to\infty}(\Omega_j^+g(\widetilde H_j)u,(\e^{\i tH}M_k\e^{-\i t\widetilde H_k}-
\e^{-\i tH}M_k\e^{\i t\widetilde H_k})g(\widetilde H_k)v)\\
&=-\lim_{\eta\downarrow0}\eta\int_0^\infty\e^{-\eta t}
(\Omega_j^+g(\widetilde H_j)u,(\e^{\i tH}M_k\e^{-\i t\widetilde H_k}-
\e^{-\i tH}M_k\e^{\i t\widetilde H_k})g(\widetilde H_k)v)\,\d t\\
&=-\lim_{\eta\downarrow0}\int_{-\infty}^\infty\e^{-\eta|t|}
(\Omega_j^+g(\widetilde H_j)u,\e^{\i tH}\i(HM_k-M_k\widetilde H_k)\e^{-\i t\widetilde H_k}
g(\widetilde H_k)v)\,\d t.
\end{align*}
By using the intertwining relations of the M\o ller operators we obtain
$$
(g(\widetilde H_j)u,(S_{jk}-\delta_{jk}\widetilde P_k^+)g(\widetilde H_k)v)=
\lim_{\eta\downarrow0}\int_{-\infty}^\infty\e^{-\eta|t|}
(\e^{-\i t\widetilde H_j}u,C\e^{-\i t\widetilde H_k}v)\,\d t,
$$
where
$$
C\equiv-\i\Omega_j^{+\ast}g(H)(HM_k-M_k\widetilde H_k)g(\widetilde H_k).
$$
The identity
\begin{align*}
g(H)(HM_k-M_k\widetilde H_k)g(\widetilde H_k)&=g_1(H)\left(
(H+\i)^{-\nu}M_k-M_k(\widetilde H_k+\i)^{-\nu}\right)
g(\widetilde H_k)\\
&-g_1(H)\left(
(H+\i)^{-\nu-1}M_k-M_k(\widetilde H_k+\i)^{-\nu-1}\right)(\widetilde H_k+\i)
g(\widetilde H_k),
\end{align*}
where $g_1(x)=g(x)(x+\i)^{\nu+1}$, and Lemma \ref{fHMLemma} allow us to conclude that
$C$ is trace-class. Furthermore, since $\|\Omega_j^{+\ast}\|\le1$, we have the estimate
\begin{align*}
\|C\|_1
&\le\|g_1(H)\|\,\|(H+\i)^{-\nu}M_k-M_k(\widetilde H_k+\i)^{-\nu}\|_1
\|g(\widetilde H_k)\|\\
&+\|g_1(H)\|\,\|(H+\i)^{-\nu-1}M_k-M_k(\widetilde H_k+\i)^{-\nu-1}\|_1\,
\|(\widetilde H_k+\i)g(\widetilde H_k)\|,
\end{align*}
and it follows that there exists a constant $c_1$ such that
\begin{equation}
\|C\|_1\le c_1\,\left(\esup_{\varepsilon\in\Delta}\,\langle\varepsilon\rangle^{\nu+1}
f(\varepsilon)\right)^2.
\label{CEstim}
\end{equation}
By Lemma \ref{TraceKernel}, $C$ has an integral kernel $c(\varepsilon',\varepsilon)$ and 
Lemma \ref{DiagKernel} allows us to write
$$
\frac{1}{2\pi}
\int g(\varepsilon)^2(u(\varepsilon),(s_{jk}(\varepsilon)
-\delta_{jk}p_k^+(\varepsilon))v(\varepsilon))_{\mathfrak h_j(\varepsilon)}\d\varepsilon=
\int(u(\varepsilon),c(\varepsilon,\varepsilon)v(\varepsilon))_{\mathfrak h_j(\varepsilon)}
\d\varepsilon,
$$
for $u\in\mathfrak F_j$ and $v\in\mathfrak F_k$, $\mathfrak F_j$, $\mathfrak F_k$ 
being dense subspaces of $\HH_{j,\ac}$ and $\HH_{k,\ac}$. By Lemma \ref{CompareKernel} 
we have
$$
\frac{1}{2\pi}\,
g(\varepsilon)^2\left(s_{jk}(\varepsilon)
-\delta_{jk}p_k^\pm(\varepsilon)\right)=c(\varepsilon,\varepsilon),
$$
for almost every $\varepsilon\in\mathbb R$, and invoking Lemma \ref{TraceKernel} we obtain
\begin{align*}
\int_{K}f(\varepsilon)^2\,\|s_{jk}(\varepsilon)-\delta_{jk}p_k^+(\varepsilon)\|_1
\,\d\varepsilon&\le
\int g(\varepsilon)^2\|s_{jk}(\varepsilon)-\delta_{jk}p_k^+(\varepsilon)\|_1\,\d\varepsilon\\
&\le2\pi\,\|C\|_1\\
&\le c_2\,\left(\esup_{\varepsilon\in\Delta}\,\langle\varepsilon\rangle^{\nu+1}
f(\varepsilon)\right)^2,
\end{align*}
and thus
\begin{align*}
\int_\Delta f(\varepsilon)^2\,\|s_{jk}(\varepsilon)-\delta_{jk}p_k^+(\varepsilon)\|_1
\,\d\varepsilon&=
\sup_{\atop{K\subset\Delta}{K\ \text{compact}}}
\int_{K}f(\varepsilon)^2\,
\|s_{jk}(\varepsilon)-\delta_{jk}p_k^+(\varepsilon)\|_1\,\d\varepsilon\\
&\le c_2\,\left(\esup_{\varepsilon\in\Delta}\,\langle\varepsilon\rangle^{\nu+1}
f(\varepsilon)\right)^2.
\end{align*}
The case where $p_k^+$ is replaced by $p_k^-$ can be handled in a similar way, starting from
Eq.~\eqref{horiz2}.\cqfd
\end{demo}

\bigskip 

Let $Q$ be a temperate charge such that $\Dom(H^\alpha)\subset\Dom(Q)$ for some $\alpha\ge0$. 
Then $Q(H+\i)^{-\alpha}$ is bounded and it follows from the identity (\ref{QConjug}) that 
$\widetilde P_k^\pm\widetilde Q_k(\widetilde H_k+\i)^{-\alpha}\widetilde P_k^\pm$ is bounded.
We thus have
$$
c_1\equiv\max_{k\in\{1,\ldots,M\}}
\esup_{\varepsilon\in\mathbb R}\,\langle\varepsilon\rangle^{-\alpha}
\|q_k^\pm(\varepsilon)\|_{\mathfrak h_k(\varepsilon)}<\infty.
$$
We shall denote by  $q_{\epsilon,k}(\varepsilon)=(1+\epsilon\varepsilon^2)^{-\alpha/2}q_k(\varepsilon)$ 
the fibers of the regularized charge
$\widetilde Q_{\epsilon,k}=\widetilde Q_k(I+\epsilon\widetilde H_k^2)^{-\alpha}$. Similarly, if for 
$j\in\{1,\ldots,M\}$, $\widetilde T_j$ denotes the generator of a $\tau_j$-invariant, gauge invariant, 
quasi-free state on $\mathcal O_j$ such that
$\Ran(\widetilde T_j)\subset\Dom(\widetilde H_j^{\alpha+\nu+1})$ then
$\widetilde H_j^{\alpha+\nu+1}\widetilde T_j$ is bounded and
$$
c_2\equiv\max_{j\in\{1,\ldots,M\}}
\esup_{\varepsilon\in\mathbb R}\,
\langle\varepsilon\rangle^{\alpha+\nu+1}\|t_j(\varepsilon)\|<\infty.
$$

Since  $s^\ast_{jk}(\varepsilon)=s_{kj}(\varepsilon)^\ast$ we have 
$$
0\le s^\ast_{jk}(\varepsilon)s_{kj}(\varepsilon)\le
\sum_{l=1}^Ms^\ast_{jl}(\varepsilon)s_{lj}(\varepsilon)=p_j^-(\varepsilon)\le I,
$$
and consequently, $\|s_{kj}(\varepsilon)\|\le\|p_j^-(\varepsilon)\|\le 1$. 
Charge conservation, Eq.~(\ref{QkConjug}), thus writes
$$
q_j^-(\varepsilon)=\sum_{l=1}^Ms^\ast_{jl}(\varepsilon)q_l^+(\varepsilon)s_{lj}(\varepsilon),
$$
from which we obtain the identity
\begin{align*}
s_{jk}^\ast(\varepsilon)q_k^+(\varepsilon)s_{kj}(\varepsilon) 
-\delta_{kj}q_j^-(\varepsilon)=\sum_{l=1}^M(\delta_{kl}-\delta_{kj})
s^\ast_{jl}(\varepsilon)q_l^+(\varepsilon)(s_{lj}(\varepsilon)-\delta_{lj}p_j^-(\varepsilon)),
\end{align*}
and then the inequality
$$
\|s_{jk}^\ast(\varepsilon)q_k^+(\varepsilon)s_{kj}(\varepsilon) 
-\delta_{kj}q_j^-(\varepsilon)\|_1\le \sum_{l=1}^M\|q_l^+(\varepsilon)\|\,
\|s_{lj}(\varepsilon)-\delta_{lj}p_j^-(\varepsilon)\|_1.
$$
We thus have
\begin{align*}
\sum_{j=1}^M
\left|\tr_{\mathfrak h_j(\varepsilon)}\left(
t_j(\varepsilon)\left\{s_{jk}^\ast(\varepsilon)
q_{\epsilon,k}^+(\varepsilon)s_{kj}(\varepsilon) 
-\delta_{kj}q_{\epsilon,j}^-(\varepsilon)\right\}\right)\right|
&\le \sum_{j,l=1}^M
\|t_j(\varepsilon)\|\,\|q_l^+(\varepsilon)\|
\,\|s_{lj}(\varepsilon)-\delta_{lj}p_j^-(\varepsilon)\|_1\\
&\le c_1 c_2\langle\varepsilon\rangle^{-\nu-1} 
\sum_{j,l=1}^M\|s_{lj}(\varepsilon)-\delta_{lj}p_j^-(\varepsilon)\|_1,
\end{align*}
and Lemma \ref{Sbound} allows us to conclude that the left hand side of this inequality belongs to 
$L^1(\mathbb R,\d\varepsilon)$. By the dominated convergence theorem, we conclude that for all
$g\in C_0^\infty(\mathbb R\setminus\Sigma_H)$ such that $0\le g\le 1$
$$
\lim_{\epsilon\downarrow 0}
\sum_{j=1}^M
\tr(\Omega_j^{-}\widetilde T_j\Omega_j^{-\ast}\Psi^{(a)}_{Q_\epsilon,g,k})
=\int g(\varepsilon)^2 \,
\tr_{\mathfrak h_j(\varepsilon)}\left(t_j(\varepsilon)\left\{s_{jk}^\ast(\varepsilon)
q_k^+(\varepsilon)s_{kj}(\varepsilon) 
-\delta_{kj}q_j^-(\varepsilon)\right\}\right)\,\frac{\d\varepsilon}{2\pi}.
$$
Similarly, if $g_n\in C_0^\infty(\mathbb R\setminus\Sigma_H)$ is a sequence such that
$0\le g_n\le1$ and $\lim_ng_n(x)=1$ for almost all $x\in\mathbb R$ we have
$$
\lim_{n}
\sum_{j=1}^M
\tr(\Omega_j^{-}\widetilde T_j\Omega_j^{-\ast}\Psi^{(a)}_{Q,g_n,k})=\int
\tr_{\mathfrak h_j(\varepsilon)}\left(t_j(\varepsilon)\left\{s_{jk}^\ast(\varepsilon)
q_k^+(\varepsilon)s_{kj}(\varepsilon) 
-\delta_{kj}q_j^-(\varepsilon)\right\}\right)\,\frac{\d\varepsilon}{2\pi},
$$
which concludes the proof of Theorem \ref{LBTheo}.

\appendix

\section{Proof of Lemma \ref{fHMLemma}}
\label{HTraceProof}
We begin by proving two auxiliary lemmas based on Hypothesis \Href{HypChi}.

\begin{Lem}\label{adjbound} Under Hypothesis $\Href{HypChi}$ we have
$$
\ad_{\widetilde H_k}^{j}(\widetilde\chi_k^{(r)})\in\B(\HH_k^{s+j/2},\HH_k^{s}),
$$
for $1\le j\le s+j\le 4\nu$, $k\in\{1,\ldots,M\}$ and $r\ge0$. Furthermore, if $s$ is an integer then
the formula
\begin{equation}
(\widetilde H_k-z)^{s}\ad_{\widetilde H_k}^j(\widetilde\chi_k^{(r)})
=\sum_{l=0}^s{s\choose l}
(-\i)^l\ad_{\widetilde H_k}^{j+l}(\widetilde\chi_k^{(r)})
(\widetilde H_k-z)^{s-l},
\label{CommutBiz}
\end{equation}
holds on $\HH_k^{s+j/2}$.
\end{Lem}

\begin{demo} By hypothesis the assertion is verified for $s=0$. By interpolation, it suffice to prove 
the assertion for integer $s$. We begin by remarking that the identity (\ref{ResolvCommutId}) implies 
that for all $z\in\Res(\widetilde H_k)$
$$
(\widetilde H_k-z)^{-s-j/2}\ad_{\widetilde H_k}^j(\widetilde\chi_k^{(r)})
=\sum_{l=0}^s{s\choose l}
\i^l(\widetilde H_k-z)^{-l-j/2}\ad_{\widetilde H_k}^{j+l}(\widetilde\chi_k^{(r)})
(\widetilde H_k-z)^{-s}.
$$
This formula is purely algebraic. However, it is justified by Hypothesis \Href{HypChi} (i) which implies, 
by duality, that $\ad_{\widetilde H_k}^{j+l}(\widetilde\chi_k^{(r)})\in\B(\HH_k,\HH_k^{-(j+l)/2})$ for
$1\le l+j\le4\nu$. For $l\in\{0,\ldots,n\}$ the estimate
\begin{align*}
\|(\widetilde H_k-z)^{-l-j/2}\ad_{\widetilde H_k}^{j+l}(\widetilde\chi_k^{(r)})&
(\widetilde H_k-z)^{-s}\|_{\B(\HH_k^{-s},\HH_k)}\\
\le\|(\widetilde H_k-z)^{-l-j/2}&\|_{\B(\HH_k^{-(j+l)/2},\HH_k^{l/2})}
\|\ad_{\widetilde H_k}^{j+l}(\widetilde\chi_k^{(r)})\|_{\B(\HH_k,\HH_k^{-(j+l)/2})}
\|(\widetilde H_k-z)^{-s}\|_{\B(\HH_k^{-s},\HH_k)},
\end{align*}
shows that $(\widetilde H_k-z)^{-s-j/2}\ad_{\widetilde H_k}^j(\widetilde\chi_k^{(r)})$ is bounded from
$\HH_k^{-s}$ into $\HH_k$. We conclude that, for $s+j\le4\nu$ we have
$\ad_{\widetilde H_k}^{j}(\widetilde\chi_k^{(r)})\in\B(\HH_k^{-s},\HH_k^{-s-j/2})$ and the assertion 
follows by duality. By taking the adjoint of the identity (\ref{ResolvCommutId}) we obtain
$$
\ad_{\widetilde H_k}^j(\widetilde\chi_k^{(r)})(\widetilde H_k-z)^{-s-j/2}
=\sum_{l=0}^s{s\choose l}
(-\i)^l(\widetilde H_k-z)^{-s}\ad_{\widetilde H_k}^{j+l}(\widetilde\chi_k^{(r)})
(\widetilde H_k-z)^{-l-j/2},
$$
which shows that (\ref{CommutBiz}) holds.\cqfd
\end{demo}

\bigskip
\begin{Lem}\label{fMMfFormLemma} Under Hypotheses \Href{HypReservoir}, \Href{HypResDyn}, and 
\Href{HypChi} we have
\begin{align*}
(H-z)^{-1}M_k^{(r)}-M_k^{(r)}(\widetilde H_k-z)^{-1}&=
\sum_{j=1}^{l-1}\i^jJ_k^{(r)}
\ad_{\widetilde H_k}^j(\widetilde\chi_k^{(r)})(\widetilde H_k-z)^{-(j+1)}\\
&+\i^{l} (H-z)^{-1}J_k^{(r)}\ad_{\widetilde H_k}^{l}(\widetilde\chi_k^{(r)})
(\widetilde H_k-z)^{-l},
\end{align*}
for all $z\in\Res(H)\cap\Res(\widetilde H_k)$, $k\in\{1,\ldots,M\}$, $r\ge0$ and $l\le2\nu$.
\end{Lem}

\begin{demo} Lemma  \ref{adjbound} allows us to write, for $j=1,\ldots,2\nu$ and for all
$u\in\HH_k^{1+j/2}$, by invoking Hypothesis \Href{HypReservoir}
  (vi) then the identity (\ref{chidomin2}) and then Hypothesis  \Href{HypChi} (ii)
\begin{align*}
v\equiv J_k^{(r)}\ad_{\widetilde H_k}^j(\widetilde\chi_k^{(r)})u&=
J_k^{(r-1)}\tilde1_k^{(r)}\ad_{\widetilde H_k}^j(\widetilde\chi_k^{(r)})u\\
&=J_k^{(r-1)}\widetilde\chi_k^{(r-1)}\tilde1_k^{(r)}
\ad_{\widetilde H_k}^j(\widetilde\chi_k^{(r)})u\\
&=J_k^{(r-1)}\widetilde\chi_k^{(r-1)}
\ad_{\widetilde H_k}^j(\widetilde\chi_k^{(r)})u.
\end{align*}
Hypothesis  \Href{HypResDyn} (ii) implies $v\in\Dom(H)$ and
$
Hv=J_k^{(r-1)}\widetilde H_k\widetilde\chi_k^{(r-1)}
\ad_{\widetilde H_k}^j(\widetilde\chi_k^{(r)})u
$.
By using Hypothesis \Href{HypChi} (ii) and the identity (\ref{chidomin2}) we may continue with
\begin{align*}
Hv&=J_k^{(r-1)}\widetilde H_k\widetilde\chi_k^{(r-1)}
\tilde1_k^{(r)}\ad_{\widetilde H_k}^j(\widetilde\chi_k^{(r)})u\\
&=J_k^{(r-1)}\widetilde H_k
\tilde1_k^{(r)}\ad_{\widetilde H_k}^j(\widetilde\chi_k^{(r)})u\\
&=J_k^{(r-1)}\widetilde H_k\ad_{\widetilde H_k}^j(\widetilde\chi_k^{(r)})u.
\end{align*}
In a similar manner, we show that
$$
J_k^{(r)}\ad_{\widetilde H_k}^j(\widetilde\chi_k^{(r)})\widetilde H_ku=
J_k^{(r-1)}\ad_{\widetilde H_k}^j(\widetilde\chi_k^{(r)})\widetilde H_ku,
$$
and we obtain
\begin{align*}
J_k^{(r)}\ad_{\widetilde H_k}^j(\widetilde\chi_k^{(r)})\widetilde H_ku
-HJ_k^{(r)}\ad_{\widetilde H_k}^j(\widetilde\chi_k^{(r)})u
&=-J_k^{(r-1)}[\widetilde H_k,\ad_{\widetilde H_k}^j(\widetilde\chi_k^{(r)})]u\\
&=\i J_k^{(r-1)}\ad_{\widetilde H_k}^{j+1}(\widetilde\chi_k^{(r)})u\\
&=\i J_k^{(r-1)}\tilde1_k^{(r)}\ad_{\widetilde H_k}^{j+1}(\widetilde\chi_k^{(r)})u\\
&=\i J_k^{(r)}\tilde1_k^{(r)}\ad_{\widetilde H_k}^{j+1}(\widetilde\chi_k^{(r)})u\\
&=\i J_k^{(r)}\ad_{\widetilde H_k}^{j+1}(\widetilde\chi_k^{(r)})u.
\end{align*}
By setting $R\equiv(H-z)^{-1}$ and $\widetilde R=(\widetilde H_k-z)^{-1}$, 
this identity allows us to write, for all $u\in\HH_k^{j/2}$,
\begin{align*}
RJ_k^{(r)}\ad_{\widetilde H_k}^j(\widetilde\chi_k^{(r)})u-
J_k^{(r)}\ad_{\widetilde H_k}^j(\widetilde\chi_k^{(r)})\widetilde Ru
&=R(J_k^{(r)}\ad_{\widetilde H_k}^j(\widetilde\chi_k^{(r)})\widetilde H_k-
HJ_k^{(r)}\ad_{\widetilde H_k}^j(\widetilde\chi_k^{(r)})
)\widetilde Ru\\
&=\i RJ_k^{(r)}\ad_{\widetilde H_k}^{j+1}(\widetilde\chi_k^{(r)})\widetilde Ru.
\end{align*}
We thus obtain the formula
$$
RJ_k^{(r)}\ad_{\widetilde H_k}^j(\widetilde\chi_k^{(r)})\widetilde R^j=
J_k^{(r)}\ad_{\widetilde H_k}^j(\widetilde\chi_k^{(r)})\widetilde R^{j+1}+
\i RJ_k^{(r)}\ad_{\widetilde H_k}^{j+1}(\widetilde\chi_k^{(r)})\widetilde R^{j+1},
$$
and by iteration,
$$
\i RJ_k^{(r)}\ad_{\widetilde H_k}(\widetilde\chi_k^{(r)})\widetilde R
=\sum_{j=1}^{l-1}\i^jJ_k^{(r)}\ad_{\widetilde H_k}^j(\widetilde\chi_k^{(r)})\widetilde R^{j+1}
+\i^{l} RJ_k^{(r)}\ad_{\widetilde H_k}^{l}(\widetilde\chi_k^{(r)})\widetilde R^{l}.
$$
We conclude by remarking that
$$
RM_k^{(r)}-M_k^{(r)}\widetilde R
=\i RJ_k^{(r)}\ad_{\widetilde H_k}(\widetilde\chi_k^{(r)})\widetilde R,
$$
{\sl c.f.} Lemma \ref{ResolvEqu}.\cqfd
\end{demo}

\bigskip

\noindent{\bf Proof of Lemma \ref{fHMLemma}.}
By differentiation of the formula in Lemma  \ref{fMMfFormLemma} with $l=2\nu$ we obtain
\begin{align*}
(H-z)^{-\ell}&M_k^{(r)}-M_k^{(r)}(\widetilde H_k-z)^{-\ell}=
\sum_{j=1}^{2\nu-1}\i^jJ_k^{(r)}
\ad_{\widetilde H_k}^j(\widetilde\chi_k^{(r)})(\widetilde H_k-z)^{-\ell-j}\\
&+\i^{2\nu}\sum_{j=0}^{\ell-1}{\ell-1\choose j} (H-z)^{-\ell+j}J_k^{(r)}
\ad_{\widetilde H_k}^{2\nu}(\widetilde\chi_k^{(r)})
(\widetilde H_k-z)^{-2\nu-j}.
\end{align*}
Thus, it suffices to show that each factor of the type
$\ad_{\widetilde H_k}^p(\widetilde\chi_k^{(r)})(\widetilde H_k-z)^{-q}$ appearing on the right hand side
of this relation is trace-class. We note that for each one of these factors we have $p\le2\nu$ and 
$q\ge\nu+p/2$.

For all $p\le2\nu$, Hypotheses \Href{HypReservoir} (ii) and \Href{HypChi} (ii) imply 
that $(I-\widetilde\chi_k^{(r-1)})\ad_{\widetilde H_k}^p(\widetilde\chi_k^{(r)})=0$.
By  \Href{HypChi} (iii) we also have
$\widetilde\chi_k^{(r+1)}\ad_{\widetilde H_k}^p(\widetilde\chi_k^{(r)})=0$. Consequently, we have
\begin{equation}
\ad_{\widetilde H_k}^p(\widetilde\chi_k^{(r)})=\varphi_k^{(r)}
\ad_{\widetilde H_k}^p(\widetilde\chi_k^{(r)}),
\label{chichiad}
\end{equation}
where $\varphi_k^{(r)}\equiv(\widetilde\chi_k^{(r-1)}-\widetilde\chi_k^{(r+1)})$. By writing
\begin{align*}
\ad_{\widetilde H_k}^p(\widetilde\chi_k^{(r)})(\widetilde H_k-z)^{-q}
&=\varphi_k^{(r)}\ad_{\widetilde H_k}^p(\widetilde\chi_k^{(r)})(\widetilde H_k-z)^{-q}\\
&=\left(\varphi_k^{(r)}(\widetilde H_k+\i)^{-\nu}\right)
\left((\widetilde H_k+\i)^{\nu}\ad_{\widetilde H_k}^p(\widetilde\chi_k^{(r)})
(\widetilde H_k-z)^{-q}\right),
\end{align*}
we remark that the first factor of the right hand side is trace-class by Hypothesis \Href{HypChi} (iv) 
while the second factor is bounded by Lemma \ref{adjbound} as soon as $q\ge \nu+p/2$.

Lemma \ref{fMMfFormLemma} and the Helffer-Sj\"ostrand formula (\ref{HSFormula}) allow us to write
$$
f(H)M_k^{(r)}-M_k^{(r)}f(\widetilde H_k)=
\sum_{j=1}^{2\nu-1}\frac{\i^j}{j!}J_k^{(r)}\ad_{\widetilde H_k}^j(\widetilde\chi_k^{(r)})
f^{(j)}(\widetilde H_k)+\mathcal R,
$$
where the remainder is given by
$$
\mathcal R=
\frac{\i^{2\nu-1}}{2\pi}\int\bar\partial\tilde f(z)(H-z)^{-1}
J_k^{(r)}\ad_{\widetilde H_k}^{2\nu}(\widetilde\chi_k^{(r)})(\widetilde H_k-z)^{-2\nu}\,\d z\wedge
\d\bar z,
$$
and where $\tilde f$ is an almost-analytic extension of $f$ of order $2\nu+1$. By writing
\begin{align*}
\ad_{\widetilde H_k}^j(\widetilde\chi_k^{(r)})f^{(j)}(\widetilde H_k)
&=\varphi_k^{(r)}\ad_{\widetilde H_k}^j(\widetilde\chi_k^{(r)})f^{(j)}(\widetilde H_k)\\
&=\left(\varphi_k^{(r)}(\widetilde H_k+\i)^{-\nu}\right)
\left((\widetilde H_k+\i)^{\nu}\ad_{\widetilde H_k}^j(\widetilde\chi_k^{(r)})
f^{(j)}(\widetilde H_k)\right),
\end{align*}
we note as before that the first factor of the right hand side is trace-class and the second factor is 
bounded.

The remainder $\mathcal R$ is treated in a similar manner with
$$
\ad_{\widetilde H_k}^{2\nu}(\widetilde\chi_k^{(r)})(\widetilde H_k-z)^{-2\nu}
=\left(\varphi_k^{(r)}(\widetilde H_k+\i)^{-\nu}\right)
\left((\widetilde H_k+\i)^{\nu}\ad_{\widetilde H_k}^{2\nu}(\widetilde\chi_k^{(r)})
(\widetilde H_k-z)^{-2\nu}\right),
$$
where the first factor of the right hand side is trace-class. Writing the second factor as
$$
(\widetilde H_k+\i)^{\nu}\ad_{\widetilde H_k}^{2\nu}(\widetilde\chi_k^{(r)})
(\widetilde H_k-z)^{-2\nu}
=\left((\widetilde H_k+\i)^{\nu}\ad_{\widetilde H_k}^{2\nu}(\widetilde\chi_k^{(r)})
(\widetilde H_k+\i)^{-2\nu}\right)
\left((\widetilde H_k+\i)^{2\nu}(\widetilde H_k-z)^{-2\nu}\right),
$$
we observe that the first factor of the right hand side is bounded by Lemma \ref{adjbound}.
The second factor is bounded by
$$
\|(\widetilde H_k+\i)^{2\nu}(\widetilde H_k-z)^{-2\nu}\|\le c|\mathrm{Im}z|^{-2\nu},
$$
for $z\in\supp\,\tilde f$. We deduce that $\mathcal R$ is trace-class.
\cqfd


\section{Proof of Lemma \ref{TraceOmega}}
\label{LocalisationProof}

In this appendix we prove the trace-norm localization Lemma \ref{TraceOmega}.
We reproduce a large part of the proof of Lemma 3.2 in \cite{AEGSS}. However, for the
reasons stated in the beginning of Section \ref{CurrentSectII}, we have to provide an alternative
proof of Lemma A.6 in \cite{AEGSS} (Lemma \ref{NotSoSimpleStuff} below) which is the key to 
the control of the trace-norm.

\subsection{Estimates in norm of $\B(\mathcal H)$}

In this section we prove two estimates, uniform in $a\ge1$ in the norm of  $\B(\mathcal H)$.

\begin{Lem}\label{BLocLemma}Under Hypotheses \Href{HypReservoir}, \Href{HypResDyn},
\Href{HypConjugate} and \Href{HypAHTrace} the following estimates hold for all
$f\in C_0^\infty(\mathbb R)$, all $0<\gamma\le1$ and all $\varphi\in C^\infty(\mathbb R)$ such 
that $\varphi'\in C_0^\infty(\mathbb R)$.
\begin{enumerate}[(i)]
\item
$$
\sup_{a\ge1}\|[f(H),\varphi(\pm A-a)](\gamma(\pm A-a)+\i)^m\|<\infty.
$$
\item
$$
\sup_{a,\alpha\ge1}\langle\alpha\rangle^{2}
\|F((\pm A-a)<-\alpha)[f(H),\varphi(\pm A-a)](\gamma(\pm A-a)+\i)^m\|<\infty.
$$ 
\end{enumerate}
\end{Lem}

\begin{demo} Without loss of generality we may suppose that for 
$R\equiv\sup\{|x|\,|\,x\in\supp\,\varphi'\}>0$ we have $\varphi(x)=0$ if $x<-R$ and 
$\varphi(x)=L\ge0$ if $x>R$. We set $A'\equiv\sigma A-a$, with $\sigma\in\{\pm\}$ and we consider
$$
\|F(A'\in\mathcal A)[f(H),\varphi(A')](\gamma A'+\i)^m\|.
$$
We must show that, uniformly in $a\ge1$ when $\alpha\to+\infty$, this expression is $O(1)$ if 
$\mathcal A=\mathbb R$ and $O(\alpha^{-2})$ if $\mathcal A=]-\infty,-\alpha[$ or, 
equivalently, if $\mathcal A=]-\infty,-R\alpha[$. We finally remark that Hypothesis \Href{HypAHTrace}
implies, via Theorem \ref{HMourre}, that $H\in C^{m+2}_\mathrm{loc}(A)$ and thus that 
$f(H)\in\B_A^{m+2}(\mathcal H)$.

\bigskip
\noindent (i) By invoking Theorem \ref{CommutExpand} we obtain the expansion
\begin{equation}
[f(H),\varphi(A')]=\sum_{j=1}^{m}\frac{(-\i\sigma)^j}{j!}
\ad_A^j(f(H))\varphi^{(j)}(A')+R_m,
\label{ThisCommutExpansion}
\end{equation}
where the remainder is given by the formula
$$
R_m=-\frac{(\i\sigma)^{m+1}}{2\pi\i}\int\bar\partial\tilde\varphi(z)
(A'-z)^{-1}\ad_A^{m+1}(f(H))(A'-z)^{-m-1}\,\d z\wedge\d\bar z,
$$
$\tilde\varphi$ being the almost-analytic extension of $\varphi$ of order $m+1$ given 
by  (\ref{quasianalyticextension}) (with $n=m+1$). We easily show, starting from the 
formula (\ref{quasianalyticextension}), that there exist constants $c_1$ and $c_2$ such that
$\supp\,\bar\partial\tilde\varphi\subset\{z=x+\i y\,|\,\langle x\rangle\le c_1+c_2|y|\}$. 
Taking into account the fact that $0<\gamma\le1$ we get that
$$
\|(A'-z)^{-m}(\gamma A'+\i)^m\|
\le\sup_{a'\in\mathbb R}\left|\frac{a'+\i}{a'-z}\right|^m\le c_3(1+|\mathrm{Im} z|^{-m}),
$$
for all $z\in\supp\,\bar\partial\tilde\varphi$. With the help of the inequality (\ref{ftildeestimate}) 
we obtain the estimate
\begin{align*}
\|R_m(A'+\i)^m\|&\le c_4\int|\bar\partial\tilde\varphi(x+\i y)|
(|y|^{-2}+|y|^{-m-2})\,\d x\d y\\
&\le c_5\sum_{j=0}^{m+3}\int\langle x\rangle^{j-2}|\varphi^{(j)}(x)|\,\d x<\infty.
\end{align*}
Since, for $j\ge1$, 
$$
\|\varphi^{(j)}(A')(\gamma A'+\i)^m\|
\le\sup_{x\in\mathbb R}|\varphi^{(j)}(x)|\sup_{x\in\supp\,\varphi'}|x+\i|^m<\infty
$$
the expansion (\ref{ThisCommutExpansion}) allows us to conclude that
$$
\sup_{a\ge1}\|[f(H),\varphi(A')](\gamma A'+\i)^m\|<\infty.
$$

\bigskip
\noindent (ii) We set $A_\alpha\equiv -\alpha^{-1}(\sigma A-a)-2$. We easily verify that
$F(A'<-3\alpha)=F(A'<-3\alpha)h(A_\alpha)$ for all $\alpha\ge1$ and that $h(A_\alpha)\varphi(A')=0$ 
for all $\alpha\ge R$. We obtain
\begin{align*}
\|F(A'<-3\alpha)[f(H),\varphi(A')](\gamma A'+\i)^m\|&\le
\|h(A_\alpha)[f(H),\varphi(A')](\gamma A'+\i)^m\|\\
&\le\|h(A_\alpha)f(H)\varphi(A')(\gamma A'+\i)^m\|\\
&\le\|[f(H),h(A_\alpha)]\varphi(A')(\gamma A'+\i)^m\|,
\end{align*}
for all $\alpha\ge R$. By applying the expansion (\ref{ThisCommutExpansion}) to $[f(H),h(A_\alpha)]$, 
we have
\begin{equation}
[f(H),h(A_\alpha)]=\sum_{j=1}^{m+1}\frac{(\i\sigma\alpha^{-1})^j}{j!}
\ad_A^j(f(H))h^{(j)}(A_\alpha)+R_{m+1},
\label{OtherCommutExpansion}
\end{equation}
where the remainder is given by the formula
$$
R_{m+1}=-\frac{(-\i\sigma\alpha^{-1})^{m+2}}{2\pi\i}\int\bar\partial\tilde h(z)
(A_\alpha-z)^{-1}\ad_A^{m+2}(f(H))(A_\alpha-z)^{-m-2}\,\d z\wedge\d\bar z,
$$
where $\tilde h$  is an almost-analytic extension of order $m+2$. The estimate
$$
\|(A_\alpha-z)^{-m}(\gamma A'+\i)^m\|\le\sup_{a'\in\mathbb R}\left|\frac{\alpha a'+\i}{a'+2+z}\right|^m
\le c_6\alpha^m(1+|\mathrm{Im} z|^{-m}),
$$
allows us to obtain, as before,
\begin{align*}
\|R_{m+1}(\gamma A'+\i)^m\|&\le c_7\,\alpha^{-2}
\int|\bar\partial\tilde h(x+\i y)|
(|y|^{-3}+|y|^{-m-3})\,\d x\d y\\
&\le c_8\alpha^{-2} \sum_{j=0}^{m+4}\int\langle x\rangle^{j-3}|h^{(j)}(x)|\,\d x\le c_9\alpha^{-2}.
\end{align*}
Since $h^{(j)}(A_\alpha)\varphi(A')=0$ for $\alpha\ge R$ and $j\ge1$, we may conclude that
$$
\|F(A'<-3\alpha)[f(H),\varphi(A')](\gamma A'+\i)^m\|\le c_{9}\alpha^{-2}.
$$
\cqfd
\end{demo}

\bigskip
We finish this section with a simple lemma.

\begin{Lem}\label{AnLemma} Under Hypotheses \Href{HypReservoir}, \Href{HypResDyn},
\Href{HypConjugate}, and \Href{HypAHTrace} we have
$$
\sup_{a\ge1,\sigma\in\{\pm\}}\|[f(H),1_k]h_\sigma^{(a)}(A)(\gamma(\sigma A-a)+\i)^m\|<\infty,
$$
for all $k\in\{0,\ldots,M\}$, $f\in C_0^\infty(\mathbb R)$ and $0<\gamma\le1$.

\end{Lem}

\begin{demo}An elementary analysis shows that
$$
\sup_{x\in\mathbb R,\sigma\in\{\pm\}}\left|\left(\frac{\gamma(\sigma x-a)+\i}{x+\i}\right)^m
h_\sigma^{(a)}(x)\right|\le \mu(a)^m\equiv
\left(\sup_{x\ge a-1}\frac{1+(x-a)^2}{1+x^2}\right)^{m/2},
$$
and that $\mu(a)\le2$ provided $a\ge 0$. The functional calculus allows us to write
$$
\sup_{a\ge1,\sigma\in\{\pm\}}\|[f(H),1_k]h_\sigma^{(a)}(A)(\gamma(\sigma A-a)+\i)^m\|
\le 2^m\, \|[f(H),1_k](A+\i)^m\|,
$$
and to reduce the proof to the assertion $[f(H),1_k](A+\i)^m\in\B(\mathcal H)$. Since
$\widetilde H_k\in\B_{\widetilde A_k}^m(\HH_k^1,\HH_k)$ by Hypothesis \Href{HypAHTrace} (ii), 
Theorem \ref{HMourre} implies that $f(H)\in\B_{A_j}^m(\mathcal H)$ for all $j\in\{0,\ldots,M\}$ and 
we may apply Lemma \ref{AfAinv}  to show that $(A_j+\i)^{-m}f(H)(A_j+\i)^m\in\B(\mathcal H)$. 
Since $1_kA_j=0$ for  $k\not=j$, Lemma \ref{ADefOK} implies 
$1_k(A_j+\i)^{-m}f(H)(A_j+\i)^m=\i^{-m}1_kf(H)(A_j+\i)^m$ from which  we conclude that
$1_kf(H)(A_j+\i)^m$ is bounded. This leads to the result that
\begin{align*}
[f(H),1_k](A+\i)^m&=\sum_{j=0}^M\left(1_jf(H)1_k(A_k+\i)^m-1_kf(H)1_j(A_j+\i)^m\right)\\
&=\sum_{\atop{j=0}{j\not= k}}^M\left(1_jf(H)1_k(A_k+\i)^m-1_kf(H)1_j(A_j+\i)^m\right),
\end{align*}
is bounded. \cqfd
\end{demo}

\subsection{The spectral multiplicity of $H$}

We now state a slightly unexpected corollary of Theorem \ref{Kernelpsi}.

\begin{Prop}\label{SpectralMult} Hypotheses \Href{HypReservoir}--\Href{HypTrace0} imply that the 
spectral multiplicity of the Hamiltonian $H$ is locally finite. More precisely, if 
\begin{equation}
U:\mathcal H_\ac\to\int^\oplus\mathfrak h(\varepsilon)\,\d\varepsilon,
\label{HspectralRep}
\end{equation}
denotes the spectral representation associated with the absolutely continuous part of $H$, then
$$
\int_\Delta\dim\mathfrak h(\varepsilon)\,\d\varepsilon<\infty,
$$
for all compact $\Delta\subset\mathbb R\setminus\Sigma_H$.
\end{Prop}

\begin{demo} Let $\Delta\subset\mathbb R\setminus\Sigma_H$. We consider the charge 
$Q\equiv I$ and the corresponding current
$$
\widetilde\Psi^{\#(a+)}_{g}\equiv\sum_{k=1}^M
\widetilde\Psi^{\#(a+)}_{g,k}=\sum_{k=1}^M
g(\widetilde H_k)\i[f(\widetilde H_k),
h_+^{(a)}(\widetilde A_k)M_k^\ast M_kh_+^{(a)}(\widetilde A_k)]g(\widetilde H_k),
$$
where $g\in C_0^\infty(\mathbb R\setminus\Sigma_H)$, $g=1$ on $\Delta$ and 
$f\in C_0^\infty(\mathbb R)$. By expanding the commutator we obtain
\begin{align*}
\widetilde\Psi^{\#(a+)}_{g,k}
&=g(\widetilde H_k)\i[f(\widetilde H_k),
h_+^{(a)}(\widetilde A_k)]\widetilde\chi_k^2h_+^{(a)}(\widetilde A_k)g(\widetilde H_k)\\
&+g(\widetilde H_k)h_+^{(a)}(\widetilde A_k)\i[f(\widetilde H_k),\widetilde\chi_k^2]
h_+^{(a)}(\widetilde A_k)g(\widetilde H_k)\\
&+g(\widetilde H_k)h_+^{(a)}(\widetilde A_k)\widetilde\chi_k^2
\i[f(\widetilde H_k),h_+^{(a)}(\widetilde A_k)]g(\widetilde H_k),
\end{align*}
and it follows from Lemma \ref{HchiLemma} that the second term of the right hand side of 
this identity is trace-class. The two remaining terms have a similar structure. Writing
$$
[f(\widetilde H_k),h_+^{(a)}(\widetilde A_k)]g(\widetilde H_k)
=\left([f(\widetilde H_k),h_+^{(a)}(\widetilde A_k)](\widetilde A_k+\i)^m\right)\left(
(\widetilde A_k+\i)^{-m}g(\widetilde H_k)\right),
$$
Hypothesis  \Href{HypAHTrace} shows that the second factor of the right hand side is trace-class.
We easily show, by following the proof of of Lemma \ref{BLocLemma} (i), that the first factor is 
bounded. The current $\widetilde\Psi^{\#(a+)}_{g,k}$ is thus trace-class and we may apply 
Theorem \ref{Kernelpsi}. The diagonal of its integral kernel is given by
$$
\psi_{g,k}(\varepsilon,\varepsilon)=\frac{1}{2\pi}g(\varepsilon)^2p_k^+(\varepsilon),
$$
and it follows from Theorem \ref{TraceKernel} (iii) that
$$
\frac{1}{2\pi}\int_\Delta
\tr_{\mathfrak h_k(\varepsilon)}(p_k^+(\varepsilon))\,\d\varepsilon\le
\frac{1}{2\pi}\int g(\varepsilon)^2
\tr_{\mathfrak h_k(\varepsilon)}(p_k^+(\varepsilon))\,\d\varepsilon
\le\|\widetilde\Psi^{\#(a+)}_{g,k}\|_1<\infty.
$$
Since the asymptotic projection $P_k^+$ commutes with $H$ it admits, in the 
decomposition (\ref{HspectralRep}) the representation
$$
(UP_k^+u)(\varepsilon)=\pi_k^+(\varepsilon)(Uu)(\varepsilon).
$$
Moreover, Lemma \ref{CompareKernel} allows us to show that for almost all $\varepsilon$ 
the operators  $p_k^+(\varepsilon)$ and $\pi_k^+(\varepsilon)$ are orthogonal projections.
Furthermore the intertwining relation $f(H)\Omega_k^+=\Omega_k^+f(\widetilde H_k)$ implies
$$
(U\Omega_k^+u)(\varepsilon)=\omega_k^+(\varepsilon)(U_ku)(\varepsilon).
$$
The relations $\Omega_k^+\Omega_k^{+\ast}=P_k^+$, 
$\Omega_k^{+\ast}\Omega_k^{+}=\widetilde P_k^+$ and Lemma \ref{CompareKernel} allow us 
to conclude that
$$
\pi_k^+(\varepsilon)=\omega_k^{+}(\varepsilon)\omega_k^{+\ast}(\varepsilon),
\qquad
p_k^+(\varepsilon)=\omega_k^{+\ast}(\varepsilon)\omega_k^{+}(\varepsilon),
$$
for almost all $\varepsilon$. In particular, we have
$$
\tr_{\mathfrak h_k(\varepsilon)}(p_k^+(\varepsilon))
=\tr_{\mathfrak h(\varepsilon)}(\pi_k^+(\varepsilon)),
$$
for almost all $\varepsilon$. It follows from the fact that $\sum_{k=1}^MP_k^+=P_\ac(H)$ that
$$
\int_\Delta\tr_{\mathfrak h(\varepsilon)}(I)\,\d\varepsilon=
\sum_{k=1}^M\int_\Delta\tr_{\mathfrak h(\varepsilon)}(\pi_k^+(\varepsilon))\,\d\varepsilon
\le 2\pi\sum_{k=1}^M\|\widetilde\Psi^{\#(a+)}_{g,k}\|_1<\infty,
$$
which, given the fact that $\dim\mathfrak h(\varepsilon)=\tr_{\mathfrak h(\varepsilon)}(I)$, 
concludes the proof.\cqfd
\end{demo}

\subsection{Trace-norm estimates}

The following lemma is the key to control the trace-norm in the proof of Lemma \ref{TraceOmega}.

\begin{Lem}\label{NotSoSimpleStuff} Under Hypotheses \Href{HypReservoir}--\Href{HypTrace0},
$$
\sup_{a\in\mathbb R,\sigma\in\{\pm\}} \|(\gamma(\sigma A-a)+\i)^{-m}g(H)\|_1<\infty,
$$
for all $g\in C_0^\infty(\mathbb R\setminus\Sigma_H)$ and $\gamma>0$ sufficiently small.
\end{Lem}

\begin{demo} We set $T\equiv(\gamma(\sigma A-a)+\i)^{-1}$ and, without loss of generality, 
we suppose that $$\sup_{x\in\mathbb R}|g(x)|\le1.$$ We begin by showing that $T^mg(H)$ is trace-class.
We have
$$
T^mg(H)=\sum_{k=0}^M(\gamma(\sigma A_k-a)+\i)^{-m}1_kg(H),
$$
where the $k=0$ term is bounded uniformly by Hypothesis \Href{HypAHTrace} (iii),
$$
\|(\gamma(\sigma A_0-a)+\i)^{-m}1_0g(H)\|_1=
\|(-\gamma a+\i)^{-m}1_0g(H)\|_1\le\|1_0g(H)\|_1.
$$

For $k\in\{1,\ldots,M\}$ Hypothesis \Href{HypReservoir} (v) and the identity (\ref{chidomin2}) imply 
that $(1_k-\chi_k^2)1_k^{(2)}=(1_k^{(2)}-\chi_k^2)1_k^{(2)}=0$. Hypothesis \Href{HypReservoir} (iii)
allows us to conclude that $(1_k-\chi_k^2)=(1_k-\chi_k^2)1_0^{(2)}$. It follows from this identity and 
from (\ref{SmoothResolv}) that
\begin{align*}
(\gamma(\sigma A_k-a)+\i)^{-m}1_kg(H)
&=M_k(\gamma(\sigma \widetilde A_k-a)+\i)^{-m}M_k^\ast g(H)
+(-\gamma a+i)^{-m}(1_k-\chi_k^2)g(H)\\
&=M_k(\gamma(\sigma \widetilde A_k-a)+\i)^{-m}g(\widetilde H_k)M_k^\ast\\
&+M_k(\gamma(\sigma \widetilde A_k-a)+\i)^{-m}(M_k^\ast g(H)-g(\widetilde H_k)M_k^\ast)\\
&+(-\gamma a+i)^{-m}(1_k-\chi_k^2)1_0^{(2)}g(H).
\end{align*}
The first term on the right hand side of this identity is trace-class by Hypothesis \Href{HypAHTrace} (i). 
Lemma \ref{fHMLemma} shows that the second term is trace-class. Finally, the last term is 
trace-class by Hypothesis \Href{HypAHTrace} (iii).

We now show that $T^mg(H)$ is uniformly bounded in $\mathcal L^1(\mathcal H)$. 
Let  $f\in C_0^\infty(\mathbb R\setminus\Sigma_H)$ such that $0\le f\le1$ and $fg=g$. 
We begin with the identity
\begin{equation}
T^{m}g(H)=f(H)T^{m}g(H)+[T^{m},f(H)]g(H).
\label{TmgId}
\end{equation}
By invoking Lemma \ref{ExactCummutExpansion} to expand the commutator of the second
term of the right hand side of this identity we obtain
$$
[T^{m},f(H)]=\sum_{j=1}^m{m\choose j}
(\i\gamma\sigma)^jT^j\ad_{A}^j(f(H))T^m\equiv BT^m.
$$
Hypothesis \Href{HypAHTrace} (ii) and Theorem \ref{HMourre} allow us to estimate
$$
\|B\|\le \sum_{j=1}^m{m\choose j}
\gamma^j\|\ad_{A}^j(f(H))\|\le C\gamma,
$$
for a constant $C$ and $\gamma$ small enough. We may thus conclude from identity (\ref{TmgId})
$$
\|T^{m}g(H)\|_1\le\|f(H)T^{m}g(H)\|_1
+C\gamma\|T^{m}g(H)\|_1,
$$
and conclude that for $\gamma<(2C)^{-1}$ we have
$$
\|T^{m}g(H)\|_1\le2\,\|f(H)T^{m}g(H)\|_1.
$$
Since $g=fg$ and $|g|\le1$ we may also write
$$
\|T^{m}g(H)\|_1\le2\,\|f(H)T^{m}f(H)\|_1
\le2\,\|f(H)|T|^{m}f(H)\|_1=2\,\tr(f(H)|T|^{m}f(H)).
$$
We consider now the spectral representation (\ref{HspectralRep}). Since the spectrum of $H$ 
is purely absolutely continuous on $\supp\,f$ we have
$$
f(H)|T|^{m}f(H)=P_\ac(H)f(H)|T|^{m}f(H)P_\ac(H),
$$
and Theorem \ref{TraceKernel} allows us to conclude that this operator has an integral kernel
$c(\varepsilon,\varepsilon')\in\mathcal L^1(\mathfrak h(\varepsilon'),\mathfrak h(\varepsilon))$
such that
$$
\tr(f(H)|T|^{m}f(H))=\int_{\supp\,f}\tr(c(\varepsilon,\varepsilon))\,\d\varepsilon.
$$
Lemma \ref{DiagKernel} implies
\begin{align}
\int (u(\varepsilon),c(\varepsilon,\varepsilon)
u(\varepsilon))_{\mathfrak h(\varepsilon)}\,\d\varepsilon
&=\lim_{\eta\downarrow0}\int_{-\infty}^\infty\e^{-\eta|t|}(u,\e^{-\i tH}f(H)|T|^mf(H)
\e^{\i tH}u)\,\d t\nonumber\\
&=\lim_{\eta\downarrow0}\int_{-\infty}^\infty\e^{-\eta|t|}\||T|^{m/2}f(H)\e^{\i tH}u\|^2\,\d t.
\label{etaformH}
\end{align}

We shall estimate this last integral. We note that for all $\alpha\ge0$ we have
$$
\||T|^{m/2}F(\gamma|\sigma A-a|>\alpha)\|\le\langle\alpha\rangle^{-m/2},
$$
and thus
\begin{align*}
\||T|^{m/2}&f(H)\e^{-\i\sigma tH}f(H)|T|^{m/2}\|\\
&\le\|F(\gamma|\sigma A-a|\le \alpha)\e^{-\i\sigma tH}f(H)^2F(\gamma|\sigma A-a|\le \alpha)\|
+3\langle\alpha\rangle^{-m/2}\\
&\le\|F(\gamma(\sigma A-a)\le\alpha)\e^{-\i\sigma tH}f(H)^2F(\gamma(\sigma A-a)\ge-\alpha)\|
+3\langle\alpha\rangle^{-m/2}.
\end{align*}
Setting $\alpha=\gamma\vartheta t/2\ge0$, Hypothesis \Href{HypAHTrace} (i) and 
Theorem \ref{HMourre} allow us to apply Proposition \ref{MainPropEst} to obtain 
$$
\|F(\gamma(\sigma A-a)\le\gamma\vartheta t/2)\e^{-\i\sigma tH}f(H)^2
F(\gamma(\sigma A-a)\ge-\gamma\vartheta t/2)\|\le c\langle\vartheta t\rangle^{-s},
$$
for constants $c$ and $s>1$, uniformly in $a$. We thus have
$$
\||T|^{m/2}f(H)\e^{-\i\sigma tH}f(H)|T|^{m/2}\|\le c_1\left(\langle t\rangle^{-s}+\langle t\rangle^{-m/2}\right),
$$
for $t\ge0$. This inequality extends to all $t\in\mathbb R$ by taking the adjoint. 
Taking into account the fact that
$$
|T|^{m/2}f(H)(H-z)^{-1}f(H)|T|^{m/2}=\pm\i\int_0^\infty\!\!
|T|^{m/2}f(H)\e^{\mp\i t(H-z)}f(H)|T|^{m/2}\,\d t,
$$
we get
$$
\sup_{z\in\mathbb C\setminus\mathbb R}\||T|^{m/2}f(H)(H-z)^{-1}f(H)|T|^{m/2}\|\le c_2,
$$
for all $a\ge1$. By invoking Theorem XIII.25 of \cite{RS4} and its corollary we obtain
$$
\int_{-\infty}^\infty\||T|^{m/2}f(H)\e^{\i tH}u\|^2\,\d t\le c_3 \,\|u\|^2,
$$
with $c_3=2c_2$. Relation (\ref{etaformH}) thus allows us to write
$$
0\le\int (u(\varepsilon),c(\varepsilon,\varepsilon)
u(\varepsilon))_{\mathfrak h(\varepsilon)}\,\d\varepsilon\le 
\int(u(\varepsilon),c_3u(\varepsilon))_{\mathfrak h(\varepsilon)}\,\d\varepsilon,
$$
from which we get $\esup_{\varepsilon\in\mathbb R}\|c(\varepsilon,\varepsilon)\|\le c_3$ and
$$
\tr_{\mathfrak h(\varepsilon)}(c(\varepsilon,\varepsilon))
\le c_3\tr_{\mathfrak h(\varepsilon)}(I),
$$
for almost all $\varepsilon$. Consequently
\begin{align*}
\|T^mg(H)\|_1&\le
2\tr(f(H)|T|^{m}f(H))\\
&=2\int_{\supp\,f}\tr(c(\varepsilon,\varepsilon))\,\d\varepsilon\\
&\le 2 c_3\int_{\supp\,f}\tr_{\mathfrak h(\varepsilon)}(I)\,\d\varepsilon\\
&\le c_4<\infty,
\end{align*}
by an application of Proposition \ref{SpectralMult}.
\cqfd
\end{demo}

\subsection{Proof of Lemma \ref{TraceOmega}}
\label{TraceOmegaDemoSubSect}

Since $h_\sigma^{(a)}(0)=0$ for $a\ge1$, Lemma \ref{ADefOK} implies
$$
h_\sigma^{(a)}(A)=\sum_{j=0}^M1_jh_\sigma^{(a)}(A_j)1_j,
$$
and thus $h_\sigma^{(a)}(A_k)=1_kh_\sigma^{(a)}(A)=h_\sigma^{(a)}(A)1_k$ which allows us to write, 
with $A'\equiv\sigma A-a$, $0<\gamma<1$ and $\mathcal A\subset\mathbb R$
\begin{align*}
\|F(&A'\in\mathcal A)[f(H),h_\sigma^{(a)}(A_k)]g(H)\|_1
=\|F(A'\in\mathcal A)[f(H),h_\sigma^{(a)}(A)1_k]g(H)\|_1\\
&\le\|F(A'\in\mathcal A)[f(H),h_\sigma^{(a)}(A)1_k](\gamma A'+\i)^m\|\,
\|(\gamma A'+\i)^{-m}g(H)\|_1.
\end{align*}
If $\gamma>0$ is sufficiently small, the second factor of the right hand side of this inequality is 
uniformly bounded for $a\in\mathbb R$ by Lemma \ref{NotSoSimpleStuff}. We must thus control 
the first factor and show that it is uniformly bounded if $\mathcal A=\mathbb R$ and that it 
decreases as $\langle\alpha\rangle^{-s}$ uniformly in $a$ if $\mathcal A=]-\infty,-\alpha[$. 

From $h_\sigma^{(a)}=h_\sigma^{(a-2)}h_\sigma^{(a)}$ we obtain
$$
h_\sigma^{(a)}(A)1_k=h_\sigma^{(a-2)}(A)h_\sigma^{(a)}(A)1_k
=h_\sigma^{(a-2)}(A)1_kh_\sigma^{(a)}(A),
$$
and thus
\begin{align*}
[f(H),1_kh_\sigma^{(a)}(A_k)]&=[f(H),h_\sigma^{(a-2)}(A)]1_kh_\sigma^{(a)}(A)
+h_\sigma^{(a-2)}(A)1_k[f(H),h_\sigma^{(a)}(A)]\\
&+h_\sigma^{(a-2)}(A)[f(H),1_k]h_\sigma^{(a)}(A),
\end{align*}
then
\begin{align*}
F(A'\in\mathcal A)&[f(H),h_\sigma^{(a)}(A)1_k](\gamma A'+\i)^m\\
&=F(A'\in\mathcal A)[f(H),h_\sigma^{(a-2)}(A)](\gamma A'+\i)^mh_\sigma^{(a)}(A)1_k\\
&+1_kh_\sigma^{(a-2)}(A)F(A'\in\mathcal A)[f(H),h_\sigma^{(a)}(A)](\gamma A'+\i)^m\\
&+h_\sigma^{(a-2)}(A)F(A'\in\mathcal A)[f(H),1_k]h_\sigma^{(a)}(A)(\gamma A'+\i)^m.
\end{align*}
In the corresponding estimate
\begin{align}
\|F(A'\in\mathcal A)&[f(H),h_\sigma^{(a)}(A_k)](\gamma A'+\i)^m\|\label{ThreeTermsId}\\
&\le\|F(A'\in\mathcal A)[f(H),h_\sigma^{(a-2)}(A)](\gamma A'+\i)^m\|\nonumber\\
&+\|F(A'\in\mathcal A)[f(H),h_\sigma^{(a)}(A)](\gamma A'+\i)^m\|\nonumber\\
&+\|h_\sigma^{(a-2)}(A)F(A'\in\mathcal A)\|\,
\|[f(H),1_k]h_\sigma^{(a)}(A)(\gamma A'+\i)^m\|,\nonumber
\end{align}
the second factor of the last term of the right hand side is uniformly bounded for $a\ge1$ 
by Lemma \ref{AnLemma}. The first factor is uniformly bounded for $a\in\mathbb R$ and 
vanishes when $\mathcal A\cap[a-3,\infty[$ is empty. In particular, for all $s>0$ there 
exists a constant $C_s$ such that
$$
\|h_\sigma^{(a-2)}(A)F(\sigma A<a-\alpha)\|\,
\|[f(H),1_k]h_\sigma^{(a)}(A)(\gamma(\sigma A-a)+\i)^m\|\le C_s\langle\alpha\rangle^{-s},
$$
for all $a,\alpha\ge1$. The two first terms of the right hand side of (\ref{ThreeTermsId}) are both 
of the form
$$
\|F(A'\in\mathcal A)[f(H),\varphi(A')](\gamma A'+\i)^m\|,
$$
where $\varphi'\in C_0^\infty(]-3,1[)$ and $0\le\varphi\le1$. Applying Lemma \ref{BLocLemma}
completes the proof.
\cqfd


\end{document}